\documentclass{aastex631}

\usepackage{graphicx}
\usepackage{natbib}
\usepackage{epsfig}
\usepackage{multirow}
\usepackage{amsmath} 
\usepackage{threeparttable}
\usepackage{stackengine} 
\usepackage{amsfonts}
\usepackage{amssymb}
\usepackage{graphicx}
\usepackage[utf8]{inputenc}

\newcommand{\dpi}{\dot{\varpi}} 		 
\newcommand{\dpisec}{\dot{\varpi}_{\rm sec}} 
\newcommand{\dom}{\dot{\Omega}} 
\newcommand{\ares}{a_{\rm res}} 
\newcommand{\dd}{^\circ~{\rm d}^{-1}} 
\newcommand{\kms}{{\rm km~s}^{-1}} 
\newcommand{\cms}{{\rm cm~s}^{-1}} 
\newcommand{\gcm}{{\rm g~cm}^{-2}} 
\newcommand{\density}{{\rm g~cm}^{-3}} 
\newcommand{\degd}{$\dd$} 
\newcommand{\degyr}{$^\circ\ {\rm yr}^{-1}$} 
\newcommand{\fpat}{$\Omega_P$} 
\newcommand{\apriori}{{\it a priori\ }}

\newcommand{\ie}{{\it i.e.,}}

\newcommand{\beq}{\begin{equation}}
\newcommand{\eeq}{\end{equation}}
\newcommand{\Cassini}{{\it Cassini}}
\newcommand{\Cas}{\Cassini\ }
\newcommand{\Gaia}{{\it Gaia}}
\newcommand{\HST}{{\it Hubble Space Telescope}}
\newcommand{\hst}{{\it HST}}
\newcommand{\Juno}{{\it Juno\ }}
\newcommand{\JWST}{{\it JWST}}
\newcommand{\Voyager}{{\it Voyager}}

\newcommand{\fw}{5.5in} 

\shorttitle{Uranus ring orbits, pole, satellites, and gravity field}
\shortauthors{French et al.}

\begin{document}
\title{The Uranus System from Occultation Observations (1977-2006): Rings, Pole Direction, Gravity Field, and Masses of Cressida, Cordelia, and Ophelia}
\author{Richard G. French}
\affil{Department of Astronomy, Wellesley College, Wellesley MA 02481 USA}

\author{Matthew M. Hedman}
\affil{Department of Physics, University of Idaho, Moscow, ID 83844 USA}

\author{Philip D. Nicholson}
\affil{Astronomy Department, Cornell University, Ithaca NY 14853 USA}

\author{Pierre-Yves Longaretti}
\affil{Institut de Plan\'etologie et d'Astrophysique de Grenoble, Grenoble, FR}

\author{Colleen A. McGhee-French}
\affil{Department of Astronomy, Wellesley College, Wellesley MA 02481 USA}

\correspondingauthor{Richard G. French}
\email{rfrench@wellesley.edu}
\begin{abstract}
From an analysis of 31 Earth-based stellar occultations and three \Voyager\ 2 occultations spanning 1977--2006 \citep{French2023a}, we determine the keplerian orbital elements of the centerlines (COR) of the nine main Uranian rings to high accuracy, with typical RMS residuals of 0.2 -- 0.4 km and 1-$\sigma$ formal errors in $a, ae,$ and $a\sin i$ of order 0.1 km, registered on an absolute radius scale accurate to 0.2 km at the 2-$\sigma$ level. The $\lambda$ ring shows more substantial scatter, with few secure detections. 
We identify a host of free and forced normal modes in several of the ring centerlines and inner and outer edges. In addition to the previously-known free modes $m=0$ in the $\gamma$ ring and $m=2$ in the $\delta$ ring, we find two additional outer Lindblad resonance (OLR) modes ($m=-1$ and $-2$) and a possible $m=3$ inner Lindblad resonance (ILR) mode in the $\gamma$ ring. No normal modes are detected for rings 6, 5, 4, $\alpha$, or $\beta$.
Five separate normal modes are forced by small moonlets: the 3:2 inner ILR of Cressida with the $\eta$ ring, the 6:5 ILR of Ophelia with the $\gamma$ ring, the 23:22 ILR of Cordelia with the $\delta$ ring, the 14:13 ILR of Ophelia with the outer edge of the $\epsilon$ ring, and the counterpart 25:24 OLR of Cordelia with the ring's inner edge. The phases of the modes and their pattern speeds are consistent with the mean longitudes and mean motions of the satellites, confirming their dynamical roles in the ring system. We find no evidence of normal modes excited by internal planetary oscillations. 
We determine the width-radius relations for nearly all of the detected modes, with positive width-radius slopes for ILR modes (including the $m=1$ elliptical orbits) and negative slopes for most of the detected OLR modes, supporting the standard self-gravity model for ring apse alignment. We find no convincing evidence for librations of any of the rings.
The Uranus pole direction at epoch TDB 1986 Jan 19 12:00 is $\alpha_P=77.311327\pm 0.000141^\circ$ and $\delta_P=15.172795\pm0.000618^\circ$. The slight pole precession predicted by \cite{Jacobson2023} is not detectable in our orbit fits, and the absolute radius scale is not strongly correlated with the pole direction.
From Monte Carlo fits to the measured apsidal precession and nodal regression rates of the eccentric and inclined rings, we determine the zonal gravitational coefficients 
$J_2=(3509.291\pm0.412)\times 10^{-6}, J_4=(-35.522\pm0.466)\times10^{-6}$, and $J_6$ fixed at $0.5\times 10^{-6}$, with a correlation coefficient $\rho(J_2,J_4)=0.9861$, for a reference radius $R=$25559 km. This result differs significantly from both earlier and more recent results \citep{Jacobson2014,Jacobson2023}, owing to our inclusion of previously neglected systematic effects, such as the offset of semimajor axes of the geometric ring centerlines from their estimated dynamical centers of mass and the significant contributions of Cordelia and Ophelia to the precession rate of the $\epsilon$ ring. Although we cannot set useful independent limits on $J_6$, we obtain strong joint constraints on combinations of $J_2,J_4,$ and $J_6$ that are consistent with our measurements. These can be used to limit the range of realistic models of the planet's internal density distribution and wind profile with depth.
The observed anomalous apsidal and nodal precession rates of the $\alpha$ and $\beta$ rings are consistent with the presence of unseen moonlets with masses and orbital radii predicted by \cite{Chancia2016}. The $\gamma$ ring's putative $m=3$ mode does not appear to be forced by a satellite, whose predicted size would be too large to have avoided prior detection. If this mode is excluded from the orbit fit, the solution for the $\gamma$ ring has a very large anomalous apsidal precession rate of unknown origin.
From the amplitudes and resonance radii of normal modes forced by moonlets, we determine the masses of Cressida, Cordelia, and Ophelia. Their estimated densities decrease systematically with increasing orbital radius and generally follow the radial trend of the Roche critical density for a shape parameter $\gamma=1.6$.
\end{abstract}
\parskip 10pt
%
\section{Introduction}
Prior to the discovery of the Uranian rings in 1977 \citep{Elliot1977,Millis1977}, Saturn was the only planet known to have rings, which in that pre-{\it Voyager} era were thought to be broad and diffuse, owing to viscous spreading associated with interparticle collisions. This simple paradigm was overturned by the detection of the narrow, eccentric, inclined Uranian rings that somehow manage to avoid circularization due to differential precession and to be sharp-edged and radially confined in the absence of direct evidence for shepherd satellites for most of the rings. The discovery of Jupiter's dusty rings by {\it Voyager 1} in 1979 \citep{Smith1979} expanded the scope and variety of rings in the solar system, and the subsequent {\it Voyager} Saturn encounters revealed a bewildering variety of complex ring structure, much of it produced by the gravitational influence of nearby satellites. The \Cassini\ mission's 13 year reconnaissance of the Saturn system greatly expanded our knowledge of the structure and dynamical environment of the rings \citep{Colwell2009, Schmidt2009, Cuzzi2009, Charnoz2009, Cuzzi2018}. From the {\it Voyager 2} flyby of Uranus in 1986, a closer view of its rings revealed the faint narrow $\lambda$ ring, sheets of dusty ring material between the narrow rings, and a gallery of nearby tiny moons \citep{Smith1986}. Subsequent observations revealed additional details of this dusty and satellite-rich system \citep{dePater2002,dePater2006,Showalter2006,dePater2013}. The census of giant plant ring systems in the solar system was extended by the detection of the Neptune's ring arcs from stellar occultation observations \citep{Hubbard1986a, Hubbard1986b, Manfroid1986} and {\it Voyager 2} images \citep{Smith1989}. Several tiny trans-Neptunian objects have rings: 10199 Chariklo's rings were discovered during a stellar occultation in 2013 \citep{Braga2014} and observed from the James Webb Space Telescope in 2022 \citep{Santos2022}, Centaur 2060 Chiron has possible ring material \citep{Ortiz2015}, and a 70 km-wide ring was detected in orbit around dwarf planet Haumea \citep{Ortiz2017}. Further afield, the search is on for evidence of ringed extra-solar planets (see \cite{Sicardy2018} for a recent discussion).

This observed wide variety of ring phenomena leaves us with many unanswered questions about their physical properties, internal structure, dynamics, origins, and evolution \citep{Esposito2018}. 
The observational basis to advance our understanding of the narrow Uranian rings in particular rests in large part on an extensive set of stellar occultation observations obtained from 1977 to 2006, described in Paper 1 \citep{French2023a}. Here, we make use of these observations
to determine accurate orbits for the Uranian rings and ring edges, the planet's pole direction and gravity field, and the masses of the moonlets Cressida, Cordelia, and Ophelia.

The paper is organized as follows: In Section 2, we summarize the Earth-based and {\it Voyager 2} Uranus ring occultation observations used for this work. Section 3 presents our geometric model of the Uranus ring system and our kinematical model for the rings and ring edges. In Section 4 we present the orbits of the ten narrow rings and ring edges and identify a host of free and forced normal modes. Then, in Section 5, we determine the direction of the planet's pole and the absolute ring radius scale, and in Section 6 we examine the widths, shapes, and masses of the rings. Next, in Section 7, we solve for the zonal gravity coefficients $J_2$ and $J_4$, taking into account several important systematic effects that have been ignored in previous analyses. In Section 8 we summarize the evidence for anomalous precession in several rings and the possibility that it is caused by nearby unseen small satellites, and in Section 9 we estimate the masses and densities of Cressida, Cordelia, and Ophelia from their associated forced normal modes in the rings. In Section 10, we discuss the broader context of the dynamics of narrow ringlets, compare the Uranus and Saturn narrow rings, and identify persistent unsolved problems. Our main conclusions are summarized in Section 11. 
%
\section{Observations}
\label{sec:observations}
The observations underlying the present work include 31 Earth-based stellar occultations by the narrow Uranian rings,
sometimes viewed with multiple telescopes and at multiple wavelengths, and three \Voyager\ occultations: one by the Radio Science Subsystem (RSS; \cite{Tyler1986,Gresh1989}) and stellar occultations of $\sigma$ Sgr and $\beta$ Per observed by the Photopolarimeter (PPS; \cite{Lane1986}) and the Ultraviolet Spectrometer (UVS; \cite{Holberg1987}). Paper 1 provides a detailed description of the Earth-based observations and a summary of the \Voyager\ data. Here, we limit ourselves to a brief overview of the observations, emphasizing characteristics of the data and event geometry that affect the accuracy of the ring measurements used for this study and the scientific interpretation of our results.

        \subsection{Earth-based stellar occultations}
For most of the 1977--2006 interval of occultation observations, the Uranus system presented a nearly pole-on view as seen from Earth, a geometry that favored the determination of the absolute radius scale of the rings while limiting the initial accuracy of their inclinations and of the planet's pole direction. 
The majority of the Earth-based occultation events were observed between 1980--1990, when Uranus was traversing the Milky Way and high-SNR occultation opportunities were frequent. After the initial discovery observations, which were made at visual wavelengths, most occultations were observed in the infrared K band ($\lambda\sim2.2\ \mu$m), where a strong methane absorption band minimizes the planet's signal within the photometric aperture. Most of the IR observations were made using InSb aperture photometers, centered on the occultation star but also containing the background signal from the rings, planet, and sky. Many events were recorded continuously in what is known as DC mode, most suitable for photometric conditions. In other cases, especially for faint stars or events with a strong background signal, observations were conducted in chopping mode, where the secondary mirror of the telescope rapidly nodded between the event star and the nearby sky, with filtering electronics recording the difference between the brightness within the aperture at these two alternate positions. 

\begin{figure}
\centerline{\resizebox{6in}{!}{\includegraphics[angle=90]{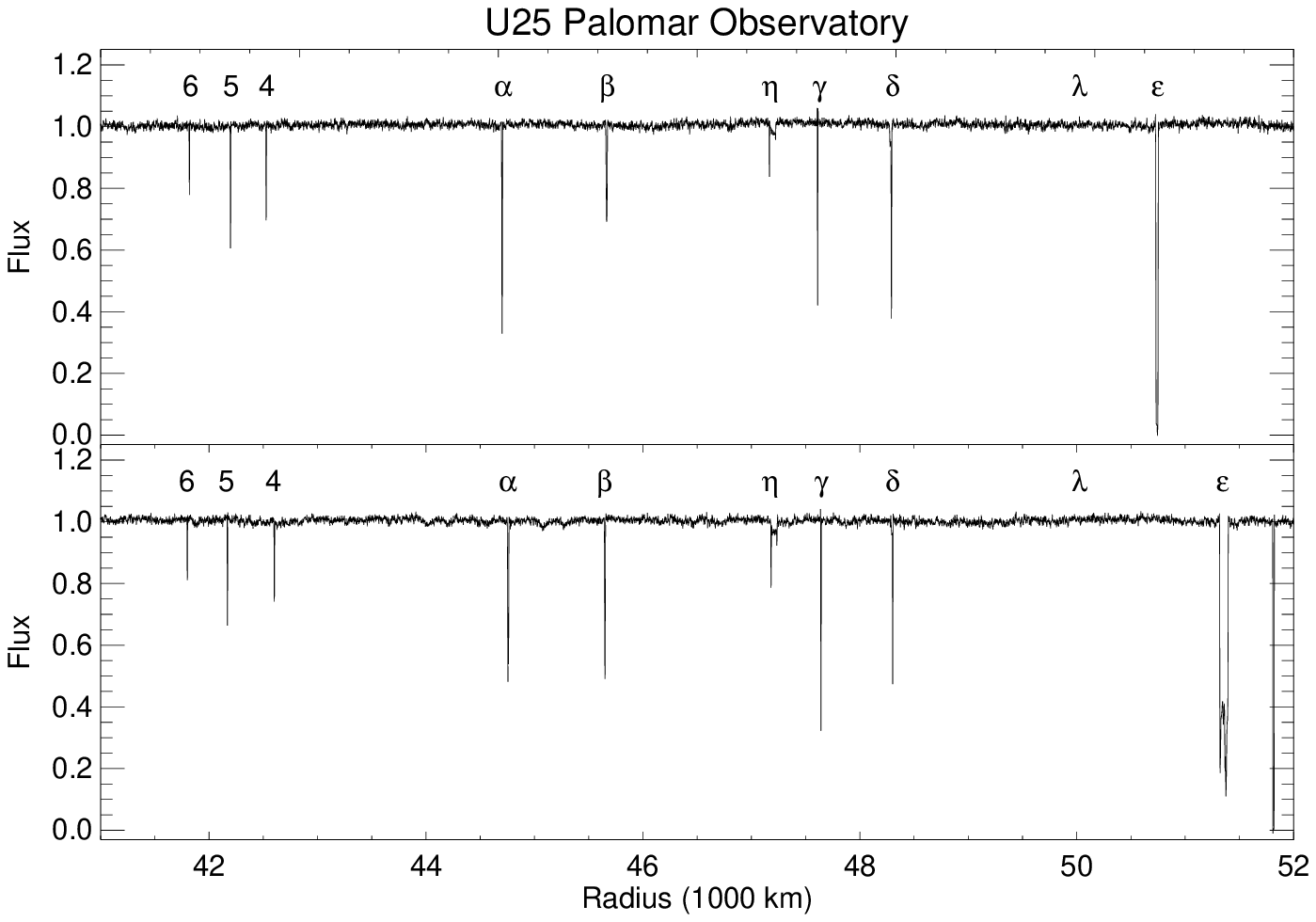}}}
\caption{Observations of the May 24, 1985 occultation of U25 from Palomar Observatory's 5~m Hale telescope at $\lambda=2.2\ \mu$m (after Fig.~4.2 \cite{Nicholson2018a}). The orbital eccentricities of several of the narrow rings are evident from their radial misalignments between ingress (top) and egress (bottom). Only the egress $\epsilon$ ring profile shows internal structure at this scale. Both the $\eta$ and $\delta$ rings have adjacent low-optical depth shoulders, not considered in this work. There is no hint of the elusive $\lambda$ ring in this very high SNR observation, indicative of its azimuthal variability in optical depth and/or its dusty nature, rendering it less evident in IR observations. }
\label{fig:U25Pal}
\end{figure}

As an example of some of the best data used in this work, {\bf Fig.~\ref{fig:U25Pal}} shows the observations of the May 24, 1985 occultation of U25 from Palomar Observatory's 5~m Hale telescope. The bright star was observed in chopping mode under photometric conditions with excellent seeing and accurate tracking, resulting in high-SNR lightcurves with stable baselines. The figure shows the normalized ingress and egress signals as a function of equatorial plane radius. The radial misalignment of several of the ingress/egress ring event pairs is due primarily to their orbital eccentricities, most notably for the outermost $\epsilon$ ring. Most ring events are sharp and unresolved at this resolution, with the exception of low optical depth companions exterior to the $\eta$ ring and interior to the $\delta$ ring (barely visible here in the ingress profile). Only the eccentric $\epsilon$ ring is radially resolved, narrower near periapse (upper panel) than near apoapse. The orbital radius of the elusive $\lambda$ ring is labeled, but it was not detected during either ingress or egress during this occultation. (The sharp dip in the signal just exterior to the egress $\epsilon$ ring is a calibration check of the normalized intensity of the occulted star.)

\begin{figure}
\centerline{\resizebox{8in}{!}{\includegraphics[angle=90]{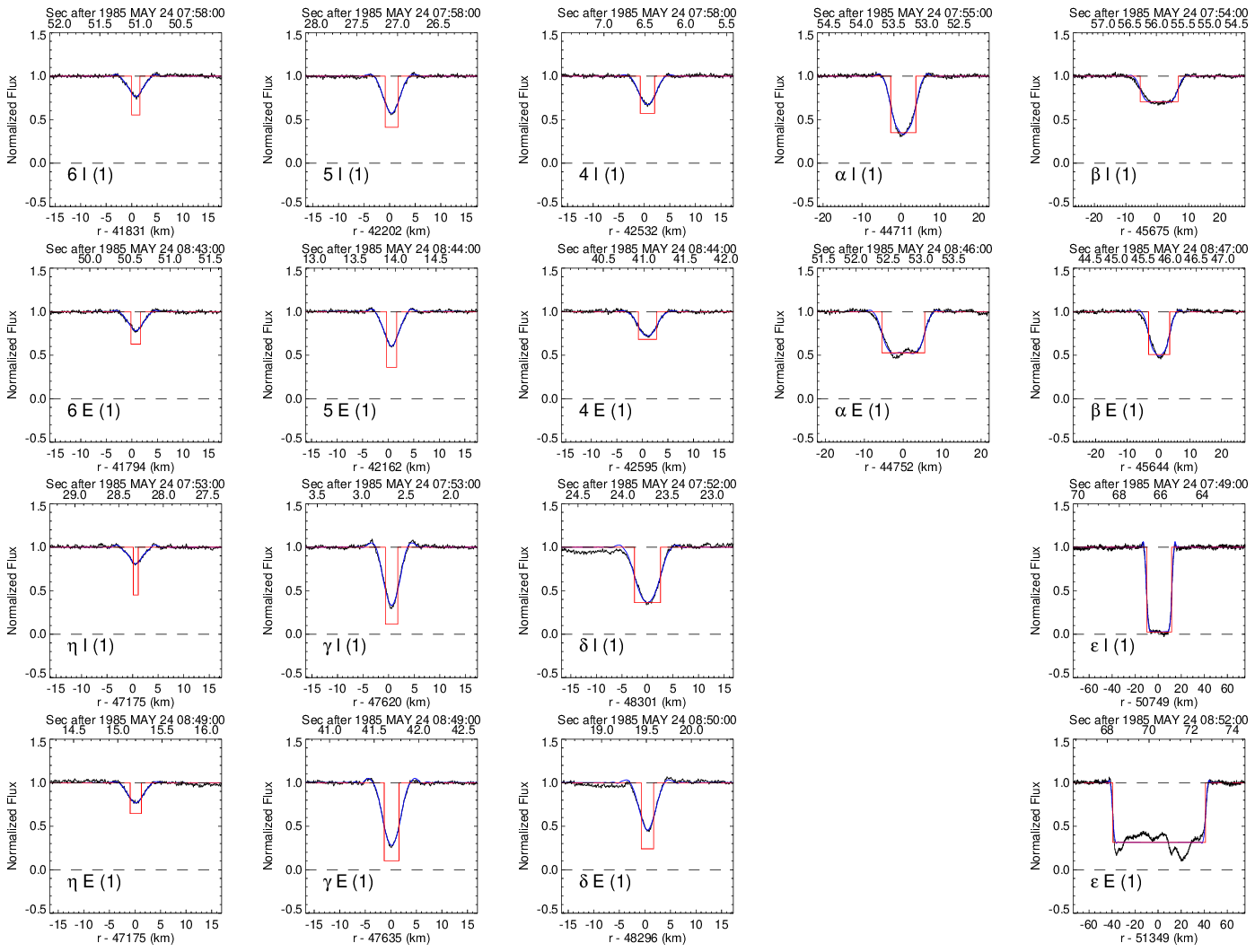}}}
\caption{Gallery of individual ring profiles from the May 24, 1985 observations of the U25 event from Palomar Observatory's 5~m Hale telescope, plotted as a function of ring plane radius (lower axis) and observed time in UTC (upper axis). The profiles are arranged in parallel rows (ingress, then egress), from the innermost ring 6 to the outermost $\epsilon$ ring. Square-well model fits are overplotted on each profile, showing the smoothing effects of diffraction and the finite angular size of the occulted star, and the geometric square-well model is shown as a box function in each panel. Rings are labeled by name, event direction (I for ingress, E for egress), and Quality Index (1 in all cases shown here -- see text for details).}
\label{fig:U25PalGallery}
\end{figure}

A gallery of the individual ring profiles for the U25 Palomar occultation is shown in {\bf Fig.~\ref{fig:U25PalGallery}}, plotted in units of normalized flux as a function of ring plane radius (lower axis) and event time (upper axis) and arranged in parallel rows for ingress and egress, increasing radially from the innermost ring 6 to the outermost $\epsilon$ ring. The gap reflects the non-detection of the $\lambda$ ring, noted above. Each profile is labeled by the ring name and event direction (I for ingress, E for egress), and a Quality Index (QI) in parentheses. As described in more detail in Paper 1, every individual Earth-based ring profile used in this study has been assigned a subjective quality index (QI), ranging from 1 (high-SNR profile with sharp ring edges well-matched by the model fit) to 4 (unreliable detection). For the observations shown, QI=1 for every ring event, but this is not the usual case -- see Paper 1 for a complete survey of all occultations. 
  
Most of the Uranian rings are at most a few km in width and are unresolved radially in the lightcurves, owing to the combined smoothing effects of Fresnel diffraction and the finite angular diameter of the occultation star. In some cases, the resolution is further limited by intrinsic instrumental time constants or by the sampling interval of the recorded data.
For this event, the Fresnel diffraction scale $F\sim\sqrt{\lambda D/2}=1.73$~km, where $\lambda$ is the observed wavelength and $D$ is the distance between the observer (or spacecraft) and the ring plane, and the projected stellar diameter at Uranus was assumed to be $d_*=5.254$ km. In this case, the stellar smoothing dominated that by diffraction, but there are visible diffraction fringes in several of the observed profiles and both effects were included in the models. (Smoothing due to the $\lambda = 2 - 2.4\ \mu$m response of the K-band filter was also taken into account.) Each profile has been fitted to a square-well model in which the ring is assumed to be sharp-edged and radially uniform in opacity. The underlying square-well model is shown as a box function in red, and the corresponding smoothed model is overdrawn on the data. In this case, the models match the observations extremely well, as indicated in the assignment of QI=1 for all profiles.

\begin{figure}
\centerline{\resizebox{4in}{!}{\includegraphics[angle=90]{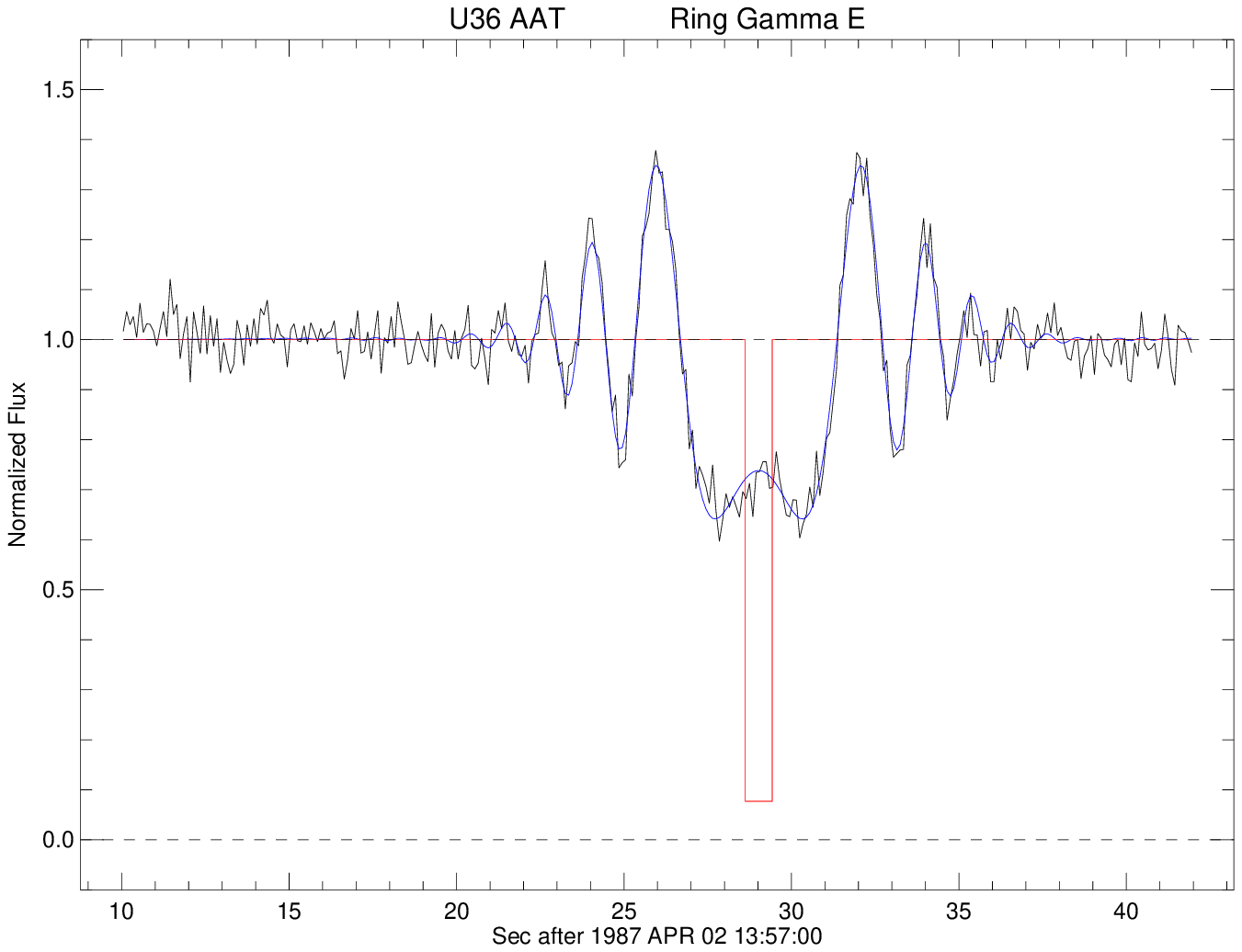}}}
\caption{The egress $\gamma$ ring profile observed from the AAT during the stellar occultation of U36 on April 2, 1987. The best-fitting square-well diffraction model is overplotted on the data, and the underlying box model of the ring is shown in red.}
\label{fig:U36AATgisqw}
\end{figure}

In the examples shown above, diffraction effects are suppressed somewhat by the significant smoothing due to the relatively large projected diameter of the occulted star at Uranus. For fainter and more distant occultation stars, however, stellar smoothing is less significant and diffraction by the ring edges is more prominent. 
{\bf Figure~\ref{fig:U36AATgisqw}} shows the premier example of this: the egress $\gamma$ ring profile observed from the Anglo-Australian Telescope (AAT) during the remarkable 4-day occultation of U36 in March/April 1987, when Uranus reversed direction at the retrograde point in its geocentric orbit and the consequent skyplane velocity of the star relative the projected ring edge was only $v_\perp=0.90\ \kms$. The projected diameter of U36 was 0.548 km, significantly smaller than the Fresnel scale $F=1.76$ km. In this example, multiple diffraction fringes are closely matched by a model of the $\gamma$ ring as a 0.73 km-wide square-well with a fractional transmission $f=0.077$. 

 For our present purposes, the key observables from the full set of individual Earth-based occultation ring profiles are the times of the midline and edges from the fitted square-well models, which we treat as our best estimates of the midline and edges of the actual ring.\footnote{The equivalent width and mean opacity are also determined from the square-well fits, but we do not make use of this information in this study.} As described below, we determine the geometry of the Uranus ring system from least-squares fits to these ring midpoint and edge location times. In our final ring orbit models, the post-fit RMS radius errors of the midpoint measurements for the nine main rings are 0.212--0.401 km and somewhat larger for the ring edges, but in all cases well below the characteristic Fresnel scale for the observations. 
 
Under these circumstances, it is appropriate to ask whether the square-well model accurately determines the widths and midlines of the narrow ring profiles that are at the heart of our investigation. We address this question in Appendix \ref{appendix:sqwell}, where we compare square-well model fits to the diffraction-limited \Voyager\ RSS ring observations with the measured widths of the corresponding spatially resolved diffraction-corrected profiles. 
 Our summary conclusion is that the square-well model demonstrably provides estimates of ring widths that are accurate at the level of a few hundred m for rings with intrinsically sharp edges, and are likely to be somewhat less accurate for rings with more gradual edges. To the extent that the fitted ring width can be viewed as equivalent to independent measurements of the inner and outer ring edges whose average defines the midline of the ring, the corresponding accuracy of the ring midline measurements is reduced by a factor of $1/\sqrt{2}$ to about 0.2~km, a bit smaller than but of the same order as the RMS error in the ring orbit fits presented below.
In cases where one edge is sharp and the other is more gradual, the two test cases examined above show no systematic sense of error, with one ring width being underestimated and the other example being overestimated. These limitations should be kept in mind when assessing the accuracy of the ring orbit model fits to the observed measurements of the ring midlines and edges.

        \subsection{Imaging observations}
Direct images of the Uranian rings provide additional information about their photometric properties that is complementary to the results from stellar occultations. Although we make no direct use of imaging data in this paper, preliminary \citep{Showalter2011} and more detailed recent \citep{Hedman2023} analysis of \Voyager\ 2 wide- and narrow-angle camera (WAC/NAC) images have revealed periodic azimuthal brightness variations in several of the rings that are indicative of normal modes, some of which are also seen in our analysis. \Voyager\ and \HST\ (\hst) observations of small Uranian moons have provided updated estimates of the ephemeris of Ophelia that we will compare to our results in Section \ref{sec:satellites} below. Ongoing and upcoming imaging observations from \JWST\ will provide additional valuable information about the ring structure as well.
%
\section{Geometric Model of the Uranus Ring System}
\subsection{General description}

We determine the orbital elements of the rings and the pole direction and gravity field of Uranus from model fits to the measured midpoints of the occultation ring profiles from the complete set of observations described in Section \ref{sec:observations}, obtained from square-well model fits for the Earth-based stellar occultations and \Voyager\ RSS data, and from direct measurement of midpoints of the spatially resolved \Voyager\ PPS stellar occultation profiles. We make use of our well-tested orbit fitting code (RINGFIT) that implements the solar system barycenter (SSBC) occultation geometry described in detail by \cite{French1993}, with minor modifications described by \cite{French2010}.\footnote{Detailed comparisons of our calculations of the event geometry of an inclined and eccentric ring occultation event with those by R.~Jacobson, using an independent code, agree at the 0.002 km level in the derived ring plane radius for a given Earth-received event time.} The code makes extensive use of the ICY interface to NASA’s NAIF SPICE toolkit \citep{Acton1996}, which provides access to planetary ephemerides and spacecraft trajectory files. We use the J2000 heliocentric reference frame, the IAU 1976 model for the Earth shape \citep{Abalakin1981}, and the ITRF93 Earth rotation model \citep{Boucher1994}. We account for general relativistic deflection by Uranus (including the effect of $J_2$) for Earth-based stellar occultations only, solving for the deflection at the time the occultation ray is closest to Uranus in the sky plane, rather than the time the occultation ray penetrates the ring plane, as implied by Eqs.~(A20) – (A23) of \cite{French1993}. (The difference amounts to only a few meters in the derived ring plane radius.) 
 
                \subsubsection{Occultation stars}
Until recently, the typical astrometric uncertainty in the parallax- and proper motion-corrected occultation star positions was substantially larger than the uncertainty in the Uranus ephemeris, and orbit fits to the rings included fitted corrections to the predicted star positions at the epoch of each occultation. With the release of the \Gaia\ EDR3 and DR3 catalogs \citep{Gaia2021,Gaia2022}, the situation has reversed, with small but measurable systematic errors in the Uranus ephemeris in {\tt ura111.bsp} and {\tt de440.bsp} exceeding the star position uncertainties. To reduce these systematic errors, the {\tt ura178} series of ephemerides (described in Section \ref{sec:ephemerides} below) was developed using the astrometric constraints provided by the full set of ring occultation events in Paper 1 for all but the $\gamma$ and $\lambda$ rings (R. Jacobson, pers. comm). We cross-referenced the predicted star positions for all Earth-based Uranus ring occultations, using VizieR \citep{Ochsenbein2000}, to obtain the J2000 \Gaia\ DR3 catalog positions given in Table 2 of Paper 1. The \Voyager\ 2 stellar occultation stars $\sigma$ Sgr and $\beta$ Per are too bright to allow accurate \Gaia\ DR3 positions, and for these stars we make use of the {\it Hipparcos} catalog positions \citep{Perryman1997}. As part of our orbit solution, we fit for corrections to selected star positions and proper motions under the assumption that the revised Uranus ephemeris is free of systematic errors.

                \subsubsection{Telescope coordinates}
                The geocentric coordinates of many of the telescopes used for the this study are listed in multiple sources that are not always in agreement. For this work, we used modern GPS coordinates, when available. In all cases, these closely matched our estimates from Google Earth, and we therefore used Google Earth to estimate the locations of telescopes not otherwise known. 
                {\bf Table \ref{tbl:obs}} lists our adopted coordinates for each telescope, including the observatory name, ID, telescope diameter, E longitude, latitude, altitude relative to the IAU 1976 Earth model, and geocentric coordinates $x,y,z$ and radius $\rho$. These were converted to a custom SPK kernel file of observatory locations ({\tt ObsCodes\_Uranus\_20220212.spk}) for use in our SPICE-based geometry calculations in our ring orbit fitting code.

\begin{table*} [ht]
\tiny
\caption{Observatory and telescope coordinates}
\label{tbl:obs} 
\centering
\begin{tabular}{l r c r r r r r r r r r r r}\hline
Observatory & ID & Diam (cm) & \multicolumn{3}{c}{E longitude} & \multicolumn{3}{c}{Latitude} & alt (m) & x (km) & y (km) & z(km) &$\rho$ (km)\\
\hline
Centro Astron. Hispano-Aleman (CAHA) & CAL & 123 & $ +357 $ & 27 & 11.675 & $ +37 $ & 13 & 15.38 & 2161 & $   5081.895 $ & $   -226.036 $ & $   3838.254 $ &   6372.515 \\
Cerro Tololo Interamerican Obs. (CTIO) & 807 & 400 & $ +289 $ & 11 & 36.917 & $ -30 $ & 10 & 10.78 & 2380 & $   1815.052 $ & $  -5213.997 $ & $  -3187.844 $ &   6375.149 \\
 & C60 & 150 & $ +289 $ & 11 & 35.700 & $ -30 $ & 10 & 09.30 & 2380 & $   1815.029 $ & $  -5214.030 $ & $  -3187.804 $ &   6375.149 \\
European Souther Observatory & ESO & 360 & $ +289 $ & 16 & 05.900 & $ -29 $ & 15 & 39.50 & 2400 & $   1838.338 $ & $  -5258.792 $ & $  -3100.341 $ &   6375.460 \\
 & ES2 & 200 & $ +289 $ & 15 & 48.000 & $ -29 $ & 15 & 28.20 & 2317 & $   1837.913 $ & $  -5259.044 $ & $  -3099.997 $ &   6375.378 \\
 & ES1 & 104 & $ +289 $ & 15 & 41.900 & $ -29 $ & 15 & 24.00 & 2321 & $   1837.780 $ & $  -5259.161 $ & $  -3099.886 $ &   6375.382 \\
NASA Infrared Telescope Facility (IRTF) & IRT & 320 & $ +204 $ & 31 & 40.932 & $ +19 $ & 49 & 34.46 & 4212 & $  -5464.326 $ & $  -2493.467 $ & $   2151.039 $ &   6379.907 \\
Las Campanas Observatory & LAS & 250 & $ +289 $ & 17 & 45.784 & $ -29 $ & 00 & 26.67 & 2270 & $   1845.370 $ & $  -5270.713 $ & $  -3075.720 $ &   6375.410 \\
 & LAV & 100 & $ +289 $ & 17 & 59.053 & $ -29 $ & 00 & 43.22 & 2270 & $   1845.627 $ & $  -5270.361 $ & $  -3076.166 $ &   6375.409 \\
Lowell Observatory & 688 & 180 & $ +248 $ & 27 & 47.436 & $ +35 $ & 05 & 49.42 & 2204 & $  -1918.477 $ & $  -4861.173 $ & $   3647.949 $ &   6373.311 \\
McDonald Observatory & 711 & 270 & $ +255 $ & 58 & 40.667 & $ +30 $ & 40 & 18.27 & 2103 & $  -1330.794 $ & $  -5328.793 $ & $   3235.718 $ &   6374.709 \\
Mount Stromlo & 414 & 190 & $ +149 $ & 00 & 31.968 & $ -35 $ & 19 & 08.72 & 770 & $  -4466.831 $ & $   2683.001 $ & $  -3667.254 $ &   6371.799 \\
Observatorio del Teide & TEE & 155 & $ +343 $ & 29 & 20.740 & $ +28 $ & 18 & 01.82 & 2395 & $   5390.292 $ & $  -1597.793 $ & $   3007.004 $ &   6375.756 \\
Palomar Observatory & 675 & 508 & $ +243 $ & 08 & 06.612 & $ +33 $ & 21 & 22.57 & 1706 & $   4678.798 $ & $     11.472 $ & $   4324.379 $ &   6371.149 \\
Pic du Midi (OPMT) & 586 & 200 & $ +000 $ & 08 & 25.764 & $ +42 $ & 56 & 14.52 & 2891 & $   1815.052 $ & $  -5213.997 $ & $  -3187.844 $ &   6375.149 \\
 & PI1 & 106 & $ +000 $ & 08 & 31.900 & $ +42 $ & 56 & 11.20 & 2862 & $   4678.846 $ & $     11.612 $ & $   4324.285 $ &   6371.120 \\
S. African Astronomical Obs. (SAAO) & SAA & 188 & $ +020 $ & 48 & 41.710 & $ -32 $ & 22 & 44.16 & 1768 & $   5041.283 $ & $   1916.170 $ & $  -3396.939 $ &   6373.809 \\
Siding Spring Observatory (AAT) & 413 & 390 & $ +149 $ & 04 & 02.208 & $ -31 $ & 16 & 31.14 & 1164 & $  -4681.032 $ & $   2805.172 $ & $  -3292.625 $ &   6373.572 \\
Siding Spring Observatory(ANU) & ANU & 230 & $ +149 $ & 03 & 44.597 & $ -31 $ & 16 & 17.88 & 1149 & $  -4680.963 $ & $   2805.674 $ & $  -3292.268 $ &   6373.559 \\
UK Infrared Telescope (UKIRT) & UKI & 380 & $ +204 $ & 31 & 46.890 & $ +19 $ & 49 & 20.70 & 4194 & $  -5464.370 $ & $  -2493.677 $ & $   2150.635 $ &   6379.890 \\
\hline
\end{tabular}
\end{table*}

        \subsection{Kinematical model for ring orbits}\label{sec:kinmodel}

The RINGFIT orbit fitting code solves for the geometric ring orbital elements that minimize the sum of squared residuals between the observed and model ring plane radii, using a standard kinematical model for all ring features. Following \cite{Nicholson2014b}, our basic model is of a precessing, inclined keplerian ellipse, specified by
\beq
r(\lambda,t) = \frac{a(1-e^2)}{1 + e \cos f},
\label{eq:kepler}
\eeq
\noindent where the true anomaly $f=\lambda - \varpi = \lambda - \varpi_0 - \dot\varpi(t-t_0)$. Here, $r, \lambda$, and $t$ are the radius, inertial longitude, and time of the observation at the ring (not at the observer), $a$ and $e$ are the ring's semimajor axis and eccentricity, $\varpi$ and $\dot\varpi$ are its longitude of periapse and apsidal precession rate, and $t_0$ is the epoch of the fit. 
For inclined rings, we include three additional parameters: $i$ (the inclination relative to the mean ring plane), $\Omega_0$ (the longitude of the ascending node) and $\dot\Omega$ (the nodal regression rate), and compute the intercept point of the
occultation ray with the specified
inclined ring plane. The zero-point for the inertial longitudes $\lambda, \varpi_0,$ and $\Omega_0$ (as well as $\delta_m$ below) is the ascending node of Uranus's equator on Earth's equator
 of J2000, where the orientation of the Uranus pole is in the direction of positive angular momentum (\ie\ $180^\circ$ from the IAU definition of the Uranus north pole). 
 The apsidal and nodal rates $\dot\varpi$ and $\dot\Omega$ can be treated as free parameters in the orbit fit, or alternatively, as in \cite{Nicholson2014b}, the expected secular rates $\dot\varpi_{sec}$ and $\dot\Omega_{sec}$ can be calculated from the combined effects of Uranus's zonal gravity harmonics $J_2, J_4, J_6, ...$ based on the results of \cite{BRL94}, and the secular precession induced by the planet's satellites according to the expressions:
\begin{equation}
\begin{split}
\dpisec = &\sqrt{\frac{GM}{a^3}} \left\{ \frac{3}{2} J_2 \left( \frac{R}{a} \right) ^2 (1+e^2-2\sin^2i) 
 - \frac{15}{4} J_4 \left( \frac{R}{a} \right) ^4\right. \\ 
& \left. +\left[\frac{27}{64} J_2^3 -\frac{45}{32} J_2 J_4 +\frac{105}{16} J_6 \right]\left( \frac{R}{a} \right) ^6 
+ \frac{1}{4}\sum_{j=1}^{M_o} \frac{m_j}{M}\alpha_{oj}^2 b_{3/2}^1(\alpha_{oj})
+ \frac{1}{4}\sum_{j=1}^{M_i} \frac{m_j}{M}\alpha_{ij}^2 b_{3/2}^1(\alpha_{ij})
\right\}
\end{split}
\label{eq:apserate}
\end{equation}

\begin{equation}
\begin{split}
\dot\Omega_{\rm sec} = & -\sqrt{\frac{GM}{a^3}} \left\{\frac{3}{2}J_2\left(\frac{R}{a}\right)^2(1+e^2-\frac{1}{2}\sin^2i) 
- \left[\frac{9}{4}J_2^2 + \frac{15}{4}J_4 \right] \left(\frac{R}{a}\right)^4\right.\\
& \left. +\left[\frac{351}{64} J_2^3 +\frac{315}{32} J_2 J_4 +\frac{105}{16} J_6 \right]\left( \frac{R}{a} \right) ^6 
+ \frac{1}{4}\sum_{j=1}^{M_o} \frac{m_j}{M}\alpha_{oj}^2 b_{3/2}^1(\alpha_{oj})
+ \frac{1}{4}\sum_{j=1}^{M_i} \frac{m_j}{M}\alpha_{ij} b_{3/2}^1(\alpha_{ij})
\right\} ,
\end{split}
\label{eq:noderate}
\eeq
where the summation is carried out over outer (subscript $o$) and inner (subscript $i$) satellites of mass $m_j$ and orbital radius $a_{oj}$ or $a_{ij}$, where for outer satellites $\alpha_{oj}=a/a_{oj}$ and for inner satellites $\alpha_{ij}=a_{ij}/a$, and $b_{3/2}^1(\alpha_{oj})$ and $b_{3/2}^1(\alpha_{ij})$ are Laplace coefficients as defined by \cite{BC61}. (Note that the Laplace coefficient factor $\alpha_{oj}$ is squared, whereas $\alpha_{ij}$ is to the first power only.)
{\bf Figure~\ref{fig:precrates}} shows radial dependence of satellite-induced secular precession in the vicinity of the ten narrow Uranian rings, with the individual contributions shown for each satellite. As we will see, the uncertainty in the fitted apse rates is as small as $4.2\times10^{-6}$ \degd\ for the $\epsilon$ ring, comparable to the individual secular contributions of Oberon, Portia, and Puck to its precession rate. 

The orbital semimajor axes $a$ and masses used to compute the secular precession rate due to the major satellites are given in {\bf Table \ref{tbl:majorsatdata}}, from 
 \cite{Jacobson2023}.\footnote{Since satellite masses are variously quoted in the literature in units of $GM_{\rm sat}$, fractional planet mass $M_{\rm sat}/M_{\rm Ur}$, and kg, we list all three for convenient comparison.} 

\begin{table*} [ht]
\caption{Major satellite orbits and masses from \cite{Jacobson2023}}
\label{tbl:majorsatdata}
\centering
\begin{tabular}{l c c c c}
\hline
Satellite & $a$ (km)  &  $GM_{\rm sat}$ (km$^3$ s$^{-2}$)   & $M_{\rm sat}/M_{Ur}\times10^{-6}$  & $M_{\rm sat}\ (\times 10^{20}$ kg) \\
\hline
Ariel &   190928. &     82.30 $\pm$      1.20 &      14.204 $\pm$       0.207 &      12.331 $\pm$       0.180 \\
Umbriel &   265981. &     86.00 $\pm$      1.50 &      14.843 $\pm$       0.259 &      12.885 $\pm$       0.225 \\
Titania &   436283. &    230.60 $\pm$      3.40 &      39.800 $\pm$       0.587 &      34.550 $\pm$       0.509 \\
Oberon &   583447. &    207.60 $\pm$      5.00 &      35.830 $\pm$       0.863 &      31.104 $\pm$       0.749 \\
Miranda &   129828. &      4.20 $\pm$      0.20 &      0.7249 $\pm$      0.0345 &      0.6293 $\pm$      0.0300 \\
Puck &    86004. &     0.1275 $\pm$     0.0425 &       0.0220 $\pm$       0.0073 &       0.0191 $\pm$       0.0064 \\
\hline
\end{tabular}
\end{table*}

The corresponding results for the minor satellites are given in {\bf Table \ref{tbl:minorsatdata}}, with semimajor axes $a$ from \cite{Jacobson98} and satellite properties from \cite{Karkoschka2001a,Karkoschka2001b}. Each satellite is modeled as a prolate spheroid with semimajor and semiminor axes $A$ and $B$, respectively, and volume $V$ given by
\beq
V = \frac{4\pi}{3}  AB^2
\eeq
using the dimensions and uncertainties in Table V of \cite{Karkoschka2001b}. We estimate the fractional mass uncertainty for an assumed satellite density from the fractional uncertainty in its volume $\sigma(V)/V$, computed from the propagated uncertainties $\sigma(\sqrt{AB})$ and $\sigma(B/A)$.

The corresponding mass estimates and uncertainties are computed for an assumed mean density of uncompressed solid water ice $\rho=0.90$ gm\ cm$^{-3}$, which can be scaled for alternate assumed densities. Also included are the directly measured masses and inferred densities of Cordelia, Ophelia, and Cressida determined from the amplitudes and inferred locations of forced resonances between these moons and ring edges (see Section \ref{sec:moondensities} below).

\begin{table*} [ht]
\begin{center}
\caption{Minor satellite orbits, dimensions, densities, and masses$^a$}
\label{tbl:minorsatdata}
\stackunder{
\hspace*{-4cm}
\centering
\begin{tabular}{l c c c c c c c c c c}
\hline
Satellite & $a$          & $\sqrt{AB}$  & $B/A$ & $A\times B$ & $V$ &  $\frac{\sigma(V)}{V}$ & $\rho$ & $GM_{\rm sat}$   & $M_{\rm sat}/M_{Ur}$  & $M_{\rm sat}$ \\
&     km  &     km &  & km$\times$km  & $\times 10^5$ km$^3$   &     &  gm cm$^{-3}$  & $\times 10^{-3}$ km$^3$ s$^{-2}$   & $\times 10^{-10}$  & $\times 10^{16}$ kg \\
\hline
Cordelia &    49752.000 & $ 21 \pm   3 $ & 0.7 $ \pm $ 0.2 & 25$\times$18 &  0.339 & 0.349 & [ 0.90] &    2.04 $ \pm $    0.71 &    3.52 $ \pm $    1.23 &    3.05 $ \pm $    1.06 \\[0.5em]
Cordelia$^b$ & & & & & &  &  1.79$^{+ 0.97}_{- 0.49}$  &    4.06 $\pm$    0.38 &    7.00 $\pm$    0.66 &        6.08 $\pm$        0.57 \\[0.5em]
Ophelia &    53764.000 & $ 23 \pm   4 $ & 0.7 $ \pm $ 0.3 & 27$\times$19 &  0.408 & 0.504 & [ 0.90] &    2.45 $ \pm $    1.24 &    4.23 $ \pm $    2.13 &    3.67 $ \pm $    1.85 \\[0.5em]
Ophelia$^b$ & & & & & &  &  0.87$^{+ 0.89}_{- 0.30}$  &    2.38 $\pm$    0.22 &    4.11 $\pm$    0.37 &        3.57 $\pm$        0.32 \\[0.5em]
Bianca &    59165.000 & $ 27 \pm   2 $ & 0.7 $ \pm $ 0.2 & 32$\times$23 &  0.709 & 0.299 & [ 0.90] &    4.26 $ \pm $    1.27 &    7.35 $ \pm $    2.20 &    6.38 $ \pm $    1.91 \\[0.5em]
Cressida &    61767.000 & $ 41 \pm   2 $ & 0.8 $ \pm $ 0.3 & 46$\times$37 &  2.638 & 0.380 & [ 0.90] &   15.85 $ \pm $    6.02 &   27.35 $ \pm $   10.40 &   23.74 $ \pm $    9.03 \\[0.5em]
Cressida$^b$ & & & & & &  &  0.70$^{+ 0.44}_{- 0.21}$  &   12.27 $\pm$    1.41 &   21.18 $\pm$    2.44 &       18.39 $\pm$        2.12 \\[0.5em]
Desdemona &    62659.000 & $ 35 \pm   4 $ & 0.6 $ \pm $ 0.2 & 45$\times$27 &  1.374 & 0.375 & [ 0.90] &    8.25 $ \pm $    3.09 &   14.25 $ \pm $    5.34 &   12.37 $ \pm $    4.63 \\[0.5em]
Juliet &    64358.000 & $ 53 \pm   4 $ & 0.5 $ \pm $ 0.1 & 75$\times$37 &  4.301 & 0.230 & [ 0.90] &   25.83 $ \pm $    5.95 &   44.59 $ \pm $   10.26 &   38.71 $ \pm $    8.91 \\[0.5em]
Portia &    66097.000 & $ 70 \pm   4 $ & 0.8 $ \pm $ 0.1 & 78$\times$63 & 12.968 & 0.148 & [ 0.90] &   77.90 $ \pm $   11.55 &  134.44 $ \pm $   19.93 &  116.71 $ \pm $   17.30 \\[0.5em]
Rosalind &    69927.000 & $ 36 \pm   6 $ & 1.0 $ \pm $ 0.2 & 36$\times$36 &  1.954 & 0.314 & [ 0.90] &   11.74 $ \pm $    3.69 &   20.26 $ \pm $    6.36 &   17.59 $ \pm $    5.52 \\[0.5em]
Belinda &    75255.000 & $ 45 \pm   8 $ & 0.5 $ \pm $ 0.1 & 64$\times$32 &  2.745 & 0.327 & [ 0.90] &   16.49 $ \pm $    5.39 &   28.46 $ \pm $    9.30 &   24.71 $ \pm $    8.07 \\[0.5em]
Puck &    86004.000 & $ 81 \pm   2 $ & 1.0 $ \pm $ 0.1 & 81$\times$81 & 22.261 & 0.078 & [ 0.90] &  133.72 $ \pm $   10.47 &  230.79 $ \pm $   18.06 &  200.35 $ \pm $   15.68 \\[0.5em]
%
%
\hline
\end{tabular}
} {
\hspace*{-2cm}
\parbox{7.6in}{
$^a$ Except as noted, masses and uncertainties computed from satellite shapes from Table V of \cite{Karkoschka2001b} and assumed density $\rho$=0.90~gm\ cm$^{-3}$.\\
$^b$ Inferred mass from the amplitude(s) of normal mode(s) forced on ring edge(s) and corresponding density computed from the tabulated satellite volume $V$. Density uncertainty computed from the combined volume and mass uncertainties. See Section \ref{sec:moondensities}.\\
}}
\end{center}
\end{table*}

\begin{figure}
\centerline{\resizebox{7in}{!}{\includegraphics[angle=90]{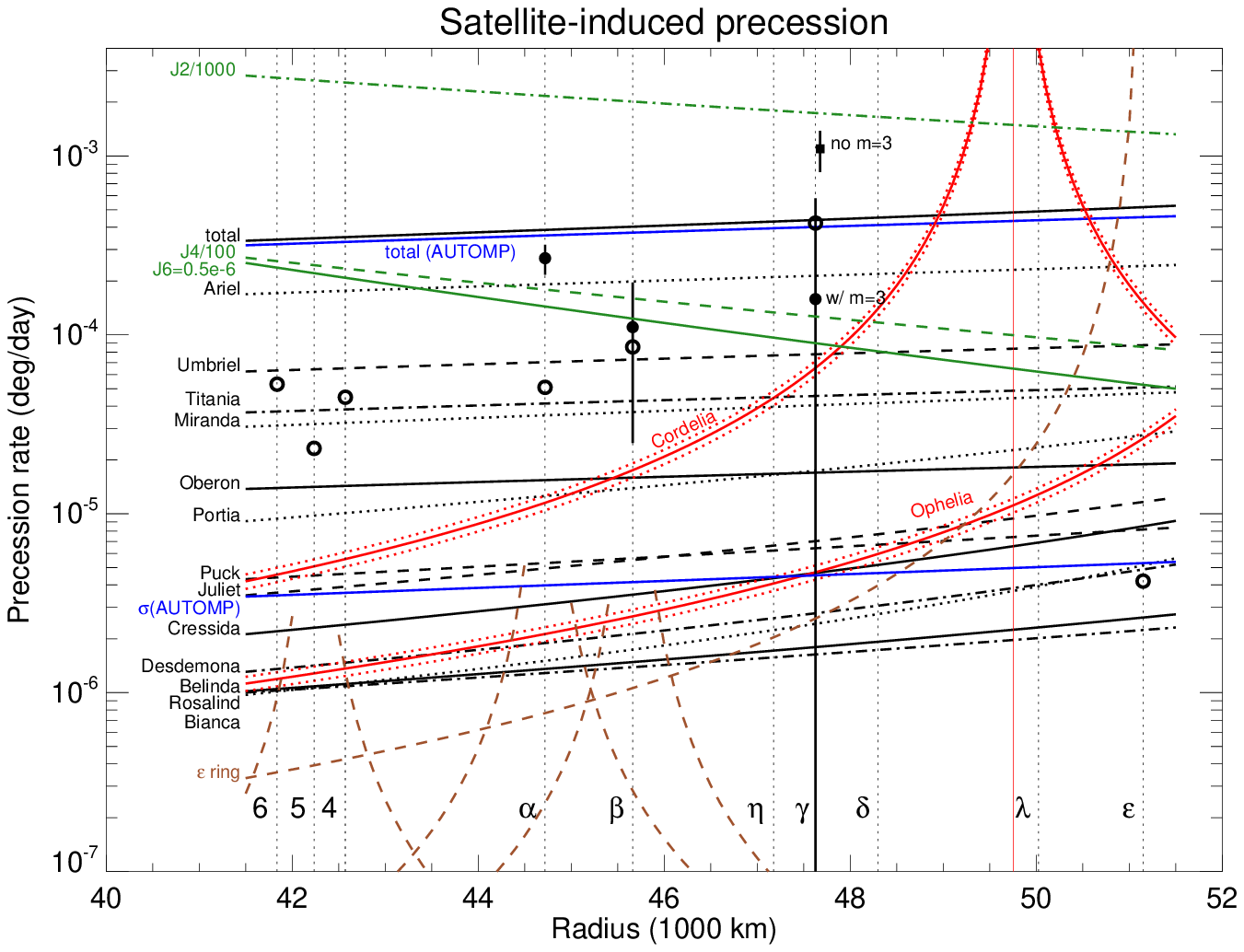}}}
\caption{The magnitude of the secular apsidal and nodal precession rates due to Uranus's satellites, plotted as a function of radius in the vicinity of the ten narrow rings. The black line labeled `total' includes all the satellites except Cordelia and Ophelia, whose individual contributions are plotted as the upper and lower solid red lines, respectively, bracketed by the uncertainties shown by dotted red lines. The blue lines labeled `total (AUTOMP)' and `$\sigma($AUTOMP)' are the sum of the the satellite-induced precession rates from the six major satellites (Ariel, Umbriel, Titania, Oberon, Miranda, and Puck), and the quadratic sum of their uncertainties, calculated from the uncertainties in their masses \citep{Jacobson2014}. The green lines show the radial dependence of the contributions to the precession rates of the rings by the gravitational coefficients $J_2$ and $J_4$, scaled to be within the vertical range of the plot, and by a representative value of $J_6=0.5\times10^{-6}$. The filled circles show the inferred anomalous precession rates for the $\alpha, \beta$, and $\gamma$ rings, with error bars indicating their 1-$\sigma$ uncertainties (see Section \ref{sec:precrates}). The filled square shows the inferred anomalous precession rate of the $\gamma$ ring for an alternate case in which the $m=3$ mode is omitted from the orbit model.
The open circles at the radii of the eccentric rings 6, 5, 4, $\alpha$, $\beta$, and $\epsilon$ ring mark the formal uncertainties in their fitted apsidal precession rates. The dashed brown lines centered on the eccentric rings show the estimated magnitude of their contributions to predicted precession rates at nearby radii, for assumed masses of the rings. In all cases, these are quite small, falling below that due to Cressida.}
\label{fig:precrates}
\end{figure}

In addition to the keplerian orbit, the orbit model allows for a combination of free or forced modes of radial distortion $\Delta r(m,\lambda,t)$ given by
\begin{equation}
\Delta r(m,\lambda,t) = -A_m\ \cos (m\theta),
\label{eq:modemodel}
\end{equation}
\noindent where
\beq
\theta = \lambda - \Omega_P(t-t_0) - \delta_m.
\eeq
\noindent 
Here, $m$ is the number of radial minima and maxima in the pattern,
$A_m$ and $\delta_m$\ are the mode's
radial amplitude and phase (the
longitude at epoch of one of the $m$ radial minima), respectively,
and the pattern speed \fpat\ is its angular rotation rate in inertial
space. As explained in more detail by \cite{Nicholson2014b}, \fpat\ is expected to be close to that of
a Lindblad resonance located at a ring particle's orbit, in which case
\begin{equation}
\Omega_P \simeq [(m-1)n + \dpisec]/m,
\label{eq:modepatspeed}
\end{equation}
\noindent where the mean motion $n$ and apsidal precession rate
$\dpisec$ 
are evaluated at the semimajor axis of the ring feature in question. A positive
value of $m$ corresponds to an inner Lindblad resonance (ILR)-type normal mode, expected at the outer edge of a ringlet, while a negative
value of $m$ corresponds to an outer Lindblad resonance (OLR)-type normal mode, expected at a ringlet's inner edge. In either case, \fpat\ is positive.
For orbits about an oblate planet, $n$ is given by Eq.~(8) of \cite{Nicholson2014b}. 
For each free mode, the additional fit parameters are $A_m$, \fpat\ and
$\delta_m$. 

Equation~(\ref{eq:modemodel}) can also be used to describe radial
perturbations forced by a Lindblad resonance with an external
satellite, in which case the
pattern speed, $m$-value, and phase are all determined by the
satellite's orbital parameters; for a first-order Lindblad resonance, \fpat\ is equal to the satellite's mean motion. The exact resonance location $\ares$\ is specified implicitly by Eq.~\ref{eq:modepatspeed}. \cite{Porco1987} were the first to compute such satellite resonance locations in the vicinity of the known Uranian rings.

Our final expression for the modeled shape of each ringlet in its orbital plane is
\beq
r_{\rm mod}(\lambda,t) = r(\lambda,t) + \sum_{i=1}^M \Delta r(m_i,\lambda,t),
\eeq
where $M$ is the number of modeled free or forced normal modes.

In addition to the kinematical orbit parameters for each ring, the least-squares fit incorporates several fitable physical and geometrical parameters, including the direction of the (possibly precessing) planetary pole, offsets to the catalog positions or proper motions of Earth-based occultation stars (or, alternatively, skyplane offsets to Uranus ephemeris), and time offsets for selected telescope observations. RINGFIT also allows for a variety of weighting schemes, including iteratively computing the weight of each ring's observations from its RMS residuals, or alternatively using the RMS residuals of each separate set of telescope observations to determine the weight of each data point.

        \subsection{Ephemerides}\label{sec:ephemerides}
For this analysis, we used the recent JPL {\tt ura178.bsp} Uranus satellite ephemerides, the {\tt peph.ura178.bsp} planetary ephemerides, and the associated \Voyager\ 2 ephemeris {\tt vgr2.ura178.bsp} at Uranus. These ephemerides were updated from the {\tt ura111.bsp, ura116.bsp, vgr2.ura111.bsp,} and {\tt de440.bsp} series \citep{Jacobson2014,Park2021}, using constraints provided by the Uranus ring occultation data in Paper 1, excluding the $\gamma$ and $\lambda$ rings \citep{Jacobson2023}. For the U0 observations, we used the SPICE kernel described in Paper 1 for the flight path of the Kuiper Airborne Observatory (KAO) during the occultation. 
The U137 and U138 Uranus ring occultations were observed using the \hst, for which we require accurate spacecraft ephemerides to compute the occultation geometry. There are two sources of \hst\ ephemerides: JPL's NAIF website\footnote{\url{https://naif.jpl.nasa.gov/pub/naif/HST/kernels/spk/}} hosts a single kernel file ({\tt hst.bsp}) that is updated regularly and contains NORAD-provided two-line elements (TLE) that have an estimated accuracy of a few km, and the Space Telescope Science Institute (STScI) hosts individual files provided by the Flight Dynamics Facility at the Goddard Space Flight Center (GSFC). The ephemeris accuracy depends primarily on the level of solar activity, which affects drag, and is estimated to be better than 300~m, under most ranges of solar activity. For our orbit fits, we used the more accurate STScI files, which we converted from their original {\tt *.orx} format to FITS files, using the IRAF task {\tt hstephem}, and then to standard NAIF SPK format using the NAIF tool {\tt mkspk}. The two custom-produced SPK files are {\tt pg3f0000r.bsp} for U137 and {\tt pg490000r.bsp} for U138.

The complete list of kernel files used in our analysis is given in {\bf Table \ref{tbl:kernels}}.

	\begin{table*} [ht]
	\begin{center} 
	\caption{Spice kernels}
	\label{tbl:kernels} 
	\begin{threeparttable}
	\centering
	\begin{tabular}{l l }\hline
	File name\tnote{a} & Description\\
	\hline 

{\tt pleph.ura178.bsp}  & Planetary ephemerides, constrained by ring data in Paper 1\\
{\tt ura178.bsp} & Major Uranus satellite ephemerides, constrained by ring data in Paper 1 \\
{\tt vgr2.ura178.bsp}& {\it Voyager 2} trajectory at Uranus, constrained by ring data in Paper 1\\   
{\tt ura115.bsp} & Minor Uranus satellite ephemerides \\                          
{\tt earthstns\_itrf93\_040916.bsp}& Geocentric coordinates of DSN groundstations    \\          
{\tt earth\_720101\_070426.bpc }& Earth rotation model   \\  
{\tt ObsCodes\_Uranus\_20220212.spk}  & Custom geocentric observer coordinates \\            
{\tt pg3f0000r.bsp } & \hst\ ephemeris for U137\\                            
{\tt pg490000r.bsp} & \hst\ ephemeris for U138 \\
{\tt urkao\_v1.bsp }& Custom ephemeris for KAO (U0)\\                             
{\tt naif0012.tls}& leap seconds \\
 	 \hline
	\end{tabular}
	\begin{tablenotes}
	\item[a] All kernels listed are available from \url{ftp://ssd.jpl.nasa.gov/pub/eph} or as part of the Uranus ring occultation support bundle on PDS. See Appendix A of \cite{French2023b}.
	\end{tablenotes}
  \end{threeparttable}
\end{center} 
\end{table*} 

\section{Ring Orbit Fits}
The most recent comprehensive model for the orbital elements of the ten narrow Uranian rings was published by \cite{French1988}, with updates provided in the review articles by \cite{French1991} and \cite{Nicholson2018a}. Subsequently, more recent occultation data were incorporated into orbit models used to determine the geometry of the possible detection of the $\lambda$ ring during the 1992 Jul 11 occultation of U103 \citep{French1996}, but no details were provided about the updated orbital elements. \cite{Jacobson2014} utilized a subset of the both published and unpublished ring occultation data from 1977--1992 in an updated model for the keplerian orbits of eight of the ten narrow rings (excluding $\gamma$ and $\delta$, each of which was known to have significant normal modes), the gravity field of the planet, and the orientation of the pole. Finally, the tabulated residuals from an unpublished orbit fit were used by \cite{Chancia2017} to estimate the mass of the moon Cressida from its perturbation on the $\eta$ ring. 

Here, we extend these analyses by utilizing the full set of observations from 1977--2006 documented in Paper 1 to determine the orbital shapes of both the ring midlines and their edges. As a first step, we updated our search for normal modes, as described below.

\subsection{Normal Modes}
\label{sec:normalmodes}
 The first evidence for normal modes in narrow ringlets was provided by \cite{French1986}, who identified an $m=2$ ILR in the shape of the $\delta$ ring and an $m=0$ OLR in the shape of the $\gamma$ ring.\footnote{\cite{French1986} interpreted the $\delta$ and $\gamma$ ring modes as being forced by unseen satellites because the pattern speeds differed measurably from the expected values for free normal modes located at the semimajor axes of the rings, but in retrospect this was based on an a value of $GM$ for Uranus that subsequently proved to be in error by five times its estimated formal error. With an improved determination of $GM$, the resonance radii of the modes closely matched the semimajor axes of the rings, as expected for free normal modes unassociated with satellites.} Subsequently, tentative detections of edge waves in the $\epsilon$ ring associated with resonances with Cordelia and Ophelia were reported by \cite{French1995}, based on a subset of the observations presented here, and \cite{Chancia2017} identified the signature of an $m=3$ mode in the $\eta$ ring, forced by Cressida.
From the wealth of \Cassini\ occultation observations, a host of free and forced normal modes have also been identified in Saturn's ringlets and sharp ring edges, sometimes in astonishing numbers: the informally-designated ``Strange" ringlet in the Cassini Division is both eccentric and inclined and has seven normal modes, and the inner edge of the Barnard Gap has 11 \citep{French2016}! (For a recent review of observations of narrow rings, gaps, and sharp edges at Saturn, Uranus, and Neptune, see \cite{Nicholson2018b} and references therein; for a detailed modern exposition of the theory of narrow rings and sharp edges, see \cite{Longaretti2018}.)

In our search for normal modes in the Uranian rings, we focus on the measured locations of the center of each ring (COR), although as noted above, dynamical arguments suggest that ILRs should be associated with outer edges of rings (OER) and OLRs with inner edges (IER). For ringlets only a few km wide, these distinctions may be unwarranted, and in any event the measurement of the COR from a square-well fit is necessarily affected by the locations of the IER and OER, each of which might be distorted by a local normal mode. Additional evidence for normal modes in the Uranian rings is provided by periodic azimuthal brightness variations in the rings in \Voyager\ \citep{Hedman2023} and \hst\ images \citep{Showalter2011}. 

Our general procedure for identifying free and forced normal modes is as follows:
For each ring feature (COR, IER, or OER), we begin with the best-fitting keplerian ellipse model and then scan the residuals over a range of pattern speeds \fpat\ and candidate wavenumbers from $m=1$ to $m=30$ for ILR-type perturbations, and from $m=0$ to $m=-30$ for OLR-type perturbations, solving for the best-fitting amplitude $A_m$ and phase $\delta_m$ at each pattern speed. The range of values scanned for \fpat\ for each mode is centered on the predicted value for the semimajor axis of the ring feature, based on Eq.~\ref{eq:modepatspeed}, and is sufficiently broad to provide a sampling of the statistical significance of a putative detection. (In Section \ref{sec:eccinc}, we make use of similar information to set detection limits on the eccentricities and inclinations of some of the rings.)
We then add the statistically significant detected modes to the kinematical model of the ring feature, fit for their amplitudes, pattern speeds and phases, and
form a new set of residuals. We repeat the frequency scanning process to search for additional weaker modes.
With the addition of successive normal modes, the RMS residual is reduced and the sensitivity to even weaker modes is increased.

We now describe the normal modes detected in this systematic search (none were identified at a statistically significant level for rings 6, 5, 4, $\alpha,\beta,$ or $\lambda$). 

        \subsubsection{The $\eta$ ring}
The very narrow core of the $\eta$ ring has no measurable eccentricity or inclination, but a normal mode scan reveals the presence of a forced $m=3$ mode associated with Cressida, as first reported by \cite{Chancia2017}. 
{\bf Figure~\ref{fig:nm_etam3COR}} shows the $\eta$ ring COR $m=3$ normal mode scan based our our complete set of observations. The pattern speed \fpat\ for the best fit is displaced from the expected value for a free normal mode, and instead matches the mean motion of Cressida, as expected for a first-order Lindblad resonance. (We present quantitative results for this and other mode searches below, once all identified modes have been incorporated into our final orbit models for ring midlines and edges.)
\begin{figure}
\centerline{\resizebox{3.75in}{!}{\includegraphics[angle=0]{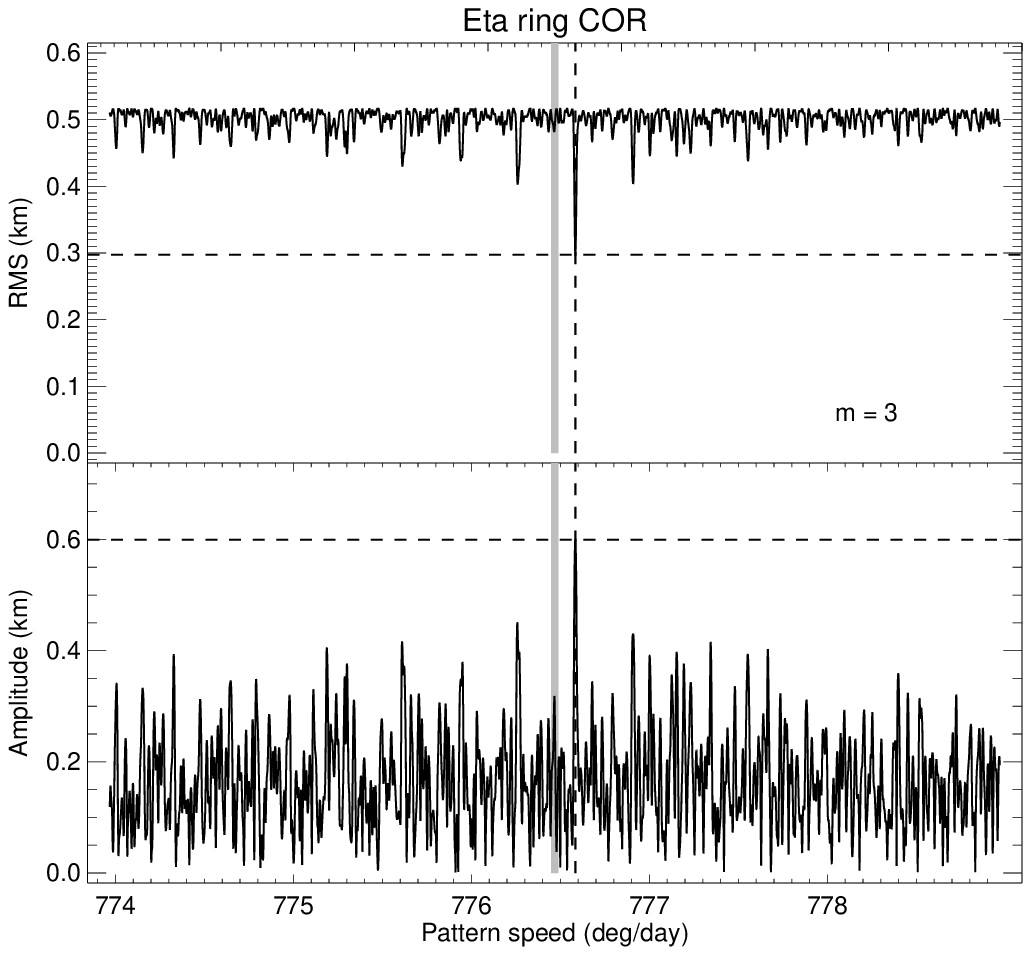}}}
\caption{Normal mode scan of the $\eta$ ring COR for $m=3$. Upper panel: RMS residual as a function of \fpat, centered on the predicted value for an $m=3$ ILR, marked by a thick vertical gray line. The best-fitting pattern speed \fpat\ (dashed vertical line) is faster than the predicted value, and matches the mean motion of Cressida. Lower panel: amplitude $A_{3}$ of the fitted $m=3$ mode for each pattern speed. Our final orbit fit gives $A_3=0.600\pm0.069$ km, very similar to the result from the normal mode scan.}
\label{fig:nm_etam3COR}
\end{figure}
No other $\eta$ ring normal modes were detected at a statistically significant level.
        \subsubsection{The $\gamma$ ring}
Unlike the $\delta$ and $\eta$ rings, the $\gamma$ ring is measurably eccentric, as demonstrated by the $m=1$ COR normal mode scan of the residuals to a circular orbit model in {\bf Fig.~\ref{fig:nm_gamma1COR}}. Here, no other modes are included, and the minimum RMS residual is quite large: 4.0 km. The best-fitting amplitude $A_1=4.94$ km from the scan, and the corresponding pattern speed marked by a vertical dashed line is very slightly faster than the predicted value for the ring's semimajor axis. We will return to this disagreement in Section \ref{sec:J2J4J6}, where we determine Uranus's gravity field from the observed apsidal and nodal precession rates of the rings.

\begin{figure}[!htbp]
\centerline{\resizebox{3.75in}{!}{\includegraphics[angle=0]{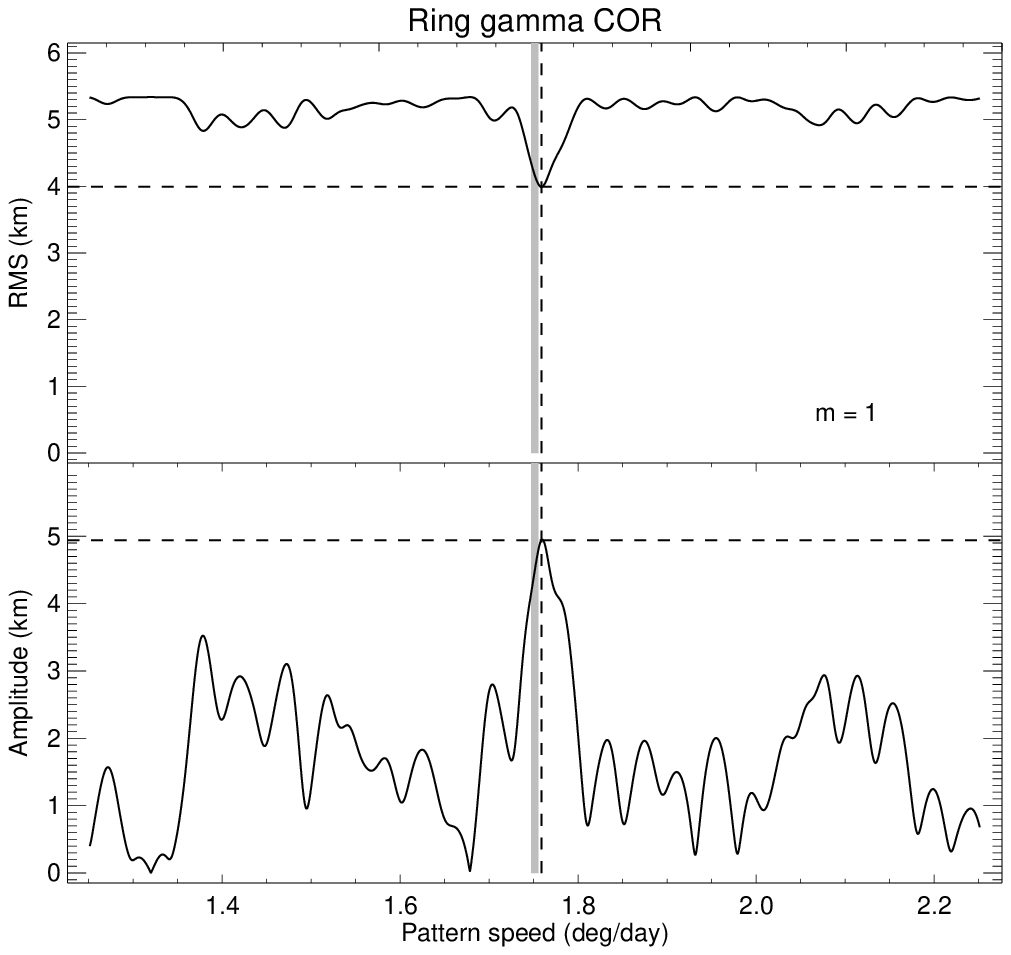}}}
\caption{Normal mode scan of the residuals to a circular model for the $\gamma$ ring COR for $m=1$.
Upper panel: RMS residual as a function of \fpat, centered on the predicted value for an $m=1$ ILR, marked by a thick vertical gray line. Lower panel: amplitude $A_1$ of the fitted $m=1$ mode for each pattern speed. The best-fitting model based on this normal mode scan has $A_1=4.78$ km, compared to the final value $A_1=5.509\pm0.076$ km when all detected normal modes are included.}
\label{fig:nm_gamma1COR}
\end{figure}

The $m=0$ normal mode amplitude is roughly equal to the $m=1$ mode amplitude, as seen in the $\gamma$ ring COR $m=0$ normal mode scan ({\bf Fig.~\ref{fig:nm_gamma0COR}}). Here, the best-fitting $m=1$ mode from the final orbit fit has been included prior to the scan. (The upper limit to the RMS residual in the upper panel in this case is 4.2 km, a bit larger than in Fig.~\ref{fig:nm_gamma1COR} owing to the slightly different best-fitting keplerian ellipses in the two cases.)

\begin{figure}[!htbp]
\centerline{\resizebox{3.75in}{!}{\includegraphics[angle=0]{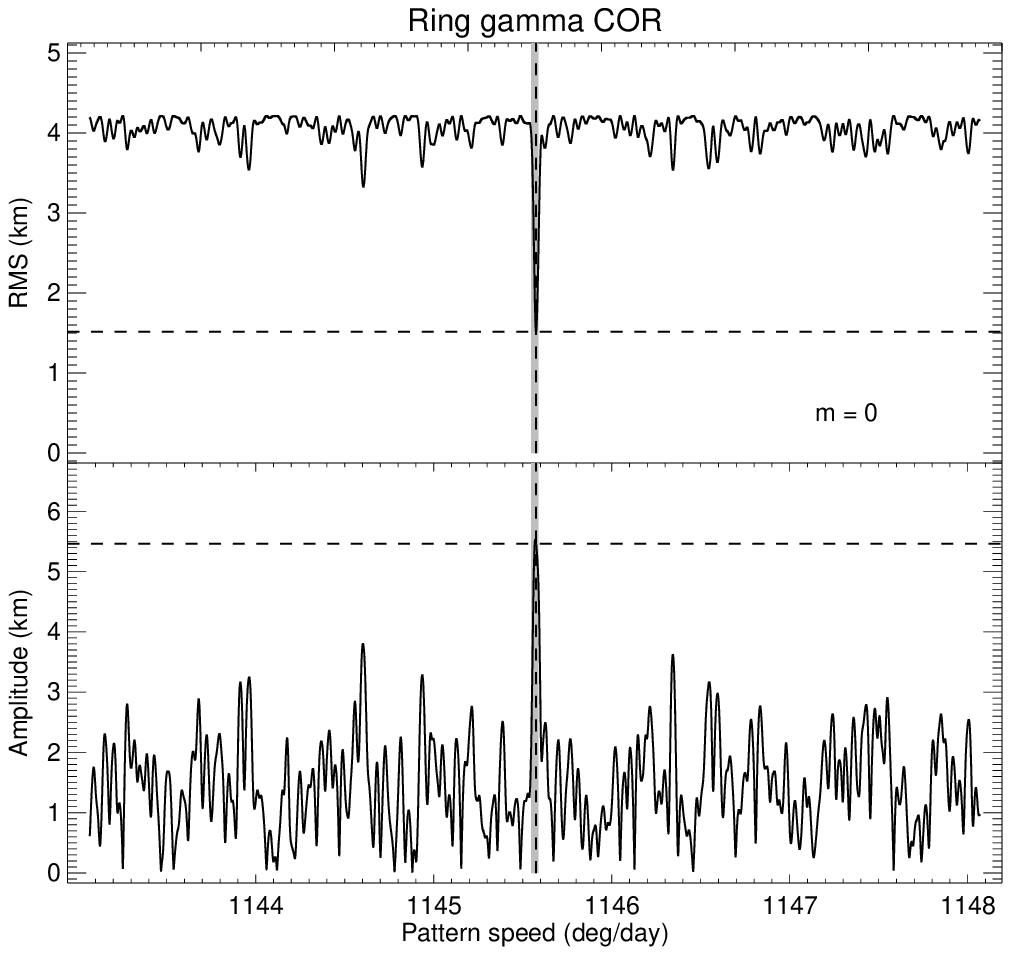}}}
\caption{Normal mode scan of the $\gamma$ ring COR for $m=0$, after inclusion of best-fitting $m=1$ mode from adopted orbit solution.
Upper panel: RMS residual as a function of \fpat, centered on the predicted value for an $m=0$ OLR, marked by a thick vertical gray line. Lower panel: amplitude $A_0$ of the fitted $m=0$ mode for each pattern speed. The best-fitting model has $A_0=5.509\pm0.076$ km from our final orbit solution. }
\label{fig:nm_gamma0COR}
\end{figure}

We extended our search for $\gamma$ ring normal modes over the range $m=-30$ to 30, and we successively added each statistically significant mode to our RINGFIT model and repeated the search for additional modes until none were found.
To illustrate the signatures of the additional detected $\gamma$ ring normal modes, the best fitting final model for all other detected modes was subtracted from the measured radii prior to performing the normal mode scans shown in each figure below. We present these in decreasing order of fitted mode amplitude $A_m$. Our search yielded the detection of the $\gamma$ ring $m=-1$ OLR. Note that Eq.~(\ref{eq:modepatspeed}) implies that $\Omega_P\sim2n\sim 2290$ \degd, much greater than the $m=1$ apse rate $\dot\varpi\simeq1.751$ \degd. The corresponding normal mode scan is shown in 
{\bf Fig.~\ref{fig:nm_gamma-1COR}}. The best-fitting pattern speed is slightly faster than that predicted for a free $m=-1$ mode at the COR, and the corresponding resonance radius is therefore somewhat interior to the midline of the ring. This is consistent with the anticipated association of OLRs with inner ring edges. The fitted amplitude $A_{-1}=1.822\pm0.067$ km from our final orbit solution. In this and subsequent cases, the normal mode scans give slightly different values for the fitted amplitudes compared to our final orbit fit because the normal mode scans allow for an additional fitted parameter of a radial offset from the ring midline or edge, while in the final orbit fit all modes are constrained to be centered on the corresponding midline or edge.

\begin{figure}[!htbp]
\centerline{\resizebox{3.75in}{!}{\includegraphics[angle=0]{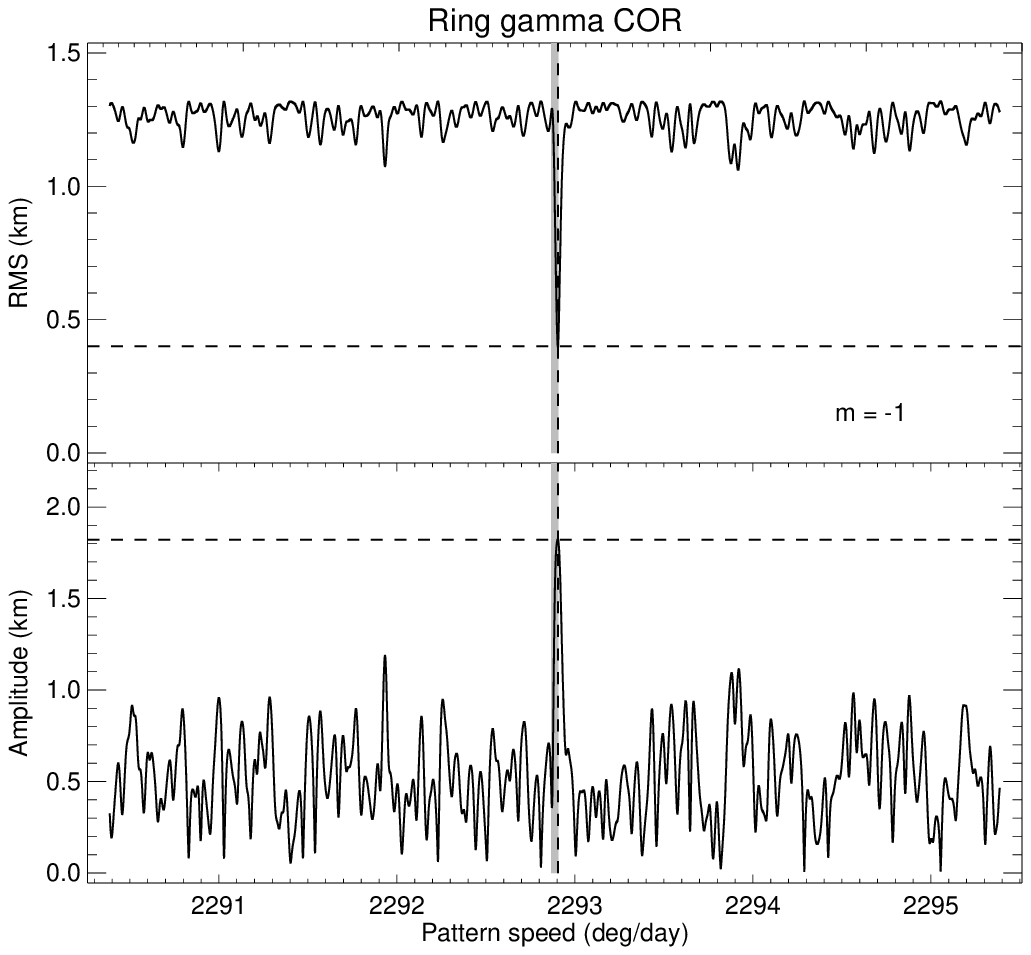}}}
\caption{Normal mode scan of the $\gamma$ ring COR for $m=-1$, after inclusion of modes $m=-2,0,1,3,$ and 6 from adopted orbit solution.
Upper panel: RMS residual as a function of \fpat, centered on the predicted value for an $m=-1$ OLR, marked by a thick vertical gray line. Lower panel: amplitude $A_{-1}$ of the fitted $m=-1$ mode for each pattern speed. The best-fitting model has $A_{-1}=1.822\pm0.067$ km from our final orbit solution.}
\label{fig:nm_gamma-1COR}
\end{figure}

We identified a second OLR in the $\gamma$ ring's midline: the $m=-2$ normal mode shown in 
{\bf Fig.~\ref{fig:nm_gamma-2COR}}. Once again, the best-fitting \fpat\ is slightly faster than the predicted value for the COR, indicating that the resonance radius is interior to the COR as expected for an OLR. This mode is somewhat weaker than the $m=-1$ mode, with a fitted amplitude $A_{-2}=0.690\pm0.076$ km from our final orbit solution.

\begin{figure}[!htbp]
\centerline{\resizebox{3.75in}{!}{\includegraphics[angle=0]{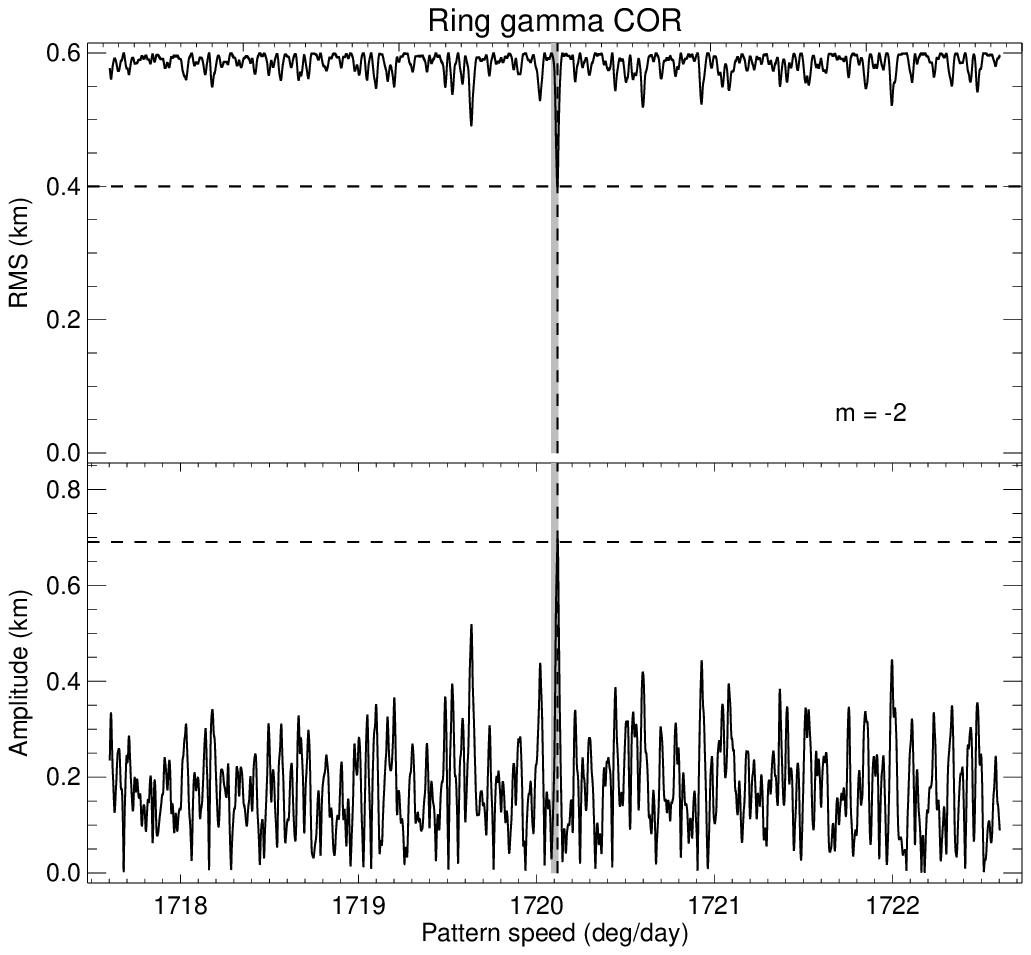}}}
\caption{Normal mode scan of the $\gamma$ ring COR for $m=-2$, after inclusion of modes $m=-1,0,1,3,$ and 6 from adopted orbit solution. Upper panel: RMS residual as a function of \fpat, centered on the predicted value for an $m=-2$ OLR, marked by a thick vertical gray line. Lower panel: amplitude $A_{-2}$ of the fitted $m=-2$ mode for each pattern speed. The best-fitting model has $A_{-2}=0.690\pm0.076$ km from our final orbit solution.}
\label{fig:nm_gamma-2COR}
\end{figure}

We also detected the $m=6$ ILR forced by Ophelia. {\bf Figure~\ref{fig:nm_gamma6COR}} shows the $\gamma$ ring COR $m=6$ normal mode scan, with a clear offset in the best-fitting \fpat\ compared to the predicted value for the ring's semimajor axis, indicating that the resonance location is interior to the centerline of the ring. The best-fitting model has $A_{6}=0.637\pm0.063$ km from our final orbit solution. Similar scans (not shown) confirm that the $m=6$ mode is also present on both ring edges. 

\begin{figure}[!htbp]
\centerline{\resizebox{3.75in}{!}{\includegraphics[angle=0]{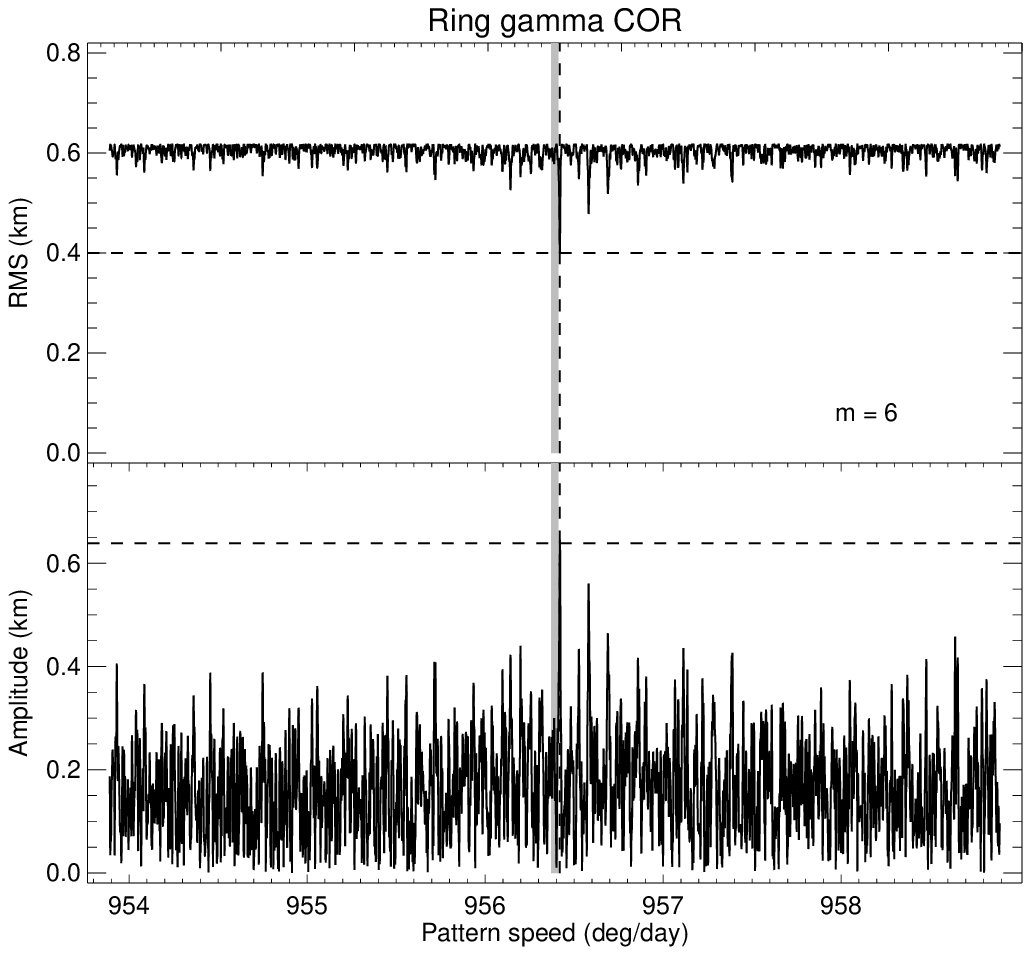}}}
\caption{Normal mode scan of the $\gamma$ ring COR for $m=6$, forced by Ophelia, after inclusion of modes $m=-2,-1,0,1,$ and 3 from adopted orbit solution. Upper panel: RMS residual as a function of \fpat, centered on the predicted value for an $m=6$ ILR, marked by a thick vertical gray line. Lower panel: amplitude $A_{6}$ of the fitted $m=6$ mode for each pattern speed. The best-fitting model has $A_{6}=0.637\pm0.063$ km from our final orbit solution.}
\label{fig:nm_gamma6COR}
\end{figure}

Finally, there is evidence for a free $m=3$ ILR with a resonance location near the outer edge of the $\gamma$ ring. The normal mode scan is shown in {\bf Fig.~\ref{fig:nm_gamma3COR}}. The best-fitting pattern speed corresponds to a resonance location exterior to the outer edge of the ringlet. There is no known satellite with the appropriate orbit to force this mode, and to assess the statistical significance of the detection, we performed a scan over a very wide range of pattern speeds. The two left panels of the figure show that the best-fitting normal mode lies very close to the predicted pattern speed and that there are no comparably strong aliases over the large range of pattern speeds considered. The upper right panel shows a histogram of the RMS residuals for all of the individual pattern speeds in the normal mode scans at left, plotted as the number of results N per bin of width 0.0005 km, with an arrow marking the value for the candidate $m=3$ mode near the ring center. The lower right panel shows a similar histogram of the best-fitting amplitude for each pattern speed, showing the number of results N per bin of width 0.0018 km. Again, the arrow marks the candidate mode, which is well above the noise threshold.

\begin{figure}[!htbp]
\centerline{\resizebox{6in}{!}{\includegraphics[angle=90]{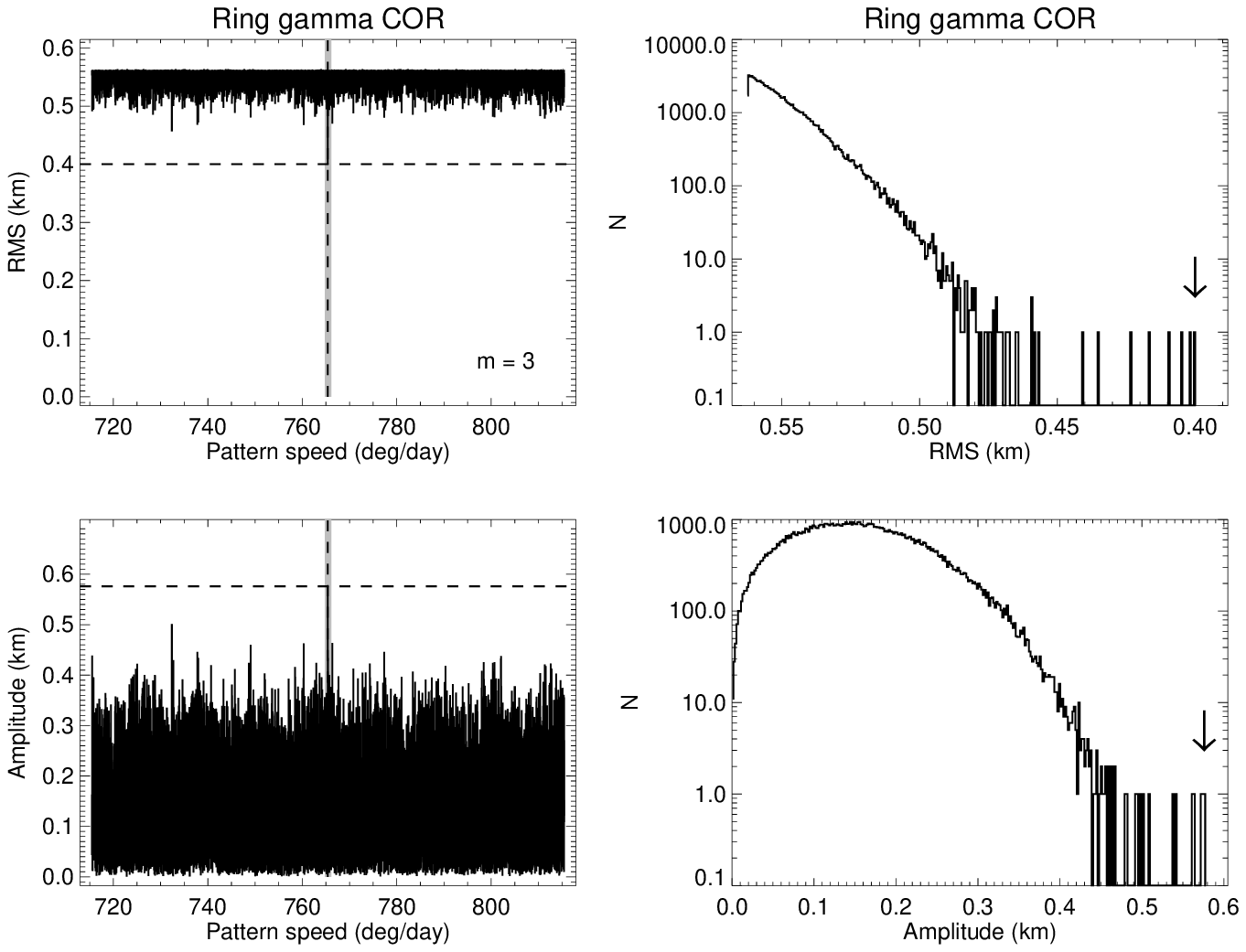}}}
\caption{Normal mode scan of the $\gamma$ ring COR for $m=3$, performed over a wide range of pattern speeds to illustrate the statistical significance of the detection near the expected pattern speed. See text for details.}
\label{fig:nm_gamma3COR}
\end{figure}

The amplitude of the $m=3$ mode is even stronger on the outer edge of the $\gamma$ ring, as shown from the normal mode scan shown in {\bf Fig.~\ref{fig:nm_gamma3OER}}, where we again have scanned over a wide range of pattern speeds. In this case (as for other ring edges), the RMS residuals are larger than for the COR normal mode scans because the fitted ring widths from the square-well model are less accurate than the ring midlines, as discussed above.
\begin{figure}[!htbp]
\centerline{\resizebox{6in}{!}{\includegraphics[angle=90]{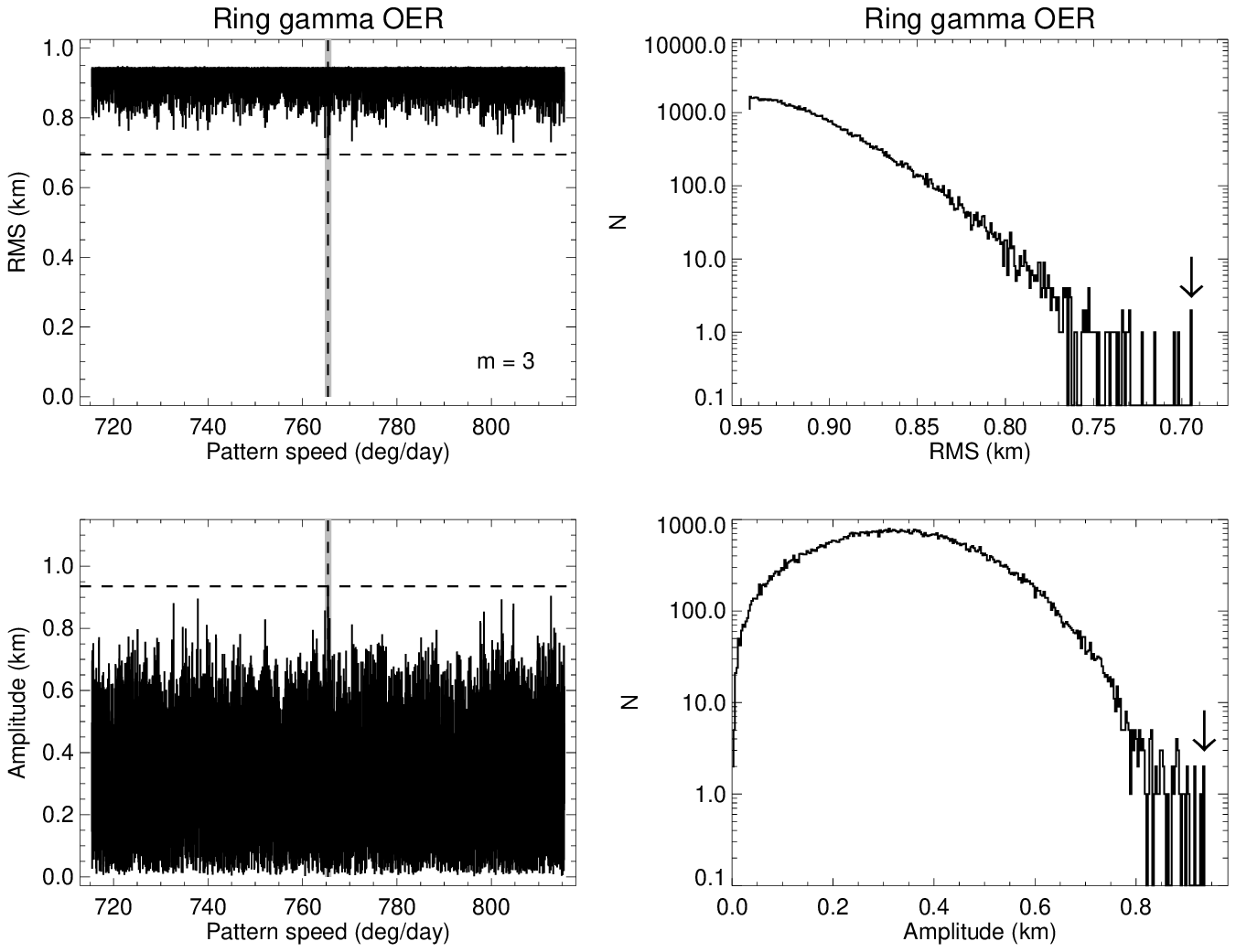}}}
\caption{Normal mode scan of the $\gamma$ ring OER for $m=3$, performed over a wide range of pattern speeds to illustrate the statistical significance of the detection near the expected pattern speed for the rings's outer edge.}
\label{fig:nm_gamma3OER}
\end{figure}
On the other hand, we found no evidence for an $m=3$ mode near the $\gamma$ ring's inner edge -- the corresponding normal mode scan is shown in {\bf Fig.~\ref{fig:nm_gamma3IER}}. As we discuss below, including the $m=3$ mode not only reduces the RMS residuals of the COR fit but also reduces the inferred anomalous precession rate of the $\gamma$ ring.

\begin{figure}[!htbp]
\centerline{\resizebox{3.75in}{!}{\includegraphics[angle=0]{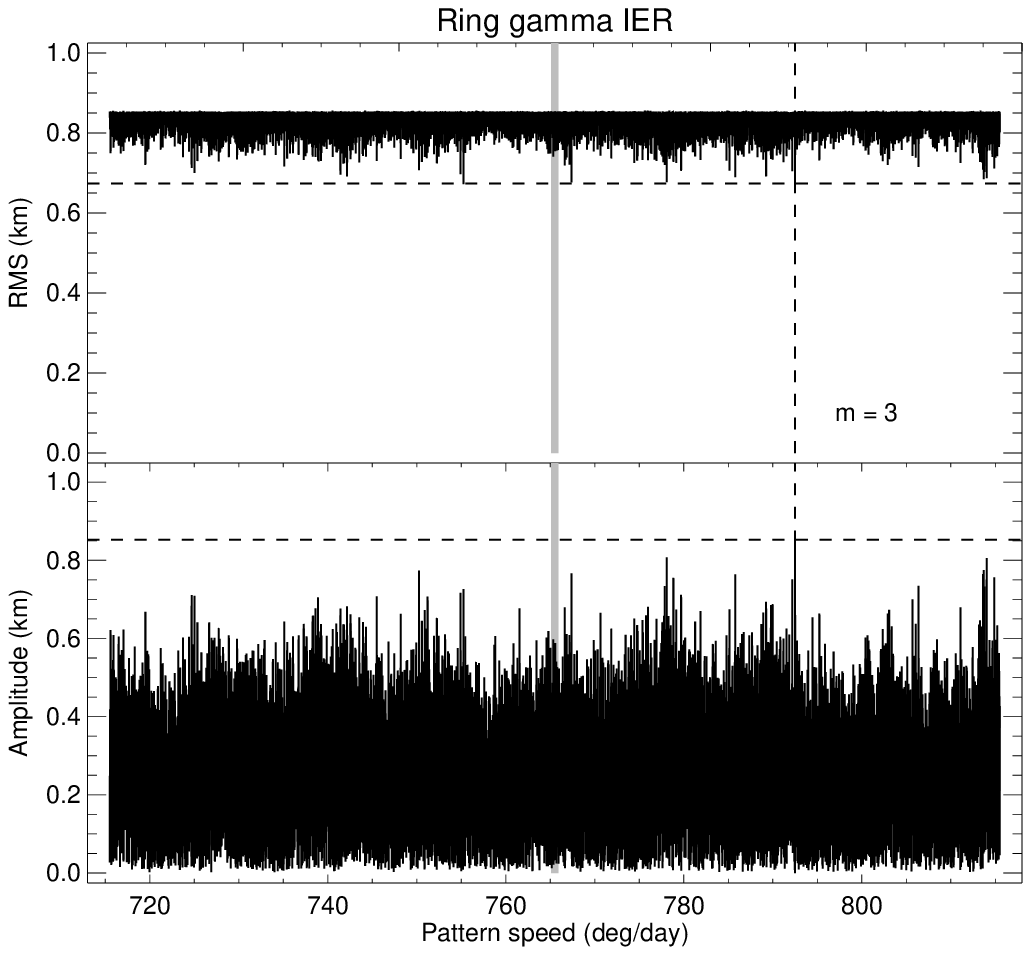}}}
\caption{Normal mode scan of the $\gamma$ ring IER for $m=3$, showing no evidence for the mode in the vicinity of the expected pattern speed for this ring edge, shown as a thick vertical gray line.}
\label{fig:nm_gamma3IER}
\end{figure}

        \subsubsection{The $\delta$ ring}
The $\delta$ ring has no detectable eccentricity or inclination, but the RMS residual of a circular orbit model is 2.35 km, nearly ten times that of the eccentric and inclined mode-free rings 6, 5, and 4, indicative of an unmodeled strong perturbation in the ring's shape. In previous orbit models, this was attributed to an $m=2$ mode, and this signature is clearly present in the 
normal mode scan shown in {\bf Fig.~\ref{fig:nm_delta2COR}}. The upper panel shows a sharp decrease in the RMS error at the predicted pattern speed for the semimajor axis of the ring, marked by a solid vertical line. The best-fitting \fpat\ is marked by a solid dashed line, which in this case coincides with the predicted value. The lower panel shows the fitted amplitude of the normal mode at each assumed pattern speed in the scan, with a best-fitting value $A_2\simeq3.2$ km.

\begin{figure}
\centerline{\resizebox{3.75in}{!}{\includegraphics[angle=0]{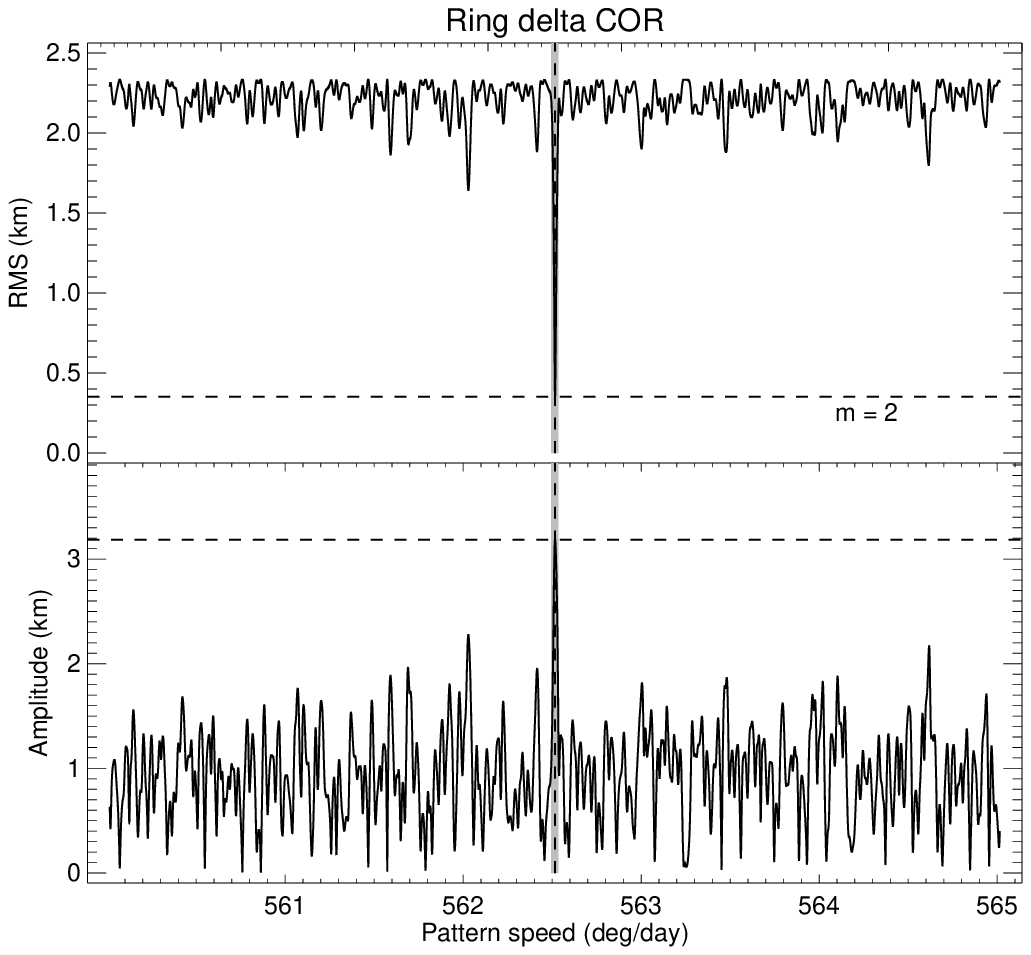}}}
\caption{Normal mode scan of the $\delta$ ring COR for $m=2$. Upper panel: RMS residual as a function of \fpat, centered on the predicted value for an $m=2$ ILR, marked by a thick vertical gray line. Lower panel: amplitude $A_2$ of the fitted $m=2$ mode for each pattern speed. The best-fitting amplitude from the normal mode scan is $A_2=3.2$ km; from our final orbit fit, $A_2=3.169\pm0.059$ km.}
\label{fig:nm_delta2COR}
\end{figure}

After including the free parameters for the $m=2$ mode of the $\delta$ ring, we formed a new set of residuals and repeated the normal mode scan for the same set of possible wavenumbers. The results of this search revealed a weak but statistically significant $m=23$ mode in the COR data, shown in {\bf Fig.~\ref{fig:nm_delta23COR}}. The RMS residual is further reduced and the fitted amplitude $A_{23}\simeq0.34$ km in the normal mode scan (similar to $A_{23}=0.339\pm0.063$ km from our final orbit fit). 
The pattern speed is slower than that expected for a free $m=23$ mode located at the semimajor axis of the COR, and the corresponding resonance radius is somewhat exterior to the ring midline. Instead, the pattern speed and phase of the observed $m=23$ mode match the orbital characteristics of Cordelia, as originally proposed by \cite{Chancia2017}, who calculated the expected normal mode amplitude $A_{23}$ for a range of assumed satellite masses and densities. Normal mode scans of $\delta$ ring IER and OER showed no detection of the $m=23$ mode on either ring edge near the expected pattern speed, but the RMS noise level in these scans was substantially greater than the observed amplitude at the ring centerline, so the non-detections are not surprising. A search for the $m=23$ mode in the width of the ring itself was also negative.
No other $\delta$ ring normal modes were detected at a statistically significant level.

\begin{figure}
\centerline{\resizebox{3.75in}{!}{\includegraphics[angle=0]{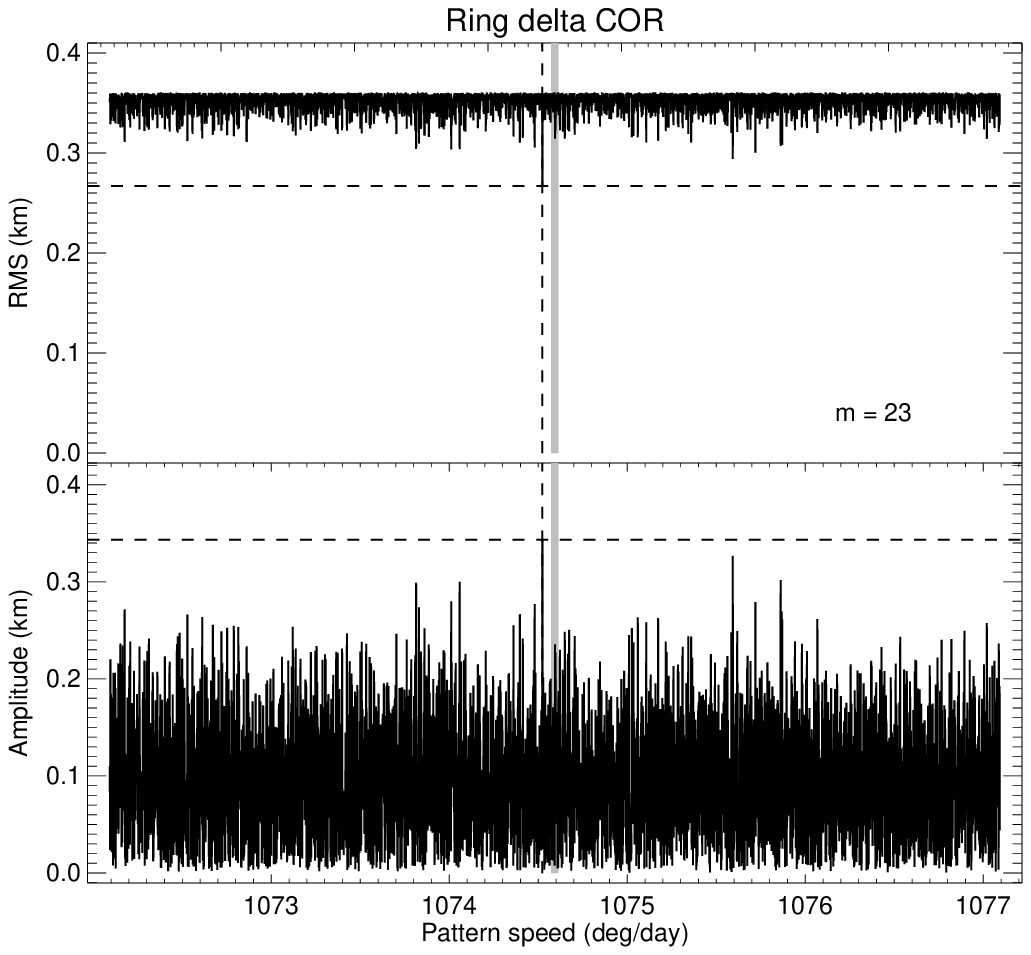}}}
\caption{Normal mode scan of the $\delta$ ring COR for $m=23$. Upper panel: RMS residual as a function of \fpat, centered on the predicted value for an $m=23$ ILR, marked by a thick vertical gray line. The best-fitting pattern speed \fpat\ (dashed vertical line) is slower than the predicted value but matches the mean motion of Cordelia $\Omega_P=1074.52\dd$. Lower panel: amplitude $A_{23}$ of the fitted $m=23$ mode for each pattern speed. The best-fitting normal mode scan value is $A_{23}=0.34$ km, close to the value from the final orbit fit of $A_{23}\simeq0.339\pm0.063$.}
\label{fig:nm_delta23COR}
\end{figure}

        \subsubsection{The $\epsilon$ ring}
        \label{sec:epsmodes}
        The inner and outer edges of the $\epsilon$ ring have long been associated with the nearby $m=-24$ OLR with Cordelia and the $m=14$ ILR with Ophelia, respectively \citep{Porco1987}. \cite{Chancia2017} showed that, for reasonable assumptions about the masses of the two satellites, the expected amplitudes of the corresponding edge waves would lie in the range $A_{-24}=0.4-1.0$ km for the inner edge (IER) and a somewhat weaker $A_{14}=0.25-0.7$ km at the OER. As noted above, \cite{French1995} found tentative evidence for these edge waves.     
{\bf Figure~\ref{fig:nm_eps14OER}} shows the results of the $m=14$ ILR normal mode scan of the $\epsilon$ ring OER. There is a convincing minimum in the RMS residuals at a pattern speed a bit faster than the value predicted for the ring edge, indicating that the resonance lies slightly interior to the outer edge. 
\begin{figure}
\centerline{\resizebox{3.75in}{!}{\includegraphics[angle=0]{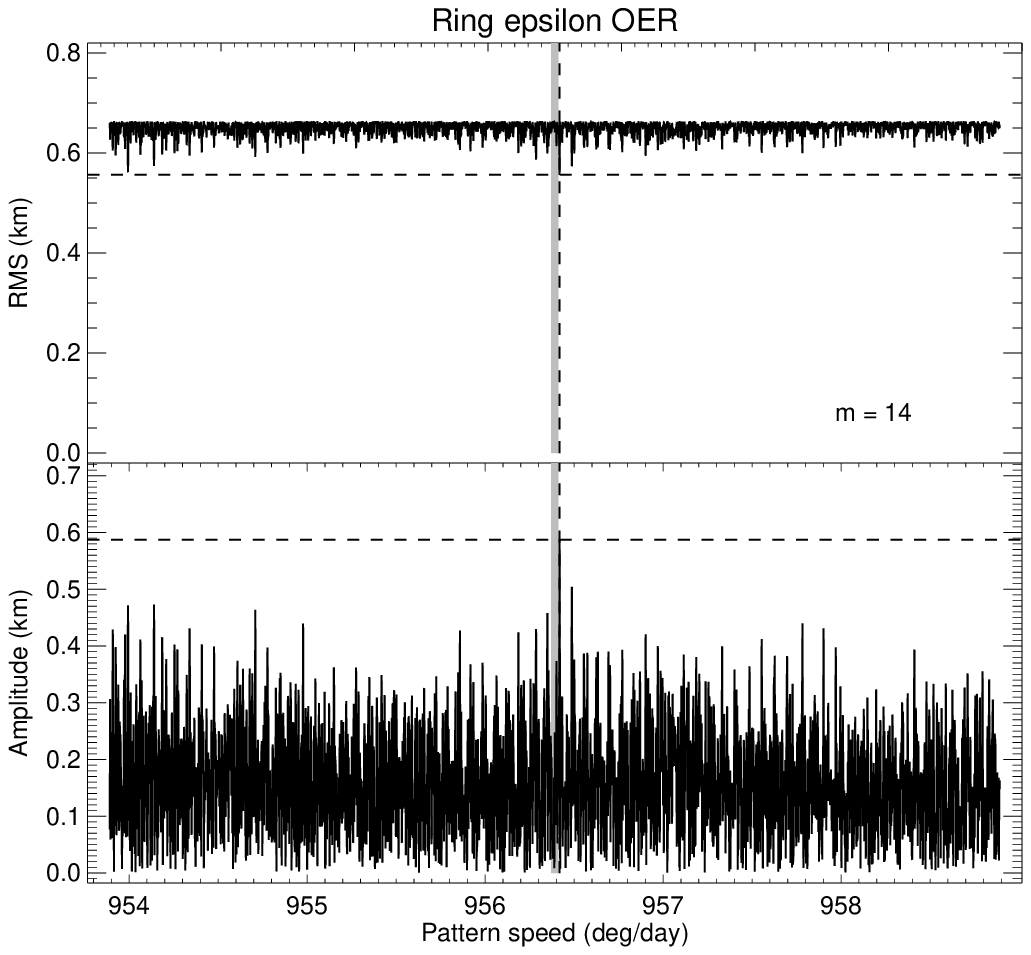}}}
\caption{Normal mode scan of $\epsilon$ ring OER for $m=14$, showing a minimum in the RMS residuals with amplitude $A_{14}\simeq0.60$~ km at a pattern speed near that of Ophelia's mean motion $\Omega_P\simeq956.42$ \degd and slightly faster than that expected at the ring's outer edge (thick gray line).}
\label{fig:nm_eps14OER}
\end{figure}
     {\bf Figure~\ref{fig:nm_eps-24IER}} shows the results of the $m=-24$ OLR normal mode scan of the $\epsilon$ ring IER. In this case, the best-fitting pattern speed is near Cordelia's mean motion $\Omega_P\simeq1074.52$ \degd and is slightly slower than that expected at the exact ring edge, indicating that the resonance is located within the ring itself, near the inner edge.

\begin{figure}
\centerline{\resizebox{3.75in}{!}{\includegraphics[angle=0]{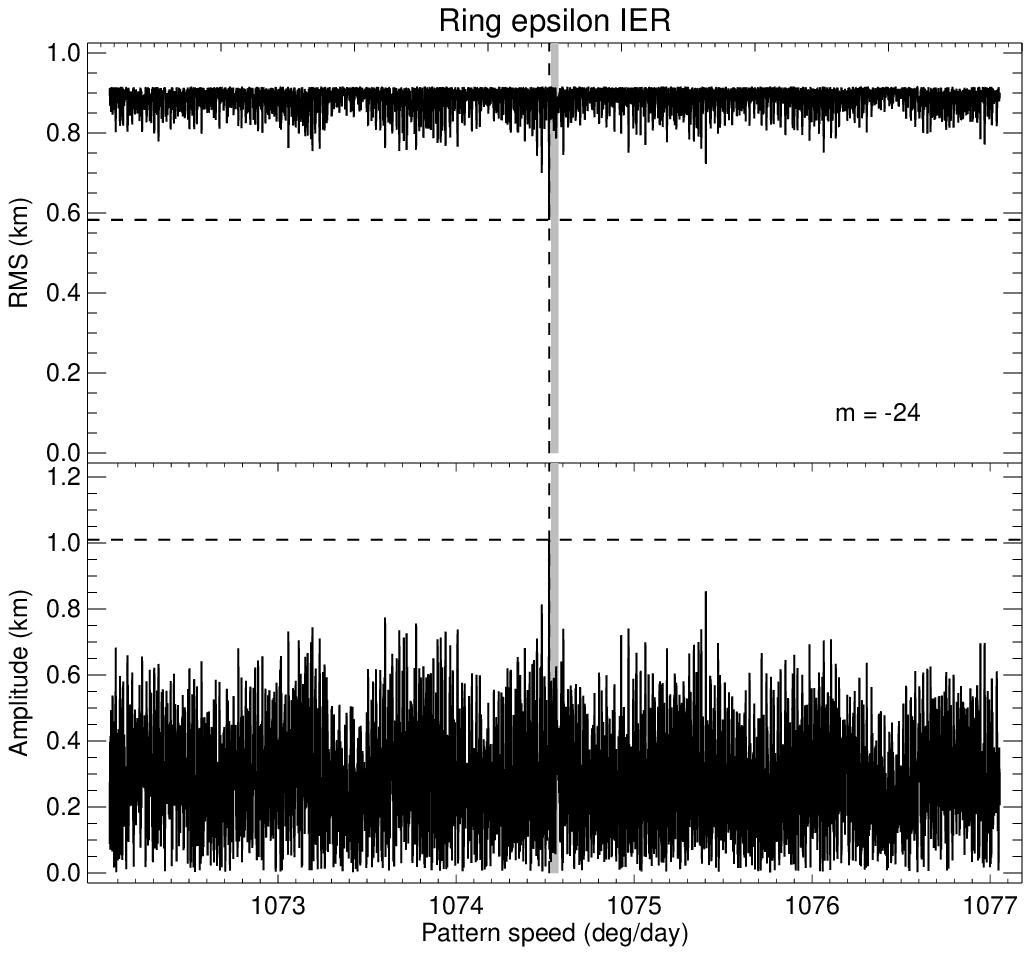}}}
\caption{Normal mode scan of $\epsilon$ ring IER for $m=-24$, showing a minimum in the RMS residuals with amplitude $A_{-24}\simeq1.0$~ km at a pattern speed near that of Cordelia's mean motion $\Omega_P\simeq1074.52$ \degd, and slightly slower than that expected at the ring's inner edge (thick gray line).}
\label{fig:nm_eps-24IER}
\end{figure}

In Section \ref{sec:satellites}, we confirm the association of these two normal modes with Cordelia and Ophelia on the basis of the fitted phases, pattern speeds, and amplitudes of the $\epsilon$ ring IER and OER edge waves compared to the expectations from satellite ephemerides, using the full set of observations reported here.

\subsubsection{Signatures of the $\epsilon$ ring edge waves in the shape of the ring midline}
        \label{sec:epsmodesCOR}
        While these normal modes are forced at the {\it edges} of the relatively wide $\epsilon$ ring and their dynamical influence might well be restricted to the immediate vicinity of the edges, the measured centerline of the ring should be offset by approximately one half of the sum of the local radial amplitudes of the two edge modes.  
We performed normal mode scans on the $\epsilon$ ring COR radius residuals for $m=14$ and $m=-24$ and found weak but secure signatures of both modes, with the measured phases and pattern speeds being in excellent agreement with the values found at the corresponding ring edges, as discussed below in Section \ref{sec:satellites}. 
         {\bf Figure~\ref{fig:nm_eps14COR}} shows the results for $m=14$ COR scan. Here, the vertical dashed line shows the best-fitting pattern speed near the value from our final orbit fit \fpat$=956.418051\pm 0.000269$~\degd, close to the mean motion of Ophelia, with an amplitude $A_{14}=0.383\pm0.071$~km, a bit more than half that at the OER, for which $A_{14}=0.590\pm0.130$~km. 
\begin{figure}
\centerline{\resizebox{4in}{!}{\includegraphics[angle=0]{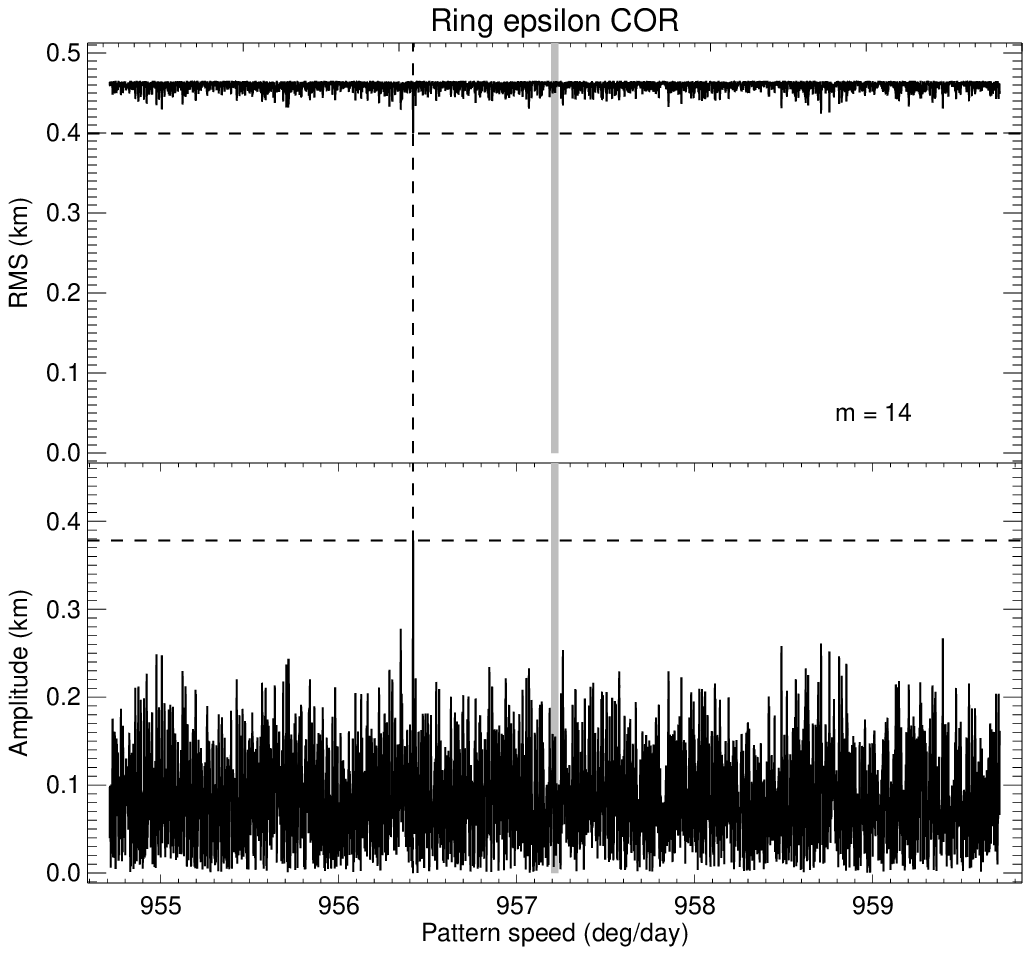}}}
\caption{Normal mode scan of the $\epsilon$ ring COR for the $m=14$ mode forced by Ophelia. The best-fitting pattern speed \fpat$=956.418051\pm0.000269$ \degd, close to the mean motion of Ophelia.}
\label{fig:nm_eps14COR}
\end{figure}
{\bf Figure~\ref{fig:nm_eps-24COR}} shows the results for the $m=-24$ COR scan. Here, the vertical dashed line is near the best-fitting \fpat$=1074.522703\pm 0.000142$ \degd, very close to the mean motion of Cordelia (see Table \ref{tbl:satmeanmo}), with an amplitude $A_{-24}=0.443\pm0.061$~km from our final orbit fit, a bit less than half that at the IER, for which $A_{-24}=1.011\pm0.107$~km. 
\begin{figure}
\centerline{\resizebox{4in}{!}{\includegraphics[angle=0]{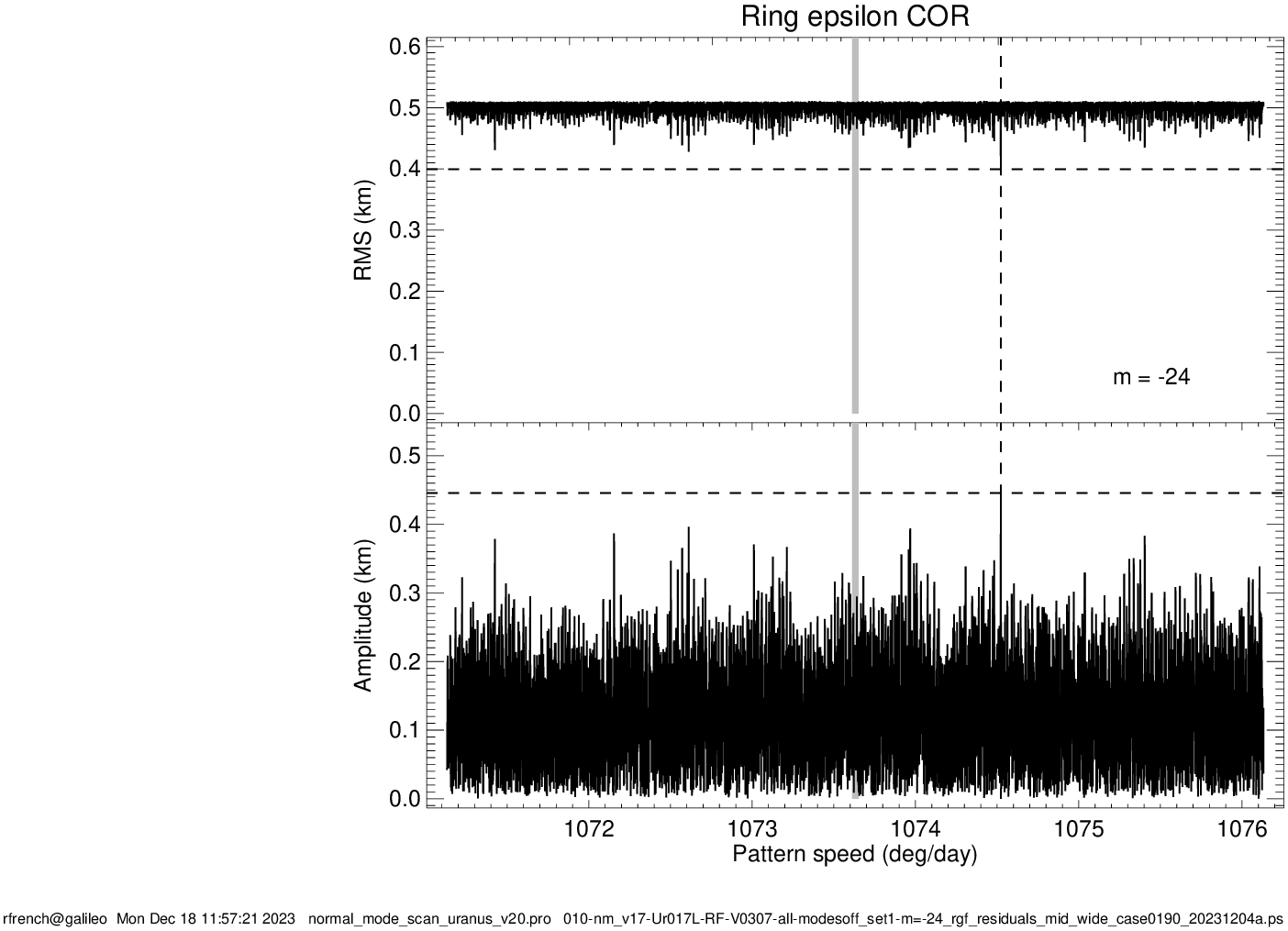}}}
\caption{Normal mode scan of the $\epsilon$ ring COR for the $m=-24$ mode forced by Cordelia. The best-fitting \fpat=$1074.522703\pm 0.000142$ \degd, close to the mean motion of Cordelia.}
\label{fig:nm_eps-24COR}
\end{figure}

        \subsubsection{Searches for other normal modes}
We performed an orbit fit that included all of the normal modes identified above and used normal mode scans for wavenumbers between $m=-30$ and 30 to search for evidence of any additional statistically significant detections in the orbit fit residuals near the predicted pattern speeds, for all ring widths, edges and midlines. None were found. \cite{Chancia2017} identified two weak first-order resonances with predicted amplitudes $A_m\sim0.1$ km for assumed moonlet densities $\rho= 0.9$~g~cm$^{-3}$: the 13:12 Cordelia resonance with the $\eta$ ring and the 2:1 Portia resonance with ring 6. Our 3-$\sigma$ detection limits for these modes are 0.204 km and 0.168 km, respectively, based on Rayleigh distribution fits to the amplitudes in the corresponding normal mode scans, placing them just out of reach for secure identification in our observations.

Internal planetary oscillations provide another possible source for normal modes in the rings. Saturn's C and B rings provide a tapestry revealing dozens of such signatures in occultation profiles \citep{Rosen91, Colwell2009, Baillie11, HN13, HN14, KronoIII, KronoIV, French2021, Hedman2022}. Although the Uranus system of narrow rings provides incomplete radial coverage for possible modes, Saturn's Maxwell ringlet -- very similar in structure to the Uranus $\epsilon$ ring -- is home to an $m=2$ mode forced by Saturn's internal oscillations, and provides an example of a narrow ringlet located precisely at the mode's resonance radius \citep{French16}. As part of our normal mode survey, we searched for evidence of distortions in the shapes of the ring midlines and edges 
near the predicted values for wavenumbers and pattern speeds excited by resonances with fundamental Uranus normal modes \citep{AHearn2022}. We found no statistically significant matches with radial amplitudes $\Delta A_m>0.2$ km, but the possibility remains that internal oscillations can reveal their presence in azimuthal variations in ring brightness seen in images, an important task for the future.

        \subsection{Adopted ring orbital elements and normal modes}
        \label{sec:orbfitCORIEROER}
        
        Having characterized a set of detected normal modes, we then solved for the ring orbital elements and Uranus system geometry iteratively and in several stages, taking into account the greater accuracy of the ring midline measurements than those of the ring edges. We first used the COR (ring midline) observations to fit for the planet's pole direction, time offsets for selected stations, and corrections to selected occultation star positions and proper motions, in addition to the orbital elements and normal modes of the nine principal rings. Next, we fitted separately for the IER/OER orbital elements and normal modes only, using the fitted widths and mid-times and assuming the system geometry, star positions, and station offset times established by the COR fit. Finally, we solved for the best circular orbit model for the $\lambda$ ring and the skyplane offset positions of the secondary stars in the multiple-star U36 occultation and the binary star U102 occultation.
         
The combined results of the separate unweighted COR and IER/OER orbit fits for the ten narrow rings are given in {\bf Tables \ref{tbl:orbel}} and {\bf \ref{tbl:normalmodes}}. For these fits, the apsidal precession and nodal regression rates $\dpi$ and $\dom$ were included as separate free parameters for the rings with measurable eccentricity and inclination, respectively, rather than being constrained by the gravitational field of Uranus and the secular precession due to the satellites. We will return to these contributions below, when we evaluate the evidence for anomalous forced precession of individual rings and set constraints on $J_2, J_4,$ and $J_6$ from the COR observations. Similarly, all normal mode pattern speeds and longitudes at epoch were included as separate free parameters, even for modes identified as being forced by Cressida, Cordelia, and Ophelia. Table \ref{tbl:orbel} lists the fitted keplerian orbital elements to $N$ ring event times for the IER/COR/OER of each ring, in increasing order of semimajor axis. (For the $\gamma$ ring COR, we include the results of the nominal fit that includes the $m=3$ mode as well as the alternate fit in which the $m=3$ mode is omitted.) All errors listed are 1-$\sigma$ formal errors from the separate unweighted least-squares fits for the COR observations and for the IER/OER data (which include the fitted ring widths from the square-well models discussed in Appendix A). Eccentricities $ae$ and inclinations $a\sin i$ in square brackets were held fixed during orbit determination. The epoch of the fit $t_0$ is TDB 1986 Jan 19 12:00, near the time of the \Voyager\ 2 encounter with Uranus, chosen to be near the mid-point of the time interval of the entire set of observations so as to minimize the correlations between the fitted apse and node rates and the longitudes at epoch.\footnote{This is the same epoch used by \cite{Jacobson2014,Jacobson2023}.}

The entries $\Delta \dot\varpi$ and $\Delta \dot\Omega$ are the differences between the observed apse and node rates and the predicted values computed for the fitted geometric semimajor axis $a$ of the given feature. The predicted rates take into account the estimated secular apse or node precession due to both major and minor satellites.\footnote{The predicted rates tabulated here are based solely on the planet's gravitational field and the satellite-induced precession. As discussed in Section \ref{sec:COO}, in solving for the gravitational field, we take into account the estimated difference between the semimajor axes of the geometric centers of the rings and their estimated radially averaged centers of mass. For a normal mode forced by a satellite, the predicted mode pattern speed corresponds to the satellite's mean motion. We discuss these forced modes in Section \ref{sec:satellites}.} The entries $\Delta a_{\dot\varpi}$ and $\Delta a_{\dot\Omega}$ are the corresponding differences for the fitted apse and node rates $\dot \varpi$ and $\dot \Omega$. In practice, we compute these iteratively to high precision, but they can be estimated to reasonable accuracy for $\Delta a_{\dot\varpi}$ and $\Delta a_{\dot\Omega}$ of order a few km from the linear approximations:
\beq
\Delta a_{\dot\varpi}\simeq \frac{\dot\varpi({\rm observed})-\dot\varpi({\rm predicted})}{d\dot\varpi/da} = \frac{\Delta\dot\varpi}{d\dot\varpi/da}
\label{eq:Delta_adotvarpi}
\eeq
and
\beq
\Delta a_{\dot\Omega}\simeq\frac{\dot\Omega({\rm observed})-\dot\Omega({\rm predicted})}{d\dot\Omega/da}= \frac{\Delta\dot\Omega}{d\dot\Omega/da}.
\label{eq:Delta_adotOmega}
\eeq
 Since $d\dot\varpi/da\propto -J_2 a^{-9/2}<0$ and $d\dot\Omega/da\propto +J_2 a^{-9/2}>0$, $\Delta\dot\varpi$ and $\Delta a_{\dot\varpi}$ have opposite signs, while $\Delta\dot\Omega$ and $\Delta a_{\dot\Omega}$ have the same signs.

Table \ref{tbl:normalmodes} lists the fitted amplitudes $A_m$, phases $\delta_m$, and pattern speeds $\Omega_P$ for all detected normal modes. For the $\gamma$ ring, we include the results of two separate fits, one that included the suspected $m=3$ mode and one in which that mode is omitted. The quantity $\Delta \Omega_P$ is the difference between the fitted pattern speed and the predicted value computed for the fitted geometric semimajor axis $a$ (Table \ref{tbl:orbel}) and our adopted gravity field of the planet and satellites; $\Delta a_\Omega$ is the corresponding difference between the geometric semimajor axis and the predicted resonance radius for the fitted pattern speed, with the linear approximation:

\beq
\Delta a_{\Omega_P}\simeq \frac{\Omega_P({\rm observed})-\Omega_P({\rm predicted})}{d\Omega_P/da}= \frac{\Delta\Omega_P}{d\Omega_P/da}.
\label{eq:Delta_aOmegap}
\eeq
From Eq.~(\ref{eq:modepatspeed}), $d\Omega_P/da\propto -(m-1)a^{-5/2}$, so the sign of $\Delta a_{\Omega_P}$ depends on both the sign of $(m-1)$ and the sign of $\Delta\Omega_P$.

We will make use of these relations in Section \ref{sec:widths}, when we examine the width, shape, and differential precession of the rings.

\begin{longrotatetable}
\begin{deluxetable}{c l c c r r r r r}
\tablecolumns{9}
\tablecaption{Uranian ring keplerian orbital elements\label{tbl:orbel}}
\tablewidth{0pt}
\tablehead{
\colhead{Ring} & 
\colhead{Feature$^{(a)}$} & 
\multicolumn{2}{c}{$a$ (km)} &
\colhead{$ae$ (km)} & 
\colhead{$\varpi_0\ (^\circ)^{(b)}$} & 
\colhead{$\dot\varpi\ (\dd$) }&
\colhead{$\Delta\dot\varpi\ (\dd)$ }&
\colhead{$\Delta a_{\dot\varpi}$ (km)}\\[-0.7em]
\colhead{} & 
\colhead{} & 
\colhead{N} &
\colhead{RMS (km)} &
\colhead{$a\sin i$ (km)} & 
\colhead{$\Omega_0\ (^\circ)$} &
\colhead{$\dot\Omega\ (\dd)$} &
\colhead{$\Delta\dot\Omega\ (\dd)$} &
\colhead{$\Delta a_{\dot\Omega}$ (km)}
}
\startdata
 $ 6 $  & IER & \multicolumn{2}{c}{ $41835.920\pm0.102$ } & $42.068\pm0.157$  & $181.733\pm0.213$  & $2.7620198\pm0.0001046$  & $-0.0003140$ & $1.352\pm0.450$  \\ 
 &  &           45     &  $ 0.830 $  & $44.747\pm0.320$  & $90.465\pm0.498$  & $-2.7569020\pm0.0001882$  & $-0.0000256$ & $-0.110\pm0.813$ \\
  & COR & \multicolumn{2}{c}{ $41837.092\pm0.096$ } & $42.499\pm0.077$  & $181.657\pm0.112$  & $2.7619579\pm0.0000516$  & $-0.0001040$ & $0.448\pm0.222$  \\ 
 &  &           50     &  $ 0.263 $  & $44.643\pm0.104$  & $89.727\pm0.272$  & $-2.7565287\pm0.0000502$  & $0.0000766$ & $0.331\pm0.217$ \\
  & OER & \multicolumn{2}{c}{ $41838.237\pm0.102$ } & $42.930\pm0.158$  & $181.526\pm0.209$  & $2.7619237\pm0.0001028$  & $0.0001277$ & $-0.550\pm0.442$  \\ 
 &  &           45     &  $ 0.603 $  & $44.446\pm0.323$  & $88.850\pm0.507$  & $-2.7560947\pm0.0001898$  & $0.0002455$ & $1.061\pm0.820$ \\
[.75em]  $ 5 $  & IER & \multicolumn{2}{c}{ $42233.577\pm0.089$ } & $80.052\pm0.148$  & $176.823\pm0.089$  & $2.6715990\pm0.0000451$  & $-0.0003241$ & $1.456\pm0.203$  \\ 
 &  &           59     &  $ 0.603 $  & $40.572\pm0.270$  & $297.500\pm0.482$  & $-2.6667586\pm0.0001679$  & $-0.0000154$ & $-0.069\pm0.757$ \\
  & COR & \multicolumn{2}{c}{ $42234.893\pm0.091$ } & $80.237\pm0.076$  & $176.819\pm0.048$  & $2.6715756\pm0.0000227$  & $-0.0000548$ & $0.246\pm0.102$  \\ 
 &  &           67     &  $ 0.212 $  & $40.951\pm0.114$  & $296.068\pm0.259$  & $-2.6663516\pm0.0000445$  & $0.0000998$ & $0.450\pm0.200$ \\
  & OER & \multicolumn{2}{c}{ $42236.261\pm0.089$ } & $80.463\pm0.148$  & $176.851\pm0.089$  & $2.6715450\pm0.0000449$  & $0.0002192$ & $-0.985\pm0.202$  \\ 
 &  &           59     &  $ 0.641 $  & $41.331\pm0.270$  & $294.460\pm0.475$  & $-2.6658983\pm0.0001677$  & $0.0002496$ & $1.125\pm0.756$ \\
[.75em]  $ 4 $  & IER & \multicolumn{2}{c}{ $42569.511\pm0.089$ } & $44.931\pm0.146$  & $256.398\pm0.183$  & $2.5978170\pm0.0000927$  & $-0.0006562$ & $3.057\pm0.431$  \\ 
 &  &           57     &  $ 0.744 $  & $23.183\pm0.313$  & $339.064\pm1.489$  & $-2.5940762\pm0.0004674$  & $-0.0005631$ & $-2.630\pm2.183$ \\
  & COR & \multicolumn{2}{c}{ $42571.124\pm0.091$ } & $45.347\pm0.076$  & $255.880\pm0.093$  & $2.5980293\pm0.0000433$  & $-0.0000977$ & $0.455\pm0.202$  \\ 
 &  &           63     &  $ 0.264 $  & $23.413\pm0.076$  & $336.828\pm0.671$  & $-2.5932795\pm0.0001121$  & $-0.0001117$ & $-0.522\pm0.524$ \\
  & OER & \multicolumn{2}{c}{ $42572.743\pm0.089$ } & $45.742\pm0.143$  & $255.294\pm0.181$  & $2.5981594\pm0.0000908$  & $0.0003798$ & $-1.769\pm0.423$  \\ 
 &  &           57     &  $ 0.727 $  & $24.329\pm0.308$  & $333.891\pm1.361$  & $-2.5925059\pm0.0004191$  & $0.0003157$ & $1.475\pm1.958$ \\
[.75em]  $ \alpha $  & IER & \multicolumn{2}{c}{ $44714.884\pm0.083$ } & $32.512\pm0.118$  & $205.885\pm0.213$  & $2.1851508\pm0.0000949$  & $-0.0007266$ & $4.229\pm0.552$  \\ 
 &  &           73     &  $ 0.542 $  & $11.853\pm0.276$  & $206.062\pm1.839$  & $-2.1831510\pm0.0005365$  & $-0.0010535$ & $-6.145\pm3.130$ \\
  & COR & \multicolumn{2}{c}{ $44718.473\pm0.086$ } & $33.916\pm0.062$  & $206.074\pm0.112$  & $2.1855003\pm0.0000510$  & $0.0002392$ & $-1.392\pm0.297$  \\ 
 &  &           81     &  $ 0.249 $  & $12.005\pm0.089$  & $203.837\pm0.894$  & $-2.1813455\pm0.0001311$  & $0.0001373$ & $0.801\pm0.765$ \\
  & OER & \multicolumn{2}{c}{ $44722.161\pm0.084$ } & $35.335\pm0.118$  & $206.243\pm0.195$  & $2.1857991\pm0.0000898$  & $0.0011714$ & $-6.819\pm0.523$  \\ 
 &  &           73     &  $ 0.543 $  & $11.676\pm0.289$  & $202.596\pm1.749$  & $-2.1799630\pm0.0005158$  & $0.0008880$ & $5.186\pm3.012$ \\
[.75em]  $ \beta $  & IER & \multicolumn{2}{c}{ $45656.770\pm0.087$ } & $18.573\pm0.128$  & $319.589\pm0.435$  & $2.0309488\pm0.0001626$  & $-0.0004852$ & $3.103\pm1.040$  \\ 
 &  &           71     &  $ 0.574 $  & $3.799\pm0.194$  & $224.376\pm8.293$  & $-2.0272655\pm0.0019207$  & $0.0007997$ & $5.128\pm12.315$ \\
  & COR & \multicolumn{2}{c}{ $45661.056\pm0.087$ } & $20.106\pm0.071$  & $317.925\pm0.222$  & $2.0308229\pm0.0000857$  & $0.0000590$ & $-0.377\pm0.548$  \\ 
 &  &           78     &  $ 0.266 $  & $4.004\pm0.070$  & $232.878\pm4.169$  & $-2.0279790\pm0.0006048$  & $-0.0005821$ & $-3.733\pm3.880$ \\
  & OER & \multicolumn{2}{c}{ $45665.286\pm0.089$ } & $21.745\pm0.131$  & $316.622\pm0.383$  & $2.0308319\pm0.0001456$  & $0.0007290$ & $-4.664\pm0.932$  \\ 
 &  &           71     &  $ 0.468 $  & $3.851\pm0.206$  & $239.903\pm7.558$  & $-2.0287602\pm0.0016596$  & $-0.0020226$ & $-12.970\pm10.649$ \\
[.75em]  $ \eta $  & IER & \multicolumn{2}{c}{ $47174.853\pm0.093$ } & [0.00]$^{(c)}$ & & & & \\
 &  &           56     &  $ 0.679 $  & [0.00]$^{(c)}$ & & & & \\
  & COR & \multicolumn{2}{c}{ $47176.009\pm0.088$ } & [0.00]$^{(c)}$ & & & & \\
 &  &           60     &  $ 0.297 $  & [0.00]$^{(c)}$ & & & & \\
  & OER & \multicolumn{2}{c}{ $47177.080\pm0.093$ } & [0.00]$^{(c)}$ & & & & \\
 &  &           55     &  $ 0.525 $  & [0.00]$^{(c)}$ & & & & \\
[.75em] \multicolumn{4}{l}{$\gamma$ ring, including $m=3$ normal mode}\\
[.75em]  $ \gamma $  & IER & \multicolumn{2}{c}{ $47624.606\pm0.103$ } & $5.016\pm0.122$  & $47.274\pm1.570$  & $1.7516631\pm0.0006611$  & $0.0001409$ & $-1.091\pm5.119$  \\ 
 &  &           76     &  $ 0.536 $  & [0.00]$^{(c)}$ & & & & \\
  & COR & \multicolumn{2}{c}{ $47626.170\pm0.089$ } & $5.306\pm0.071$  & $48.720\pm0.858$  & $1.7514650\pm0.0004210$  & $0.0001448$ & $-1.121\pm3.260$  \\ 
 &  &           83     &  $ 0.401 $  & [0.00]$^{(c)}$ & & & & \\
  & OER & \multicolumn{2}{c}{ $47627.865\pm0.100$ } & $5.470\pm0.130$  & $47.145\pm1.458$  & $1.7524301\pm0.0005903$  & $0.0013286$ & $-10.286\pm4.572$  \\ 
 &  &           76     &  $ 0.692 $  & [0.00]$^{(c)}$ & & & & \\
[.75em] \multicolumn{4}{l}{$\gamma$ ring, excluding $m=3$ normal mode}\\[.75em]
 $ \gamma $  & IER & \multicolumn{2}{c}{ $47624.606\pm0.129$ } & $5.016\pm0.153$  & $47.274\pm1.964$  & $1.7516631\pm0.0008271$  & $0.0001409$ & $-1.091\pm6.405$  \\ 
 &  &           76     &  $ 0.536 $  & [0.00]$^{(c)}$ & & & & \\
  & COR & \multicolumn{2}{c}{ $47626.289\pm0.094$ } & $5.314\pm0.074$  & $47.573\pm0.881$  & $1.7524471\pm0.0003578$  & $0.0011422$ & $-8.842\pm2.771$  \\ 
 &  &           83     &  $ 0.490 $  & [0.00]$^{(c)}$ & & & & \\
  & OER & \multicolumn{2}{c}{ $47627.987\pm0.122$ } & $5.524\pm0.157$  & $46.394\pm1.775$  & $1.7536847\pm0.0006681$  & $0.0025990$ & $-20.112\pm5.175$  \\ 
 &  &           76     &  $ 0.895 $  & [0.00]$^{(c)}$ & & & & \\
[.75em]  $ \delta $  & IER & \multicolumn{2}{c}{ $48297.775\pm0.082$ } & [0.00]$^{(c)}$ & & & & \\
 &  &           73     &  $ 0.608 $  & [0.00]$^{(c)}$ & & & & \\
  & COR & \multicolumn{2}{c}{ $48300.227\pm0.082$ } & [0.00]$^{(c)}$ & & & & \\
 &  &           80     &  $ 0.273 $  & [0.00]$^{(c)}$ & & & & \\
  & OER & \multicolumn{2}{c}{ $48302.752\pm0.083$ } & [0.00]$^{(c)}$ & & & & \\
 &  &           72     &  $ 0.683 $  & [0.00]$^{(c)}$ & & & & \\
[.75em]  $ \lambda $  & COR & \multicolumn{2}{c}{ $50026.557\pm1.314$ } & [0.00]$^{(c)}$ & & & & \\
 &  &            8     &  $ 3.477 $  & [0.00]$^{(c)}$ & & & & \\
[.75em]  $ \epsilon $  & IER & \multicolumn{2}{c}{ $51120.014\pm0.077$ } & $386.530\pm0.107$  & $307.089\pm0.016$  & $1.3632690\pm0.0000070$  & $-0.0028286$ & $30.128\pm0.075$  \\ 
 &  &           79     &  $ 0.583 $  & [0.00]$^{(c)}$ & & & & \\
  & COR & \multicolumn{2}{c}{ $51149.279\pm0.081$ } & $405.894\pm0.062$  & $307.076\pm0.009$  & $1.3632575\pm0.0000042$  & $-0.0001081$ & $1.154\pm0.045$  \\ 
 &  &           89     &  $ 0.396 $  & [0.00]$^{(c)}$ & & & & \\
  & OER & \multicolumn{2}{c}{ $51178.588\pm0.079$ } & $425.242\pm0.111$  & $307.061\pm0.015$  & $1.3632618\pm0.0000070$  & $0.0026240$ & $-28.026\pm0.075$  \\ 
 &  &           77     &  $ 0.560 $  & [0.00]$^{(c)}$ & & & & \\
\enddata
\tablenotetext{(a)}{\ \ \ COR: center of ring. RunIDs: ringfit\_v1.9\_Ur018M-RF-V0351-URA178-COR-v2 and ringfit\_v1.9\_Ur018M-RF-V0351-URA178-COR-lambda.  
IER: inner edge of ring, OER: outer edge of ring. RunID for all edges: ringfit\_v1.9\_Ur018M-RF-V0351-URA178-IER-OER-v4.}
\tablenotetext{(b)}{\ \ The epoch is TDB 1986 Jan 19 12:00. The zero-point for inertial longitudes is the ascending node of Uranus's equator on Earth's equator of J2000, where the orientation of the Uranus pole is in the direction of positive angular momentum (\ie\ $180^\circ$ from the IAU definition of the Uranus north pole).}
\tablenotetext{(c)}{\ \  Eccentricities $e$ and inclinations $a\sin i$ in square brackets were held fixed during orbit determination.}
\end{deluxetable}
\end{longrotatetable}
 
\begin{longrotatetable}
\begin{deluxetable}{c l r r r r r r}
\tablecolumns{8}
\tablecaption{Uranian ring normal modes \label{tbl:normalmodes}}
\tablewidth{0pt}
\tablehead{
\colhead{Ring} & 
\colhead{Feature$^{(a)}$} & 
\colhead{$m$} &
\colhead{$A_m$ (km)} & 
\colhead{$\delta_m\ (^\circ)$} & 
\colhead{$\Omega_p\ (\dd)$} &
\colhead{$\Delta\Omega_p$\ ($\dd$)} &
\colhead{$\Delta a_p$ (km)}
}
\startdata
 $ \eta $  & IER & $       3$ & $0.452\pm0.122$  & $77.208\pm5.798$  & $776.585539\pm0.002490$  & $0.086574$  & $-3.499\pm0.101$  \\ 
  & COR & $       3$ & $0.600\pm0.069$  & $75.901\pm2.359$  & $776.584048\pm0.001164$  & $0.113697$  & $-4.595\pm0.047$  \\ 
  & OER & $       3$ & $0.677\pm0.124$  & $75.946\pm3.819$  & $776.584059\pm0.001818$  & $0.140201$  & $-5.666\pm0.073$  \\ 
[.75em] \multicolumn{4}{l}{$\gamma$ ring, including $m=3$ normal mode}\\
[.75em]  $ \gamma $  & IER & $       0$ & $6.137\pm0.138$  & $309.462\pm1.178$  & $1145.576306\pm0.000538$  & $-0.050871$  & $1.411\pm0.015$  \\ 
  &   & $       6$ & $0.590\pm0.110$  & $33.405\pm2.208$  & $956.418079\pm0.000877$  & $-0.022759$  & $0.754\pm0.029$  \\ 
  &   & $      -2$ & $1.099\pm0.146$  & $35.175\pm2.940$  & $1720.120475\pm0.001212$  & $-0.071814$  & $1.325\pm0.022$  \\ 
  &   & $      -1$ & $1.656\pm0.120$  & $31.309\pm4.318$  & $2292.907872\pm0.002002$  & $-0.098006$  & $1.357\pm0.028$  \\ 
  & COR & $       0$ & $5.509\pm0.076$  & $311.695\pm0.750$  & $1145.576845\pm0.000352$  & $0.006033$  & $-0.167\pm0.010$  \\ 
  &   & $       3$ & $0.577\pm0.073$  & $54.989\pm2.254$  & $765.399832\pm0.001152$  & $-0.065364$  & $2.706\pm0.048$  \\ 
  &   & $       6$ & $0.637\pm0.063$  & $32.518\pm1.171$  & $956.419622\pm0.000474$  & $0.025958$  & $-0.861\pm0.016$  \\ 
  &   & $      -2$ & $0.690\pm0.079$  & $26.843\pm2.762$  & $1720.117977\pm0.001217$  & $0.010437$  & $-0.193\pm0.022$  \\ 
  &   & $      -1$ & $1.822\pm0.067$  & $25.852\pm2.169$  & $2292.904186\pm0.001054$  & $0.011240$  & $-0.156\pm0.015$  \\ 
  & OER & $       0$ & $4.702\pm0.128$  & $313.066\pm1.574$  & $1145.576675\pm0.000698$  & $0.066941$  & $-1.857\pm0.019$  \\ 
  &   & $       3$ & $0.935\pm0.129$  & $56.867\pm2.520$  & $765.397669\pm0.001028$  & $-0.026590$  & $1.101\pm0.043$  \\ 
  &   & $       6$ & $0.892\pm0.117$  & $28.027\pm1.363$  & $956.419529\pm0.000560$  & $0.076982$  & $-2.552\pm0.019$  \\ 
  &   & $      -1$ & $1.955\pm0.122$  & $25.258\pm3.578$  & $2292.903174\pm0.001739$  & $0.132603$  & $-1.836\pm0.024$  \\ 
[.75em] \multicolumn{4}{l}{$\gamma$ ring, excluding $m=3$ normal mode}\\[.75em]
 $ \gamma $  & IER & $       0$ & $6.137\pm0.172$  & $309.462\pm1.474$  & $1145.576306\pm0.000673$  & $-0.050871$  & $1.411\pm0.019$  \\ 
  &   & $       6$ & $0.590\pm0.138$  & $33.405\pm2.763$  & $956.418079\pm0.001097$  & $-0.022759$  & $0.754\pm0.036$  \\ 
  &   & $      -2$ & $1.099\pm0.183$  & $35.175\pm3.678$  & $1720.120475\pm0.001517$  & $-0.071814$  & $1.325\pm0.028$  \\ 
  &   & $      -1$ & $1.656\pm0.150$  & $31.309\pm5.403$  & $2292.907872\pm0.002505$  & $-0.098006$  & $1.357\pm0.035$  \\ 
  & COR & $       0$ & $5.450\pm0.079$  & $311.233\pm0.762$  & $1145.576581\pm0.000376$  & $0.010044$  & $-0.279\pm0.010$  \\ 
  &   & $       6$ & $0.698\pm0.066$  & $30.504\pm1.075$  & $956.419653\pm0.000442$  & $0.029566$  & $-0.980\pm0.015$  \\ 
  &   & $      -2$ & $0.635\pm0.083$  & $27.010\pm3.104$  & $1720.120139\pm0.001183$  & $0.019028$  & $-0.351\pm0.022$  \\ 
  &   & $      -1$ & $1.807\pm0.070$  & $26.105\pm2.325$  & $2292.903886\pm0.001134$  & $0.019506$  & $-0.270\pm0.016$  \\ 
  & OER & $       0$ & $4.724\pm0.160$  & $313.481\pm1.841$  & $1145.575881\pm0.000844$  & $0.070553$  & $-1.958\pm0.023$  \\ 
  &   & $       6$ & $1.057\pm0.140$  & $26.147\pm1.358$  & $956.419580\pm0.000563$  & $0.080719$  & $-2.676\pm0.019$  \\ 
  &   & $      -1$ & $1.966\pm0.151$  & $22.643\pm4.393$  & $2292.903576\pm0.002122$  & $0.141832$  & $-1.964\pm0.029$  \\ 
[.75em]  $ \delta $  & IER & $       2$ & $2.349\pm0.105$  & $171.686\pm1.527$  & $562.516712\pm0.000580$  & $-0.042324$  & $2.415\pm0.033$  \\ 
  & COR & $       2$ & $3.169\pm0.059$  & $170.383\pm0.666$  & $562.516205\pm0.000258$  & $0.000146$  & $-0.008\pm0.015$  \\ 
  &   & $      23$ & $0.339\pm0.063$  & $6.844\pm0.472$  & $1074.523021\pm0.000189$  & $-0.072580$  & $2.173\pm0.006$  \\ 
  & OER & $       2$ & $4.035\pm0.105$  & $171.248\pm0.885$  & $562.516313\pm0.000351$  & $0.044481$  & $-2.539\pm0.020$  \\ 
[.75em]  $ \epsilon $  & IER & $     -24$ & $1.011\pm0.107$  & $2.916\pm0.259$  & $1074.522889\pm0.000099$  & $-0.034140$  & $1.082\pm0.003$  \\ 
  & COR & $      14$ & $0.383\pm0.071$  & $14.105\pm0.628$  & $956.418015\pm0.000269$  & $-0.798192$  & $28.425\pm0.010$  \\ 
  &   & $     -24$ & $0.443\pm0.061$  & $2.828\pm0.346$  & $1074.522703\pm0.000142$  & $0.888501$  & $-28.177\pm0.005$  \\ 
  & OER & $      14$ & $0.590\pm0.130$  & $14.904\pm0.704$  & $956.418119\pm0.000298$  & $0.024894$  & $-0.887\pm0.011$  \\ 
\enddata
\tablenotetext{(a)}{\ \ \ COR: center of ring. RunIDs: ringfit\_v1.9\_Ur018M-RF-V0351-URA178-COR-v2 and ringfit\_v1.9\_Ur018M-RF-V0351-URA178-COR-lambda.  
IER: inner edge of ring, OER: outer edge of ring. RunID for all edges: ringfit\_v1.9\_Ur018M-RF-V0351-URA178-IER-OER-v4.}
\tablenotetext{(b)}{\ \ The epoch is TDB 1986 Jan 19 12:00. The zero-point for inertial longitudes is the ascending node of Uranus's equator on Earth's equator of J2000, where the orientation of the Uranus pole is in the direction of positive angular momentum (\ie\ $180^\circ$ from the IAU definition of the Uranus north pole). All longitudes reduced to the  minimum value of $\lambda \mod 360^\circ/|m|$.}
\end{deluxetable}
\end{longrotatetable}

\subsection{Offset times for selected events}
As discussed in Paper 1, the accuracy of the absolute timing of individual data sets is highly dependent on the availability of accurate time standards at each observatory, the methods used to incorporate time signals into the data streams, and the intrinsic and sometimes variable time delays introduced by filtering electronics and recording systems. For several occultations observed with multiple telescopes, there are significant systematic offsets between the predicted and observed event times. We incorporate these into our orbit fit by including these station offset times as free parameters, as given in 
{\bf Table \ref{tbl:timeoffsets}}. For each event, the absolute timing was anchored by observations from one or more telescopes that were assumed to have the most reliable time reference. Details of the timing for each of the Earth-based observations are included in Paper 1. 

For the \Voyager\ occultations, we fitted for along-track spacecraft time offsets relative to the nominal \Voyager\ trajectory for the RSS occultation and the $\beta$ Per stellar occultation. The JPL solution for the \Voyager\ trajectory in {\tt vgr2.ura178.bsp} assumed an offset time $dt=0$ for the $\sigma$ Sgr stellar occultation, and instead solved for an offset to the {\it Hipparcos} catalog position of this multiple star system. We have followed this prescription in our orbit fits as well.

	\begin{table*} [ht]
	\begin{center} 
	\caption{Fitted station offset times}
	\label{tbl:timeoffsets} 
	\begin{threeparttable}
	\centering
	\begin{tabular}{l l r }\hline
	Event & Station & Offset (s)  \\
	\hline 
U12 & ESO (2m) &  $  -0.073 \pm   0.016 $  \\
 & Las Campanas (IR) &  $   0.099 \pm   0.022 $  \\
 & Las Campanas (vis) &  $   0.114 \pm   0.019 $  \\
U14 & ESO (1m) &  $  -0.091 \pm   0.007 $  \\
 & Las Campanas (IR) &  $   0.064 \pm   0.006 $  \\
 & Pic du Midi &  $   3.717 \pm   0.009 $  \\
 & Pic du Midi (1m) &  $   0.760 \pm   0.012 $  \\
 & Teide (ingress) &  $   0.040 \pm   0.010 $  \\
 & Teide (egress) &  $   0.467 \pm   0.010 $  \\
U25 & McDonald Obs. &  $  -0.024 \pm   0.006 $  \\
U36A & IRTF &  $  -8.769 \pm   0.262 $  \\
U103 & ESO (2m) &  $   0.055 \pm   0.013 $  \\
U103 & CTIO &  $  -0.035 \pm   0.013 $  \\ 
U134 & SAAO (egress) &  $   0.475 \pm   0.170 $  \\
U137 & HST &  $   0.635 \pm   0.008 $  \\
U144 & CAHA (ingress) &  $   0.428 \pm   0.068 $  \\
 & CAHA (egress) &  $   0.724 \pm   0.240 $  \\
Vgr2 RSS & DSS-43 &  $  -0.013 \pm   0.011 $  \\
Vgr2 $\sigma$ Sgr & PPS &  $ [0.000] $  \\ 
Vgr2 $\beta$ Per & PPS &  $  -0.061 \pm   0.016 $  \\
 	 \hline
	\end{tabular}
  \end{threeparttable}
\end{center} 
\end{table*} 

\subsection{Orbit fit residuals}
The residuals of the separate COR and IER/OER fits summarized in Tables \ref{tbl:orbel} and \ref{tbl:normalmodes} are shown in {\bf Fig.~\ref{fig:residuals}}. For the COR observations shown in the upper panel, the RMS residuals for the nine main rings range from 0.212 km for ring 5 to 0.401 km for the $\gamma$ ring including the $m=3$ mode and 0.490 km when the $m=3$ mode is absent. For the IER/OER observations, the RMS residuals range from 0.468 km for $\beta$ OER to 0.830 km for ring 6 IER, and 0.895 km for the $\gamma$ ring OER when the $m=3$ mode is excluded from the fit. In all, 651 data points were included in the COR fit, for an RMS per degree of freedom of 0.348 km. In comparison, the IER/OER fit included 1174 data points and an RMS per degree of freedom of 0.655 km. As noted previously, the COR measurements are intrinsically more accurate, which prompted the separate fits for the ring midlines and the ring edges.

\begin{figure}
\centerline{\resizebox{4in}{!}{\includegraphics[angle=0]{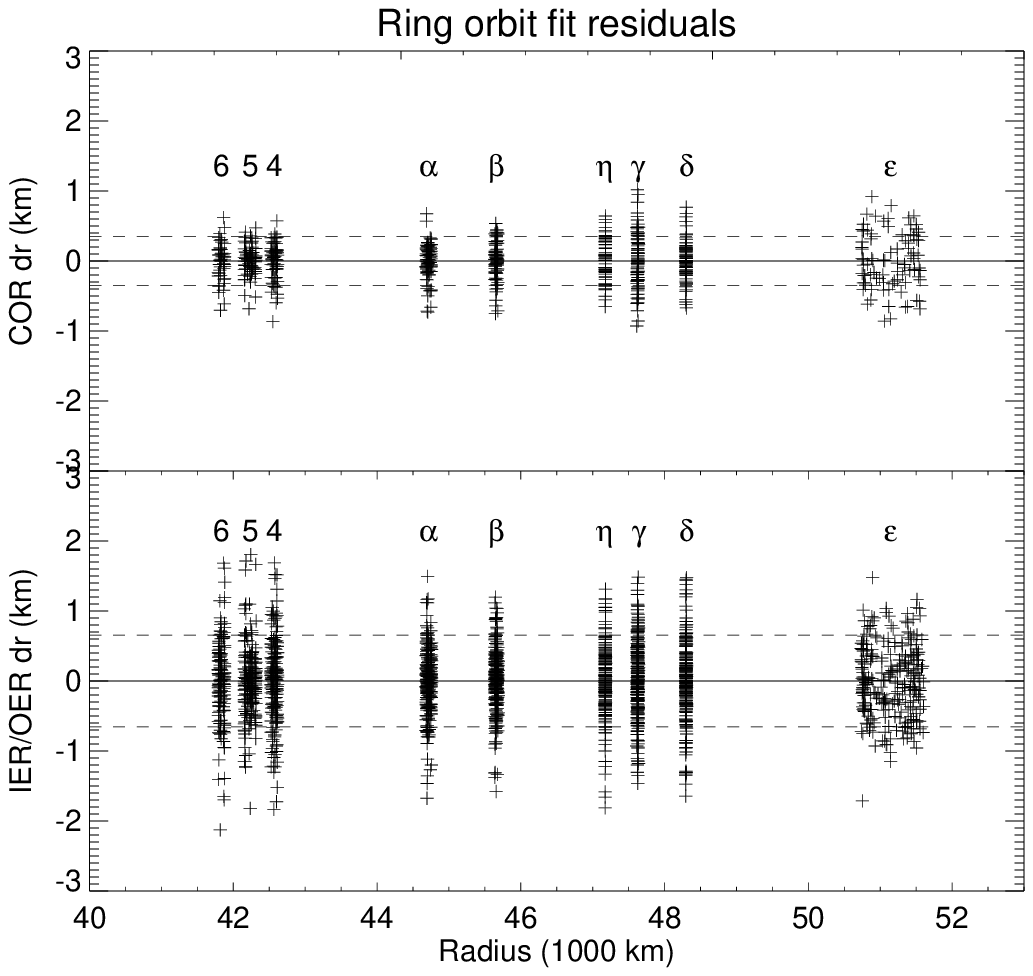}}}
\caption{Ring plane residuals of the COR (top panel) and IER/OER (bottom panel) fits in Tables \ref{tbl:orbel} and \ref{tbl:normalmodes}, including the $\gamma$ ring $m=3$ normal mode and excluding the $\lambda$ ring. The dashed lines in each panel show the RMS residuals per degree of freedom for each fit.}
\label{fig:residuals}
\end{figure}

{\bf Table \ref{tbl:residualsbyevent}} provides a more detailed view of the distribution of the COR fit RMS residuals by occultation event and observatory code. The total number of data points for each event is listed as N$_{\rm tot}$, and for multi-station events, N gives the number of data points per station. 

\startlongtable
\begin{deluxetable*} {lclrrc}
\tablecolumns{6}
\tabletypesize{\small}
\tablecaption{RMS residuals by event and observatory}
\label{tbl:residualsbyevent} 
\tablewidth{0pt}
\tablehead{
\colhead{Star}  & \colhead{Date} & \colhead{Obs} & \colhead{N$_{\rm tot}$} & \colhead{N} & 
\colhead{RMS (km)}\\
}
\startdata
U0 & 1977-03-10 & KAO &  16 &     &  0.350 \\
U2 & 1977-12-23 & TEN &   3 &     &  0.028 \\
U5 & 1978-04-10 & 304 &  16 &     &  0.421 \\
U9 & 1979-06-10 & 304 &   7 &     &  0.476 \\
U11 & 1980-03-20 & 807 &   6 &     &  0.193 \\
U12 & 1980-08-15 &     &  47 &     &  0.208 \\
 &            & 807 &     &  13 &  0.205 \\
 &            & ESO &     &  17 &  0.164 \\
 &            & LAS &     &   8 &  0.257 \\
 &            & LAV &     &   9 &  0.237 \\
U13 & 1981-04-26 & 413 &  18 &     &  0.190 \\
U14 & 1982-04-22 &     &  87 &     &  0.324 \\
 &            & 586 &     &   8 &  0.480 \\
 &            & 807 &     &  18 &  0.274 \\
 &            & ES1 &     &  14 &  0.162 \\
 &            & LAS &     &  16 &  0.305 \\
 &            & LAV &     &  10 &  0.451 \\
 &            & PI1 &     &   5 &  0.363 \\
 &            & TEE &     &   8 &  0.338 \\
 &            & TEN &     &   8 &  0.237 \\
U15 & 1982-05-01 & 414 &  17 &     &  0.371 \\
U16 & 1982-06-04 & 675 &  18 &     &  0.278 \\
U17B & 1983-03-25 & SAA &  13 &     &  0.179 \\
U23 & 1985-05-04 &     &  30 &     &  0.350 \\
 &            & 711 &     &   9 &  0.299 \\
 &            & 807 &     &  18 &  0.379 \\
 &            & TEN &     &   3 &  0.306 \\
U25 & 1985-05-24 &     &  54 &     &  0.259 \\
 &            & 675 &     &  18 &  0.239 \\
 &            & 711 &     &  18 &  0.178 \\
 &            & 807 &     &  18 &  0.335 \\
U28 & 1986-04-26 & 568 &  17 &     &  0.289 \\
U34 & 1987-02-26 & 568 &  16 &     &  0.348 \\
U36A & 1987-04-02 &     &  25 &     &  0.264 \\
 &            & 413 &     &   3 &  0.217 \\
 &            & 568 &     &   5 &  0.332 \\
 &            & 807 &     &   8 &  0.314 \\
 &            & ANU &     &   1 &  0.058 \\
 &            & IR2 &     &   2 &  0.149 \\
 &            & UKI &     &   6 &  0.190 \\
 U1052  & 1988-05-12 & 568 &  10 &     &  0.288 \\
U65 & 1990-06-21 & 568 &  15 &     &  0.328 \\
U83 & 1991-06-25 & 568 &  18 &     &  0.265 \\
U84 & 1991-06-28 & 568 &  18 &     &  0.222 \\
 U102A  & 1992-07-08 & 568 &   6 &     &  0.385 \\
U103 & 1992-07-11 &     &  27 &     &  0.281 \\
 &            & 675 &     &  12 &  0.230 \\
 &            & ES2 &     &  15 &  0.316 \\
 U9539  & 1993-06-30 & 807 &  18 &     &  0.281 \\
U134 & 1995-09-09 &     &  18 &     &  0.273 \\
 &            & SA1 &     &   9 &  0.284 \\
 &            & SAA &     &   9 &  0.261 \\
U137 & 1996-03-16 &     &  22 &     &  0.336 \\
 &            & 568 &     &  18 &  0.359 \\
 &            & HST &     &   4 &  0.206 \\
U138 & 1996-04-10 &     &  18 &     &  0.167 \\
 &            & 675 &     &   9 &  0.165 \\
 &            & HST &     &   9 &  0.168 \\
U144 & 1997-09-30 &     &  15 &     &  0.445 \\
 &            & CAE &     &   5 &  0.450 \\
 &            & CAI &     &   5 &  0.495 \\
 &            & SAA &     &   5 &  0.382 \\
U149 & 1998-11-06 &     &  12 &     &  0.369 \\
 &            & 568 &     &   6 &  0.395 \\
 &            & 688 &     &   6 &  0.342 \\
 U0201  & 2002-07-29 & 675 &  12 &     &  0.411 \\
 U0602  & 2006-09-20 & 568 &  14 &     &  0.367 \\
\enddata
\end{deluxetable*}

         \subsection{Limits on eccentricity and inclination}
        \label{sec:eccinc}
In our adopted final orbit model, several of the narrow rings have no detectable eccentricity and/or inclination. Nevertheless, we can set upper limits on these quantities from a statistical analysis of the patterns of orbit fit radial residuals for a given ring. We illustrate this procedure with the $\eta$ ring. The top panel of {\bf Fig.~\ref{fig:eta_ecc}} shows the results of a series of least-squares fits to the residuals in the measurements of the $\eta$ ring radius after subtraction of the best-fitting circular orbit and $m=3$ normal mode, over a range of precession rates centered on the predicted apsidal rate appropriate for the ring's fitted semimajor axis. The free parameters for each fit are the eccentricity and mean anomaly of the best-fitting ellipse with the assumed precession rate. The top panel shows the RMS residuals of each fit as a function of the assumed pattern speed, and the second panel shows the amplitude $ae$ of the best-fitting ellipse. The solid vertical lines mark the expected precession rate for an eccentric ring with the actual radius of the $\eta$ ring. If the ring were measurably eccentric, we would expect to see a sharp dip in the RMS residuals and a peak in the fitted amplitude, centered on this pattern speed. Instead, the best-fitting eccentric model for the $\eta$ ring is marked by the vertical dashed lines, far removed from the physically significant expected precession rate. The bottom panel of the figure shows a histogram of the distribution of fitted amplitudes ($\Delta a = ae$) from the middle panel. The overplotted solid line shows the best-fitting Rayleigh distribution to this histogram, appropriate for a random one-sided distribution:
\beq
N(\Delta a) = N_0 \frac{\Delta a}{\sigma^2} \exp{(-\Delta a^2/2\sigma^2),} 
\label{eq:eta_ecc}
\eeq
\noindent where we fit for $N_0$ and $\sigma$. In this case, $\sigma(\Delta ae)=0.063$ km, which is comparable to the 1-$\sigma$ formal uncertainties of the rings with measurable eccentricities (Table \ref{tbl:orbel}). Setting the eccentricity detection limit at 2-$\sigma$, the corresponding upper limit to the $\eta$ ring's eccentricity is $ae\lesssim0.126$ km.

\begin{figure}
\centerline{\resizebox{4in}{!}{\includegraphics[angle=0]{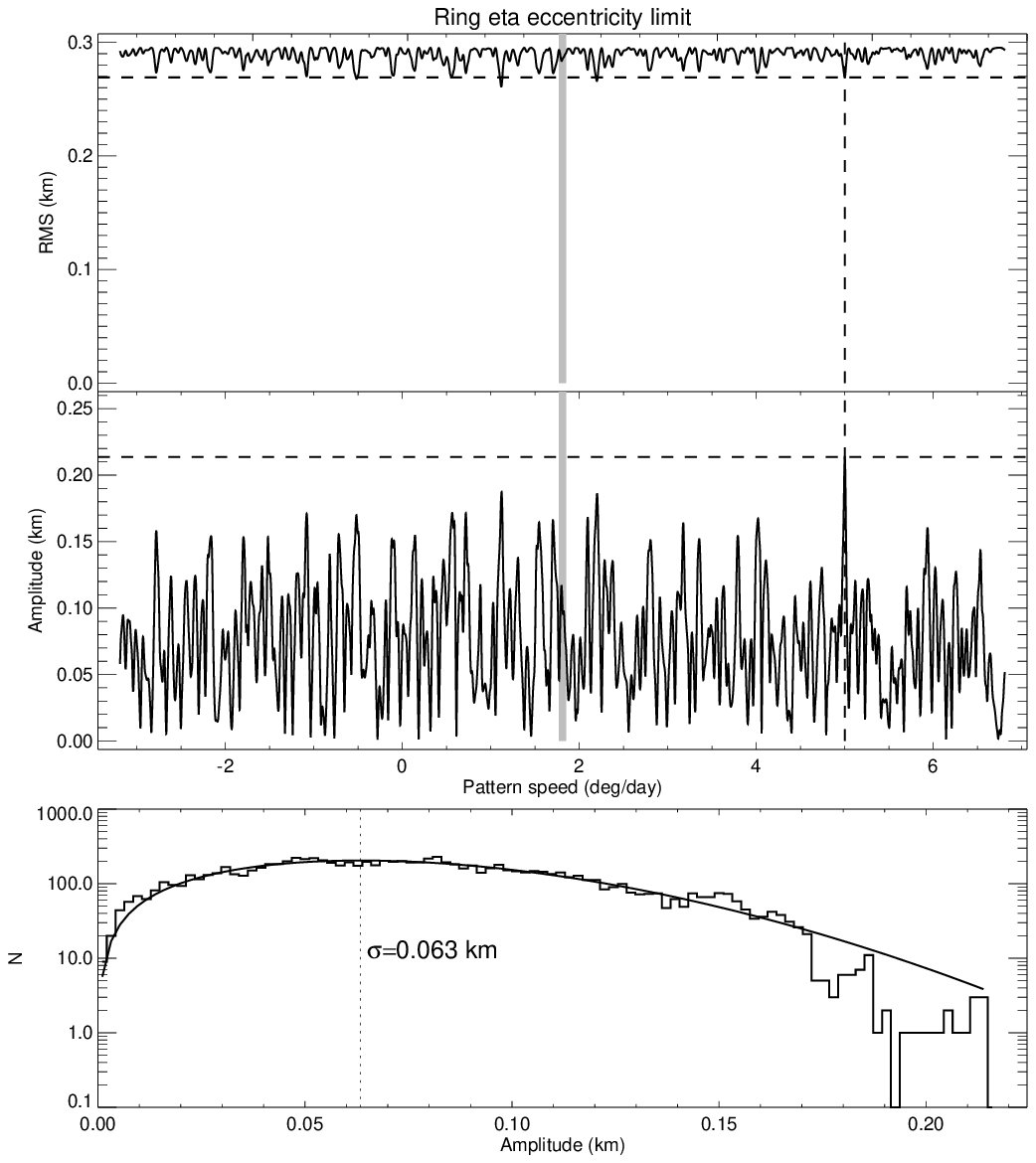}}}
\caption{Statistical limit on the eccentricity of the $\eta$ ring. The upper two panels show the null results of an $m=1$ normal mode scan, with no detectable dip in the RMS residuals near the predicted apsidal precession rate for the observed semimajor axis of the ring, shown as a thick vertical gray line. The bottom panel shows a histogram of the amplitudes given by the individual least-squares fits for each assumed pattern speed, along with the best-fitting Rayleigh distribution with $\sigma=0.063$ km.}
\label{fig:eta_ecc}
\end{figure}

Using the same approach to estimate a detection limit for inclination, {\bf Fig.~\ref{fig:eta_inc}} shows a scan over nodal regression rates for the $\eta$ ring, where we follow the prescription given by Eqs.~(8)--(11) of \cite{French2016} to relate the observed radial residual $\Delta r$ to the predicted radial displacement corresponding to the local vertical displacement of an inclined ring. Once again, there is no indication of a measurable inclination at the expected pattern speed (marked by a thick vertical gray line), with $\sigma(\Delta a\sin i)=0.145$ km. Adopting the same 2-$\sigma$ detection threshold, the upper limit to the inclination of the $\eta$ ring is $a\sin i\lesssim0.290$ km.

\begin{figure}
\centerline{\resizebox{4in}{!}{\includegraphics[angle=0]{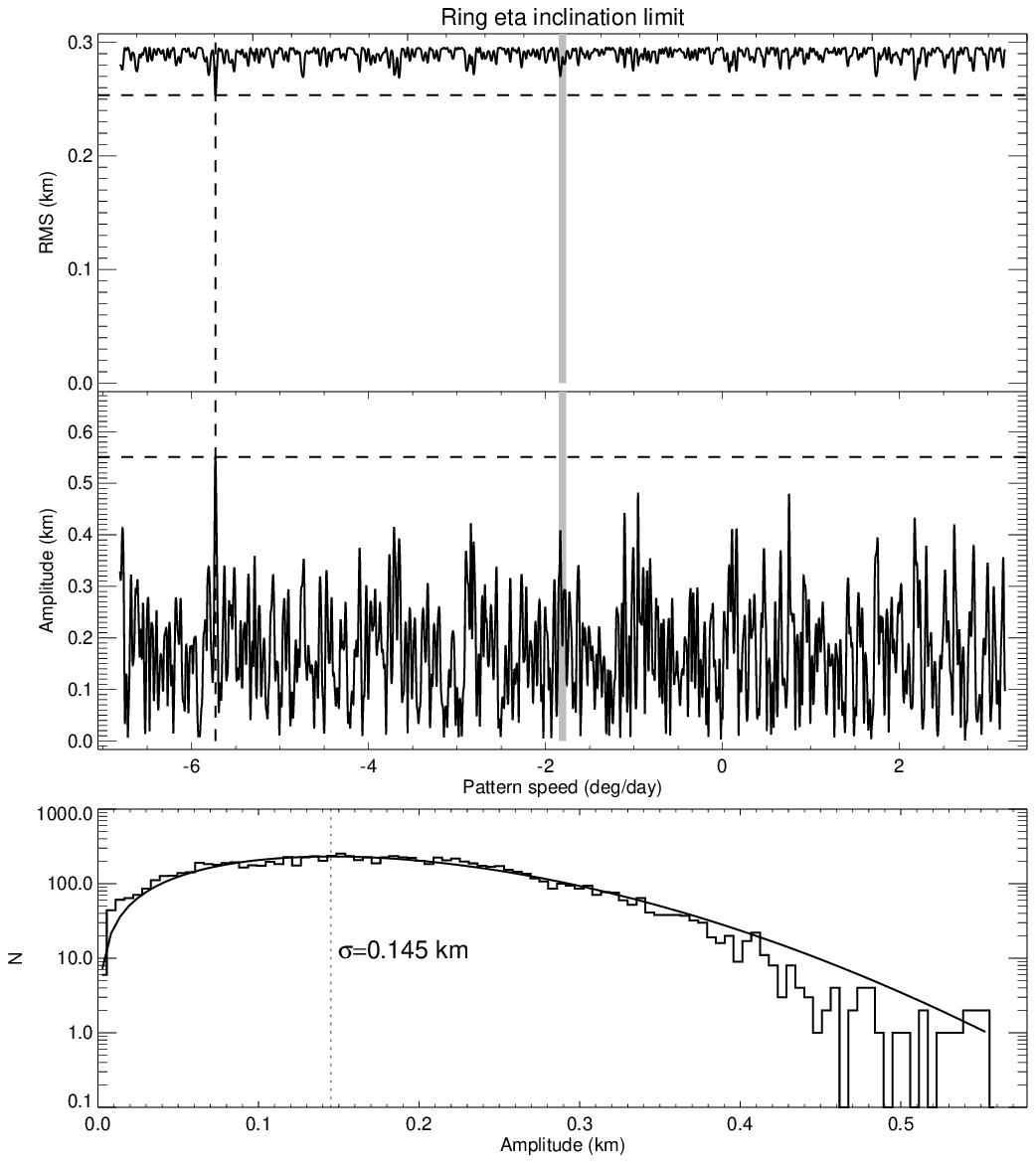}}}
\caption{Statistical limit on the inclination of the $\eta$ ring. The upper two panels show the null results of an $m=1$ inclination scan, with no detectable dip in the RMS residuals near the predicted nodal regression rate for the observed semimajor axis of the ring, marked by a thick vertical gray line. The bottom panel shows a histogram of the amplitudes given by the individual least-squares fits for each assumed pattern speed, along with the best-fitting Rayleigh distribution with $\sigma=0.145$ km.}
\label{fig:eta_inc}
\end{figure}

 Proceeding in a similar fashion for the other rings, we obtain the results in {\bf Table \ref{tbl:eccinc}} for the 2-$\sigma$ upper limits on the eccentricities and inclinations of all rings with no detected $m=1$ mode and/or inclination.

	\begin{table*} [ht]
	\begin{center} 
	\caption{2-$\sigma$ eccentricity/inclinations  limits}
	\label{tbl:eccinc} 
	\begin{threeparttable}
	\centering
	\begin{tabular}{c c c }\hline
	Ring &  $ae$ (km) & $a\sin i$ (km) \\
	\hline 
%
$\eta$ & 0.126 & 0.290\\  
$\gamma$ & -- & 0.286\\ 
$\delta$ & 0.088 & 0.284\\ 
$\epsilon$ &-- & 0.158\\  
 	 \hline
	\end{tabular}
  \end{threeparttable}
\end{center} 
\end{table*}

\subsection{The $\lambda$ ring}
         \label{sec:lambda}
The $\lambda$ ring differs from the other nine narrow Uranian rings in the apparent wavelength dependence of its equivalent width \citep{French1991} and in having an enhanced brightness in the forward scattering direction compared to the other narrow rings \citep{Smith1986,Ockert1987}. From a comparison of {\it Voyager} UVS and PPS stellar occultations ($\lambda=0.11~\mu$m and 0.27~$\mu$m, respectively) and Earthbased IR observations at $\lambda=2.2~\mu$m, \cite{Kangas1987} modeled the wavelength dependence of the equivalent depth of the fitted square-well profiles by assuming a two-component population of particles large compared to the observed wavelength and of smaller particles that scattered in the Mie regime with $\tau\propto 1/\lambda$. They derived an upper limit of 6\% for the contribution of large particles to the total optical depth at $\lambda=0.11\ \mu$m. Collectively, these results suggest that the $\lambda$ ring is primarily composed of micron-sized dust. However, it is also azimuthally variable, having arcs and clumps \citep{Ockert1987,Colwell1990,Showalter1995}, some of which might have larger particles detectable in the IR. A detection at infrared wavelengths was reported for the U103 occultation from CTIO and possibly from ESO as well \citep{French1996}. \cite{Kangas1989} identified several additional candidate $\lambda$ ring events from an analysis of Earth-based occultations at $\lambda=2.2\ \mu$m, and to these we add a likely detection from the \hst\ observations of the U138 occultation at $\lambda = 0.362 - 0.705\ \mu$m (see Paper 1 for observational details).

 For the orbit fit presented here, we include the two secure \Voyager\ PPS UV occultation detections (we omit the lower-SNR UVS observations for the same event, to avoid double-counting), five Earth-based IR candidate events, and the likely U138 \hst\ detection, for a total of eight possible $\lambda$ ring events, listed in {\bf Table \ref{tbl:lambda}}. For each event, we include the observed radius $r_{\rm obs}$, the residual $dr$ relative to our adopted circular orbit fit, the ring profile equivalent width $EW$, and the wavelength of the observations. Radial profiles and square-well model fits to the six Earth-based candidate detections are shown in {\bf Fig.~\ref{fig:lambda}}. 

\begin{table*} [ht]
\caption{Candidate $\lambda$ ring occultation events.}
\label{tbl:lambda}
\begin{threeparttable}
\begin{tabular}{l l c c c r c c}\hline
Ring & Obs & Dir & UTC & $r_{\rm obs}$ (km) &  $dr$ (km) & $EW$ (km) & $\lambda$ ($\mu$m) \\
\hline
U23 & 807 & E & 1985-05-04T06:05:50.0905 &   50022.817 &  $      -3.740 $  &  $   0.27 $ & 2.20 \\
$\sigma$ Sgr & PPS & I & 1986-01-24T05:30:42.3000 &   50025.105 &  $      -1.452 $  & 0.19 & 0.27 \\
$\sigma$ Sgr & PPS & E & 1986-01-24T08:06:57.1900 &   50022.528 &  $      -4.030 $  & 0.19 & 0.27 \\
U28 & 568 & I & 1986-04-26T13:27:23.3009 &   50032.187 &  $       5.630 $  &  $   0.28 $ & 2.20 \\
U28 & 568 & E & 1986-04-26T15:11:51.8047 &   50029.089 &  $       2.532 $  &  $   0.23 $ & 2.20 \\
U103 & 807 & I & 1992-07-11T07:59:42.0300 &   50026.778 &  $       0.221 $  &  $   0.32 $ & 2.20 \\
U103 & ES2 & I & 1992-07-11T07:59:43.5982 &   50023.378 &  $      -3.179 $  &  $   0.48 $ & 2.20 \\
U138 & HST & E & 1996-04-10T12:15:28.1639 &   50030.576 &  $       4.018 $  &  $   0.30 $ & 0.362--0.705 \\
\hline
\end{tabular}
\end{threeparttable}
\end{table*}

\begin{figure}
\centerline{\resizebox{5.5in}{!}{\includegraphics[angle=90]{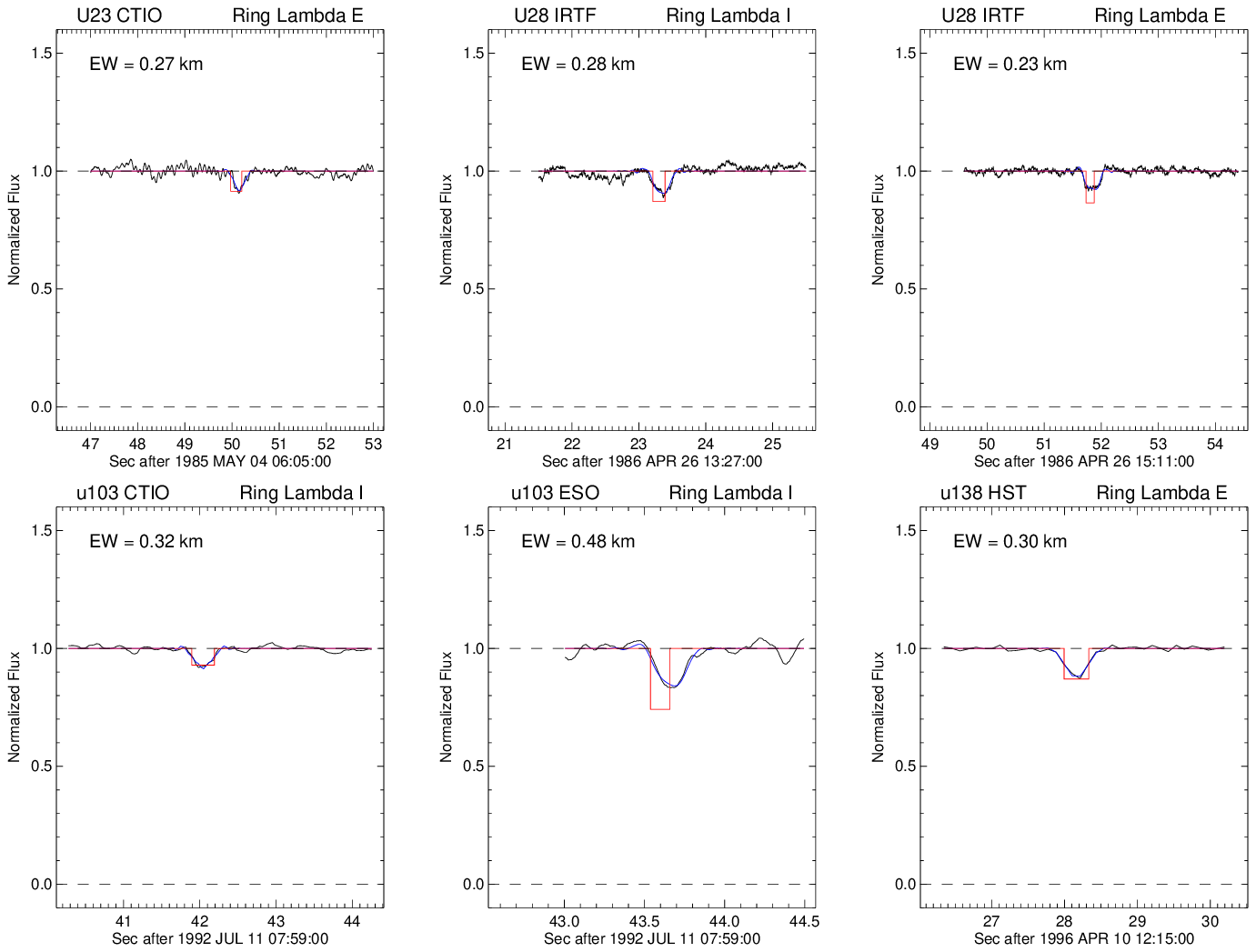 }}}
\caption{Candidate Earth-based occultation detections of the $\lambda$ ring. (The substantial offset of the U103 ESO model curve relative to the square well shown in red is due to the large instrumental time constant for that dataset -- see Paper 1.)}
\label{fig:lambda}
\end{figure}
       
No satisfactory keplerian elliptical model fitted these measurements, and our circular orbit fit gives $a=50026.557\pm 1.314$ km and an RMS residual of 3.477 km (Table \ref{tbl:orbel}), compared to the \cite{Jacobson2014} value $a=50024.16\pm0.96$, which was based on the combined \Voyager\ PPS and UVS detections during the $\sigma$ Sgr occultation and the {\tt vgr2.ura111.bsp} ephemeris.
A separate fit using the {\tt vgr.ura178.bsp} ephemeris that included only the two \Voyager\ points yielded $a=50023.816\pm1.289$ km, close to the \cite{Jacobson2014} result. With the addition of the \hst\ detection, which we regard as convincing (in part because it was observed at wavelengths fairly close to the PPS sensitivity of $\lambda=0.27\ \mu$m), we obtained $a=50026.069\pm2.37$ km with an RMS residual of 3.363 km.
       
Overall, the orbit fits to the $\lambda$ ring show considerably more scatter than the other rings, even when restricted to the convincing \Voyager\ detections. The \Voyager\ images show evidence of an $m=6$ pattern in the ring’s radial position with an amplitude of $4.6\pm0.9$ km \citep{Showalter1995} that could be compatible with this excess scatter, but the occultation data are too sparse to confirm the existence of this pattern at this point. In the absence of additional secure observations, its semimajor axis should probably be regarded as uncertain at the level of a few km. 
     
\subsection{Multiple-star occultation geometry}
\label{sec:secondaries}
Among the Uranus occultations reported in Paper 1, the ring profiles of two occultations revealed that they involved multiple-star systems. 
The occultation of U36 by Uranus and the rings was a remarkable multi-day event, lasting from 1987 Mar 30 through Apr 2 and occurring while Uranus was at the end of its retrograde loop as seen from Earth \citep{Elliot1987}. Besides the ring occultations of the primary star, additional secondary events involving three additional stars were observed, as described in detail in Paper 1. Similarly, star U102A was found to have a binary companion, U102B. We solved for the skyplane offsets $\Delta f$ (east) and $\Delta g$ (north) of secondary stars U36B and U36C relative to the primary star U36A (only one ring event was observed for U36D, preventing a unique determination of its offset position), and of star U102B relative to star U102A, from separate orbit fits to the secondary ring event times, with the results shown in {\bf Table~\ref{tbl:secondaries}}. %
\begin{table*} [ht]
\caption{Skyplane positions of secondary stars}
\label{tbl:secondaries}
\begin{center}
\begin{tabular}{l r r}\hline
Star & $\Delta f$ (km) & $\Delta g$ (km)  \\
\hline
U36B & $+9.15 \pm 0.32 $& $+20.74\pm 0.77$\\
U36C & $-128.86 \pm 0.32 $ & $+138.53\pm 0.77$\\
U102B & $+287.28 \pm 0.42 $ & $-283.02\pm 1.80$\\ 
\hline
\end{tabular}
\end{center}
\end{table*}

\subsection{Star positions, proper motions and planetary ephemeris offsets}
\label{sec:starpositions}
Under the assumption that the {\it Gaia} DR3 star positions at the catalog epoch of TDB 2016 Jan 1 12:00 are more accurate than the proper motion-corrected positions, in our nominal orbit fit we solved for corrections to the catalog values of the proper motions of the Earth-based occultation stars. To provide alternative representations of the corrections to the relative positions of the star and planet for each occultation, we performed two additional fits: first, we assumed that the {\tt ura178} series of ephemerides were exact and used the unmodified star catalog proper motions, solving for corrections to the predicted star positions for each occultation; next, we used the unmodified catalog values to compute the proper motion and parallax-corrected star positions for each observation, and instead fitted for skyplane offsets $f_0$ and $g_0$ (east and north) of Uranus relative to the ephemeris position for each occultation. All three fits returned virtually identical pole directions and ring orbital elements, as expected.
{\bf Table~\ref{table:stars}} lists the results of these three separate fits, and includes for each star the proper motion correction in RA ($d \alpha \cos \delta/dt$) and Dec ($d \delta/dt$), corrections to the catalog positions $d\alpha$ and $d\delta$, and sky-plane ephemeris offsets $f_0$ and $g_0$, along with their correlation coefficient $\rho$ (which was identical for the three fits). Except for a few outliers, mostly associated with multiple star systems, the fitted star offsets $d\alpha$ and $d\delta$ for most of the remaining events are on the order of a few mas. Similarly, the sky-plane ephemeris offsets $f_0$ and $g_0$ are well under 100 km, confirming that the substantial systematic drift in time of the {\tt ura111} Uranus ephemeris that amounted to several hundred km in the sky plane by the time of the final ring occultation in 2006 has been effectively eliminated in the {\tt ura178} ephemeris.\footnote{The {\it Gaia DR3} catalog ID for each star is included in Table 2 of Paper 1.}

\begin{table*} [ht]
\scriptsize
\begin{center} 
\caption{Fitted corrections to proper motions, star positions, and planet ephemeris}
\label{table:stars} 
\stackunder{
\hspace*{-2cm}
\centering
\begin{tabular}{l  | r  r  | r r | r r |  r}
\hline
& \multicolumn{2}{c|}{Proper motion fit} & \multicolumn{2}{c|}{Star offset fit}&\multicolumn{2}{c|}{Ephemeris offset fit} & \\
Star &  $d\alpha \cos \delta/dt$ (mas/yr)  & $d\delta/dt$ (mas/yr)  &  \multicolumn{1}{c}{$d\alpha$ (mas)} & $d\delta$ (mas ) & \multicolumn{1}{c}{$f_0$ (km)} & \multicolumn{1}{c |}{$g_0$ (km)} & \multicolumn{1}{c}{$\rho$\tnote{a}}\\
\hline 
U0  &  $   0.04476 \pm    0.00028$ & $   0.01458 \pm    0.00049$ & $   -1.7370 \pm     0.0108$ & $   -0.5660 \pm     0.0189$ &  $    22.583 \pm      0.141$ & $     7.357 \pm      0.246 $ & $ -0.009 $ \\
U2  &  $  -0.05768 \pm    0.00127$ & $   0.08987 \pm    0.03230$ & $    2.1930 \pm     0.0483$ & $   -3.4165 \pm     1.2273$ &  $   -30.649 \pm      0.678$ & $    47.658 \pm     17.144 $ & $ -0.952 $ \\
U5  &  $   0.06834 \pm    0.00034$ & $  -0.03155 \pm    0.00038$ & $   -2.5782 \pm     0.0130$ & $    1.1901 \pm     0.0142$ &  $    33.119 \pm      0.167$ & $   -15.291 \pm      0.182 $ & $ -0.236 $ \\
U9  &  $  -0.02314 \pm    0.00062$ & $   0.04092 \pm    0.00258$ & $    0.8459 \pm     0.0227$ & $   -1.4966 \pm     0.0943$ &  $   -10.931 \pm      0.293$ & $    19.324 \pm      1.217 $ & $  0.889 $ \\
U11  &  $   0.02566 \pm    0.00092$ & $  -0.01070 \pm    0.00039$ & $   -0.9183 \pm     0.0330$ & $    0.3828 \pm     0.0140$ &  $    12.094 \pm      0.434$ & $    -5.043 \pm      0.184 $ & $ -0.163 $ \\
U12  &  $  -0.02100 \pm    0.00029$ & $  -0.03987 \pm    0.00087$ & $    0.7429 \pm     0.0104$ & $    1.4103 \pm     0.0307$ &  $   -10.111 \pm      0.141$ & $   -19.197 \pm      0.418 $ & $  0.621 $ \\
U13  &  $   0.00444 \pm    0.00023$ & $   0.00634 \pm    0.00066$ & $   -0.1540 \pm     0.0080$ & $   -0.2202 \pm     0.0228$ &  $     1.997 \pm      0.104$ & $     2.852 \pm      0.295 $ & $  0.280 $ \\
U14  &  $   0.13015 \pm    0.00019$ & $   0.08204 \pm    0.00052$ & $   -4.3854 \pm     0.0063$ & $   -2.7647 \pm     0.0176$ &  $    57.299 \pm      0.082$ & $    36.120 \pm      0.231 $ & $  0.150 $ \\
U16  &  $   0.00340 \pm    0.00024$ & $  -0.00384 \pm    0.00052$ & $   -0.1142 \pm     0.0081$ & $    0.1289 \pm     0.0176$ &  $     1.481 \pm      0.106$ & $    -1.673 \pm      0.228 $ & $  0.104 $ \\
U15  &  $   0.03359 \pm    0.00023$ & $  -0.00179 \pm    0.00067$ & $   -1.1310 \pm     0.0078$ & $    0.0602 \pm     0.0226$ &  $    14.717 \pm      0.102$ & $    -0.785 \pm      0.295 $ & $  0.208 $ \\
U17B  &  $   0.02862 \pm    0.00031$ & $  -0.00861 \pm    0.00051$ & $   -0.9378 \pm     0.0103$ & $    0.2822 \pm     0.0167$ &  $    12.579 \pm      0.138$ & $    -3.787 \pm      0.224 $ & $  0.337 $ \\
U23  &  $   0.03614 \pm    0.00024$ & $   0.04487 \pm    0.00032$ & $   -1.1082 \pm     0.0074$ & $   -1.3756 \pm     0.0100$ &  $    14.650 \pm      0.098$ & $    18.184 \pm      0.132 $ & $  0.175 $ \\
U25  &  $   0.30159 \pm    0.00022$ & $   0.16083 \pm    0.00029$ & $   -9.2306 \pm     0.0068$ & $   -4.9224 \pm     0.0088$ &  $   121.117 \pm      0.089$ & $    64.590 \pm      0.115 $ & $  0.043 $ \\
U28  &  $   0.04119 \pm    0.00027$ & $  -0.05310 \pm    0.00051$ & $   -1.2227 \pm     0.0081$ & $    1.5763 \pm     0.0150$ &  $    16.341 \pm      0.108$ & $   -21.063 \pm      0.201 $ & $  0.183 $ \\
U34  &  $   0.42907 \pm    0.00042$ & $  -0.28186 \pm    0.00523$ & $  -12.3768 \pm     0.0121$ & $    8.1289 \pm     0.1507$ &  $   174.952 \pm      0.171$ & $  -114.803 \pm      2.128 $ & $  0.845 $ \\
U36A  &  $   0.31706 \pm    0.00029$ & $  -0.36079 \pm    0.00056$ & $   -9.1173 \pm     0.0083$ & $   10.3772 \pm     0.0160$ &  $   125.254 \pm      0.114$ & $  -142.893 \pm      0.220 $ & $ -0.078 $ \\
U1052  &  $   0.00355 \pm    0.00089$ & $  -0.04157 \pm    0.00037$ & $   -0.0981 \pm     0.0247$ & $    1.1490 \pm     0.0102$ &  $     1.314 \pm      0.331$ & $   -15.392 \pm      0.137 $ & $  0.049 $ \\
U65  &  $   0.04912 \pm    0.00042$ & $   0.00392 \pm    0.00053$ & $   -1.2539 \pm     0.0108$ & $   -0.1001 \pm     0.0136$ &  $    16.736 \pm      0.144$ & $     1.336 \pm      0.181 $ & $  0.192 $ \\
U83  &  $   0.04739 \pm    0.00044$ & $   0.07387 \pm    0.00051$ & $   -1.1620 \pm     0.0109$ & $   -1.8112 \pm     0.0125$ &  $    15.566 \pm      0.146$ & $    24.263 \pm      0.168 $ & $  0.205 $ \\
U84  &  $   0.05320 \pm    0.00067$ & $   0.03366 \pm    0.00036$ & $   -1.3038 \pm     0.0164$ & $   -0.8252 \pm     0.0089$ &  $    17.460 \pm      0.220$ & $    11.051 \pm      0.119 $ & $  0.193 $ \\
U102A  &  $  -0.57815 \pm    0.00051$ & $   0.49276 \pm    0.00250$ & $   13.5766 \pm     0.0119$ & $  -11.5716 \pm     0.0587$ &  $  -182.414 \pm      0.160$ & $   155.471 \pm      0.789 $ & $ -0.409 $ \\
U103  &  $   0.02529 \pm    0.00097$ & $   0.05789 \pm    0.00036$ & $   -0.5939 \pm     0.0229$ & $   -1.3590 \pm     0.0085$ &  $     7.981 \pm      0.307$ & $    18.263 \pm      0.115 $ & $  0.324 $ \\
U9539  &  $   0.00114 \pm    0.00031$ & $  -0.04155 \pm    0.00172$ & $   -0.0257 \pm     0.0070$ & $    0.9353 \pm     0.0387$ &  $     0.347 \pm      0.094$ & $   -12.624 \pm      0.523 $ & $ -0.239 $ \\
U134  &  $  -0.01846 \pm    0.00286$ & $   0.01964 \pm    0.01047$ & $    0.3749 \pm     0.0580$ & $   -0.3988 \pm     0.2127$ &  $    -5.187 \pm      0.800$ & $     5.504 \pm      2.932 $ & $  0.983 $ \\
U137  &  $  -0.01975 \pm    0.00051$ & $  -0.02044 \pm    0.00370$ & $    0.3909 \pm     0.0100$ & $    0.4045 \pm     0.0733$ &  $    -5.768 \pm      0.148$ & $    -5.964 \pm      1.081 $ & $ -0.873 $ \\
U138  &  $  -0.02093 \pm    0.00072$ & $  -0.01165 \pm    0.00272$ & $    0.4130 \pm     0.0142$ & $    0.2300 \pm     0.0537$ &  $    -5.981 \pm      0.206$ & $    -3.332 \pm      0.779 $ & $ -0.818 $ \\
U144  &  $  -0.05555 \pm    0.00208$ & $   0.03560 \pm    0.00405$ & $    1.0136 \pm     0.0380$ & $   -0.6498 \pm     0.0740$ &  $   -14.230 \pm      0.532$ & $     9.126 \pm      1.037 $ & $  0.905 $ \\
U149  &  $  -0.05626 \pm    0.00056$ & $  -0.06285 \pm    0.00189$ & $    0.9649 \pm     0.0095$ & $    1.0781 \pm     0.0324$ &  $   -13.951 \pm      0.138$ & $   -15.582 \pm      0.469 $ & $ -0.280 $ \\
U0201  &  $  -0.69338 \pm    0.00185$ & $   0.43748 \pm    0.00102$ & $    9.3095 \pm     0.0248$ & $   -5.8740 \pm     0.0136$ &  $  -128.665 \pm      0.343$ & $    81.183 \pm      0.189 $ & $  0.566 $ \\
U0602  &  $  -0.53994 \pm    0.00258$ & $  -0.36195 \pm    0.00403$ & $    5.0120 \pm     0.0240$ & $    3.3595 \pm     0.0374$ &  $   -69.474 \pm      0.332$ & $   -46.570 \pm      0.519 $ & $ -0.278 $ \\
 	 \hline
	\end{tabular}
} {
\hspace*{-1cm}
\parbox{6.6in}{
$^a$ Correlation coefficient.
}}
\end{center} 
\end{table*} 

\section{Uranus Pole Direction and Ring Plane Radius Scale}
\label{sec:pole}

We determined the uncertainty in the Uranus pole direction and ring plane radius scale from a series of test fits, described below. The results are included in {\bf Table~\ref{tbl:radiuspole}} and shown in {\bf Figs.~\ref{fig:pole}} and {\bf\ref{fig:radiusscale}}. 

\subsection{Uranus pole direction}

\begin{itemize}
\item Fit 1: Nominal pole direction and ring orbital elements. This fit included the COR measurements of Earth-based and \Voyager\ observations of the nine principal rings, resulting in the adopted orbital elements and normal modes given in Tables \ref{tbl:orbel} and \ref{tbl:normalmodes}. The uncertainties in the pole direction are the formal 1-$\sigma$ errors from the unweighted least squares fit to the ring observations, plotted as the black error ellipse in the Fig.~\ref{fig:pole}. 
\item Fit 2:  Adopted pole direction, including Voyager trajectory uncertainties. Fit 1 assumed that the \Voyager\ {\tt vgr2.ura178.bsp} ephemeris was exact and did not take into account the formal uncertainties in the spacecraft position that were estimated as part of the ephemeris solution. We conducted a series of Monte Carlo simulations in which we randomly displaced the nominal spacecraft position by the corresponding estimated uncertainty during each of the three separate occultations (RSS, $\sigma$ Sgr, and $\beta$ Per) to derive an error ellipse representing the pole direction uncertainties arising solely from the estimated \Voyager\ ephemeris errors. We convolved this probability density function with the pole direction error ellipse from Fit 1 to derive a modified 1-$\sigma$ error ellipse and correlation coefficient $\rho(\alpha_P,\delta_P)$ that accounted for the combined uncertainties in our orbit fit and the systematic uncertainties in the spacecraft trajectory. Overall, the $\Voyager$ trajectory uncertainty has a relatively small effect on the final error budget: the fitted pole direction is the same as in Fit 1, but the error estimates have increased modestly from $\sigma(\alpha_P)=0.000128^\circ$ to $0.000141^\circ$, and from $\sigma(\delta_P)=0.000472^\circ$ to $0.000618^\circ$. This result is shown as a red error ellipse in Fig.~\ref{fig:pole} and represents our best estimate of the final pole direction and its 1-$\sigma$ error. 
\item Fit 3: Precession of the Uranus pole direction.
For Fit 1, we assumed that the Uranus pole direction is fixed in time, but it is predicted to experience slight precession caused by the weak periodic torques supplied by the sun and the planet's satellites. \cite{Jacobson2023} developed a trigonometric series representation of these contributions to the pole direction over time, derived from numerical integrations of the {\tt ura178} satellite ephemerides. 

Over the course of the 29-year interval of the ring occultation observations considered here, the corresponding predicted direction of the pole varied periodically by $\pm0.0002^\circ$ in $\alpha_P$ and $\pm 0.00015^\circ$ in $\delta_P$, following the dominant short-period term in the series with a period of 17.79 years (6494 days), comparable to the leading terms $E_1$ and $I_1$ in Miranda's eccentricity and inclination arguments (\cite{Laskar1986}, Table 3). For Fit 3, we adopted the trigonometric representation of the pole precession as a function of time, fitting for the pole direction at the TDB 1986 Jan 19.5 (\ie\ Jan 19 12:00) reference epoch assumed for the ring orbital elements. The brick-red oscillatory path labeled {\tt ura178} is the pole direction given by this trigonometric series over the interval 1600--2600, beginning at the filled circle at the lower left of the path. The {\tt ura178} pole direction over the interval of the occultation observations 1977--2006 is shown as a thicker circular line near the center of the figure.
The semimajor axes of the rings and the RMS residuals for this fit were virtually identical to those in Fit 1, indicating that the predicted pole precession has little effect on the derived ring system geometry, as was also found by \cite{Jacobson2014,Jacobson2023}. The error ellipse for Fit 3 is slightly displaced from the Fit 1 error ellipse. Based on this insensitivity of our results to the inclusion of pole precession, we adopt a fixed pole direction of Fits 1 and 2 for our preferred orbit solution, with the error bars from Fit 2.
\item Fit 4: Earth-based observations only. This fit included only the Earth-based ring occultation data and reveals the importance of the unique geometry of the $\Voyager$ occultations in helping to constrain the pole direction and radius scale. The corresponding elongated solid blue error ellipse in Fig.~\ref{fig:pole} reflects the strong correlation $\rho(\alpha_P,\delta_P)=-0.88$ resulting from the limited range of viewing geometry of the ring plane available from the ensemble of Earth-based observations. Although the $\Voyager$ data contributed only 38 ring measurements (compared to 613 Earth-based measurements), they significantly reduce the estimated error in the pole direction (and radius scale, as we show below). Note that the error ellipses for Fits 1 and 2 are nearly centered on the Fit 5 error ellipse, reflecting the fact that the $\Voyager$ ephemeris was constrained by both the Earth-based and spacecraft ring observations.
\item Fit 5: \cite{Jacobson2023} solution. This pole solution was determined as part of the development of the {\tt ura178} series of ephemerides that made use of nearly all the Uranus ring occultation data in Paper 1, with the exception of the $\gamma$ and $\lambda$ rings. Extensive detailed comparisons of fits using only the COR ring data showed virtually perfect agreement between Jacobson's test cases (personal communication) and our Fit 1 results, a robust confirmation of the validity of our two independent orbit fitting codes. In contrast to Fits 1--4, the Fit 5 pole solution was based on a comprehensive fit to extensive historical astrometric and spacecraft navigational data, resulting in a somewhat larger error ellipse (displaced slightly from our adopted Fit 1 pole) that incorporates uncertainties in these additional data sources and provides a more conservative estimate of the accuracy of the pole determination.
\item Fit 6: \cite{Jacobson2014} solution. The \cite{Jacobson2014} pole direction at the J2000 epoch ($\alpha=77.310\pm0.002^\circ$ and $\delta=15.172\pm0.002^\circ$) was derived using a subset of the COR ring occultation observations included here, excluding the $\gamma$ and $\delta$ rings and post-1991 data. The \Voyager\ occultation geometry was calculated from the
{\tt vgr2.ura111.bsp} ephemeris, which was developed using extensive spacecraft navigation and satellite astrometry observations, but no ring occultation data. The \cite{Jacobson2014} pole is shown as the dotted brown error ellipse with an assumed correlation $\rho(\alpha_P,\delta_P)=0$. The systematic offset from the Fit 1 pole position is due primarily to differences between the updated {\tt vgr2.ura178.bsp} ephemeris and the earlier spacecraft ephemeris, as well as to the use by \cite{Jacobson2014} of keplerian ring orbital elements that modeled the $\eta$ ring as inclined and eccentric, excluded the $\gamma$ and $\delta$ rings, and neglected the contributions of normal modes to the $\eta$ and $\epsilon$ ring shapes.\footnote{In contrast, the \cite{Jacobson2023} solution for the Uranus system geometry fitted for the same normal modes in the ring midlines as in our orbit fits, and closely matched our keplerian ring orbital elements.}
\item Fit 7: \cite{French1991} solution. The \cite{French1991} pole direction at the B1950 epoch ($\alpha=76.5969\pm0.0034^\circ$ and $\delta=15.1117\pm0.0033^\circ$) was derived using a subset of the ring occultation observations included here, excluding the post-1991 data, and based on an early \Voyager\ ephemeris. The result (precessed to the J2000 frame) is shown as the dotted green error ellipse with an assumed correlation $\rho(\alpha_P,\delta_P)=0$. The fitted pole direction is remarkably close to the Fit 1 result, although with a much larger error estimate. The similarity in pole directions for Fits 1 and 7 is due in part to the largely overlapping data sets and to a likely similarity in the two \Voyager\ ephemerides.
\end{itemize}

\begin{deluxetable}{c r c c r r r r l}
\tablecolumns{9}
\tabletypesize{\scriptsize} 
\tablecaption{Uranus pole direction and ring plane radius scale
\label{tbl:radiuspole}}
\tablewidth{0pt}
\tablehead{
\colhead{Fit} & 
\colhead{$\alpha_P$ (deg)} & 
\colhead{$\delta_P$ (deg)} & 
\colhead{$\rho^{(a)}$}&
\colhead{$\Delta\alpha_P$ (deg)}&
\colhead{$\Delta\delta_P$ (deg)}&
\colhead{$\langle\Delta{\rm a}\rangle$ (km)} & 
\colhead{$\sigma(\Delta{\rm a})$ (km)} &
\colhead{Description}
}
\startdata
1 &  $ 77.311327 \pm  0.000128 $  &  $ 15.172795 \pm  0.000472 $  &  $ -0.22 $  &  ---   &   --- & --- & ---  & nominal pole direction \\
2 &  $ 77.311327 \pm  0.000141 $  &  $ 15.172795 \pm  0.000618 $  &  $ -0.42 $  &  ---   &   --- & --- & ---  & w/ Vgr trajectory errors \\
3 &  $ 77.311210 \pm  0.000128 $  &  $ 15.172762 \pm  0.000471 $  &  $ -0.22 $  &  $ -0.000117 $  &  $ -0.000033 $  &  $ -0.001 $  &  0.003 & {\tt ura178} pole \\
4 &  $ 77.311246 \pm  0.000274 $  &  $ 15.173393 \pm  0.002013 $  &  $ -0.88 $  &  $ -0.000081 $  &  $  0.000597 $  &  $  0.278 $  &  0.027 & Earth-based only \\
5 &  $ 77.311200 \pm  0.000400 $  &  $ 15.172400 \pm  0.001700 $  &  $ -0.39 $  &  $ -0.000127 $  &  $ -0.000395 $  &      ---   &   ---  & Jacobson (2023) \\
6 &  $ 77.310000 \pm  0.003000 $  &  $ 15.172000 \pm  0.002000 $  &     [1.0]      & $ -0.001327 $  &  $ -0.000795 $  &  $ -0.015 $  &  0.091 & Jacobson (2014) \\
7 &  $ 77.310877 \pm  0.003400 $  &  $ 15.174564 \pm  0.003300 $  &     [1.0]      & $ -0.000450 $  &  $  0.001768 $  &  $ -0.055 $  &  0.092 & French et al. (1991) \\
\enddata
\tablenotetext{(a)}{\ \ Correlation coefficient $\rho(\alpha_P,\delta_P)$}
\end{deluxetable}

\begin{figure}
\centerline{\resizebox{6in}{!}{\includegraphics[angle=90]{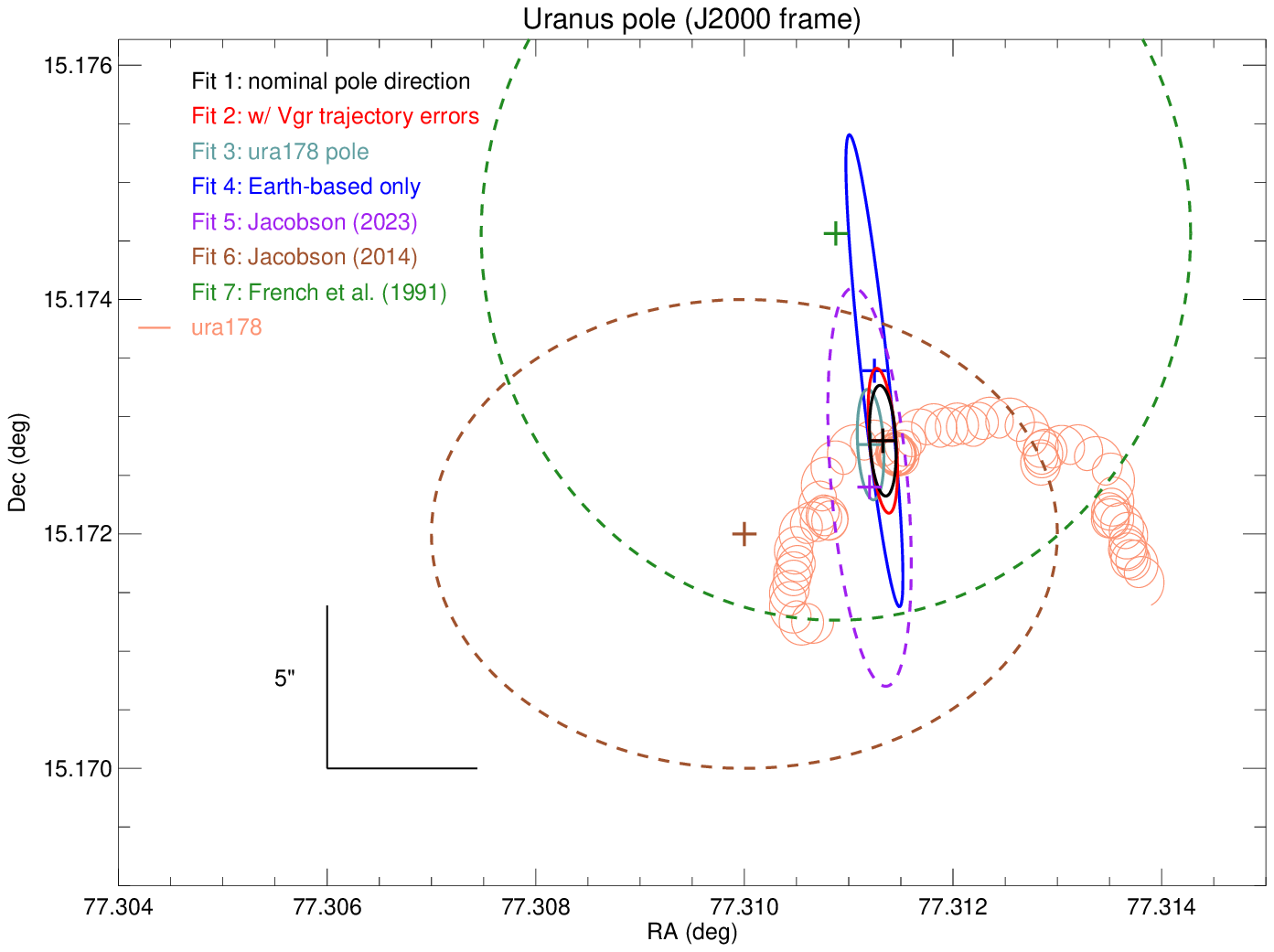}}}
\caption{Uranus pole direction in the J2000 reference frame from the fits listed in Table \ref{tbl:radiuspole}. Ellipses are shown as solid lines for Uranus ring and pole fits performed using the data in Paper 1, and as dotted lines for previously published results. See text for details of each fit.}
\label{fig:pole}
\end{figure}
\subsection{Ring plane radius scale}
\label{sec:radscale} 
A secure dynamical association between satellite resonances and the narrow Uranian rings depends on the accuracy of the absolute radius scale. Figure~\ref{fig:radiusscale} shows the differences $\Delta a$ in the semimajor axes of the nine main rings between the fits listed in Table~\ref{tbl:radiuspole} and our adopted geometric solution (Fits 1 and 2 have identical pole direction and ring orbital elements, and differ only in the uncertainty in the pole direction), plotted as a function of orbital radius and displaced horizontally for clarity. (Note that $\Delta a$ should not be confused with $\Delta a_{\dot\varpi}, \Delta a_{\dot\Omega},$ and $\Delta a_{\Omega_P}$ in Tables ~\ref{tbl:orbel} and \ref{tbl:normalmodes}.) Table \ref{tbl:radiuspole} includes $\langle \Delta a\rangle$, the average offset in fitted semimajor axes from the Fit 1 values, and $\sigma(\Delta a)$, the standard deviation of the offsets. The error bars on each point in the figure correspond to the individual 1-$\sigma$ uncertainties in the determination of the semimajor axis for each ring in each fit. The Fit 1 error bars are those in our adopted orbit fit in Table \ref{tbl:orbel}. As described above, for Fit 2 we incorporated the uncertainties in the \Voyager\ ephemeris into the overall error estimate for the pole direction. From this same series of Monte Carlo simulations, the RMS scatter in the fitted semimajor axes of the rings due to spacecraft trajectory uncertainties was quite small, varying from 0.033 km for ring 6 to 0.027 for the $\epsilon$ ring. The error bars shown (in red) for Fit 2 were computed by combining these small errors in quadrature with the formal errors in Fit 1. The results are so similar to those in Fit 1 that we have not modified the error bars for the orbital elements in Table \ref{tbl:radiuspole} to include the very small contribution due to spacecraft ephemeris uncertainties. The fitted semimajor axes and orbital elements from Fit 3, which included pole precession, are virtually identical to those in Fits 1 and 2, with $\langle \Delta a\rangle=-0.001$ km and $\sigma(\Delta a)=0.003$ km.

\begin{figure}
\centerline{\resizebox{6in}{!}{\includegraphics[angle=90]{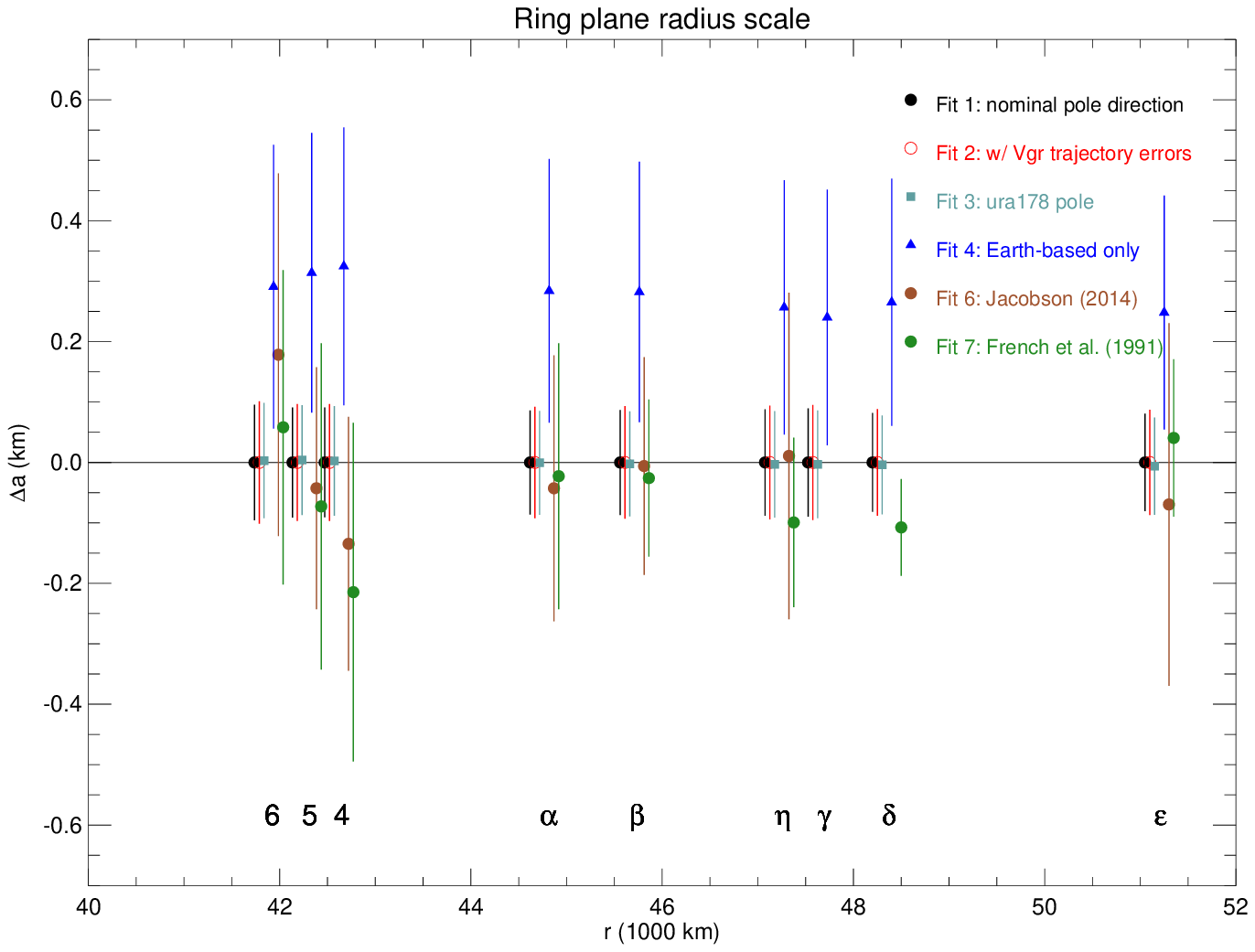 }}}
\caption{Ring plane radius scale. The differences $\Delta a$ in the semimajor axes of the nine main rings between fits listed in Table~\ref{tbl:radiuspole} and our adopted geometric solution are shown as a function of orbital radius, displaced horizontally for clarity. The error bars on each point are the individual 1-$\sigma$ uncertainties in the determination of the semimajor axis for each ring in each fit.}
\label{fig:radiusscale}
\end{figure}

Fit 4 (shown as blue squares in Fig.~\ref{fig:radiusscale}), based solely on Earth-based data, deserves special attention. The fitted semimajor axes are systematically higher ($\langle \Delta a\rangle=0.278$ km), and with substantially larger error bars than in Fits 1 and 2 that included the \Voyager\ data. The systematic offset in radius scale is directly related to the corresponding offset in the fitted pole directions between the same two fits shown in Fig.~\ref{fig:pole}, and reflects the strong correlation between the absolute radius scale and the assumed pole direction. This is understandable on a geometric basis insomuch as the separation of ingress and egress ring event times depends on the projection of the ring orbits into the skyplane resulting from the assumed orientation of the pole.

Finally, the radius scales of two previous determinations that included \Voyager\ observations, but based on different spacecraft ephemerides, are very similar to our adopted solution: the Fit 6 \citep{Jacobson2014} semimajor axes differed from the Fits 1 and 2 values on average by only $\langle \Delta a\rangle=-0.015$ km, with an RMS scatter $\sigma(\Delta a)=0.092$ km, and the corresponding Fit 7 \citep{French1991} results were $\langle \Delta a\rangle=-0.053$ km and $\sigma(\Delta a)=0.093$ km.

Based on the demonstrated importance of the \Voyager\ observations to constraining the pole direction (Fig.~\ref{fig:pole}), the resultant reduction in error estimates for the ring radii, and the similarity of the present and previously-published determinations of the radius scale based on independent \Voyager\ ephemerides, we estimate that our adopted radius scale (Fits 1 and 2) is accurate to $\sim$0.2 km at the 2-$\sigma$ level.
%
\section{Ring Widths, Shapes, and Masses}
\label{sec:widths}

We now turn to an examination of the widths, shapes, and masses of the rings, based on the centerline (COR) and edge (IER/OER) fits given in Tables \ref{tbl:orbel} and \ref{tbl:normalmodes}. 
Since differential precession of an eccentric ringlet would result in streamline crossing on a dynamically short timescale compared to the presumed age of the ring, it is commonly assumed that narrow rings precess uniformly, enforced either by the collective effects of collisions and self-gravity or by external forcing from a satellite (see \cite{Longaretti2018} for a detailed review of these and other proposed mechanisms). The self-gravity explanation was originally proposed by \cite{GT79}, motivated by the observed elliptical shape of the $\epsilon$ ring. 

\cite{BGT83} developed a two-streamline self-gravity formalism that \cite{Nicholson2018b} applied to observations of narrow rings, and it provides a useful starting point for interpreting the observed variation of width with ring radius.\footnote{We express these in terms of the {\it observed} widths and shapes of the normal modes at the actual ring edges and centerlines, rather than in terms of the spacing and eccentricities of the two streamlines, as assumed by \cite{BGT83} and \cite{Longaretti2018}. These alternate approaches differ by a factor of two in the definitions of several important geometric quantities. We identity these differences where they occur.}
In this model, differential precession is balanced by self-gravity for a free $m = 1$ mode (\ie\ not forced by a satellite) when the following equilibrium condition (Eq.~(10.109), \cite{Longaretti2018} holds:
\beq
\frac{\delta \epsilon}{\epsilon} = \frac{21\pi}{4}J_2\biggl(\frac{R}{a}\biggr)^2\frac{M_{\rm Ur}}{M_r}\biggl(\frac{\delta a'}{a}\biggr)^3\frac{1}{H(q_e^2)},
\label{eq:PY109}
\eeq
where $\epsilon$ and $\delta\epsilon$ are the mean eccentricity and the difference in eccentricities of the two streamlines, $a$ and $\delta a'$ are the mean value and difference in the streamline semimajor axes, $J_2$ is the second-order gravitational harmonic coefficient with reference radius $R$, $M_{\rm Ur}$ is the planet mass, $M_r$ is the ring mass,  $q_e=a \delta\epsilon/\delta a'$, and
\beq
H(q_e^2) = \frac{1 - (1-q_e^2)^{1/2}}{q_e^2(1-q_e^2)^{1/2}}.
\label{eq:Hq2}
\eeq
Relating the two-streamline quantities $\epsilon, \delta\epsilon$ and $\delta a'$ to the observed eccentricities and semimajor axes of the ring edges, we have
\beq
\epsilon = e =  e_{\rm COR},
\eeq
\beq
\delta e =e_{\rm OER}-e_{\rm IER} \approx 2\delta\epsilon,
\label{eq:deltae}
\eeq
\beq
\delta a = a_{\rm OER} - a_{\rm IER}\approx 2\delta a',
\label{eq:deltaa}
\eeq 
and
\beq
a= a_{\rm COR}
\eeq
The eccentricity gradient $q_e$ is identical for both the two-streamline model and the actual ring observations:
\beq
q_e = \frac{a \delta\epsilon}{\delta a'}=\frac{a\delta e/2}{\delta a/2} = \frac{a\delta e}{\delta a}.
\label{eq:qe}
\eeq
The variation in the observed width with orbital radius around the elliptical ring is $dW/dr\simeq a \delta e/ae=\delta e/e=2a\delta\epsilon/ae=2\delta\epsilon/\epsilon$. Explicitly including the factors of two in Eqs.~(\ref{eq:deltae}) and (\ref{eq:deltaa}) when translating from the two-streamline quantities $\delta \epsilon$ and $\delta a'$ to their observational counterparts $\delta e$ and $\delta a$, we have
\beq 
\frac{dW}{dr} = \frac{\delta e}{e} = \frac{2\delta\epsilon}{\epsilon}.
\label{eq:selfgrav}
\eeq
(The dimensionless quantity $dW/dr\approx \delta e/e$ should not be confused with $q_e=a \delta e/\delta a$.)

For $m\neq 1$, the corresponding result is 
\beq
\frac{dW_m}{dr}=\frac{\delta e_m}{e_m}=\frac{2\delta\epsilon_m}{\epsilon_m},
\label{eq:selfgravm}
\eeq
where $a\delta e_m$ is the difference in the mode amplitudes between the outer and inner ring edges and $a e_m$ is the COR mode amplitude $A_m$. (Note that the eccentricity gradient for $m\neq 1$ is $q_{em}=a\delta_{em}/\delta a$.)
For modes with $m\leq0$, the self-gravity model predicts that the slope of the width-radius relation will be negative.
In the derivation of Eqs.~(\ref{eq:selfgrav}) and (\ref{eq:selfgravm}), an underlying assumption of the two-streamline formalism is that the 
fractional difference in mode amplitude $e_m$ is small, which we will see is not necessarily true for the $\eta,\gamma,$ and $\delta$ rings.

An additional quantity of interest is the mean gradient in periapse longitude, given by
\beq
q_{\varpi}={ -a e\delta\varpi_{0}}/{\delta a},
\label{eq:qvarpi}
\eeq
where $\delta\varpi_{0}=\varpi_{\rm OER} - \varpi_{\rm IER}$ is the difference between the outer and inner ring edge pericenter longitudes.
Eqs.~(\ref{eq:qe}) Eq.~(\ref{eq:qvarpi}) together define a generalized $q$:
\beq
\begin{split}
q \cos \gamma & = q_{e}\\
q \sin\gamma & = q_{\varpi},
\end{split}
\eeq
where $\gamma$ is the phase lag of the minimum width of the ring relative to pericenter. Similar expressions apply for the case $m\neq 1$.  The ring width varies as
\beq
W(f) = \delta a[1-q\cos(f-\gamma)],
\eeq
where $f=\lambda-\varpi$ is the true anomaly of the ring's centerline.
As noted by \cite{Nicholson2018b}, $\tan \gamma \simeq -e\delta\varpi/\delta e$, and thus the phase lag can be much larger than the pericenter offset $\delta\varpi$ if $\delta e\ll e$.
For a single mode, we require that $q<1$ to avoid streamline crossing, although in the simultaneous presence of multiple modes this restriction may not apply to each mode individually. 

In the idealized case of a uniformly precessing ring, the fitted apse rates of the IER, COR, and OER would be equal:
\beq
\dot\varpi(a_{\rm IER}) = \dot\varpi(a_{\rm COR}) = \dot\varpi(a_{\rm OER}).
\eeq
The inner edge of an eccentric ringlet must thus precess more slowly than expected for an isolated particle in orbit with semimajor axis $a_{\rm IER}$, and the opposite is the case at the outer edge. Recalling the definitions from Section \ref{sec:orbfitCORIEROER} given in Eqs.~(\ref{eq:Delta_adotvarpi}) -- (\ref{eq:Delta_aOmegap}) (\ie\ that the fit parameter $\Delta\dot\varpi$ is the difference between the observed precession rate and that predicted from the planet's gravity harmonics and satellite precession contributions at the edge in question), uniform precession predicts that at the IER, one would expect $\Delta\dot\varpi<0$ and $\Delta a_{\dot\varpi}>0$. Similarly, for the OER one would expect $\Delta\dot\varpi>0$ and $\Delta a_{\dot\varpi}<0$. For $m\neq1$, the corresponding conditions at the IER are $\Delta\Omega_P<0$ and $\Delta a_{\Omega_P}>0$ and for inclined ringlets, $\Delta\dot\Omega>0$ and $\Delta a_{\dot \Omega}>0$. 

Under equilibrium conditions, where the inner and outer ring edges precess at the same rate as the COR, the simple self-gravity model predicts that $\Delta a_{\dot\varpi}({\rm COR})=0 $ and thus that 
\beq
\begin{split}
\Delta a_{\dot\varpi}({\rm IER}) - \Delta a_{\dot\varpi}({\rm COR}) &= \Delta a_{\dot\varpi}({\rm IER})  = a_{\rm COR} - a_{\rm IER} >0\\
\Delta a_{\dot\varpi}({\rm OER}) - \Delta a_{\dot\varpi}({\rm COR}) &= \Delta a_{\dot\varpi}({\rm OER}) = a_{\rm COR} - a_{\rm OER}<0.
\end {split}
\eeq

With this dynamical framework in mind, we next describe the shapes, widths, width-radius relations, and comparative precession rates across each ring obtained from the fitted keplerian and normal mode elements from Tables \ref{tbl:orbel} and \ref{tbl:normalmodes}.
{\bf Table \ref{tbl:widths}} lists the following quantities for each ring: mean semimajor axis $\bar a$, width $\overline W=\delta a$, wavenumber $m$, mode amplitude $A_m$, difference in mode amplitudes between the inner and outer edges $a\delta e_m = A_m({\rm OER}) - A_m({\rm IER})$, pericenter longitude difference between the outer and inner edges $\delta \varpi_{0m}$, gradients $q_{em}$ and $q_{\varpi m}$, width-radius relation $dW_m/dr =a\delta e_m/\bar a e_m$, and $\delta\Omega_P$, the observed difference in pattern speeds between the outer and ring edges, defined for $m=1$ as
\beq
 \delta\Omega_P =\delta\dot\varpi =  \dot\varpi({\rm OER}) - \dot\varpi(\rm{ IER})
\eeq
 and for $m\neq1$ as
 \beq
 \delta\Omega_P = \Omega_P({\rm OER}) - \Omega_P({\rm IER}).
 \eeq

\begin{deluxetable}{c c r r r r r r r r r }
\tablecolumns{11}
\tablecaption{Width-radius results\label{tbl:widths}}
\tablewidth{0pt}
\tablehead{ 
\colhead{Ring} & 
\colhead{$\bar a$} & 
\colhead{${\overline{W}}$} & 
\colhead{$m$} & 
\colhead{${A_m }$} &  
\colhead{$a\delta e_m$} &  
\colhead{${\delta\varpi_{0m}}$} &  
\colhead{${q_{em}}$} &  
\colhead{${q_{\varpi m}}$} &   
\colhead{$dW_m/dr$} &  
\colhead{${\delta\Omega_p}$} \\[-1em] 
\colhead{ } & 
\colhead{ km} & 
\colhead{km} & 
\colhead{ } & 
\colhead{km} &  
\colhead{km} &  
\colhead{deg} &  
\colhead{ } &  
\colhead{ } &   
\colhead{ } &  
\colhead{\degyr}    
}
\startdata
6 & 41837.092 &     2.316 &   1 & $  42.499 $  & $   0.862 $  & $  -0.207 $  & $   0.372 $  & $   0.077 $  & $   0.020 $  & $  -0.035 $  \\
5 & 42234.893 &     2.684 &   1 & $  80.237 $  & $   0.412 $  & $   0.028 $  & $   0.153 $  & $  -0.004 $  & $   0.005 $  & $  -0.020 $  \\
4 & 42571.124 &     3.231 &   1 & $  45.347 $  & $   0.810 $  & $  -1.103 $  & $   0.251 $  & $   0.277 $  & $   0.018 $  & $   0.125 $  \\
$\alpha$ & 44718.473 &     7.277 &   1 & $  33.916 $  & $   2.823 $  & $   0.359 $  & $   0.388 $  & $  -0.139 $  & $   0.083 $  & $   0.237 $  \\
$\beta$ & 45661.056 &     8.516 &   1 & $  20.106 $  & $   3.172 $  & $  -2.967 $  & $   0.372 $  & $   1.105 $  & $   0.158 $  & $  -0.043 $  \\
$\eta$ & 47176.009 &     2.227 &   3 & $   0.600 $  & $   0.224 $  & $  -1.262 $  & $   0.101 $  & $   0.127 $  & $   0.374 $  & $  -0.540 $  \\
$\gamma$ & 47626.170 &     3.258 &   1 & $   5.306 $  & $   0.454 $  & $  -0.129 $  & $   0.139 $  & $   0.018 $  & $   0.086 $  & $   0.280 $  \\
 & & &   0 & $   5.509 $  & $  -1.435 $  & $   3.605 $  & $  -0.440 $  & $   1.587 $  & $  -0.261 $  & $   0.135 $  \\
 & & &   6 & $   0.637 $  & $   0.302 $  & $  -5.378 $  & $   0.093 $  & $   0.499 $  & $   0.475 $  & $   0.530 $  \\
 &   &      & $  -1$  & $   1.822 $  & $   0.299 $  & $  -6.051 $  & $   0.092 $  & $   0.556 $  & $   0.164 $  & ---  \\
 &    &     & $  -2$  & $   0.690 $  & $  -1.099 $  & --- & $  -0.337 $  & --- & $  -1.592 $  & $   4.742 $  \\
 &    &     & $   3$  & $   0.577 $  & $   0.935 $  & --- & $   0.287 $  & --- & $   1.622 $  & $  -0.790 $  \\
$\delta$ & 48300.227 &     4.977 &   2 & $   3.169 $  & $   1.686 $  & $  -0.438 $  & $   0.339 $  & $   0.148 $  & $   0.532 $  & $  -0.146 $  \\
 &  & &  23& $   0.339 $ &  --- & --- & --- & --- & --- & --- \\
$\epsilon$ & 51149.279 &    58.574 &   1 & $ 405.894 $  & $  38.712 $  & $  -0.028 $  & $   0.661 $  & $  -0.019 $  & $   0.095 $  & $  -0.003 $  \\
\enddata
\end{deluxetable}


        \subsection{Rings 6, 5, 4, $\alpha$, and $\beta$}
        \label{sec:654ab}
We begin with the five inner rings, each of which has a measurable eccentricity and inclination but no detected free or forced normal modes. In {\bf Fig.~\ref{fig:widths6beta}}, each column shows key characteristics for a given ring. 
In the top row, $\Delta r(\lambda)= r(\lambda) - a_{\rm COR}$ is the difference between the model ring radius $r(\lambda)$ of the $m=1$ mode (computed from Eq.~(\ref{eq:kepler}) at the epoch of the orbit fit) and the mean semimajor axis $a_{\rm COR}$, plotted as a function of inertial longitude $\lambda$ at epoch for each ring's inner edge (blue), midline (green), and outer edge (red). 
\begin{figure}
\centerline{\resizebox{\textwidth}{!}{\includegraphics[angle=90]{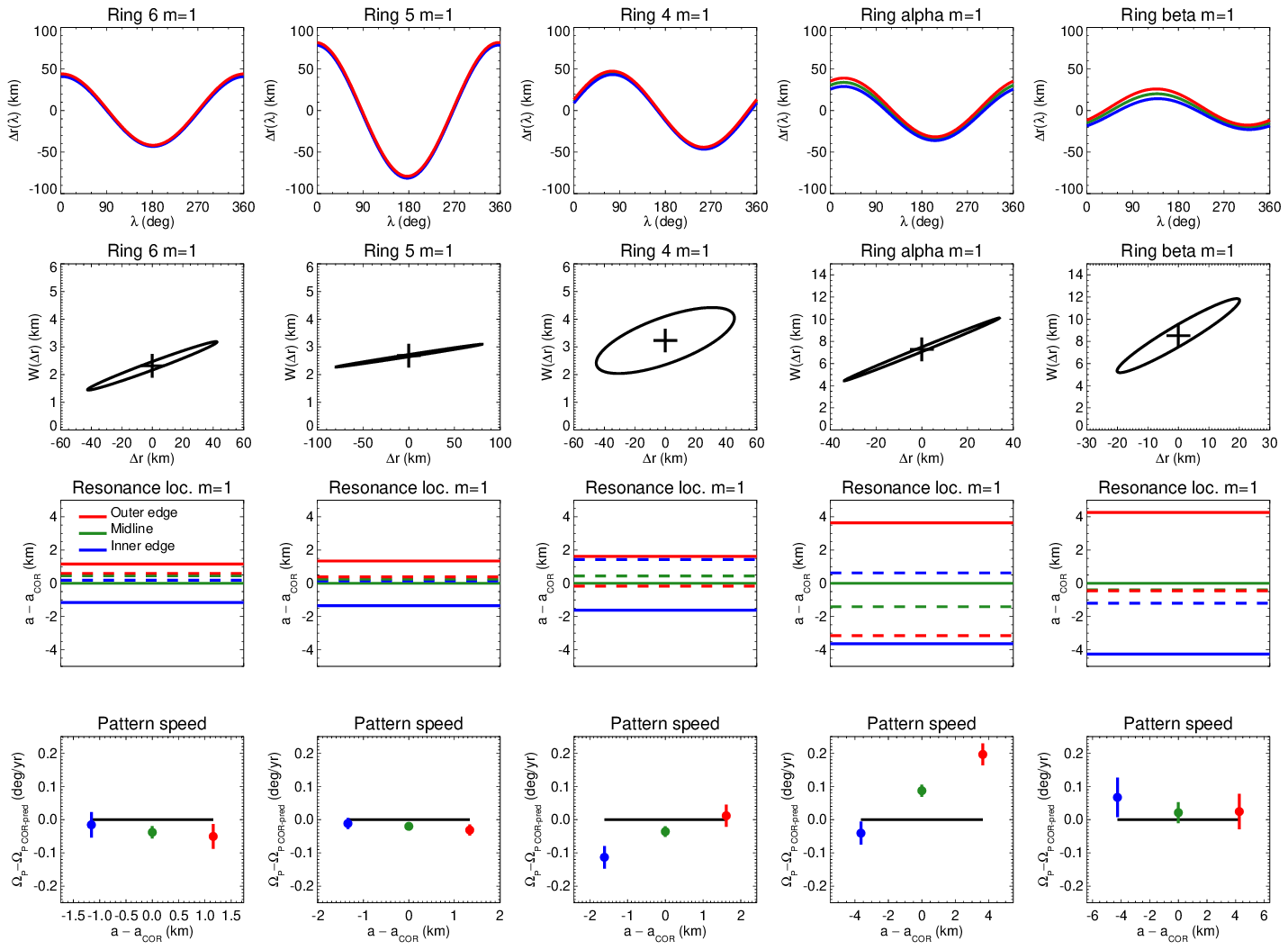}}}
\caption{Width variations of rings 6, 5, 4, $\alpha$, and $\beta$. The top row shows $\Delta r(\lambda)$, the variation of ring radius with inertial longitude $\lambda$ at epoch for the IER, COR, and OER (blue, green, and red, respectively. The second row shows the corresponding variation of ring width $W(\Delta r)$, and the third row shows the predicted resonance radii for each detected mode (shown dashed) relative to ring's semimajor axis $a_{\rm COR}$, compared to the locations of the ring edges (shown as solid lines). The final row shows the differences in the observed pattern speeds of the inner and outer edges, and that of the COR, relative to the {\it predicted} COR value (based on the Uranus gravity field and satellite contributions to the precession rates), plotted as a function of semimajor axis relative to $a_{\rm COR}$.}
\label{fig:widths6beta}
\end{figure}
In the second row, the ring widths $W(\Delta r)$ are plotted as a function of $\Delta r= r_{\rm COR} - a$. 
When the longitudes of the periapses of the two edges are aligned, $W(r)$ is a straight line with slope $dW/dr=\delta e/e$,  but for misaligned apses, the shape is an ellipse. All five inner rings show a positive slope, as predicted for a self-gravitating eccentric ring by Eq.~(\ref{eq:selfgrav}). The apses are most closely aligned for rings 6, 5, and $\alpha$, with more significant misalignments for rings 4 and $\beta$, as indicated by the elliptical shapes of the width-radius relations.

The third row of Fig.~\ref{fig:widths6beta} shows the semimajor axes (dashed lines) calculated from the fitted pattern speeds of the normal modes (in this case, the periapse precession rates) at the ring edges and midlines, compared to the observed ring midline and edge semimajor axes themselves, shown as solid lines. The computed semimajor axes (generically referred to as ``resonance radii'' to include the case of modes forced by satellites) based on the combined precessional effects of the gravity field of the planet and the known satellites closely match the ring midlines for all but ring $\alpha$, which together with ring $\beta$ precess somewhat faster than predicted for the mean radii of these rings. This anomalous precession may be due to small unseen satellites, as discussed below in Section \ref{sec:precrates}.
 
Finally, the fourth row compares the observed pattern speeds of the IER, COR, and OER relative to the {\it predicted} pattern speed of the COR, plotted as a function of semimajor axis. In the ideal case of a uniformly precessing ring, the precession rates $\dot\varpi$ of the IER, COR, and OER should be equal. Systematic offsets between the observed and predicted COR precession rates are in some cases due to forced satellite modes, where the forcing frequency is close to but not equal to the COR precession rate, and is also affected in some cases by the difference between the geometric midline of a ring and its radial center of mass, an issue addressed in Section \ref{sec:COO}.
Statistically 
significant differences in the pattern speeds at the ring edges such as those seen for rings 4 and $\alpha$ could be indications of libration or circulation of a normal mode, as has been observed at the outer edge of Saturn's B ring \citep{Spitale2010,Nicholson2014a,French2023b}. We explore these possibilities in Section \ref{sec:librations}.

        \subsection{The $\eta$ and $\delta$ rings}
       {\bf Figure~\ref{fig:widthsetadelta}} shows the corresponding results for the $\eta$ and $\delta$ rings. In the top panel, the edge and midline locations are plotted as a function of $m\lambda$, since there are $m$ complete normal mode cycles around the circumference of the ring. The $\eta$ ring is nearly constant in width, with a very small but positive $dW/dr$, and has nearly aligned mode pericenters. The $m=3$ normal mode is clearly driven by the 3:2 ILR with Cressida, as discussed by \cite{Chancia2017} and confirmed here by the resonance locations for the edges and midline being about 5 km interior to the ring and well-matched to the Cressida resonance radius calculated from the satellite mean motion of \cite{Showalter2006}, labeled as SL06. The fitted $m=3$ pattern speeds for the IER, COR, and OER agree within the error bars, but are controlled by the forcing frequency of Cressida, rather than by the gravity field of the planet, resulting in the large vertical offset in the left panel in the fourth row of the figure.
       
The  $\delta$ ring $m=2$ mode has a substantial width variation, with nearly identical calculated resonance radii for the IER, COR, and OER falling very near to the ring midline. It has a positive width-radius relation, as expected from Eq.~\ref{eq:selfgravm} for a self-gravitating ring with free normal mode $m>1$, with nearly aligned pericenters at the ring midline and edges. The fitted $m=2$ pattern speeds for the IER, COR, and OER agree within the error bars, as shown in the middle panel in the fourth row of the figure.

The $\delta$ ring $m=23$ mode is relatively weak, with $A_m=0.342\pm0.063$~km, and is detectable only in the ring midline (see Table \ref{tbl:normalmodes}) -- the outer and inner ring edges are shown as dashed lines in the upper right panel of the figure, and $W_m(\Delta r)$ is unobserved. The resonance location for the observed pattern speed matches the \cite{Jacobson98} value for the mean motion of Cordelia (labeled Jac98 in the figure), and lies just inside the outer edge of the $\delta$ ring. Once again, there is a large vertical offset between the observed pattern speed and the predicted COR pattern speed, since the $m=23$ mode is controlled by the forcing frequency of Cordelia. We will make use of the measured amplitude and resonance location to estimate Cordelia's mass in Section \ref{sec:satellites}.
\begin{figure}
\centerline{\resizebox{\fw}{!}{\includegraphics[angle=0]{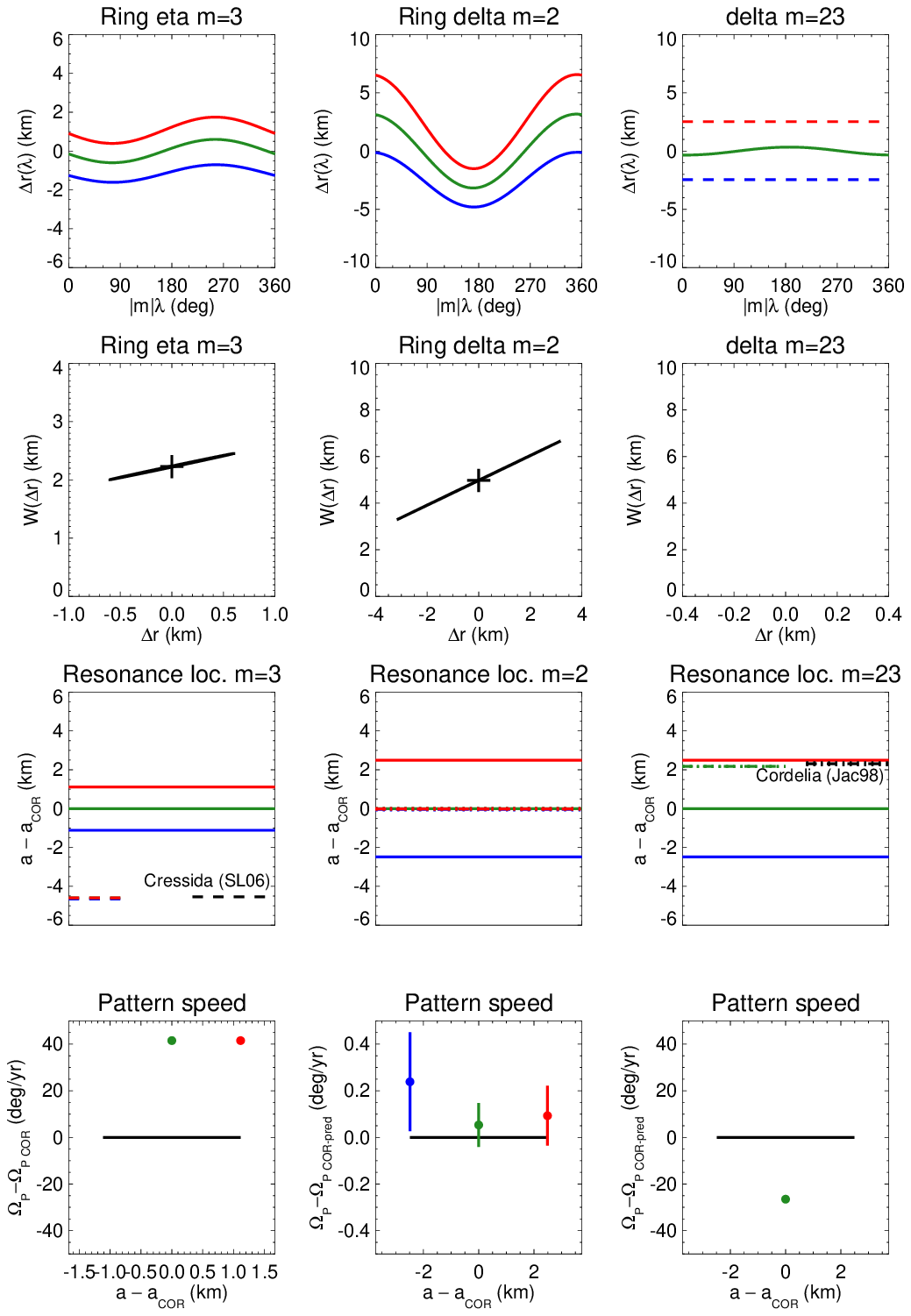}}}
\caption{Width variations of rings $\eta$ and $\delta$, with the same layout as for Fig.~\ref{fig:widths6beta}. 
 The predicted resonance location for Cressida, assuming the mean motion given by \cite{Showalter2006} (labeled SL06), is shown as a dashed line for the $\eta$ ring $m=3$ mode, and lies $\sim5$ km interior to the ring's midline. The predicted resonance location for Cordelia, assuming the mean motion given by \cite{Jacobson98} (labeled Jac98), is shown as a dashed line for the $\delta$ ring $m=23$ mode, and lies just interior to the outer edge of the ring. The $m=23$ mode was not detected in the shape of the IER or OER of the $\delta$ ring. 
 }
\label{fig:widthsetadelta}
\end{figure}
        
        \subsection{The $\gamma$ ring}
The $\gamma$ ring has the most complex shape of all the Uranian rings, being distorted by six distinct normal modes. The contributions of each mode to the ring width and shape are shown in {\bf Figure~\ref{fig:widthsgamma}}, with the following notable features:
\begin{itemize}
\item The $m=1$ mode has a large COR amplitude $A_1=5.306\pm0.071$ km and well-aligned apses, but it contributes rather little to the overall width variation of the ring itself, with a very small but positive slope $dW_m/dr=0.086$. The outer edge of the ring shows an anomalously high $m=1$ precession rate: from Table \ref{tbl:orbel} (for the model including the $m=3$ normal mode) %
the resonance radius computed from the fitted apse rate for the OER lies %
8.591 km interior to the COR semimajor axis, as shown in the left panel of row three of Fig.~\ref{fig:widthsgamma}.
\item The $m=0$ mode has a similarly large COR amplitude $A_0=5.511\pm0.076$ km, but unlike the $m=1$ mode, it is narrowest at apoapse and widest at periapse, resulting in a negative value of $dW_m/dr$ and slightly misaligned periapse phases, producing the elliptical shape of the $W_m(\Delta r)$ plot. The resonance location closely matches the semimajor axis of the ring. The negative slope of the width-radius relation is consistent with the self-gravity model prediction for 
an outer Lindblad resonance (OLR) with $m\le0$ (see Eq.~(\ref{eq:selfgravm})). The width of the $\gamma$ ring is primarily controlled by the $m=0$ mode and not the $m=1$ mode, even though both modes have comparable $ae\sim5$ km. This observation supports the physical argument posited by \cite{Nicholson2018b} that self-gravity needs to overcome only the relatively slow differential precession for an $m=1$ perturbation, whereas for $m\neq 1$ it is necessary to overcome the much greater gradient in mean motion.
\item The weaker $m=6$ mode is driven by the 6:5 ILR with Ophelia, with edge and midline pattern speeds matching the \cite{Jacobson98} value for the satellite mean motion within its error bars (labeled Jac98 in the figure). The more accurate recent Ophelia mean motion (Robert French, personal communication -- labeled RF23) has a predicted resonance location that closely matches that of the observed $m=6$ pattern speed, lying within the ring itself. The proximity of the Ophelia 6:5 resonance to the inner edge of the $\gamma$ ring is surprising and the opposite of standard theories of resonant shepherding. A potential analog to this structure is the ringlet ER1 in Saturn's inner C ring. The outer edge of this ringlet falls near a resonance with a planetary normal mode \citep{French2021}, which is also the opposite of expectations since the pattern speed of this normal mode is faster than the ring's orbital motion. Both these rings may have both their edges shepherded by their respective resonances, as predicted by \cite{BGT86}.
\item The $m=-1$ mode has a small positive $dW_m/dr= 0.164$, contradicting the expectation that for an isolated OLR, the slope should be negative. Given the multiple simultaneous modes in the $\gamma$ ring and the small amplitude of this mode, this apparent discrepancy may not be dynamically significant.
\item The $m=-2$ mode is undetectable on the outer edge of the ring, and is strongest on the inner edge. In this case, $dW_m/dr=-1.592$, negative as expected for this OLR, and second in magnitude of all measured modes in any of the rings, with the exception of the $m=3$ mode described next.
\item The $m=3$ mode was detected only on the outer edge and midline of the ring, with a positive $dW_m/dr=1.622$, the largest value for any of the modes identified here. The fitted apses of the COR and OER $m=3$ modes are well-aligned and match closely in pattern speed. There is no known satellite that forces this mode. On the other hand, if this is a free mode, its resonance radius is quite far removed from the ring midline, unlike the other free modes identified in the Uranus system. The pattern speed of the $m=3$ structure is much less than that expected for fundamental planetary normal modes \citep{AHearn2022} and is significantly larger than the planet's rotation rate \citep{Desch1986, Warwick1986, Helled2010}. Thus this pattern is unlikely to be generated by either a fundamental planetary normal mode or a persistent anomaly carried around by the planet's winds similar to those that produce waves in Saturn's rings \citep{Hedman2022}. If this signal turns out to be real and generated by the planet, it therefore would have to be due a more complex and possibly mixed planetary mode.
A final possibility is that this putative detection is instead a random pattern emerging from the co-addition of five other modes to the shape of the ring. 
We will see below in Section \ref{sec:precrates}, however, that exclusion of this mode from the ensemble of $\gamma$ ring modes results in a very large anomalous $m=1$ precession rate with no identifiable source. Furthermore, the RMS residuals for the $\gamma$ ring COR fit when the mode is absent increase from 0.401 km to 0.490 km, the largest of any of the rings; for the OER, the increase is from 0.692 km to 0.895 km (Table \ref{tbl:orbel}). At this point, the origin of the $m=3$ mode remains mysterious.
\end{itemize}
\begin{figure}
\centerline{\resizebox{\textwidth}{!}{\includegraphics[angle=90]{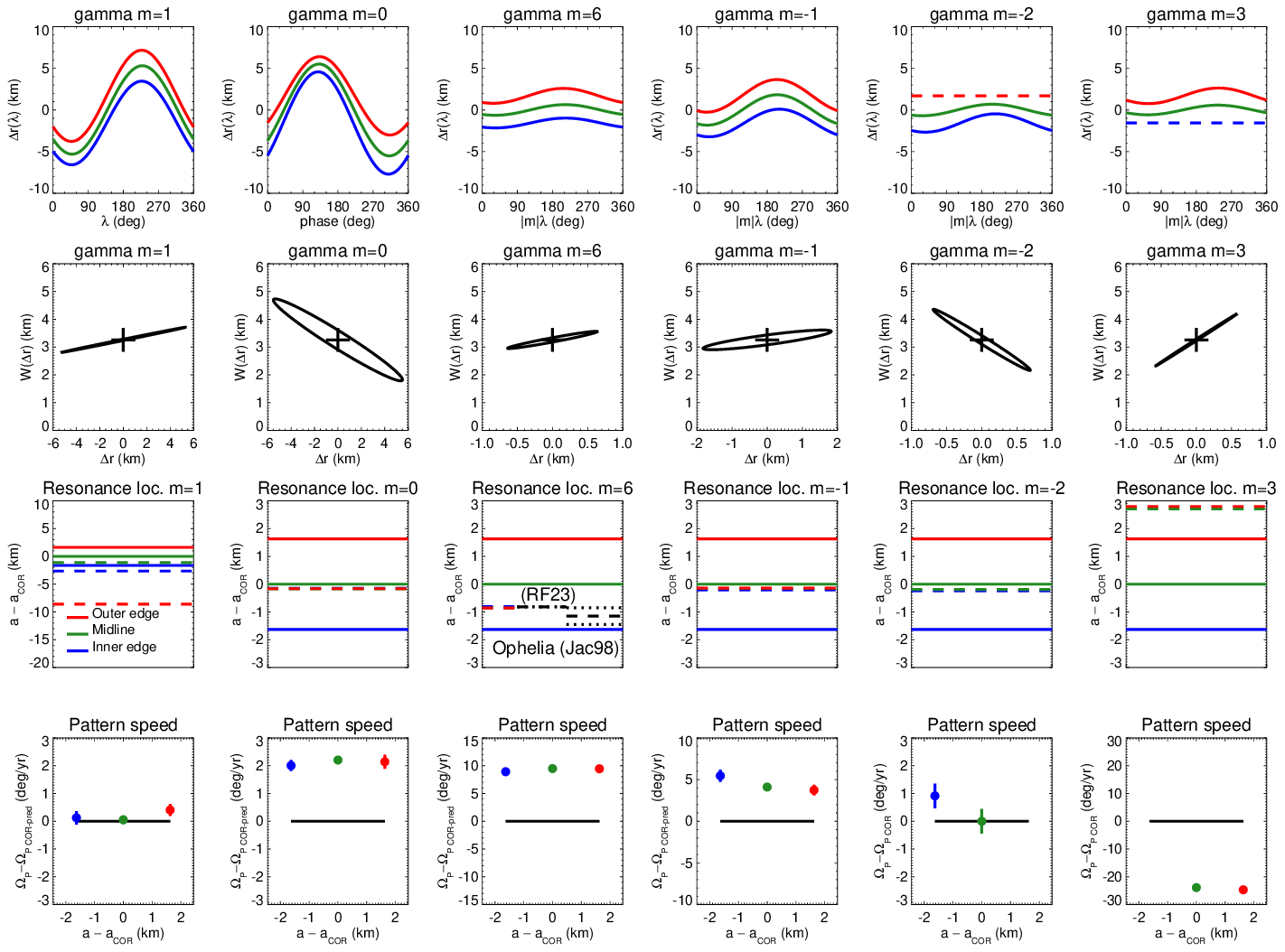}}}
\caption{Width variations of the $\gamma$ ring due to six normal modes, with the same layout as for Fig.~\ref{fig:widths6beta}. 
The predicted resonance location for Ophelia, assuming the mean motion given by \cite{Jacobson98} (labeled Jac98), is shown as a dashed line (bounded by its uncertainty) for the $m=6$ mode, and lies within the ring, near its inner edge. The more accurate recent Ophelia mean motion (Robert French, personal communication -- labeled RF23) has a predicted resonance location matching that of the observed $m=6$ pattern speed. The $m=-2$ mode was not detected in the shape of the OER, and the $m=3$ mode was not detected in the shape of the IER. The final row shows the differences in the observed pattern speeds of the inner and outer edges, and that of the COR, relative to the {\it predicted} COR value (based on the Uranus gravity field and satellite contributions to the precession rates), plotted as a function of semimajor axis relative to $a_{\rm COR}$. Note the different vertical axis ranges for the final row of plots.}
\label{fig:widthsgamma}
\end{figure}

        \subsection{The $\epsilon$ ring}
The $\epsilon$ ring is the widest and most eccentric of the Uranian rings, with a well-defined width-radius relation ($dW_m/dr=0. 095$) shown in 
{\bf Fig.~\ref{fig:widthseps}}. Its width varies by almost a factor of five, from 19.9 km at periapse to 97.3 km at apoapse. The periapses of the inner and outer edges are aligned to within $0.028^\circ$ and the resonance radius for the $m=1$ mode closely matches the midline of the ring, which is not surprising since the $\epsilon$ ring establishes a tie point for the determination of the planet's gravity field that governs its precession rate. The very small systematic offset between the observed IER, COR, and OER pattern speeds and the predicted COR value is a consequence of the radial offset of the geometric center of the ring from the ring center of mass, discussed in Section \ref{sec:COO}. The resonance radius of the $m=-24$ OLR lies within the ring 1.082 km outside the ring's IER, near to the $m=-24$ resonance of Cordelia computed from the mean motion given by \cite{Jacobson98}, as shown at upper right. Similarly, the resonance radius of the $m=14$ ILR also lies within the ring 0.887 km inside the ring's OER, near to the mean motion resonance of Ophelia computed from the \cite{Jacobson98} mean motion, as shown at lower right. The more accurate recent Ophelia mean motion (Robert French, personal communication -- labeled RF23) 
closely matches the observed $m=14$ pattern speed.
\begin{figure}
\centerline{\resizebox{\fw}{!}{\includegraphics[angle=90]{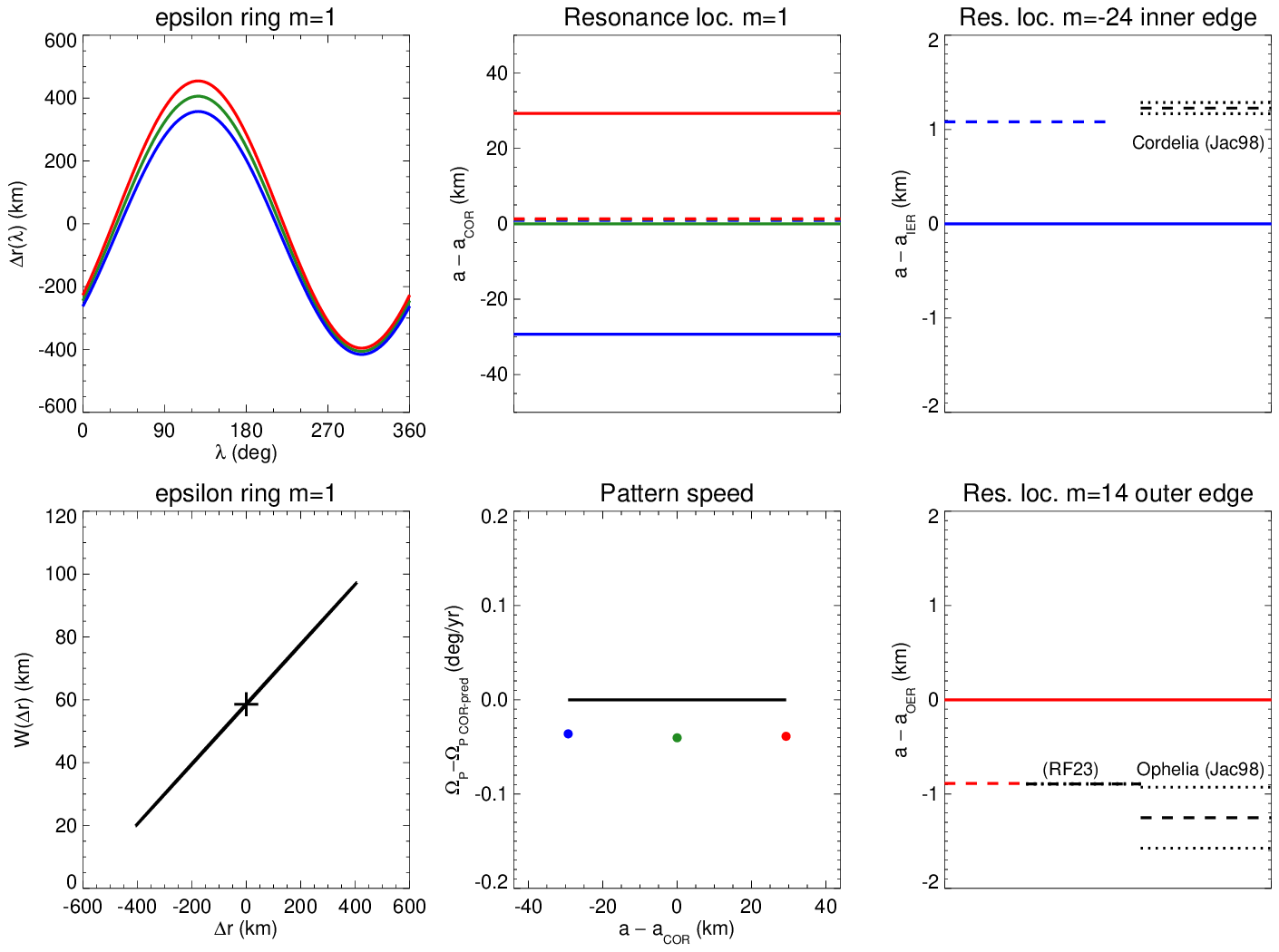}}}
\caption{Width variations of the $\epsilon$ ring. Upper left: $\Delta r(\lambda)$, the variation of ring radius with inertial longitude $\lambda$ at epoch for the IER, COR, and OER (blue, green, and red, respectively. Lower left: the corresponding variation of ring width $W(\Delta r)$. Upper middle: the predicted resonance radii corresponding to the observed apse rates at each ring edge relative to ring's semimajor axis $a_{\rm COR}$. Lower middle: the difference in the observed pattern speeds of the IER, COR, and IER relative to the predicted COR pattern speed, plotted as a function of semimajor axis relative to $a_{\rm COR}$. Upper right: the resonance location of the $m=-24$ mode (shown as a dashed blue line) relative to the inner edge of the $\epsilon$ ring (solid blue line), computed from the fitted $m=-24$ pattern speed of the IER. The corresponding resonance location computed from the \cite{Jacobson98} mean motion of Cordelia (labeled Jac98) is shown as a black dashed line, bounded by its uncertainties shown as dotted lines. Lower right: the resonance location of the $m=14$ mode (shown as a dashed red line) relative to the outer edge of the $\epsilon$ ring (solid red line), computed from the fitted $m=14$ pattern speed of the OER. The corresponding resonance location computed from the \cite{Jacobson98} mean motion of Ophelia is shown as a black dashed line labeled Jac98, bounded by its uncertainties shown as dotted lines. The more accurate recent Ophelia mean motion (Robert French, personal communication -- labeled RF23) has a predicted resonance location that closely matches that of the observed $m=14$ pattern speed. }
\label{fig:widthseps}
\end{figure}

 \subsection{Ring masses, surface densities, and a search for librations} 
 \label{sec:librations}
 In Section \ref{sec:654ab}, we compared the pattern speeds of the inner and outer edges of the eccentric rings, and in several instances there appear to be significant differences that could be 
be indications of librations, as has been observed at the outer edge of the B ring \citep{Spitale2010,Nicholson2014a,French2023b}. 
Notably, $\Delta\dot\varpi$ for ring 4 and the $\alpha$ ring, and $\Delta\varpi_0$ for the $\beta$ ring, appear to be significantly different from zero (see Table \ref{tbl:orbel} and Fig.~\ref{fig:widths6beta}), although neither is statistically significant for the $\epsilon$ ring, where the experimental uncertainties are a factor of 10 smaller.

From the two-streamline self-gravity model of eccentric rings, \cite{BGT83} identified a characteristic libration frequency $\Omega_{\rm sg}$ given by
\beq
\Omega_{\rm sg}=\frac{n}{\pi}\frac{M_r}{M_{\rm Ur}}\biggl(\frac{a}{\delta a'}\biggr)^2 H(q_e^2)=\frac{4n}{\pi}\frac{M_r}{M_{\rm Ur}}\biggl(\frac{a}{\delta a}\biggr)^2 H(q_e^2),
\label{eq:Omegasg}
\eeq
where $n\simeq \sqrt{GM_{\rm Ur}/a^3}$ is the ring particle mean motion and the factor of 4 in the right hand expression arises from the difference between $\delta a'$ and $\delta a$ (Eq.~(\ref{eq:deltaa})). The function $H(q_e^2)$ is defined in Eq.~\ref{eq:Hq2}.

Assuming that self-gravity prevents differential precession due primarily to $J_2$ and combining Eqs.~(\ref{eq:PY109}) and (\ref{eq:Omegasg}), the libration period for an eccentric ringlet $P_{\rm lib} = 2\pi/\Omega_{\rm sg}$ is given by
\beq
P_{\rm lib} =\frac{8\pi}{21} \frac{ {a}^{9/2} } {J_2 R^2 (GM_{\rm Ur})^{1/2}}\frac{q_e}{ae}.
\label{eq:Plib}
\eeq

Alternatively, this can written more simply as
\beq
P_{\rm lib} = \Biggl(\frac{2\pi}{\frac{d\dot\varpi}{da}\delta a}\Biggr)\frac{\delta e}{e} = P_{\Delta \dot\varpi} \frac{\delta e}{e},
\eeq
where $\frac{d\dot\varpi}{da}\delta a$ is just the differential precession frequency across the ring and the factor in parentheses is the corresponding differential precession period $P_{\Delta \dot\varpi}$ of the inner and outer ring edges. Since $\delta e/e \ll 1$ for most of the eccentric rings, the predicted libration periods are considerably shorter than the differential precession periods.

Substituting numerical values appropriate for Uranus into Eq.~(\ref{eq:Plib}), 
\beq
P_{\rm lib} = 960   \biggl(\frac{a}{50,000\ {\rm km}}\biggr)^{9/2} \biggl( \frac{10\ \rm km}{ae}\biggr) q_e{\rm\ yr}.
\label{eq:libpd}
\eeq

From the two-streamline self-gravity model, the mass of an eccentric ringlet (\ie\ with a free $m=1$ mode) is given by 
\beq
M_r=\frac{21\pi}{16}J_2\biggl(\frac{R}{a}\biggr)^2\biggl(\frac{e}{\delta e}\biggr)\biggl(\frac{\delta a}{a}\biggr)^3\frac{M_{\rm Ur}}{H(q_e^2)}.
\eeq
For a ringlet with a free mode with $m\neq1$, the corresponding relations are
\beq
P_{\rm lib} =\frac{4\pi}{3(m-1)n} \frac{q_e}{e_m} =3.1 \frac{1}{(m-1)} \biggl(\frac{a}{50,000\ {\rm km}}\biggr)^{5/2} \biggl( \frac{10\ \rm km}{ae_m}\biggr) q_e{\rm\ yr}
\eeq
and
\beq
M_r=\frac{3\pi(m-1)}{8}\biggl(\frac{e_m}{\delta e_m}\biggr)\biggl(\frac{\delta a}{a}\biggr)^3\frac{M_{\rm Ur}}{H(q_e^2)}.
\label{eq:Mselfgravm}
\eeq

As noted by \cite{GT79}, the two-streamline self-gravity ring mass estimate differs from the results of more careful treatments that use a large number of streamlines $N\gg2$. Assuming that $q_e$ is constant across the ring and that $\delta e/e\ll 1$, numerical results show that the many-streamline ring mass from the standard self-gravity (SSG) model converges almost exactly to 
\beq
M_{\rm SSG} =\frac{1}{2}M_r
\eeq
for all of the eccentric Uranus rings. 

The corresponding mean surface mass density from the many-streamline standard self-gravity model is given by 
\beq
\Sigma_{\rm {\rm SSG}} = \frac{M_{\rm SSG}}{ 2\pi   a \delta a}=3.18 \biggl(\frac{M_{\rm SSG}}{10^{14} {\ \rm kg}}\biggr)\biggl( \frac{50,000\ {\rm km}}{a}\biggr)\biggl(\frac{10 {\rm\ km}}{\delta a}\biggr)\ \gcm.
\eeq

The standard self-gravity model has been criticized on a number of grounds. For example, the derived surface densities for the $\alpha$ and $\beta$ rings are quite low, and inconsistent with the inferred large particle sizes for these rings \citep{French1991}. Furthermore, the observed variations in the radial optical depth profile of the $\epsilon$ ring over a range of true anomalies disagree with model predictions \citep{Graps1995}. \cite{GP87} pointed out that gas drag due to the extended hydrogen atmosphere of Uranus poses a severe problem for shepherding of the $\alpha$ and $\beta$ rings unless their masses have been seriously underestimated. Finally, the model ignores potentially important dynamical effects, such as the requirement for ring shepherding that there be substantial collisional dissipation near ring edges \citep{BGT82}. 

Motivated in part by these concerns, \cite{Chiang2000} developed a heuristic collisional self-gravity (CSG) model for apse alignment of narrow eccentric rings that includes the pressure-induced accelerations resulting from interparticle collisions near ring edges. Their model assumes that $\delta e/e \ll 1$, that $q_e$ is constant across the ring, and that the collisional pressure gradient is confined to a localized edge zone of radial scale $\lambda\simeq c_b/n$ (with dispersion velocity $c_b$ and mean motion $n$) over which the ring surface density falls sharply to zero. The combined contributions to the precession rate across the ring due to planetary oblateness, ring self-gravity, and collisions uniquely determine the ring's radial surface density profile and total mass, for an assumed radial scale $\lambda$ near the ring edge. 
 By including the estimated effect of collisions, \cite{Chiang2000} found surface densities for the $\alpha,\beta,$ and $\epsilon$ rings as large as $\Sigma=75-100\ \gcm$, larger by more than a factor of 10 than the SSG values. Subsequently, \cite{Mosqueira2002} modified the treatment of the strong density gradients near ring edges in the \cite{Chiang2000} model, but with generally similar results. Qualitatively, both of these CSG approaches result in large ring masses, concentrated near the ring edges, significantly exceeding the SSG values. 
 
Using the width-radius results from Table \ref{tbl:widths}, we have computed the masses and surface densities for each of the rings with a dominant free normal mode, using both the SSG and CSG models, as shown in {\bf Table \ref{tbl:ringmasses}}, along with the libration periods $P_{\rm lib}$ predicted from the two-streamline SSG model. For the SSG calculations, we used N=500 streamlines. We adopted the computationally simpler \cite{Chiang2000} solution to illustrate the collisional self-gravity approach, for a representative dispersion velocity of $c_b=1\ \cms$, corresponding to $\lambda=$ 35 -- 50 m over the ring system. (For the CSG calculations, we assumed a very large number of streamlines -- N=20,000 -- to ensure that the narrow transition zone near the ring edge was adequately sampled to provide accurate surface density values.) As these authors note, both $M_{\rm CSG}$ and $\Sigma_{\rm CSG}$ depend rather sensitively on the width of the ring edge transition region (or, equivalently, on the dispersion velocity), scaling approximately as ${c_b}^{1.5}$ in their model. In contrast, the \cite{Mosqueira2002} results are more weakly dependent on $\lambda$ but nevertheless typically confine most of the ring mass to a small region near each edge.


\begin{table*} [ht]
\begin{center} 
\caption{Ring masses, surface densities, and libration periods}
\label{tbl:ringmasses} 
\begin{threeparttable}
\centering
\begin{tabular}{l c c r c c}
\hline
Ring &  $M_{\rm SSG}\times 10^{14}$ (kg)$^a$ & $\Sigma_{\rm SSG} \ (\gcm)^a$ &  $P_{\rm lib}$ (yr)$^b$ & $M_{\rm CSG}\times 10^{14}$ (kg)$^c$ & $\Sigma_{\rm CSG} \ (\gcm)^d$\\
\hline
6  &      0.04 &      0.58 &     37.6 & 6.78 & 111.37\\
5  &      0.23 &      3.20 &      8.5 & 7.51 & 105.47 \\
4  &      0.11 &      1.23 &     25.6 & 8.13 & 94.05 \\
$\alpha$  &      0.19 &      0.93 &     66.3  & 12.46 & 60.96\\
$\beta$  &      0.15 &      0.60 &    118.1 & 13.52 & 55.33 \\
$\delta^d$  &      3.84 &     25.42 &      3.0 \\
$\epsilon$  &     32.70 &     17.37 &     17.1 & 69.89 & 37.13\\
\hline
\end{tabular}
 \begin{tablenotes}
 	\item[a] Computed using N= 500 streamlines in the standard self-gravity model (SSG), neglecting collisional effects.
	\item[b] Computed from Eq.~(\ref{eq:libpd}).
	\item[c] Computed using N= 20,000 streamlines in the collisional self-gravity (CSG) model of \cite{Chiang2000} for an assumed dispersion velocity $c_b=1\ \cms$. As noted in the text, the ring mass and surface density scale as ${c_b}^{1.5}$.
	\item[d] Computed from the two-streamline self-gravity solution in Eq.~(\ref{eq:Mselfgravm}) for $m=2$.
\end{tablenotes}
\end{threeparttable}
\end{center} 
\end{table*}

To illustrate the general features of the SSG and CSG results, we show in {\bf Fig.~\ref{fig:CG2000ring5}} a set of radial surface density profiles computed for the $\alpha$ ring at quadrature, using N=500 streamlines. The surface density $\Sigma(a)$ is shown on a logarithmic scale at left. Under the assumption that the ring optical depth is proportional to the surface mass density, we show at right the corresponding radial optical depth profiles on a linear scale, normalized at the point of highest surface mass density. The legend at right applies to both panels. Notice first that the collisionless self-gravity profile (\ie\ the SSG model) has a convex shape, peaking at the midline of the ring, with a very low maximum surface mass density $\Sigma_{\rm SSG}  \simeq 1\ \gcm$. The \cite{Chiang2000} results for $c_b=1, 2$ and $4\ \cms$ (labeled CG2000) share the same general U-shape, but with substantially greater surface density than the SSG model, increasing as $c_b$ becomes larger and eventually reaching an average value $\Sigma_{\rm CSG} >100\ \gcm$, but even larger values in excess of $1000\ \gcm$ near the ring edges. (In comparison, even the most opaque regions of Saturn's B ring have surface densities in the range 100-140 $\gcm$ \citep{Hedman2016}.) The corresponding optical depth profiles at right are completely different from the bowl-shaped SSG profile, showing narrow opaque ring edge regions that increase in width with increasing $c_b$ (or, equivalently, increasing $\lambda$), with much lower optical depth in the central region of the ring. To illustrate the modifications introduced by \cite{Mosqueira2002}, we used their Eq.~(20) representation of the collisional precession term to compute their CSG model, using $N=500$ streamlines and $c_b=2\ \cms$ (labeled ME2002). The derived surface density profile shows more more massive edge regions than the \cite{Chiang2000} models, with lower mass density in the interior of the ring near the midline. All of the CSG results shown have optical depth profiles that differ substantially from both the SSG model and (perhaps more significantly) are clearly a poor match to the observed shapes of the \Voyager\ optical depth profiles (see Fig.~\ref{fig:cooA}).

Reviewing the entries in Table \ref{tbl:ringmasses}, the width-radius results for rings 6, 5, and 4 provide the first opportunity to estimate the masses and surface densities of these very narrow and morphologically similar rings. The SSG results yield very low but disparate surface densities $\Sigma_{\rm SSG}$ between 0.58--3.17 $\gcm$, roughly comparable to the $\alpha$ and $\beta$ ring values. The CSG surface densities are larger by factors of $\sim$ 100, ranging from $\Sigma_{\rm CSG}$ = 94 to 111 $\gcm$, the largest of any of the rings shown. (We exclude the $\gamma$ ring from the table because of the confounding and uncertain influence of its many normal modes.) The CSG surface densities for the $\alpha$ and $\beta$ rings are in the range $\Sigma_{\rm CSG}$= 50--60 $\gcm$, similar to the results obtained by \cite{Chiang2000}, although the detailed models share the same issue in producing predicted radial optical depth profiles that differ substantially from the observations. The $\delta$ ring SSG surface density is computed from the $m=2$ normal mode properties, rather than from the keplerian ellipse term in its orbital elements, yielding a value in between the $\epsilon$ ring SSG and CSG surface density values. For this very broad ringlet, the SSG and CSG models give broadly similar results, but very different predicted shapes for the radial optical depth profile.

In assessing the importance of collisional self-gravity to the narrow Uranian rings, we note that the physical assumptions of the \cite{Chiang2000} model have a direct influence on the magnitude of the surface density enhancement relative to the standard self-gravity model. The most relevant of these relates to the magnitude of the very narrow scale $\lambda$ over which the ring surface density drops to zero at the ring edge. More specifically, they assume that $\lambda \sim c_b/n$ where $c_b$ is the ring velocity dispersion in the region of width $w$ perturbed by the satellite responsible for the edge confinement, and $n$ is the local mean motion. There is no explicit justification for this choice of $\lambda$ in \cite{Chiang2000}, but it is consistent with the following arguments:
\begin{itemize}
\item {On the one hand, $\lambda$ cannot be smaller than $c_b/n$, otherwise the Rayleigh stability criterion would be violated (the pressure gradient would be large enough to change the specific angular momentum distribution from stable to unstable \citep{Lin1993}. If for some reason the drop in surface density at the edge occurred on a yet smaller scale, the ring edge would become unstable and broaden again until stability is restored and this minimum scale satisfied.}
\item {On the other hand, choosing the smallest possible scale $\lambda$ maximizes the pressure gradient at the edge and, as a consequence, the mass increase of a narrow ring at a given geometry with respect to the SSG model.}
\end{itemize}

Maximizing the mass of the ring allowed \cite{Chiang2000} to assess whether the CSG model is able to overcome the mass deficit problem pointed out by \cite{GP87}. But although explaining sharp ring edges was a puzzle when they were first discovered, making the edge as sharp as possible is not a dynamical necessity; ultimately, the scale $\lambda$ is determined by (unknown) detailed properties of the ring stress tensor. As seen in Figs.~\ref{fig:coo6}, \ref{fig:coo5}, and \ref{fig:coo4}, there is no observational reason to assume that the inner edges of rings 6, 5, and 4 are as sharp as assumed by \cite{Chiang2000}, and the outer edges, while generally sharper, may be gradual on the scale of tens of meters.
Given our current state of understanding, these points indicate that it is probably best to view the SSG surface density as a minimum value and the CSG one as a maximum one.

\begin{figure}
\centerline{\resizebox{\textwidth}{!}{\includegraphics[angle=0]{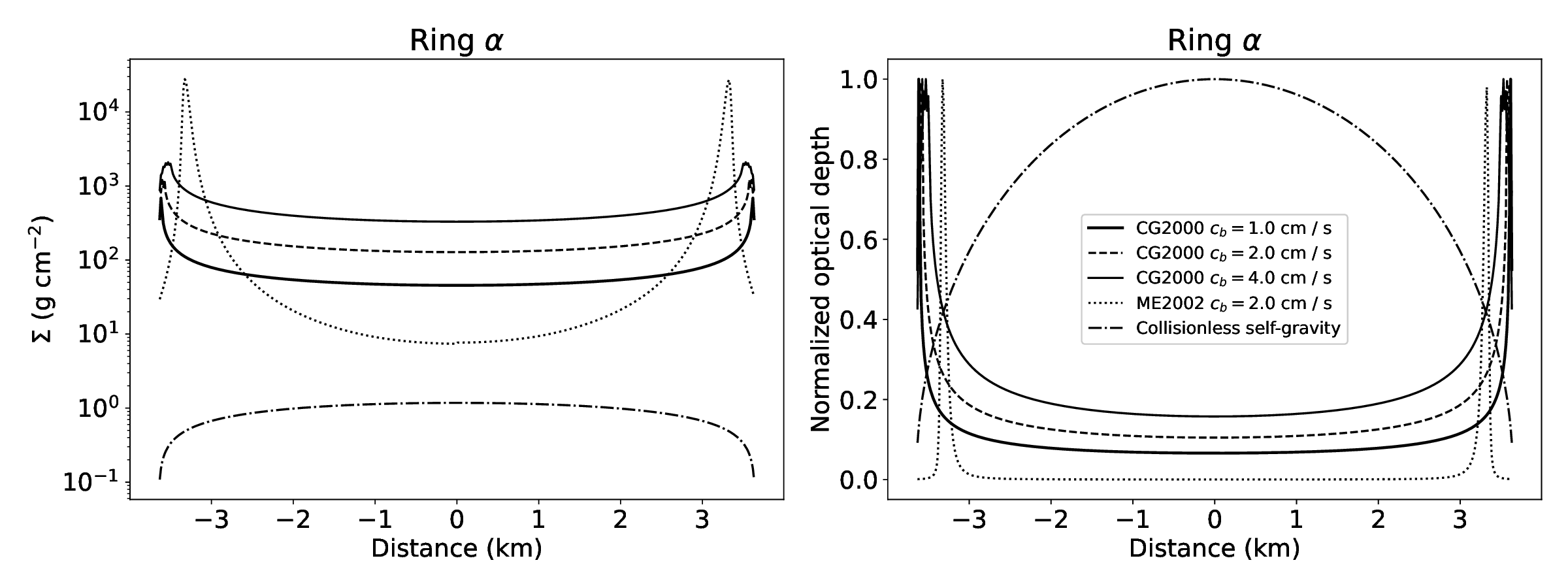}}}
\caption{Surface density profiles for the $\alpha$ ring at quadrature, shown on a logarithmic scale and left and normalized by their maximum values at right on a linear scale to show their predicted normal optical depth profiles. The collisionless self-gravity model (SSG) has a convex shape in both panels and a very low mean surface density $\Sigma_{\rm SSG}<1\gcm$. Three CSG profiles computed using the \cite{Chiang2000} model with dispersion velocities $c_b=1,2$, and $4\ \cms$ show much larger mean surface densities, increasing $\propto {c_b}^{1.5}$, with characteristic U-shaped optical depth profiles at right. The alternate collisional model of \cite{Mosqueira2002} is included for $c_b=2\ \cms$, showing qualitatively similar high surface densities confined to the ring edges.}
\label{fig:CG2000ring5}
\end{figure}

Turning now to the predicted libration periods in Table \ref{tbl:ringmasses}, note that for most rings they are decades long, making any such libration challenging to detect over the rather sparsely populated 30-year span of the occultation observations. At the other extreme, the $m=2$ mode of the $\delta$ ring has a predicted libration period of only 3 years, which is comparable to the typical interval between observations and thus susceptible to aliasing. Using the algorithm of 
\cite{Nicholson2014a}, we performed a series of fits for each ring, similar to normal mode scans, in which we specified a grid of possible libration periods between 1 and 100 years, solved for the best-fitting amplitude and phase of the libration for each period, and tabulated the RMS residuals for each fit. None of the scans returned a convincing, unique fit with a minimum RMS for a single libration period, and there was considerable aliasing for libration periods shorter than $\sim5$ years owing to the sparse sampling of the observations on this timescale. 

Looking in more detail at specific examples, Fig.~\ref{fig:widths6beta} shows that the inner and outer edges of rings 4 and $\alpha$ have fitted apse rates that differ by $\Omega_P({\rm OER})-\Omega_P({\rm IER})=\delta \Omega_P\sim0.1-0.2$\ \degyr\ with OER and IER apses misaligned by $\delta\varpi=-1.10\pm0.26^\circ$ for ring 4 
 and by $+0.36\pm0.29^\circ$ for the $\alpha$ ring. 

These can be compared to the apse misalignment expected over one quarter of a libration period $\Delta\varpi_{\rm lib}=\Delta \Omega_P P_{\rm lib}/4$. For ring 4, $\Delta\varpi_{\rm lib}=0.8^\circ$, of the same order as $\Delta\varpi$, meaning that the observed apse misalignment is of the same order as the range of excursions expected if the ring is librating. For the $\alpha$ ring, on the other hand, $\Delta\varpi_{\rm lib}=3.9^\circ$, more than an order of magnitude larger than the observed apse misalignment. If this ring is librating with a long period compared to the span of the observations, the probability of observing the apses so closely aligned would be rather small. If the ring is librating, $\Delta \Omega_P$ should be at its maximum when the apses are aligned. The $\epsilon$ ring has both well-aligned apses ($\Delta\varpi=-0.028^\circ$ from Table~\ref{tbl:orbel}) and very small $\Delta \Omega_P <0.0026$\ \degyr. Assuming simple harmonic motion, the amplitude of the libration $\Delta \phi \sim \Delta \Omega_P/\Omega_{\rm sg}= \Delta\Omega_P P_{\rm lib}/2\pi\sim0.007^\circ$. That is, if the $\epsilon$ ring is indeed librating with a period of 17.1 yr, the fitted edge apses and apse rates imply that the amplitude of libration is very small $(\sim0.01^\circ)$, below the level of detectability in our libration scans.

In a more systematic attempt to set lower limits on librations, we make use of the fit results in Table \ref{tbl:ringmasses} and proceed in the follow steps:

\begin{itemize}
\item[1.] Assume that $\delta\varpi = A_L\cos[2\pi(t-t_0)/P_L]$,
       where  $P_L$ is the (unknown) libration period
       and $A_L$ is the (unknown) libration amplitude.
\item[2.] It follows that $\delta\dot\varpi = -\frac{2\pi A_L}{P_L}\sin[2\pi(t-t_0)/P_L].$
\item[3.] Assume the nominal values of $P_L$ given in Table \ref{tbl:ringmasses}, based on the
      standard SSG model.
\item[4.] Use the observed value of either $\Delta\varpi$ {\it or}  $\Delta\dot\varpi$
      to estimate a lower limit on $A_L$. The larger amplitude $A_L$ wins.
\end{itemize}

Following  this prescription, we find that for the $\alpha$ ring, $\delta\dot\varpi$ dominates and $A_L > 2.4^\circ$.
      For the $\beta, \delta$, and $\epsilon$ rings, $\delta\varpi$ dominates
         and $A_L >  3.0^\circ, 0.4^\circ$, and $0.03^\circ$, respectively.
      For rings 6, 5 and 4 the two limits are comparable (within a factor
         of 2), with $A_L >  0.2^\circ, 0.03^\circ$, and $1.1^\circ$, respectively. 
         
In principle, more conclusive results could be obtained by subdividing the observations into several time intervals and mapping the variations of the fitted values of $\delta \Omega_P$ and $\delta \varpi$ over time to see if their phases matched the pattern expected for librations. This approach proved unsuccessful, however, owing to the uneven distribution of observations over time and the relatively small number of observations of each ring. We are left with the conclusion that any librations, if present, are at the margin of detectability.
%
\section{Uranus gravity field}
\label{sec:gravity}

In the orbit fits for the eccentric and inclined rings given in Table \ref{tbl:orbel}, we treated the apsidal precession and nodal regression rates  as kinematical parameters without any dynamical constraints. We now use the fitted apse and node rates to constrain the lowest-order zonal harmonic coefficients $J_2, J_4,$ and $J_6$ of the Uranus gravity field. 
An immediate challenge is that the radial range of the measurements spans less than 10,000 km, or equivalently, ${R}/{a}$ in Eq.~(\ref{eq:apserate}) varies only between 0.61 and 0.50 from the ring 6 to the $\epsilon$ ring, whereas to leading order the radial dependence of the $J_n$ terms is a sequence of steepening power laws: the $J_2$ term $\propto a^{-7/2}, J_4\propto a^{-11/2},$ and $J_6\propto a^{-15/2}$. As a result, $J_2, J_4,$ and $J_6$ are strongly correlated, and in particular the ring occultation data alone cannot yield an independent estimate of $J_6$, although the fit results provide coupled constraints on the trio $J_2, J_4,$ and $J_6$. In this section, we identify additional systematic and random errors that affect the determination of the gravity field, quantify their effects on the derived $J_n$, and provide a prescription to compute the difference in standard deviations between our best-fitting model and other model gravity fields described by the first three even gravity harmonics.

\subsection{Ring midlines, centers of opacity, and centers of mass}
\label{sec:COO}
In addition to the challenge of a restricted radial range of observed apse and node rates noted above, a second complication that has not been previously addressed is that the apse and node rates in Table \ref{tbl:orbel} are usually assumed to refer to the semimajor axes of the geometric ring midlines (COR), which may not correspond to the dynamical radius, or semimajor axis of the average center of mass (COM) of the rings. This is of particular concern for the $\epsilon$ ring, which has manifestly non-uniform internal radial structure, as seen in Fig.~\ref{fig:U25PalGallery}, but it applies to the other rings as well, as can be seen from the high-resolution \Voyager\ RSS ring profiles. 
{\bf Figure~\ref{fig:coo6}} shows the ingress and egress RSS optical depth profiles of ring 6. The ingress profile is near apoapse with true anomaly $f=144.4^\circ$, while the egress profile is almost exactly at periapse, with $f=2.7^\circ$. Under the simplifying approximation that optical depth is proportional to surface mass density, and thus using the center of opacity (COO) as a surrogate for the ring's COM, we define the radii of the inner and outer ring edges $r_{\rm IER}$ and $r_{\rm OER}$ by the vertical dashed lines, chosen by eye, and determine the radius of the center of opacity $r_{\rm COO}$ by integrating the radial normal optical depth profile $\tau(r)$ using
\beq
r_{\rm COO} = \frac{\int_{r_{\rm IER}}^{r_{\rm OER}} \tau(r)rdr}{\int_{r_{\rm IER}}^{r_{\rm OER}} \tau(r)dr}.
\eeq

\begin{figure}
\centerline{\resizebox{5in}{!}{\includegraphics[angle=90]{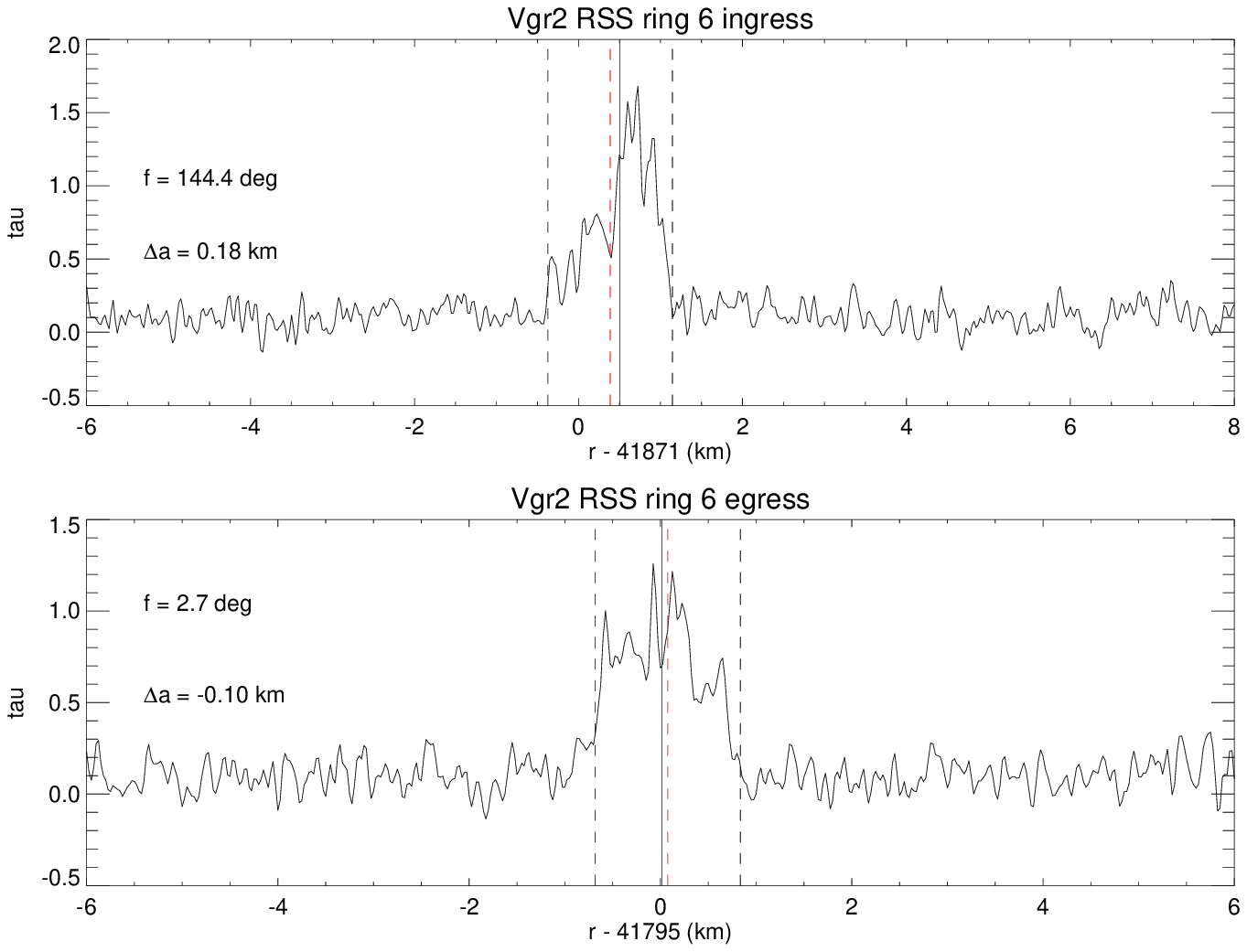}}}
\caption{Ring 6 center of opacity (COO) estimated from \Voyager\ RSS optical depth profiles. The derived values of $r_{\rm COO}$ are shown by the solid vertical blue line near the middle of each profile, but on opposite sides of the geometric center (COR) for ingress and egress, marked by the vertical dashed red lines mid-way between the two edges.}
\label{fig:coo6}
\end{figure}

The derived values of $r_{\rm COO}$ for the two RSS ring 6 observations are shown by the solid vertical blue line near the middle of each radial profile. For both ingress and egress, this is offset from the midline of the ring, marked by the vertical dashed red line mid-way between the two edges. We then determine the equivalent offset of the ring semimajor axis center of opacity $\Delta a_{\rm COO}$ from the relation
\beq 
\Delta a_{\rm COO} = \biggl(\frac{r_{\rm COO} - r_{\rm IER}}{r_{\rm OER} - r_{\rm IER}} - \frac{1}{2}\biggr) \overline W
\eeq
where $\overline W$ is the mean width of the ring defined by the difference in semimajor axes of the OER and IER fits in Table \ref{tbl:orbel}. Numerically, $\Delta a_{\rm COO} = +0.18$ km and $-0.10$ km for the ring 6 ingress and egress profiles, respectively. 

In this instance, it is clear that the internal structure of ring 6 is quite different at the two locations sampled by the RSS occultations, which may not be representative of the structure at other ring longitudes. In any event, there are no streamline models for the internal structure of any of the rings that accurately capture the ring structure at all longitudes, and we use the RSS observations to estimate the uncertainty of the ring COO or COM as well as its systematic offset from the geometric ring center determined from square-well fits to the ensemble of ring profiles.

We proceed similarly in Appendix \ref{appendix:COO} for the other eccentric rings, making additional use of the best Earth-based radial profiles for the $\epsilon$ ring in instances where it is wide enough to reveal internal structure. The derived centers of opacity are given in {\bf Table \ref{tbl:rCOO}}, which lists
 $a_{\rm COR}$, the unweighted average values for 
$\Delta a_{\rm COO}$ and their RMS variation, the statistical significance $\Delta a_{\rm COO}/\sigma(\Delta a_{\rm COO})$, and finally of $a_{\rm COO}$, with an uncertainty given by the sum in quadrature of $\sigma(a_{\rm COR})$ and $\sigma(\Delta a_{\rm COO})$.
\begin{table*} [ht]
\begin{center} 
\caption{Ring center of opacity}
\label{tbl:rCOO} 
\begin{threeparttable}
\centering
\begin{tabular}{c c c c c}\hline
Ring &  $a_{\rm COR}$ (km) & $\Delta a_{\rm COO}$ (km) &{\large $\frac{\Delta a_{\rm COO}}{\sigma(\Delta a_{\rm COO})}$ }& $a_{\rm COO}$ (km) \\
\hline 
6 & $41837.092 \pm     0.096 $ & $    0.040 \pm     0.193 $ & $    0.208 $ & $41837.132 \pm     0.215 $ \\
5 & $42234.893 \pm     0.091 $ & $    0.169 \pm     0.100 $ & $    1.686 $ & $42235.061 \pm     0.135 $ \\
4 & $42571.124 \pm     0.091 $ & $    0.404 \pm     0.087 $ & $    4.620 $ & $42571.528 \pm     0.126 $ \\
$\alpha$ & $44718.473 \pm     0.086 $ & $    0.172 \pm     0.132 $ & $    1.309 $ & $44718.645 \pm     0.157 $ \\
$\beta$ & $45661.056 \pm     0.087 $ & $    0.322 \pm     0.274 $ & $    1.175 $ & $45661.378 \pm     0.288 $ \\
$\gamma$ & $47626.170 \pm     0.089 $ & $    0.119 \pm     0.125 $ & $    0.954 $ & $47626.289 \pm     0.153 $ \\
$\epsilon$ & $51149.279 \pm     0.081 $ & $    1.134 \pm     0.488 $ & $    2.324 $ & $51150.414 \pm     0.495 $ \\
\hline
\end{tabular}
\end{threeparttable}
\end{center} 
\end{table*}

The mean offset $\Delta a_{\rm COO}$ is positive for all seven eccentric rings but statistically significant at the 2-$\sigma$ level only for ring 4 and the $\epsilon$ ring. We will make use of both the offsets and their uncertainties in our final determination of the Uranus gravity field.

As an illustration of the effect of the offset $\Delta a_{\rm COO}$ on the computed precession rate, note that at the radius of the $\epsilon$ ring, $d\dot\varpi/da =  -9.355\times {10^{-5}}$ \ \degd~km$^{-1}$, and the corresponding change in apse rate for $\Delta a_{\rm COO}=1.134$ km from Table \ref{tbl:rCOO} is $\Delta \dot\varpi=-1.06\times 10^{-4}$\ \degd. To put this in context, the formal error in the apse rate $\sigma(\dot\varpi)$ for the $\epsilon$ COR from Table \ref{tbl:orbel} is just $4.2\times {10^{-6}}$ \ \degd, and thus the radial change $\Delta a_{\rm COO}$ corresponds to change of $\Delta \dot\varpi/\sigma(\dot\varpi)\sim-1.06\times 10^{-4} /4.2\times {10^{-6}}$, or a 25$\sigma$ change in the predicted apse rate. Note that for the $\epsilon$ ring COR in Table \ref{tbl:orbel}, $\Delta a_{\dot\varpi}=1.154$ km $\simeq \Delta a_{\rm COO}=1.134$ km, demonstrating that the adopted gravity field takes into account the systematic difference between the ring's COR and COO. 

{\bf Figure~\ref{fig:precrates2}} (top panel) shows the observed ring apse and node rates from Table \ref{tbl:orbel}, plotted as solid dots, compared to the smooth curve computed from our adopted gravity model (Fit 18 in Table ~\ref{tbl:Jn}), described below in Section \ref{sec:J2J4J6}. The second panel shows the difference $\Delta \dot\varpi$ between the individually fitted apse rates from Table \ref{tbl:orbel} (with their error bars) and the rates predicted by the adopted gravity model (solid dots), and from individual alternative fits (shown as open circles) that ignored the apse and node rates of each ring in turn when fitting for $J_2$ and $J_4$. (Both series of fits account for the offsets $\Delta a_{\rm COO}$.) The bottom panel shows $\Delta \dot\Omega$ for the node rates of the measurably inclined rings. The observed apse and node rates generally match the predicted values, with the possible exceptions of the $\alpha, \beta,$ and $\gamma$ ring apse rates. In Section \ref{sec:precrates}, we will examine the evidence for anomalous precession rates of these rings and evaluate possible explanations.
\begin{figure}
\centerline{\resizebox{5in}{!}{\includegraphics[angle=0]{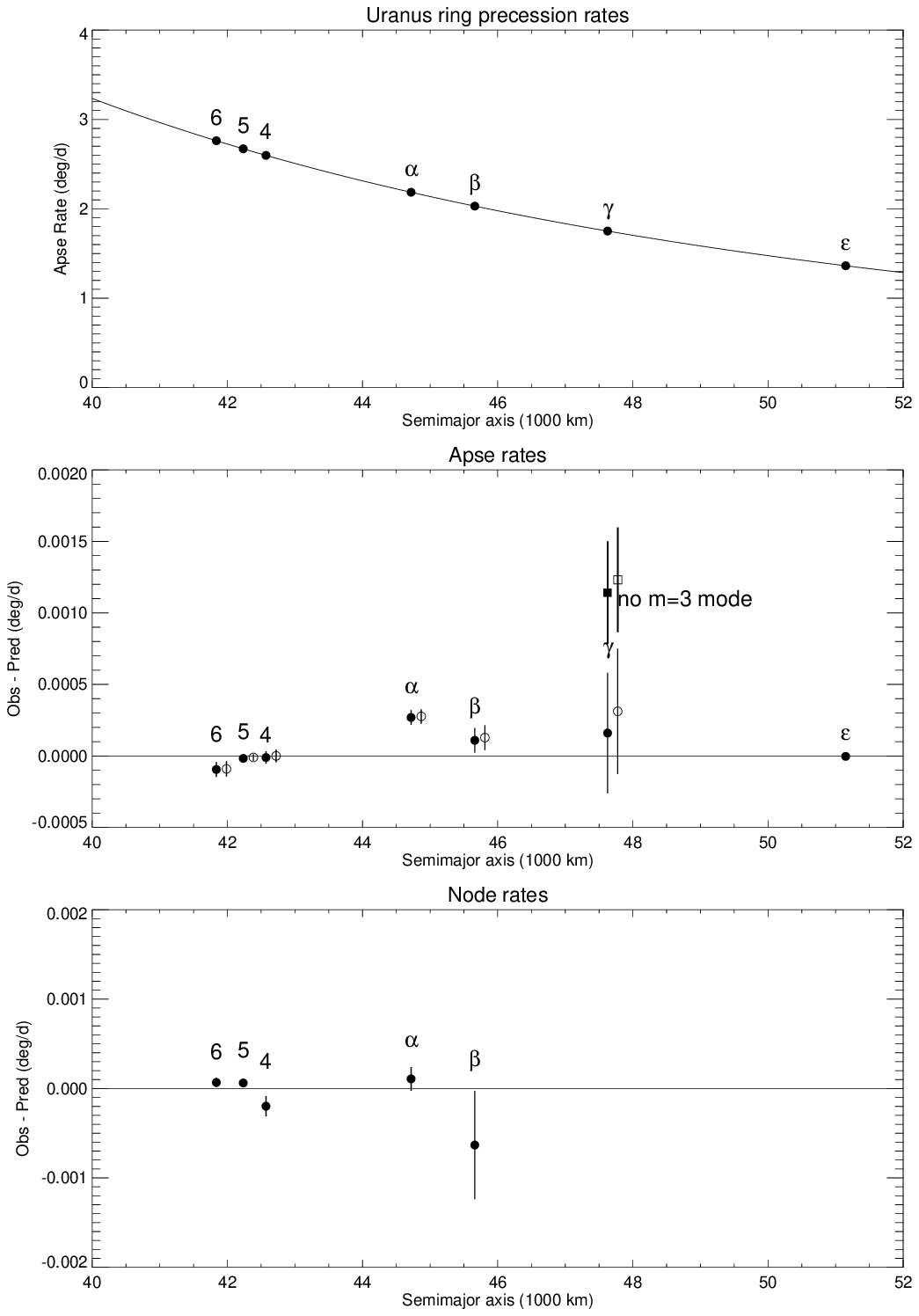}}}
\caption{Uranus ring apse and node precession rates. The top panel shows the observed apse rates of the eccentric rings, overplotted with the predicted radial variation in apse rate computed from the adopted gravity model (Fit 15 in Table \ref{tbl:Jn}). The middle panel shows 
the differences between the observed and predicted apse rates $\Delta\dot\varpi$ for each ring's center of opacity (COO), (unlike the Table \ref{tbl:orbel} values, which apply to the ring's geometric centers (COR)), plotted as filled symbols. The open symbols show the estimated anomalous precession based on separate fits for $J_2$ and $J_4$ that excluded the contribution of individual rings to the gravity field solution. These are listed in Table \ref{tbl:anomprecrates}. Rings $\alpha, \beta,$ and possibly $\gamma$ appear to have anomalously high precession rates compared to the predictions from the gravity model. For the $\gamma$ ring orbit fit that excluded the $m=3$ mode, the anomalous precession is even higher (shown as a filled and open squares). The bottom panel shows the node rate difference $\Delta\dot\Omega$ for the inclined rings.}
\label{fig:precrates2}
\end{figure}

\subsection{Determining the planetary gravity field from ring precession rates}
\label{sec:J2J4J6}
Historically, the most accurate estimates of the Uranus gravity field have been derived from the observed precession rates of the rings \citep{French1988,Jacobson2014}. With the inclusion of all ring occultation data between 1977--2006, the uncertainties in the ring apsidal and nodal rates have been substantially reduced compared to prior determinations. As we will show, however, both the mean values of $J_2$ and $J_4$ and their realistic uncertainties from ring occultation data are significantly affected by the assumed value of $J_6$ and by previously ignored systematic and random errors in: (1) the estimates of the ring centers of opacity (COO), (2) the absolute radius scale, and (3) the estimated contributions of both major and minor satellites to the ring precession rates. In this section, we review previous gravity field determinations and estimate the influence of these error sources on the derived gravity harmonics. Although we cannot obtain an independent estimate of $J_6$, we derive a set of observationally constrained linear relations between $J_2, J_4,$ and $J_6$ based on their strong mutual correlations in the ring orbit fits. 
\subsubsection{Previous determinations of the Uranus gravity field}\label{sec:prevgrav}
\cite{French1988} determined $J_2$ and $J_4$ from orbit fits to ring occultation observations from 1977--1986, with the results listed as Fit 1 in {\bf Table~\ref{tbl:Jn}}, converted from their assumed reference radius of $R=26200$ km to $R=25559$ km. \cite{Jacobson2014} solved for $J_2$ and $J_4$ from a comprehensive analysis of all available \Voyager\ navigation data and selected ring occultation measurements, and in a separate solution from the ring data alone. These results are listed as Fits 2 and 3, respectively.\footnote{The correlation coefficient $\rho(J_2,J_4)$ for these fits was provided by R. Jacobson, personal communication.} \cite{Jacobson2014} assumed that $J_6=0$, but the quoted uncertainties in $J_2$ and $J_4$ incorporated an \apriori uncertainty $\sigma(J_6)=1.0\times10^{-6}$, as well as \apriori uncertainties on other parameters of the fit including the absolute radius scale and the pole direction. The resulting ``rings only" solution (Fit 3) is shown in {\bf Fig.~\ref{fig:J2J4}} as the large solid-line brown ellipse. 

More recently, \cite{Jacobson2023} estimated $J_2$ and $J_4$ as part of a global solution for the Uranus ephemeris and system properties, based on a comprehensive analysis of historical astrometry, navigation data, and ring occultation observations. He set $J_6=(0.58\pm0.12)\times 10^{-6}$, based on a suite of Uranus interior models by \cite{Neuenschwander2022}, and the satellite contribution to the precession of the rings was included for Ariel, Umbriel, Titania, Oberon, Miranda, and Puck only, ignoring the minor satellites. (The assumed masses of these six satellites are included in Table \ref{tbl:majorsatdata}.) The \cite{Jacobson2023} results are shown Fig.~\ref{fig:J2J4} as Fit 4, where the error bars include estimated uncertainties in the astrometry and navigation data used for the fit.

\begin{table*} [ht]
\scriptsize
\begin{center} 
\caption{Uranus gravity parameters}
\label{tbl:Jn} 
\centering
\begin{tabular}{l l c c  c l}
\hline
Fit & $J_2\times 10^{-6}$ & $J_4\times 10^{-6}$ & $J_6\times 10^{-6}$ &  $\rho(J_2,J_4)$ &  Note\\
\hline
1 & $ 3513.23 \pm 0.34 $ & $-30.32 \pm 4.73$ & -- &-- & \cite{French1988} (converted to $R=25559$ km)\\
2 & $ 3510.7 \pm 0.7 $ & $-34.2 \pm 1.3$ & $0.0\pm1.0$  & 0.9784 & \cite{Jacobson2014} adopted solution\\
3 & $ 3510.5 \pm 1.3 $ & $-34.4 \pm 1.3$ & $0.0\pm1.0$  & 0.9784 & \cite{Jacobson2014} ``rings only" solution\\
4 & $ 3510.465\pm 0.058 $ & $-34.145 \pm 0.082$& $0.58\pm 0.12$ &  0.9887 & \cite{Jacobson2023} ({\tt ura178} ephemeris)\\
5 & $ 3510.464\pm 0.066 $ & $-34.158 \pm 0.092$& [0.58] &  0.9384 & COR -- 6 satellites (AUTOMP)\\
6 & $ 3510.474 \pm 0.056 $ & $-34.135 \pm 0.081$ & [0.58] &  0.9853 &  COR wtd fit to apse/node rates (AUTOMP)\\
7 & $ 3511.175 \pm 0.065 $ & $-33.492 \pm 0.092$ & [0.50] &  0.9384 &  COR + COO offsets (Table \ref{tbl:rCOO}) -- 6 satellites (AUTOMP)\\
8 & $ 3510.975 \pm 0.065 $ & $-34.030 \pm 0.092$ & [0.0] &  0.9384 &  Fit 7 but $J_6=0\times10^{-6}$\\
9 & $ 3511.374 \pm 0.065 $ & $-32.954 \pm 0.092$ & [1.0] &  0.9384 &  Fit 7 but $J_6=1\times10^{-6}$\\
10 & $ 3511.201 \pm 0.065 $ & $-33.528 \pm 0.092$ & [0.50] &  0.9384 &  Fit 7 with $\Delta a=+0.2$ km for all rings\\
11 & $ 3510.723 \pm 0.065 $ & $-33.957 \pm 0.092$ & [0.50]  &0.9384 &  Fit 7 including all but Cordelia and Ophelia\\
12 & $ 3509.291 \pm 0.067 $ & $-35.522 \pm 0.094$ & [0.50]  &0.9383 &  Fit 7 incl. all sats, basis for Monte Carlo Fit 15\\
13 & $ 3513.217 \pm 0.065 $ & $-31.669 \pm 0.091$ & [0.50]  &0.9384 &  Fit 7 excluding all satellites\\
14 & $ 3509.337 \pm 0.065 $ & $-35.438 \pm 0.091$ & [0.50]  &0.9387 &  COO, all satellites, rings 6,5,4 $\epsilon$ for $J_n$\\
[0.75em]15 & $ 3509.291 \pm 0.412 $ & $-35.522  \pm 0.466$ & [0.50]  &0.9861 &  Adopted solution: Fit 12 w/ composite error estimates\\[0.75em]
16 & $ 3509.291 \pm 0.385 $ & $-35.522 \pm 0.433$ & [0.50]      &0.9957 &  Monte Carlo errors from COO fits to apse/node rates only\\
17 & $ 3509.291 \pm 0.134 $ & $-35.522 \pm 0.147$ & [0.50]      &0.9997 &  Monte Carlo errors from satellite mass uncertainties only\\
18 & $ 3509.291 \pm 0.026 $ & $-35.522 \pm 0.036$ & [0.50]      &0.1412 &  Monte Carlo errors from radius scale uncertainty only\\
  		[0.75em]Quantity& Value & & & & Note\\
\hline 
$GM_{\rm U}$  & $5793950.300$ km$^3$ s$^{-2}$& & &  &\cite{Jacobson2023}\\
$R$  & 25559 km&  & & & Reference radius for $J_n$\\
$dJ_4/dJ_2$ & $+1.134$ & &   &   &  slope of error ellipse of adopted solution\\
$dJ_4/dJ_2$ & $+2.69540$ & &   &   & slope of line connecting COO $J_6=(0,1)\times10^{-6}$\\
$dJ_2/dJ_6$ & $+0.39909$ & &   &   & over range $J_6=(0,1)\times10^{-6}$\\
$dJ_4/dJ_6$ & $+1.07570$ & &   &   &   over range $J_6=(0,1)\times10^{-6}$\\
$d J_2/d\Delta a$ & $+0.130 \times 10^{-6}$ km$^{-1}$ & & &  & from Fits 7 and 10\\ 
$d J_4/d\Delta a$ & $-0.180 \times 10^{-6}$ km$^{-1}$ & & &  & from Fits 7 and 10\\
\hline
\end{tabular}
\end{center} 
\end{table*}

\begin{figure}
\centerline{\resizebox{\fw}{!}{\includegraphics[angle=90]{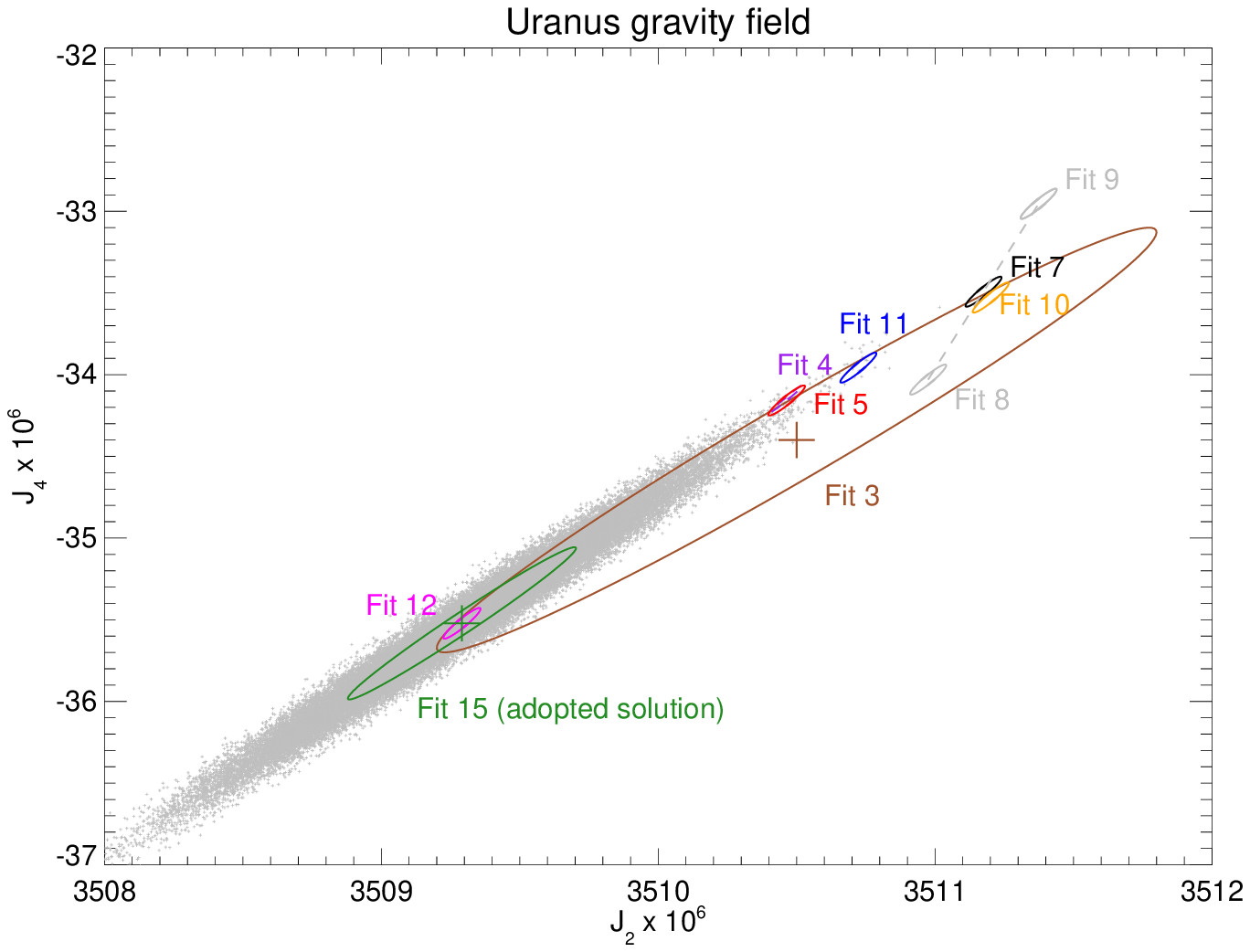}}}
\caption{Results of a suite of fits for $J_2$ and $J_4$, under a variety of assumptions about the value of $J_6$, the contributions of major and minor satellites to the ring precession rates, the differences between the ring COR and COO, and the absolute radius scale of the rings. The fit numbers correspond to the results in Table \ref{tbl:Jn}. The adopted fit (Fit 15) includes error estimates derived from the Monte Carlo set of solutions marked by the gray dots, representing the range of variation in $J_2$ and $J_4$ associated with estimated error sources. See text for details.}
\label{fig:J2J4}
\end{figure}

\subsubsection{Uranus gravity field from COR ring orbital elements}
To provide an initial estimate of $J_2$ and $J_4$ from the complete set of COR measurements used to derive the orbital elements in Tables \ref{tbl:orbel} and \ref{tbl:normalmodes}, we performed Fit 5, using the same fit parameters as our adopted COR fit except that the ring apse and node rates were calculated from the best-fitting values of $J_2$ and $J_4$, a fixed value for $J_6$, and the secular precession forced by the satellites included only Ariel, Umbriel, Titania, Oberon, Miranda, and Puck, omitting the estimated contributions of the smaller moons displayed in Fig.~\ref{fig:precrates}. To provide a direct comparison with \cite{Jacobson2023}, we let $J_6=0.58\times 10^{-6}$. This solution is plotted as the red error ellipse in Fig.~\ref{fig:J2J4}. The fitted values for $J_2$ and $J_4$ are very similar to the \cite{Jacobson2023} result (Fit 4), showing that these two independent solutions based largely on the same set of observations yield nearly identical results. 

Next, in Fit 6, we determined $J_2$ and $J_4$ from a weighted fit to the apse and node rates in Table \ref{tbl:orbel}, again assuming $J_6=0.58\times 10^{-6}$. As seen in Table \ref{tbl:Jn}, the results are similar to Fit 5, but with slightly smaller error bars and a larger correlation coefficient. The fitted values and error bars are quite close to the \cite{Jacobson2023} results in Fit 4 as well. This confirms that the primary constraints on $J_2$ and $J_4$ in the full RINGFIT results are the apse and node rates, while the other free parameters have less influence on the fitted values and uncertainties of the gravity parameters.

In a separate fit, not included in Table \ref{tbl:Jn}, we allowed $J_6$ to be an additional free parameter, but the resulting formal errors in all three gravity parameters were very large, the retrieved value of $J_6$ was implausibly large, and the correlations between the parameters were nearly singular. 

\subsubsection{Uranus gravity field from COO ring orbital elements}
To explore the effects on the derived gravity field of applying the ring radius corrections associated with the COO measurements, in Fit 7 we applied the offsets $\Delta a_{\rm COO}$ from Table \ref{tbl:rCOO} to the fitted ring semimajor axes when computing the precession rates and let $J_6=0.5\times 10^{-6}$, with all other fit conditions as in Fit 5. The result is shown as a black error ellipse and is significantly displaced from the COR solution (Fit 5),
demonstrating the sensitivity of $J_2$ and $J_4$ to the differences between the geometric ring centers (COR) and their dynamical centers (COO). 

\subsubsection{Sensitivity of $J_2$ and $J_4$ to the assumed value of $J_6$}
Although our experiment with Fit 5 described above demonstrates that we cannot fit simultaneously for $J_2,J_4,$ and $J_6$, we can nevertheless quantify the sensitivity of the results to assumed (fixed) values of $J_6$. In Fits 8 and 9, we let $J_6=(0.0,1.0)\times 10^{-6}$ respectively, applied the COO offsets as before, and solved for $J_2$ and $J_4$. The results are shown as the labeled gray error ellipses in Fig.~\ref{fig:J2J4}, bracketing the COO Fit 7 that assumed $J_6=0.5\times10^{-6}$. From this pair of fits, we find the partial derivatives $dJ_2/dJ_6$ and $dJ_4/dJ_6$ listed at the bottom of Table \ref{tbl:Jn}. The ratio of these gives the slope of the line connecting the COO $J_6=(0.0,1.0)\times 10^{-6}$
solutions, shown as the gray dashed line in the figure and listed in Table \ref{tbl:Jn}.

\subsubsection{Sensitivity of $J_2$ and $J_4$ to the absolute radius scale}
To quantify the effect of a possible systematic error in the overall absolute ring radius scale, in Fit 10 we repeated Fit 7, but assumed a systematic radius scale error of $\Delta a=+0.2$ km (our 2-$\sigma$ estimate, as shown in Section \ref{sec:radscale}). The result of the fit is shown as the orange error ellipse in Fig.~\ref{fig:J2J4}, only slightly displaced from Fit 7, indicating that a radius scale error at this level has only a modest effect on the derived gravity field. This is evident in the small magnitude of the differentials $d J_2/d\Delta a$ and $d J_4/d\Delta a$ included at the bottom of Table \ref{tbl:Jn}.

\subsubsection{Effects of minor satellites on the derived gravity field}

Previous published determinations of $J_2$ and $J_4$ have taken into account the secular precession due to Ariel, Umbriel, Titania, Oberon, Miranda, and Puck, but have ignored the contributions of the smaller moonlets. As shown in Fig.~\ref{fig:precrates},
several of these minor satellites each have predicted secular precession rates for several of the eccentric rings that are larger than those due to Oberon and Puck, and in particular 
Cordelia and Ophelia are estimated to contribute jointly to precession rate of the $\epsilon$ ring at a level of $\dot\varpi>10{^{-4}}$~\degd, greater than that of any of the other satellites except Ariel. To quantify the effect of the minor satellites, we first performed Fit 11, similar to Fit 7 except for the inclusion of all moons except Cordelia and Ophelia. The fitted values of $J_2$ and $J_4$ are significantly offset from the Fit 7 values, with the results shown as the blue error ellipse in Fig.~\ref{fig:J2J4}. Then, in Fit 12 we added the predicted precession due to tiny Cordelia and Ophelia, with the resultant values of $J_2$ and $J_4$ shown as the magenta error ellipse, shifted even further from the Fit 7 values that used only the six most massive moons. For comparison, in Fit 13 we ignored the effects of all satellites when computing the ring precession rates. The resulting values are off-scale to the upper right of the figure. The differences between Fits 7, 11, 12 and 13 demonstrate the sensitivity of the derived values of $J_2$ and $J_4$ to the assumed minor satellite masses and the uncertainties in the masses of both major and minor satellites. We evaluate these effects below.

\subsection{The influence of the $\alpha, \beta,$ and $\gamma$ rings on the derived gravity field}

 Figure~\ref{fig:precrates2} shows that the apse rates of the $\alpha, \beta,$ and $\gamma$ rings in the COR fit for $J_2$ and $J_4$ are systematically faster than predicted from the gravity solution. To quantify the influence of these three rings on the derived gravity field, we performed Fit 14, in which we applied COO corrections, included the effects of all major and minor satellites, but allowed the apse and node rates of the $\alpha, \beta,$ and $\gamma$ rings to be free parameters, rather than being constrained by $J_2$ and $J_4$. Only rings 6, 5, 4, and $\epsilon$ were so constrained. As shown in Table \ref{tbl:Jn},  
 the fitted values for $J_2$ and $J_4$ differed only slightly from the Fit 12 results, indicating that the possibly anomalous precession rates of the $\alpha, \beta,$ and $\gamma$ rings do not produce a significant systematic shift in the derived gravity field. 

\subsubsection{Effects of random and systematic errors on the derived gravity field}
So far, we have separately estimated the sensitivity of the derived values of $J_2$ and $J_4$ to the assumed value of $J_6$, to differences between ring centerlines and centers of opacity (COR and COO, respectively), to possible systematic errors in the absolute radius scale, and to the contributions of small moons to the ring precession rates. All of these systematic effects have associated uncertainties that affect the robustness of the fitted gravity field. To quantify the effects of both random and systematic errors on the derived gravity field, we performed a series of Monte Carlo fits that sampled a random distribution of estimated uncertainties in key quantities that affect the best-fitting solution for $J_2$ and $J_4$. We used the conditions of Fit 12 (including all satellites and the COO offset) as our nominal best fit for $J_2$ and $J_4$. Then, 
for each of 50,000 fits, we applied separate offsets to the semimajor axes of each ring, drawn randomly from a normal distribution with the corresponding individual standard deviation in the COO semimajor axis 
$\sigma(\Delta a_{\rm COO})$. To account for the uncertainty in the absolute radius scale, we applied an additional single random systematic offset $\Delta a$ to the semimajor axes of {\it all} of the rings, drawn from a normal distribution with a standard deviation of 0.2 km. Finally, to account for uncertainties in the masses of both major and minor satellites when computing their contributions to each ring's secular precession rate, we applied offsets to the individual satellite masses drawn randomly from a normal distribution with a standard deviation for each major satellite taken from \cite{Jacobson2023} and for the minor satellites from Table \ref{tbl:minorsatdata}, assuming our derived masses for Cressida, Cordelia, and Ophelia. 

We determined $J_2$ and $J_4$ for each Monte Carlo instance from a least squares fit to the entire system geometry, with the apse and node rates of the eccentric and inclined rings being determined from the combined effects of the modeled gravity field and satellite contributions, assuming $J_6=0.5\times 10^{-6}$ for all fits. The ensemble of fitted values of $J_2$ and $J_4$ represents the probability density function (PDF) associated with the applied random errors.
 To determine the corresponding error ellipse, we first determined the mean values of $J_2$ and $J_4$ and their standard deviations from the distribution of their fitted values. Next, we estimated the correlation coefficient $\rho(J_2,J_4)$ by determining its value such that the corresponding 1-$\sigma$ error ellipse for $\sigma(J_2)$ and $\sigma(J_4)$ enclosed $e^{-1/2} \simeq 0.6065$ of the fitted values, as appropriate for the complementary cumulative probability distribution of a bivariate normal distribution (see Appendix C). 
The results are given in Table \ref{tbl:Jn} as Fit 15, our adopted solution and error bars. The mean values for $J_2$ and $J_4$ match those from Fit 12, as expected, but with much larger uncertainties, owing to the incorporation in the Monte Carlo runs of the estimated random errors in $\Delta a_{\rm COO}$, $\Delta a$, and the satellite contributions to the ring precession rates. 

To quantify the various contributions to the total error budget, we performed three separate Monte Carlo runs. For Fit 16, we included only the estimated errors due to standard deviations in the COO semimajor axes $\sigma(\Delta a_{\rm COO})$. For Fit 17, we included only the 
contributions of the uncertainty in satellite masses to the error budget. Finally, in Fit 18, we included only the uncertainty in the absolute radius scale. The results are given in Table \ref{tbl:Jn}. In all cases, the mean values of $J_2$ and $J_4$ matched those of our adopted solution, as expected, and the quoted error bars correspond to the standard deviations in the fitted values $J_2$ and $J_4$ from the 50,000 individual random samples for each series. These fits show that the dominant source of error in the Monte Carlo runs is associated with the uncertainty in the locations of the ring centers of opacity (\ie\ the random errors in $\Delta a_{\rm COO}$). On the other hand, the substantial displacement of the best-fitting $J_2$ and $J_4$ that included only the AUTOMP satellites from our adopted solution accentuates the importance of including the effects of the minor satellites when computing the predicted precession rates of the rings. 

\subsubsection{Adopted gravity field solution and uncertainties}
Our adopted solution for the Uranus gravity field is from Fit 15:
$J_2=(3509.291\pm0.412)\times 10^{-6}, J_4=(-35.522\pm0.466)\times 10^{-6}$, and $\rho(J_2,J_4)=0.9861$. These represent the best-fitting values of $J_2$ and $J_4$ from Fit 12, which used the entire set of ring observations, corrected for the estimated differences between COO and COR values, including the contribution to ring precession of both major and minor satellites, and assume $J_6=0.5\times 10^{-6}$. The formal errors $\sigma(J_2)$ and $\sigma(J_4)$ and correlation $\rho(J_2,J_4)$ for Fit 12 account for random errors of the observations alone, whereas the Monte Carlo results summarized in Fit 15 account as well for the estimated uncertainties in $\Delta a_{\rm COO}$, $\Delta a$, and the satellite contributions to the ring precession rates.

A striking characteristic of the results in Table \ref{tbl:Jn} and evident in Fig.~\ref{fig:J2J4} is the small size of formal error ellipses in $J_2$ and $J_4$ from the individual fits, compared to the much larger differences in $J_2$ and $J_4$ that stem from a range of assumptions about the value of $J_6$, the satellites included in the calculation of secular precession, the choice of COR or COO for ring semimajor axes, the absolute radius scale, and realistic uncertainties in these quantities. Some of these differences can be predicted on the basis of the observed strong correlations between $J_2$, $J_4$, and $J_6$, as we show below. Nevertheless, the large size of the error ellipse of our final solution compared to the formal errors of the individual fits underscores the consequences of our limited knowledge of the internal radial structure of the rings that determines their dynamical semimajor axes, as well as uncertainties in the masses of the minor satellites, especially Cordelia and Ophelia, which primarily affect the $\epsilon$ ring's predicted precession rate, a key constraint on the gravity field determination.

\subsubsection{Correlations between $J_2$, $J_4$, and $J_6$}
The strong correlations between $J_2, J_4,$ and $J_6$ can be used to restrict the range of models for the interior structure and differential rotation of Uranus that are consistent with the ring occultation solution for the planet's gravity field. 
In Appendix \ref{appendix:grav}, we evaluate these results numerically and provide a prescription to compute the difference in standard deviations between a model gravity field described by the first three gravity harmonic coefficients
$\{J'_{2}, J'_{4}, J'_{6}\}$ and our adopted solution. 

\section{Anomalous precession rates}
Our determination of the gravity field of Uranus relies on a model of the radial dependence of ring apsidal precession and nodal regression rates that includes the effects of the planet and the known satellites. From an examination of the post-fit residuals in apse rate for individual rings, we can set limits on possible unseen moonlets, such as those that have been proposed to account for quasi-periodic optical depth variations in the $\alpha$ and $\beta$ ring \Voyager\ RSS optical depth profiles \citep{Chancia2016}. Figure \ref{fig:precrates2} shows small positive residuals for the apse rate of these two rings, and also for the $\gamma$ ring. Here, we determine the range of moonlet masses and orbital radii that could account for these residuals, and set limits on the anomalous precession in the vicinity of the eccentric rings.

We estimated the detection limits for anomalous precession rates of the eccentric rings from a series of fits in which we solved for full set of geometric and orbital parameters as in our adopted solution, except that we let each eccentric and possibly inclined ring in turn have its apse and node rates fitted as free parameters but solved for $J_2$ and $J_4$ from the other rings. The differences between the fitted values and the apse rates computed from Fit 12 in Table~\ref{tbl:Jn} (accounting for $\Delta a_{\rm COO}$) are listed in
{\bf Table \ref{tbl:anomprecrates}}, which summarizes our measured values and upper limits for anomalous precession. These are plotted as open circles in the middle panel of Fig.~\ref{fig:precrates2}.

\begin{table*} [ht]
\begin{center} 
\caption{Limits on anomalous precession rates}
\label{tbl:anomprecrates} 
\centering
\begin{tabular}{c r l}\hline
Ring &   $\Delta\dot\varpi \ (10^{-5}$ \degd)  & Description \\
\hline 
6 &  $   -9.07 \pm    5.31 $ & No detected anomalous precession \\
5 &  $   -1.04\pm     2.30 $ & No detected anomalous precession\\
4 &  $   +0.02\pm     4.48 $ & No detected anomalous precession \\
$\alpha$ &  $  27.66 \pm     4.99 $  & Consistent with \cite{Chancia2016} wake model\\
$\beta$ &  $   12.79 \pm     8.74 $  & Consistent with \cite{Chancia2016} wake model\\
$\gamma$ &  $  31.21\pm     43.84 $  & $m=3$ mode included in orbit model\\
$\gamma$ &  $  123.16 \pm     36.54 $  & $m=3$ mode not included in orbit model\\
$\epsilon$ & -- & Not determinable from orbit model\\
\hline
\end{tabular}
\end{center} 
\end{table*}

\label{sec:precrates}

\subsection{Rings 6, 5, and 4}
The precession rates of rings 6, 5, and 4 all contributed to the determination of the gravity field, so it is not surprising that there are no detected anomalous precession rates for these rings, although a single outlier among them would perhaps have been detectable.

\subsection{The $\alpha$ ring}
 \cite{Chancia2016} proposed that wakes observed in the \Voyager\ RSS $\alpha$ ring radial profiles could be produced by a moonlet with orbital radius $a_s= 44825^{+22}_{-12}$ km (compared to the $\alpha$ ring COR semimajor axis $a=44718.470\pm0.086$ km) and mass $M_{sI}=(3^{+4}_{-2}) \times 10^{14}$ kg (or $M_{sI}/M_{\rm Ur}=(3.46^{+4.61}_{-2.30}) \times 10^{-12}$) from the ingress profile and $M_{sE}=(1.0^{+18}_{-0.6}) \times 10^{14}$ kg (or $M_{sE}/M_{\rm Ur}=(1.15^{+20.7}_{-2.65}) \times 10^{-12}$) from the egress profile. 
 In {\bf Fig.~\ref{fig:alpha_anomprec}}, the solid diagonal line indicates the satellite mass, as a function of the orbital distance exterior to the ring COR, that would be required to produce the estimated anomalous precession rate 
 $ \Delta\dot\varpi = (27.66\pm4.99)\times10^{-5}$ \degd\ of the $\alpha$ ring, from Table \ref{tbl:anomprecrates}. 

 The dotted lines bound the uncertainty. The dashed diagonal line (visible at upper right) gives the corresponding results for a moonlet interior to the ring. The relative locations of the rings exterior to the $\alpha$ ring are shown as vertical dashed lines, and the masses of representative satellites are plotted as horizontal dotted lines, along with an estimate for the $\epsilon$ ring mass ratio $M_\epsilon/M_{\rm Ur}=7.57\times 10^{-11}$ \citep{Nicholson2014b}. Selected moonlet radii $r$ are shown for an assumed average density $\rho=1\ \density$. For example, a Cordelia-sized moon orbiting 500 km from the $\alpha$ ring could produce the observed anomalous precession rate.
 
 For comparison, the \cite{Chancia2016} estimates of the hypothetical moonlet's mass and distance from the $\alpha$ ring (using the more reliable RSS ingress profile) are shown by the labeled dot, with error bars that overlap the range of masses inferred from the measured anomalous precession rate of the $\alpha$ ring. This provides supportive evidence for an unseen icy moonlet with a radius $\gtrsim$ 4 km $\simeq$100 km exterior to the $\alpha$ ring.
 
\begin{figure}
\centerline{\resizebox{5in}{!}{\includegraphics[angle=90]{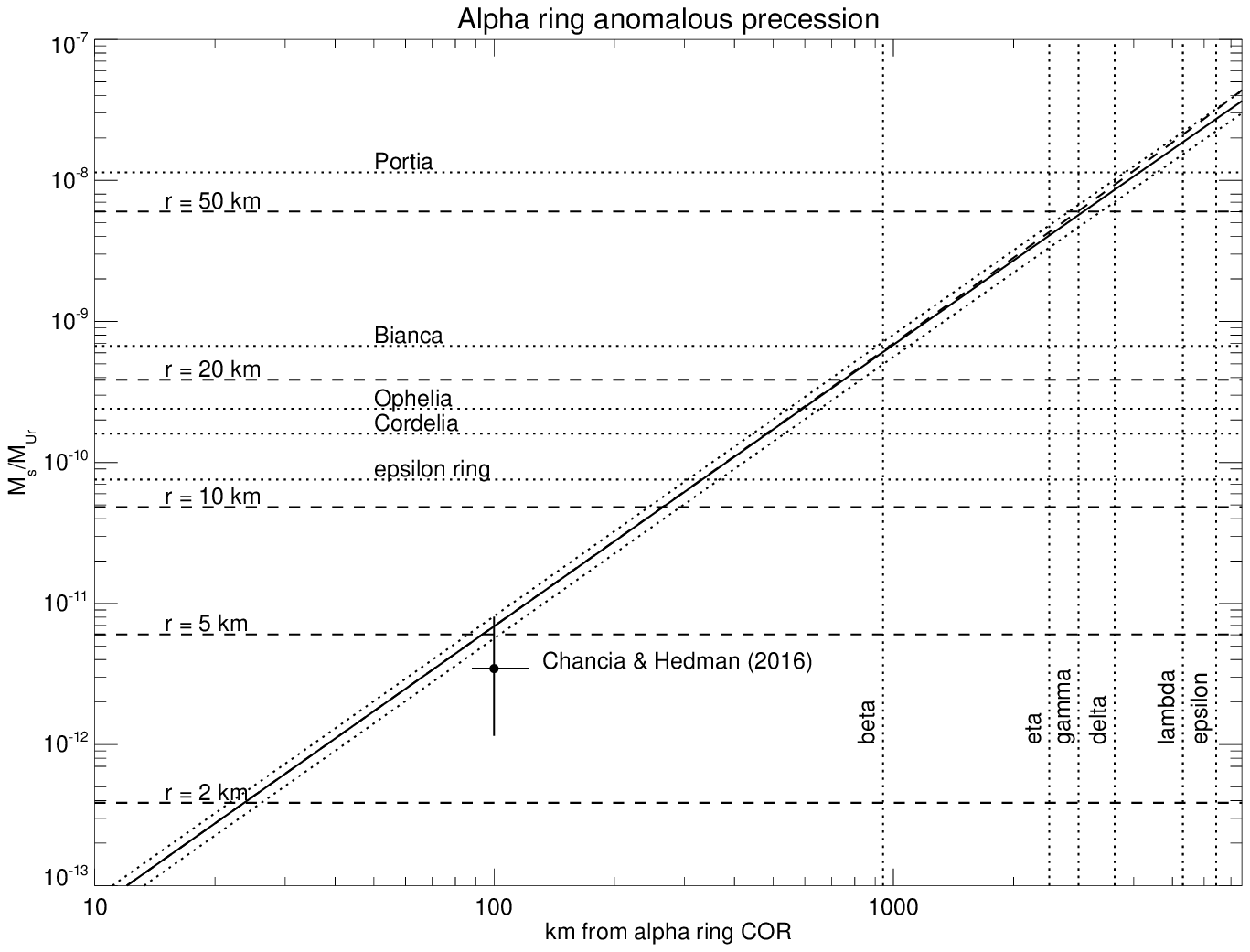}}}
\caption{Satellite size and mass required to match the observed anomalous precession rate of the $\alpha$ ring, plotted as a function of orbital distance from the semimajor axis of the ring's COR and shown as a solid diagonal line, bounded by dotted lines indicating its uncertainty. The dashed black line that deviates from the solid line at the upper right corresponds to satellite orbits interior to the $\alpha$ ring's semimajor axis. Representative masses and radii of moonlets and the $\epsilon$ ring are shown, assuming $\rho=1\ \density$. The locations of the rings exterior to the $\alpha$ ring are marked by vertical dotted lines. The black dot with error bars marks the hypothetical satellite mass and orbital radius proposed by \cite{Chancia2016} to account for wakes observed in the \Voyager\ RSS $\alpha$ ring radial profiles. The predictions from the observed anomalous precession rate match within the error bars.}
\label{fig:alpha_anomprec}
\end{figure}   
                \subsection{The $\beta$ ring}
Similarly, \cite{Chancia2016} proposed that wakes observed in the $\beta$ ring radial profiles could be produced by a moonlet with orbital radius $a_s= 45738^{+8}_{-4}$ km (compared to the $\beta$ ring COR semimajor axis $a=45661.051\pm0.087$ km) and mass $M_{sI}=(0.5^{+0.3}_{-0.2}) \times 10^{14}$ kg (or $M_{sI}/M_{\rm Ur}=(0.58^{+0.35}_{-0.23}) \times 10^{-12}$) from the ingress RSS profile and $M_{sE}=(0.7^{+0.4}_{-0.2}) \times 10^{14}$ kg (or $M_{sE}/M_{\rm Ur}=(0.81^{+0.46}_{-0.23}) \times 10^{-12}$) from the egress RSS profile. In {\bf Fig.~\ref{fig:beta_anomprec}}, the solid diagonal line indicates the satellite mass, as a function of the orbital distance exterior to the ring COR, that would be required to produce the estimated anomalous precession rate 
$\Delta\dot\varpi = (12.79\pm8.74)\times10^{-5}$ \degd~of the $\beta$ ring. 
The dotted lines bound the uncertainty. The dashed diagonal line (visible at upper right) gives the corresponding results for a moonlet orbiting interior to the ring. The relative locations of the rings exterior to the $\beta$ ring are shown as vertical dashed lines.
 
 For comparison, the \cite {Chancia2016} estimate of the hypothetical moonlet mass and distance from the $\beta$ ring from the more reliable RSS ingress profile is shown by the labeled dot, with error bars that overlap the range of masses inferred from the estimated anomalous precession rate of the ring. As with the $\alpha$ ring, this provides supportive evidence for an unseen moonlet the produced the wake signature in the ring profile, in this case with a radius $\gtrsim$ 3 km $\sim$ 80 km exterior to the $\beta$ ring.

\begin{figure}
\centerline{\resizebox{5in}{!}{\includegraphics[angle=90]{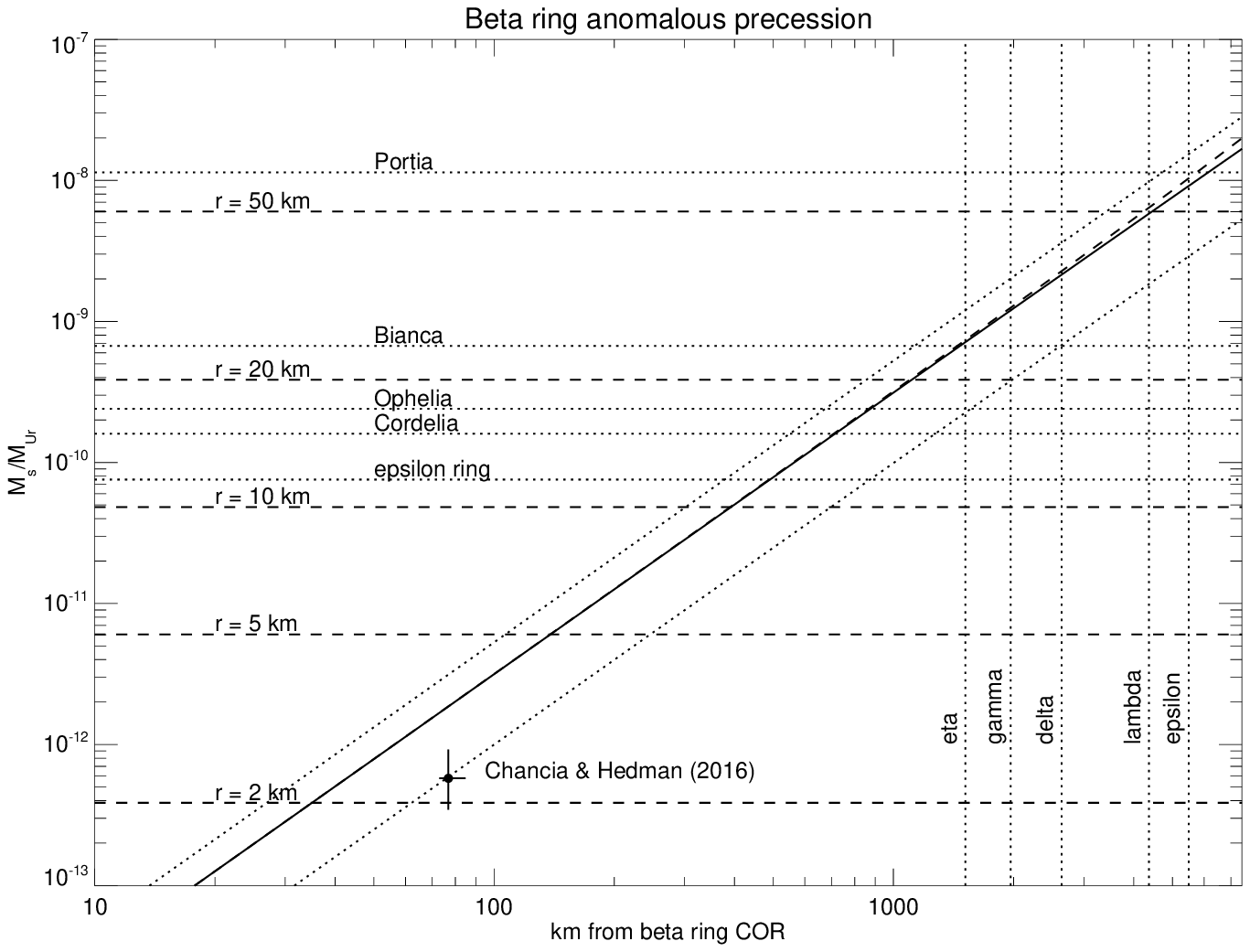}}}
\caption{Satellite size and mass required to match the observed anomalous precession rate of the $\beta$ ring, plotted as a function of orbital distance from the semimajor axis of the the ring's COR and shown as a solid diagonal line, bounded by dotted lines indicating its uncertainty. The dashed black line that deviates from the solid line at the upper right corresponds to satellite orbits interior to the $\beta$ ring's semimajor axis. Representative masses and radii of moonlets and the $\epsilon$ ring are shown, assuming $\rho=1\ \density$. The locations of the rings exterior to the $\beta$ ring are marked by vertical dotted lines. The black dot with error bars marks the hypothetical satellite mass and orbital radius proposed by \cite{Chancia2016} to account for wakes observed in the \Voyager\ RSS $\beta$ ring radial profiles. The predictions from the observed anomalous precession rate match within the error bars.}
\label{fig:beta_anomprec}
\end{figure}

                \subsection{The $\gamma$ ring}
Finally, we examine the evidence for possible unseen moonlets in the vicinity of the $\gamma$ ring. The putative $m=3$ normal mode resulted in a small positive but uncertain anomalous precession rate 
$\Delta \dot\varpi=(14.48\pm 42.10)\times10^{-5}$ \degd~(Table \ref{tbl:orbel}), consistent with zero.
When the $m=3$ mode is excluded from the fit, the anomalous precession has a substantially larger value: 
from Table \ref{tbl:anomprecrates}, $\Delta \dot\varpi=(123.16\pm36.54)\times10^{-5}$ \degd, plotted as a filled square in Fig.~\ref{fig:precrates2}.
{\bf Figure~\ref{fig:gamma_anomprec}} shows the satellite masses required to account for both of these possibilities. The upper solid and dashed diagonal lines correspond to the fit that excluded the $m=3$ mode, giving the inferred satellite mass as a function of distance exterior to or interior to the $\gamma$ ring COR, respectively.
When the $m=3$ mode is included in the fit, the predicted satellite masses are smaller, with the nominal estimates based on the measured anomalous precession rate of the $m=1$ mode being shown as the red solid and dashed lines, defined as before, with the upper limit shown as the red dotted line. (There is no corresponding lower limit, since 
$\sigma(\Delta \dot\varpi)$ exceeds its measured value.) 

The amplitude and location of the putative $m=3$ mode provide an independent constraint on the mass and location of a hypothetical satellite that forces the mode. The resonance location for the $m=3$ mode is $\sim 1.101$ km exterior to the outer edge of the ring, and if the mode is forced by a first-order ILR with an unseen satellite, its mean motion would be 
$\Omega_P({\rm OER})=n=765.3977$ \degd, 
with a corresponding semimajor axis $a=62366.66$ km, a few hundred km interior to the orbit of Desdemona. Based on the amplitude of the COR mode $A_3=0.577\pm0.073$ km and the distance from resonance $\Delta a_P=2.706\pm 0.048$ km, the required satellite mass would be $m_{\rm sat}/M_{\rm Ur}=12.5\times10^{-10}$, or about twice the mass of Cordelia (see Section \ref{sec:satellites}). This is too large to have escaped previous detection in \Voyager\ images  and would probably produce a detectable perturbation in Cordelia's orbit. Furthermore, it is too small to account for the $m=1$ anomalous precession. On this basis, we reject the hypothesis that the $m=3$ mode is forced by an external satellite. If instead the $m=3$ mode is unforced or entirely absent, and we assume that the lower radius limit of detectable satellite is $\sim10$ km, the inferred $m=1$ anomalous precession rate could be provided by a hidden moonlet orbiting within 150--250 km on either side of the $\gamma$ ring, assuming $\rho=1\ \density$.
         
\begin{figure}
\centerline{\resizebox{5in}{!}{\includegraphics[angle=90]{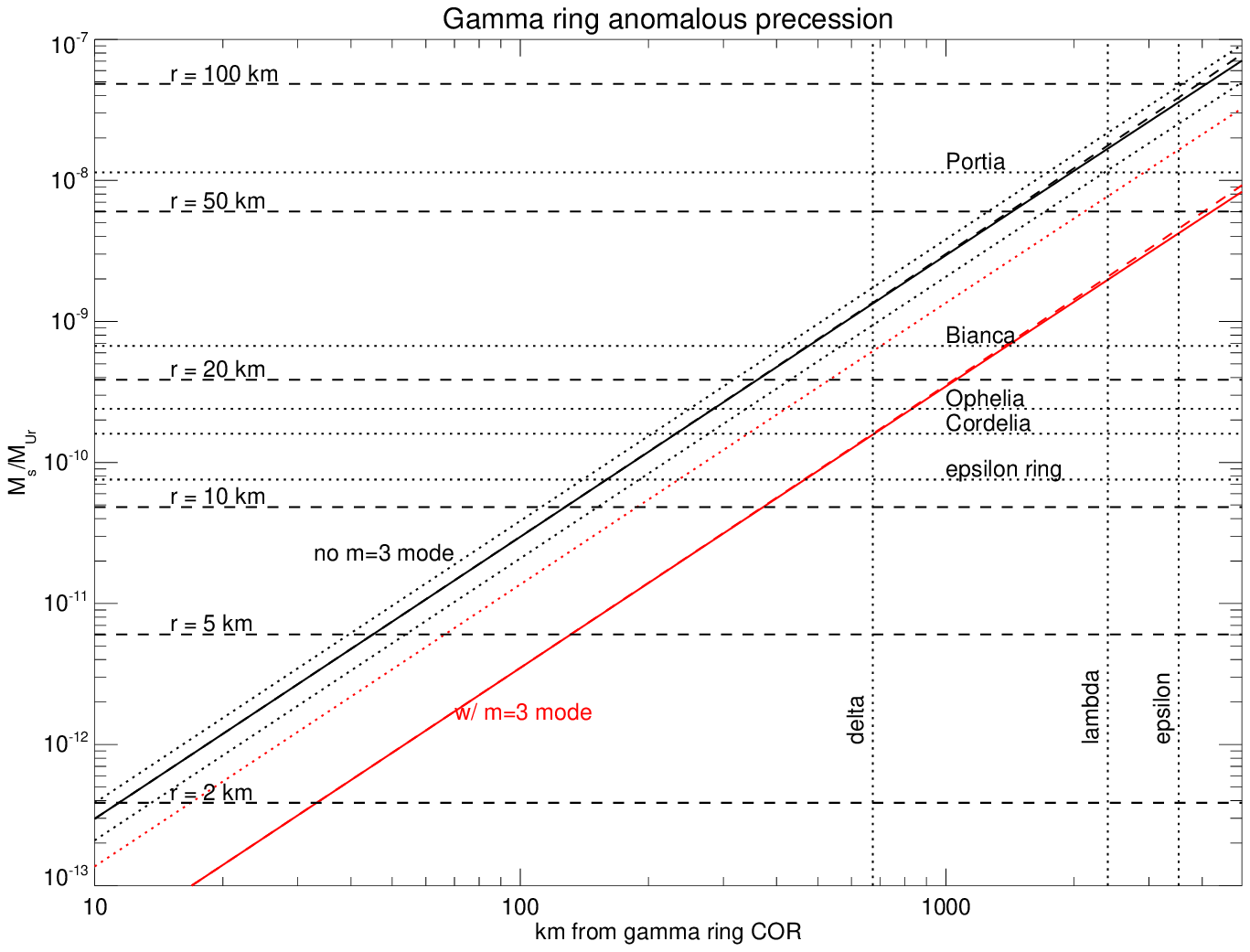}}}
\caption{Satellite size and mass required to match the observed anomalous precession rate of the $\gamma$ ring, plotted as a function of orbital distance from the ring's COR. The orbit fit that excluded the $m=3$ mode had a very high anomalous precession rate, resulting in corresponding large satellite masses, as shown by the solid black diagonal line, bounded by dotted lines denoting its uncertainty. For the alternative case that included the $m=3$ mode, the anomalous precession rate was lower, resulting in smaller predicted satellites, as indicated by the solid red diagonal line, bounded by its upper uncertainty as a red dotted line. The dashed black and red lines that deviate from the solid lines at the upper right correspond to satellite orbits interior to the $\gamma$ ring's semimajor axis. Representative masses and radii of moonlets and the $\epsilon$ ring are shown, assuming $\rho=1\ \density$. The locations of the rings exterior to the $\gamma$ ring are marked by vertical dotted lines.}
\label{fig:gamma_anomprec}
\end{figure}

                \subsection{The $\epsilon$ ring}
The $\epsilon$ ring precession rate was the main tie point for the inferred gravity field, and so no measurable anomalous precession was detectable.              

\section{Moonlet Orbits, Masses, and Densities}
\label{sec:satellites}
Five of the detected normal modes in the Uranian rings appear to be forced by nearby satellite resonances: the $\eta$ ring 3:2 ILR with Cressida, the $\gamma$ ring 6:5 ILR and the $\epsilon$ OER 14:13 ILR with Ophelia, and the $\delta$ ring 23:22 ILR and the $\epsilon$ IER 24:25 OLR with Cordelia. In this section, we examine the satellite-driven modes detected at ring midlines or edges, compare their pattern speeds and phases with the mean motions and orbital longitudes at epoch of the three satellites, and estimate the masses and densities of the satellites from the amplitudes of the modes.

\subsection{Satellite-driven normal modes}
First-order Lindblad resonances of wavenumber $m$ located near narrow rings produce radial forces that excite normal modes: $m$-lobed distortions in the ring particle orbits that rotate with a pattern speed matching the satellite's mean motion. In the idealized case where ring material can be viewed as independent test particles, the periapse of the distortion is aligned with the mean longitude of the satellite for orbits interior to the resonant radius, with the apoapse being so aligned for particles orbiting exterior to the resonance \citep{Murray1999}. Examples in the Saturn system include the Titan 1:0 ILR in the C ring, where the orientations of a host eccentric features on either side of the resonance match these expectations (see Fig.~19 \cite{Nicholson2014b}), and the Mimas 2:1 ILR near the outer edge of the B ring, where eccentric features in the Cassini Division exterior to the resonance all have their apoapses closely aligned with Mimas (see Fig.~25 \cite{French2016}). 

The situation is more complicated for an eccentric narrow ring affected by nearby resonances when the ring's self-gravity is taken into account (see \cite{Longaretti2018} for an extensive review). In this case, streamline models show that the torque balance at the ring edges can result in shepherding, which serves to account for both the sharp edges of a narrow ring and the absence of differential precession across the ring. As applied to the $\epsilon$ ring, this model predicts that the outer edge is confined by the Ophelia 14:13 ILR and the inner edge is confined by the Cordelia 24:25 OLR, each producing an $m-$lobed edge wave. 
In the absence of dissipation, the forced mode (periapse/apoapse) is predicted to be aligned with the satellite mean longitude for an (ILR/OLR), respectively, and when dissipation is present, 
the relevant apse (lags/leads) the satellite mean longitude in angle for an (ILR/OLR), respectively. (In Appendix D, we justify these statements on dynamical grounds.) 

Finally, both theoretical and numerical models show that under special circumstances a single shepherd satellite can confine a narrow ring \citep{Goldreich1995, Hanninen1994, Hanninen1995, Lewis2011}, but the applicability of these results to the Uranian rings has not been explored in detail.

The characteristics of the forced normal modes identified in the Uranian rings are summarized in
{\bf Table \ref{tbl:satmodes}}. For each mode, we include from Tables \ref{tbl:orbel} and \ref{tbl:normalmodes} the fitted semimajor axis $a$ of the feature, the wavenumber $m$, its amplitude $A_m$, the pattern speed $\Omega_m$, and the distance of the predicted resonance radius from the observed ring feature $\Delta a_P = a_{\rm res}-a$. Also listed are the mean satellite longitude at epoch $\lambda_{\rm sat}$ from the {\tt ura111} ephemeris and the longitude difference 
$\Delta\lambda = \delta'_m  - \lambda_{\rm sat}$, where ${\delta'_m}$ is defined modulo $360^\circ/m$ as the inertial longitude at epoch relative to nearest apse, as identified in the final column.\footnote{For an $m$-lobed normal mode, the phase offset of the mode apse and the satellite longitude is given by $m\Delta\lambda$.} Error bars reflect the uncertainties in $\delta'_m$ and do not include the very small uncertainties in the satellite longitude at epoch (see Table 4, \cite{Jacobson98}). Following \cite{Chancia2017}, we estimate the satellite mass $M_{\rm sat}/M_{\rm Ur} $ from the mode amplitude $A_m$ in the vicinity of a Lindblad resonance and the distance from resonance $|\Delta a_P|$ using Eq.~(10.22) from \cite{Murray1999}:
\beq
\label{eq:satmass}
A_m = \frac{2\alpha a^2(M_{\rm sat}/M_{\rm Ur})|f_d|} {3(m-1)|\Delta a_P|},
\eeq
\noindent where $\alpha= a/a_s$ (the ratio of the ring and satellite semimajor axes) and $f_d$ is the Laplace factor, given by \cite{Murray1999}. (As noted by \cite{Chancia2017}, $\frac{2\alpha |f_d|}{m-1}$ varies between 1.5 and 1.6, depending on $m$.) This relationship is based on test-particle approximation that is strictly applicable only in the evanescent region of the mode and at least 1/4 wavelength from the resonance, and so it is probably best thought of as an approximate scaling estimate, rather than as a dynamically rigorous result for all of the forced modes considered here.

\begin{table*} [ht]
\tiny
\caption{Satellite-driven normal modes}
\label{tbl:satmodes} 
\centering
\begin{tabular}{c l c r r c c r r r r  l}\hline
Satellite & Ring & $a$ & $\Delta a_p$ & $m$ & $A_m$  &$m_{\rm sat}/M_{\rm Ur} $& \multicolumn{1}{c}{$\Omega_p$}  &  \multicolumn{1}{c}{${\delta'_m}^{(a)}$} &  \multicolumn{1}{c}{${\lambda_{\rm sat}}^{(b)}$} & \multicolumn{1}{c}{$\Delta \lambda^{(c)}$} & apse\\
&   &   km & km && km &$\times 10^{-10}$& \multicolumn{1}{c}{$\dd$}  & \multicolumn{1}{c}{deg}  & \multicolumn{1}{c}{deg} & \multicolumn{1}{c}{deg} &  \\
\hline 
Cressida & $\eta$ (COR) & 47176.009 & $   -4.595 $ & $ 3 $ & $  0.600 \pm  0.069 $ & $  21.18 \pm   2.44 $ & $  776.584048 \pm 0.001164 $ & $  15.90 \pm 2.36 $ &  17.44 & $  -1.54 \pm 2.36 $ & apoapse \\
Ophelia & $\gamma$ (IER) & 47624.606 & $    0.754 $ & $ 6 $ & $  0.590 \pm  0.110 $ & $  $ & $  956.418079 \pm 0.000877 $ & $ 303.40 \pm 2.21 $ & 298.08 & $   5.33 \pm 2.21 $ & apoapse \\
 & $\gamma$ (COR) & 47626.170 & $   -0.861 $ & $ 6 $ & $  0.637 \pm  0.063 $ & $   4.23 \pm   0.42 $ & $  956.419622 \pm 0.000474 $ & $ 302.52 \pm 1.17 $ & 298.08 & $   4.44 \pm 1.17 $ & apoapse \\
 & $\gamma$ (OER) & 47627.865 & $   -2.552 $ & $ 6 $ & $  0.892 \pm  0.117 $ & $  $ & $  956.419529 \pm 0.000560 $ & $ 298.03 \pm 1.36 $ & 298.08 & $  -0.05 \pm 1.36 $ & apoapse \\
 & $\epsilon$ (COR) & 51149.279 & $   28.425 $ & $ 14 $ & $  0.383 \pm  0.071 $ & $  $ & $  956.418015 \pm 0.000269 $ & $ 296.96 \pm 0.63 $ & 298.08 & $  -1.12 \pm 0.63 $ & periapse \\
 & $\epsilon$ (OER) & 51178.588 & $   -0.887 $ & $ 14 $ & $  0.590 \pm  0.130 $ & $   3.67 \pm   0.81 $ & $  956.418119 \pm 0.000298 $ & $ 297.76 \pm 0.70 $ & 298.08 & $  -0.32 \pm 0.70 $ & periapse \\
Cordelia & $\delta$ (COR) & 48300.227 & $    2.173 $ & $ 23 $ & $  0.339 \pm  0.063 $ & $   5.58 \pm   1.09 $ & $ 1074.523021 \pm 0.000189 $ & $  69.45 \pm 0.47 $ &  70.00 & $  -0.55 \pm 0.47 $ & periapse \\
 & $\epsilon$ (IER) & 51120.014 & $    1.082 $ & $ -24 $ & $  1.011 \pm  0.107 $ & $   7.83 \pm   0.83 $ & $ 1074.522889 \pm 0.000099 $ & $  70.42 \pm 0.26 $ &  70.00 & $   0.41 \pm 0.26 $ & apoapse \\
 & $\epsilon$ (COR) & 51149.279 & $  -28.177 $ & $ -24 $ & $  0.443 \pm  0.061 $ & $  $ & $ 1074.522703 \pm 0.000142 $ & $  70.33 \pm 0.35 $ &  70.00 & $   0.33 \pm 0.35 $ & apoapse \\
\hline
\end{tabular}
\begin{itemize}
\item[a] The epoch is 1986 Jan 19 12:00 (TDB). $\delta'_m$ is the observed longitude of the mode's periapse or apoapse, as specified in the final column. It is ambiguous by additive multiples of $360^\circ/m$, adjusted to minimize the magnitude of $\Delta \lambda$.
\item[b] Mean longitude of the satellite at the epoch, from the {\tt ura111} ephemeris.
\item[c] $\Delta\lambda = \delta'_m  - \lambda_{\rm sat}$. Error bars reflect the uncertainties in $\delta'_m$ and do not include the uncertainties in the satellite ephemerides (see Table 4, \cite{Jacobson98}).
\end{itemize}
\end{table*}

All of the normal modes in Table \ref{tbl:satmodes} have pattern speeds that are very close to those of the satellites associated in the resonances. {\bf Table \ref{tbl:satmeanmo}} lists the weighted average pattern speeds of all identified modes associated with Cressida, Ophelia, and Cordelia, as well as satellite mean motions from published and unpublished sources. 
\begin{deluxetable}{l l r}
\tablecolumns{3}
\tablecaption{Satellite mean motions\label{tbl:satmeanmo}}
\tablehead{
\colhead{Satellite} & 
\colhead{Source} & 
\colhead{$\Omega_P$ (\degd)} \\[-0.9em]}
\startdata
Cressida & This work & $776.584048 \pm  0.001164$                      \\
& \cite{Jacobson98} & $776.582414  \pm  0.000022$                       \\
& \cite{Showalter2006} & $776.582789  \pm  0.000059$  \\
& Robert French (pers. comm.) & $776.5826393    \pm  0.0000053$  \\
[.5em]Ophelia & This work & $956.418404  \pm  0.000171$   \\
& \cite{Jacobson98} & $956.42833  \pm  0.00908$ \\
& Robert French (pers. comm.) & $956.418293     \pm  0.000037$  \\
[.5em]Cordelia & This work & $1074.522858 \pm 0.000075$  \\
& \cite{Jacobson98}  & $1074.51832  \pm  0.00187$ \\
\enddata
\end{deluxetable}

\subsubsection{Cressida}
The $m=3$ ILR forced by Cressida in the $\eta$ ring was first identified by \cite{Chancia2017}, based on a ring orbit fit to a subset of the data used in our current analysis. The Cressida resonance lies $\Delta a_P=-4.595$ km interior to the ring COR.
We find $A_m=0.600\pm0.069$ km, $\Omega_P=776.584048\pm0.001164\dd$, 
$\delta'_m=15.90\pm2.36^\circ$, and $\Delta \lambda = -1.54\pm 2.36^\circ $ from 60 $\eta$ ring COR measurements, comparable to but with smaller error bars than the \cite{Chancia2017} values of $A_m=0.667\pm0.113$ km, $\Omega_P=776.58208\pm0.00169\dd$, 
 and $\Delta \lambda = -2.72 \pm 6.12^\circ $ from 49 measurements. 
 
 The pattern speed of the 3:2 ILR in the $\eta$ ring is reasonably consistent with previous values for Cressida's mean motion, differing by $1.4\sigma$ from the \cite{Jacobson98} value, by $1.1\sigma$ from the \cite{Showalter2006} value, and by $1.2\sigma$ from the unpublished value (Robert French, pers. comm) based on an analysis of \hst\ images.
 
These results are consistent with the test particle interpretation of the response of the $\eta$ ring to the Cressida resonance, which is too distant to produce single-sided shepherding. Since the resonance is interior to the ringlet, the normal mode apoapse is predicted to be aligned with the satellite, as observed, with no statistically significant lead or lag in longitude.

\subsubsection{Cordelia and Ophelia}
The $m=6$ ILR Ophelia resonance has long been recognized as being located near to the $\gamma$ ring \citep{Porco1987}, and 
as described in Section \ref{sec:normalmodes} and tabulated in Table \ref{tbl:orbel}, we detected the $m=6$ mode at the ring midline and both edges, with the measured amplitude $A_m$ increasing systematically from $0.591\pm0.110$ km at the IER, to $0.637\pm 0.063$ km at the COR, and finally to $0.891\pm0.117$ km at the OER. The resonance radius lies $\Delta a_P=-2.552$ km interior to the ring OER, where the mode amplitude is greatest. The mode apoapse leads Ophelia by $\Delta\lambda=+5.33\pm2.21^\circ$ at the ring's IER, by $+4.44\pm1.17^\circ$ at the ring's COR, and at the OER, where the $m=6$ mode amplitude the largest, the apoapse is nearly aligned with Ophelia: $\Delta\lambda=-0.05\pm1.36^\circ$.

The dynamics of the $\gamma$ ring are complex, with the simultaneous presence of five free normal modes and the Ophelia 6:5 ILR forced mode that is located within the ring itself (Fig.~\ref{fig:widthsgamma}). The near-alignment of the $m=6$ mode apoapse with Ophelia is as expected if the ring material responds as isolated test particles to the satellite's forcing, but this neglects the possibly significant role of self-gravity for this narrow ringlet of high average optical depth. The apparently statistically-significant phase offsets of the mode at the ring's inner edge and midline have no obvious explanation. 

As shown previously in Sections \ref{sec:epsmodes} and \ref{sec:epsmodesCOR}, normal mode scans revealed the presence of the $m=14$ ILR with Ophelia in the $\epsilon$ OER and COR, and the $m=-24$ OLR with Cordelia in the $\epsilon$ IER and COR. The final values for the mode amplitudes, phases, and pattern speeds when included in our adopted orbit fit are given in Table \ref{tbl:orbel} and in more detail in Table \ref{tbl:satmodes}. The periapse of the $m=14$ OER mode lags behind Ophelia by $\Delta\lambda=-0.55\pm0.47^\circ$ , as expected for a normal mode at the outer edge of the ring, and the apoapse of the $m=-24$ IER mode leads Cordelia by $\Delta\lambda=0.41\pm0.26^\circ$, again in keeping with the dynamical arguments presented in Appendix D for a forced normal mode at the inner edge of a ring in the presence of dissipation.

The $m=23$ ILR Cordelia resonance was first associated with the $\delta$ ring by \cite{Porco1987}. We detected the signature of this resonance in our normal mode scans of the $\delta$ ring COR, although the fitted mode amplitude is only $A_{23}=0.339\pm0.063$ km, too small to be detectable in either of the noisier ring edge (IER/OER) orbit fit residuals. The resonance is located slightly outside of the ring COR ($\Delta a_P=2.173$ km). The test-particle model predicts that the mode periapse should be aligned with Cordelia, as observed. There is a marginally-detectable lag in the orientation of periapse behind Cordelia at the 1-$\sigma$ level: $\Delta\lambda = -0.55\pm0.54^\circ$, perhaps due to dissipation in the ring.

Comparisons of the pattern speeds and orbital longitudes of the normal modes and the $\tt ura115$ ephemeris values for Ophelia and Cordelia are shown in {\bf Fig.~\ref{fig:OphCord}}. The top left panel shows the fitted pattern speeds and their error bars for the $\gamma$ ring IER/COR/OER $m=6$ mode measurements and the $\epsilon$ ring COR and OER $m=14$ results, compared to the \cite{Jacobson98} value. The weighted mean value of our measurements is shown by the solid green line, bracketed by the estimated uncertainty. The dashed blue lines show the rather large 1-$\sigma$ uncertainty in Ophelia's mean motion from \cite{Jacobson98}. All of the measured pattern speeds are systematically below the {\tt ura115} ephemeris value by $\simeq 0.01$ \degd. The weighted mean pattern speed for the normal modes of 
      $\Omega_P=956.418404\pm0.000171$ \degd\ is just $\sim0.65 \sigma$ from the unpublished value of $\Omega_P=956.418293\pm0.000037$ \degd\ shown in red, derived from \hst\ observations (Robert French, personal communication). The lower left panel shows the longitude differences $\Delta\lambda$ between the measured phases of the detected modes and the mean longitude of Ophelia at the epoch. The $\gamma$ ring $m=6$ normal mode apoapse leads the satellite for the IER and COR, but aligns with the satellite at the OER. On the other hand, the periapse of the $\epsilon$ ring $m=14$ ILR at the ring's outer edge lags behind Ophelia, as expected in the presence of dissipation.
\begin{figure}
\centerline{\resizebox{\fw}{!}{\includegraphics[angle=90]{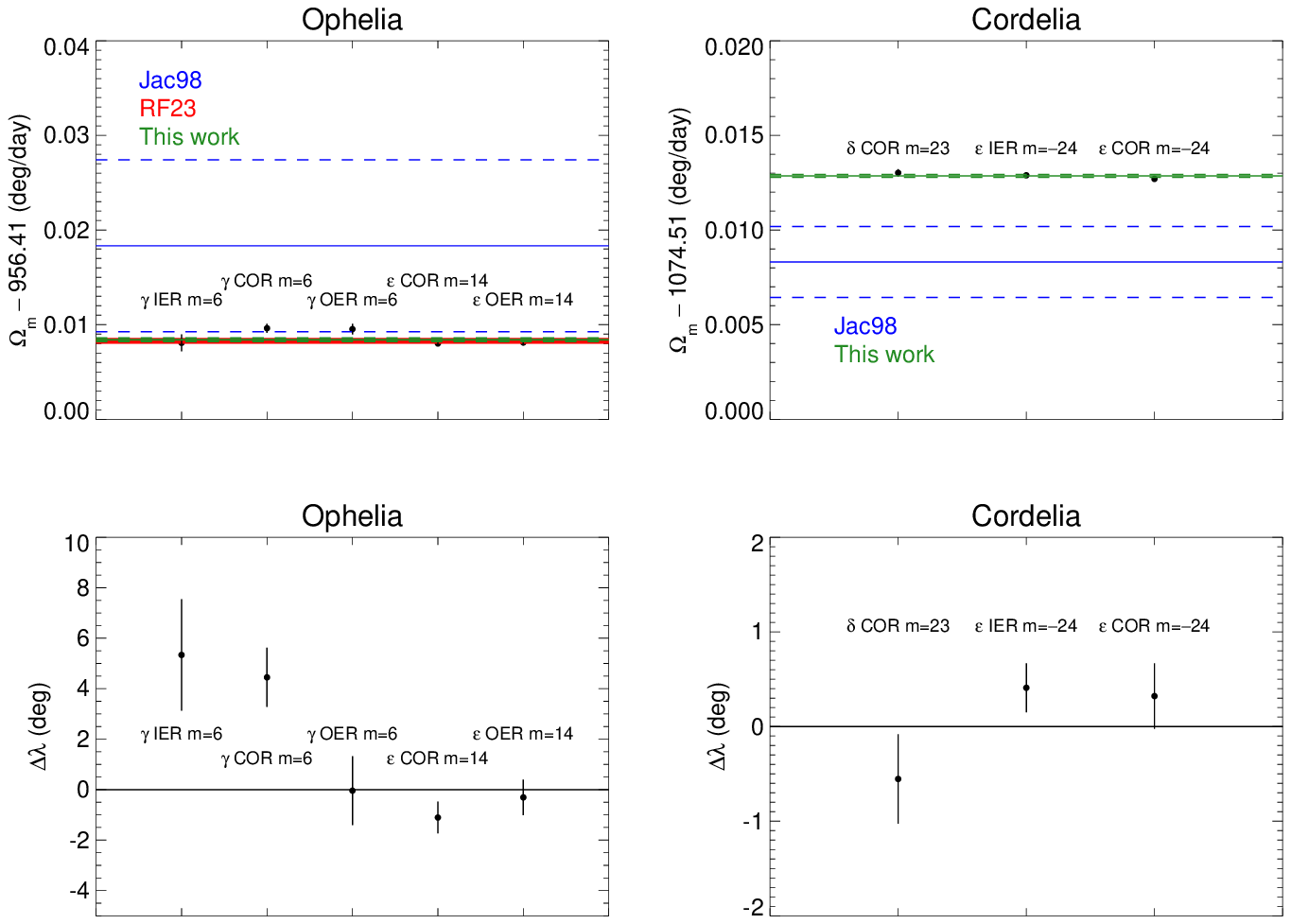}}}
\caption{Pattern speed and phase comparison for the observed normal modes forced by Ophelia and Cordelia. The upper left panel shows the observed pattern speeds $\Omega_m$ for each of five labeled modes associated with Ophelia, plotted as filled circles with error bars. The solid horizontal lines mark the mean motions of Ophelia (bounded by their uncertainties, shown dashed) reported by \cite{Jacobson98} (shown in blue and labeled Jac98) and Robert French (pers. comm.) shown in red and labeled RF23, compared to the weighted average of the mode pattern speeds from this work, shown in green. The upper right panel compares the \cite{Jacobson98} and our results for Cordelia. The bottom panels show the longitude differences $\Delta\lambda$ between the fitted periapse or apoapse longitudes and the satellite mean longitudes at epoch from the {\tt ura115} satellite ephemeris (from Table \ref{tbl:satmodes}).}
\label{fig:OphCord}
\end{figure}

The right column of Fig.~\ref{fig:OphCord} shows the comparable results for Cordelia. In this case, the best fitting pattern speeds for the $\delta $ ring COR $m=23$ mode and the $\epsilon$ ring $m=-24$ mode values for the IER and COR are in excellent mutual agreement, although they lie above the \cite{Jacobson98} values, which have much larger error bars. In the lower right panel, the longitude of the apoapse of the $\delta$ ring $m=23$ normal mode lags slightly behind Cordelia, while the $\epsilon$ ring's inner edge mode associated with the Cordelia 24:25 OLR leads the satellite, as expected in the presence of dissipation.
\subsection{Moonlet masses and densities}
\label{sec:moondensities}
The densities of Cressida, Ophelia, and Cordelia can be estimated from the ratio of their masses given in Table \ref{tbl:satmodes} to their volumes, calculated from the prolate satellite dimensions from Table IV of \cite{Karkoschka2001a}. %
These can be compared to the Roche critical density $\rho_{\rm Roche}$ 
\beq
\rho_{\rm Roche} = \frac{4\pi\rho_p} {\gamma(a/R_p)^3}= \frac{3 M_p}{\gamma a^3},
\eeq
\noindent where $\rho_p, R_p$ and $M_p$ are the planet's density, radius, and mass, $a$ is the distance from the planet, and $\gamma$ is a dimensionless parameter associated with the mass distribution and shape of the satellite \citep{Tiscareno2013}. An object without material strength and density less than $\rho_{\rm Roche}$ would be pulled apart, and thus the critical density curve represents a tidally governed lower limit to density of gravitationally bound aggregates. Following \cite{Porco2007} and \cite{Tiscareno2013}, we assume $\gamma=1.6$, corresponding to the case when the satellite's mass is assumed to be uniform in density and distributed into the shape of its Roche lobe.

The computed satellite densities are plotted as a function of orbital radius in {\bf Fig.~\ref{fig:roche}}. 
Cordelia and Ophelia each have two separate resonances that yield mutually-consistent densities with overlapping error bars computed from the uncertainties in the fitted mode amplitudes alone, although the realistic uncertainty in their average value (plotted in gray) is dominated by the contribution of the much larger satellite volume uncertainty. Even taking these larger error estimates into account, the calculated satellite densities appear to decrease with orbital radius by about a factor of two from innermost Cordelia to outermost Cressida. For comparison, the Roche critical density is plotted as a solid line, assuming that the shape parameter $\gamma=1.6$.
It increases inward quite rapidly, from $0.5\ \density$ at 2.6 $R_p$ to over $2.0\ \density$ at ring 6, supporting the view that the Uranian rings are more rocky and less porous than Saturn's rings \citep{Tiscareno2013}. For smaller values of $\gamma$, the critical density curve would lie above the measured densities and require that the moonlets have material strength. The observed densities follow the overall trend of the Roche critical density curve for $\gamma=1.6$, which may reflect the formation process and possible orbital migration of these small moons. 
 
\begin{figure}
\centerline{\resizebox{7in}{!}{\includegraphics[angle=90]{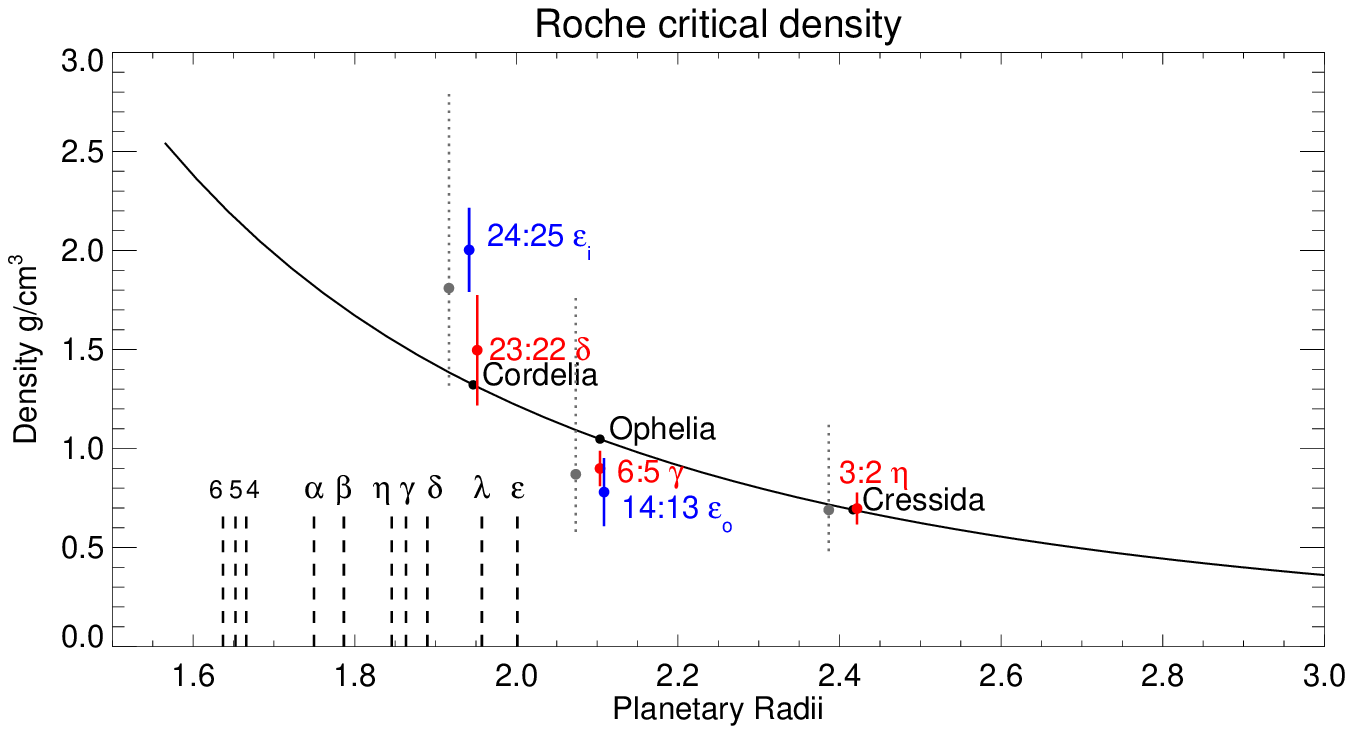}}}
\caption{Densities of Cordelia, Ophelia, and Cressida computed from fitted normal mode amplitudes $A_m$, distances from resonance $\Delta a_P$, and satellite dimensions from \cite{Karkoschka2001b}. Here, we use the COR amplitudes for the $\eta, \delta$ and $\gamma$ ring resonances, since the ring midline fits have smaller formal errors than the edge amplitudes that are based on noisier data. The error bars on the individually marked resonances reflect only the contribution of the fractional errors of $A_m$ to the overall error budget, which is dominated by the satellite volume uncertainties. The gray dots are the weighted mean densities for each satellite and the dotted error bars correspond to the combined uncertainties in the mode amplitudes and the satellite volumes (Table \ref{tbl:minorsatdata}). The Roche critical density is plotted as a solid line for shape parameter $\gamma=1.6$ and $R_p=25559$ km, with filled circles at the orbital radii of the satellites. Dashed vertical lines are labeled with the locations of the ten narrow rings. }
\label{fig:roche}
\end{figure}

\section{Discussion}

\subsection{Ring orbits, width variations, and masses}

The most secure results from nearly three decades of occultation observations of the narrow rings of Uranus are their orbital shapes and width variations. Although the number of individual occultation observations for a given ring is at most just 89 (well below the $\sim$150--260 for Saturn's rings from \Cassini\  occultation observations), the long time-baseline and remarkable geometric accuracy of the orbit reconstructions enables the detection of normal modes and width variations of the rings on a scale of $\sim0.2$ km for the Uranian rings. The internal structure of the narrowest of the rings is largely unresolved in Earth-based observations, preventing the accurate reconstruction of streamline models that could provide detailed information about self-gravity wakes, ring confinement mechanisms, and signatures of unseen satellites. 

Standard self-gravity (SSG) models predict unrealistically small ring masses and surface densities and fail to match the actual variations in $q_e(r)$ observed in the rings, but they successfully predict the slopes of the observed width-radius relations for nearly all of the inner and outer Lindblad resonances detected in the rings. Collisional self-gravity models (CSG) predict much larger ringlet masses and surface densities more consistent with expectations based on particle size estimates, but they sequester the bulk of the mass in narrow regions near the ring edges, predicting U-shaped radial optical depth profiles that are a poor match to high-resolution \Voyager\ occultation observations. More work is needed to ascertain whether the differences between the various predicted profiles and the observed profiles can be explained as the result of variations in the effective opacity of the ring material across each ring, or if they instead require a more complex dynamical model that produces a more uniform optical depth profile. Here, it might prove fruitful to compare the Uranus results with surface densities of Saturn's rings in regions of comparable optical depth. 

The accurate determination of the orbits and width-radius relations for the detected normal modes provides strong evidence for the overall balance between differential precession due to planetary oblateness and to self-gravity in preserving the apse alignment of the inner and outer edges of the eccentric ringlets. Only a subset of the Uranus rings have excited normal modes, but the $\gamma$ ring's five free modes puts it in the class of several of Saturn's rings and ring edges that have as many or more normal modes. The excitation mechanism behind the observed modes is poorly understood, although it seems likely that non-linear coupling between modes might be responsible for the simultaneous presence of many modes, as seems to be the case for the complex shape of the outer edge of the B ring \citep{French2023b}. 

While we found no statistically significant evidence for normal modes driven by internal planetary oscillations, it is important to note that the patterns generated by satellite resonances on the $\epsilon$, $\delta$ and $\gamma$ rings are relatively subtle, and finding resonances with planetary normal modes is more challenging because their pattern speeds cannot be accurately predicted in advance. This lack of detections therefore does not rule out the possibility  that planetary normal modes are  responsible for confining some of these rings. Still, these data, combined with more detailed analyses of the internal structures and longitudinal brightness variations within each of these rings, should help constrain the properties of the planet's internal normal modes.

\subsection{Ring-moonlet connections}

Among the host of narrow rings in the solar system, only the $\epsilon$ ring is securely established as being confined by shepherd satellites, with the inner satellite Cordelia having a 24:25 OLR with the ring's inner edge and the outer satellite Ophelia having a 14:13 ILR with the outer edge \citep{Porco1987}.
The amplitudes, pattern speeds, and phases of the edge waves match dynamical expectations and provide estimates of the masses of the confining satellites. The resonance radii of both edge modes are located within the ring, $\sim1$ km from the corresponding ring edge. In the future, it will be interesting to compare these values with numerical simulations.

In addition to their roles as shepherd satellites for the $\epsilon$ ring, Cordelia and Ophelia each are the source of an additional forced normal mode in the rings: the 23:22 ILR of Cordelia with the $\delta$ ring and the 6:5 ILR of Ophelia with the $\gamma$ ring. Cressida's 3:2 ILR with the $\eta$ ring provides a third example. The orientations of the excited normal modes in all three cases match a simple test-particle interpretation of the ring response to the satellite forcing (although the complex dynamics of the $\gamma$ ring remain to be investigated in detail). An example of this mechanism in the Saturn system is provided by the Barnard Gap outer edge, which is perturbed by a Prometheus 5:4 ILR. Here,  the apoapse of the mode is aligned with the satellite, as expected from the test-particle model \citep{French2016}.

The internal structure of several narrow ringlets has been interpreted as evidence for satellite wakes generated by unseen moonlets near the $\delta$ ring \citep{Horn1988} and the $\alpha$ and $\beta$ rings \citep{Chancia2016}. Although our study offers no additional supporting evidence for a $\delta$ ring moonlet, our measurements of anomalous precession rates in both the $\alpha$ and $\beta$ rings are consistent with the mass estimates for the 
proposed unseen moonlets near $\alpha$ and $\beta$ rings, making them just below the level of detectability in \Voyager\ or \hst\ images. Direct detections of these moons may have to await a future mission to the Uranus system, but these small moons could produce localized disturbances in nearby rings that could be detectable in \hst\ or \JWST\ images. 

A provocative result of our study is the striking trend of moonlet densities with orbital distance from Uranus. Taken at face value, the densities of Cordelia, Ophelia, and Cressida follow the systematic trend of the Roche critical density for $\gamma=1.6$, rising from 0.71 $\density$\ for outermost Cressida to double that value for innermost Cordelia. This finding is consistent with these moons having relatively little internal strength, which is compatible with the overall architecture of the Uranian ring-moon system \citep{Tiscareno2013}. This strikes an interesting contrast with the Neptune system, whose innermost moons must have sufficient internal strength to survive at their current locations \citep{Tiscareno2013, Brozovic2020}

Cordelia's high density is particularly interesting because it is comparable to the densities of the much larger moons Ariel (1.54 $\density$), Umbriel (1.52 $\density$), Titania (1.65 $\density$) and Oberon (1.67 $\density$), and is actually larger than the density of Miranda (1.18 $\density$) \citep{Thomas1988, Jacobson2014}. This exceptionally high density likely has implications for the origins and evolution of these small moons. For example, according to standard tidal theory, Cordelia should be migrating inwards, so perhaps Cordelia represents the dense core of a larger moon whose outer layers have been stripped away by tidal forces. In this scenario, the $\epsilon$ ring would be composed of the last bits of material lost from Cordelia's surface.

\subsection{Uranus gravity field, internal structure, and winds}
Accurate measurements of the gravity fields of giant planets can provide powerful constraints on their internal structure and the depth and nature of their wind profiles, as has been shown recently for Saturn and Jupiter from \Cas and \Juno measurements of high-order gravitational moments \citep{Iess2018,Iess2019,Dietrich2021,Kulowski 2021,Militzer2022,Galanti2022}. Uranus has very strong winds at the cloud level, with features at $\pm60^\circ$ latitude moving around the planet over 200 m s$^{-1}$ faster than the those near the equator \citep{Hammel2001,Hammel2005,Sromovsky2005}. If these winds penetrate far enough into Uranus' interior, they could significantly influence the planet's gravity field. \cite{Kaspi2013} concluded that the difference between the observed value of $J_4$ and the predicted value based on relatively simple interior models was sufficient to constrain the dynamics to the outermost 0.15\% of the total mass of Uranus, corresponding to a relatively thin weather layer no more than about 1,000 kilometers deep. More recently, \cite{Movshovitz2022} showed that the moment of inertia (MoI) is tightly correlated with the gravity field beyond $J_2$, and that to improve interior models and the MoI determination, the planet's rotation period must be known to comparable precision.
\cite{Neuenschwander2022} demonstrated that knowledge of $J_6$ to 10\% accuracy is required to constrain wind depths on Uranus: relatively high values of $J_6 = (59.9–69.04) \times 10^{-8}$
 can only be explained by
the existence of deep winds with a penetration depth of more than
250 km, whereas lower values 
$J_6 =
(46.12–53.76) \times 10^{-8} $ do not allow
for deep winds with a penetration depth as large as 1100 km. Significantly, they also demonstrated that more accurate values of $J_2$ and $J_4$ 
 can further constrain the density distribution and narrow the range of predicted
solutions in $J_6, J_8,$ and MoI of Uranus and Neptune. Finally, \cite{Soyuer2023} conclude from model studies that the scale height of the zonal winds on Uranus is below $\sim 0.03 R_U$ (770 km), somewhat shallower than found by \cite{Kaspi2013}.

Nearly all of these modeling efforts for the interior structure and dynamics of Uranus have made use of the \cite{Jacobson2014} values for $J_2$ and $J_4$ listed under Fits 2 and 3 in Table \ref{tbl:Jn}, but they have not yet exploited the strong degree of correlation between the measured values of $J_2$ and $J_4$ as an additional constraint. Although we are unable to set useful limits on $J_6$ alone, as we have shown in Table \ref{tbl:Jn} and Fig.~\ref{fig:J2J4}, our adopted gravity field solution based on a more complete set of ring observations provides tighter limits not only on $J_2$ and $J_4$, but also on allowed combinations of values of $J_2, J_4,$ and $J_6$, after accounting for systematic uncertainties in the center of opacity of the rings, the absolute radius scale, and the masses of the major and minor satellites. The prescription in Appendix \ref{appendix:grav} can be used to restrict the range of density distributions and wind models for Uranus that are consistent with the ring occultation results, although significant advances in our understanding of the planet's internal structure will probably require high-precision gravity measurements possible only from an orbiting spacecraft \citep{Movshovitz2022}.

\subsection{Open questions and future prospects}
Decades after their discovery, the Uranian rings continue to raise vexing dynamical puzzles. What confines the narrow rings? How can they maintain their eccentricities and inclinations in the presence of differential apsidal and nodal precession rates? What are their connections with the planet's host of small moons? How recently were they formed, and what is the time scale for significant evolution of ring systems?
The findings presented in this paper demonstrate the richness and complexity of the Uranian ring system, and its potential to illuminate other aspects of the Uranus system. Already, these rings have provided constraints of the gravity field of Uranus itself, estimates the mass densities of several known satellites, and evidence for a number of still-unseen moons. Future observations by earth-based telescopes or missions to the Uranus system could therefore provide key insights into multiple aspects of Ice Giant Systems \citep{OWL}.

On the one hand, detailed studies of the structure and dynamics of the Uranian rings will provide additional insights into the interiors of Ice Giants. Continued monitoring of the rings' orbital evolution will yield increasingly tight constraints on the higher-order components of the planet's gravitational field. At the same time, if any structures in the rings can be attributed to resonances with oscillations or asymmetries in the planetary normal modes \citep{AHearn2022}, those structures will provide novel information about the internal structures of Ice Giants, which are still largely unconstrained \citep{Helled2022}.

Meanwhile, further studies of these rings should clarify whether small moons do in fact exist near the rings. The sizes, locations and densities of such moons will constrain the conditions under which solid material is more likely to accrete or fragment, which is relevant to not only the Uranian rings and moons, but also the formation of moons and planets more generally. 

%
\section{Conclusions}
Our main results are summarized below:
\begin{itemize}
\item
From an analysis of 31 Earth-based stellar occultations and three \Voyager\ 2 occultations spanning 1977--2006 (Paper 1), we have determined the keplerian orbital elements of the centerlines (COR) of the nine main Uranian rings to high accuracy, with typical RMS residuals of 0.2 -- 0.4 km and 1-$\sigma$ formal errors in $a, ae,$ and $a\sin i$ of order 0.1 km, registered on an absolute radius scale accurate to 0.2 km at the 2-$\sigma$ level. The $\lambda$ ring shows more substantial scatter, with limited secure detections. 

\item
We characterized a host of free and forced normal modes in several of the ring centerlines and inner and outer edges (IER/OER). In addition to the previously-known free modes $m=0$ in the $\gamma$ ring and $m=2$ in the $\delta$ ring, we have identified two additional OLR modes ($m=-1$ and $-2$) and a possible $m=3$ mode in the $\gamma$ ring. No normal modes were detected for rings 6, 5, 4, $\alpha$, or $\beta$.

\item The origin of the $\gamma$ ring's putative $m=3$ mode is mysterious. Its pattern speed is much slower than oscillation frequencies expected from internal oscillations of the planet, and it is unlikely to be due to an unseen satellite, whose predicted size would be too large to have avoided prior detection. If it is an unforced mode, it is unusual in having its resonance radius more than 1 km exterior to the ring's outer edge. If the mode is omitted from the orbit solution, the $\gamma$ ring has a very large anomalous apsidal precession rate of unknown origin.

\item Five separate normal modes are forced by small moonlets: the 3:2 inner Lindblad resonance (ILR) of Cressida with the $\eta$ ring, the 6:5 ILR of Ophelia with the $\gamma$ ring, the 23:22 ILR of Cordelia with the $\delta$ ring, the 14:13 ILR of Ophelia with the outer edge (OER) of the $\epsilon$ ring, and the counterpart 25:24 OLR of Cordelia with the ring's inner edge. The phases of the modes and their pattern speeds are consistent with the mean longitudes and mean motions of the satellites, confirming their dynamical roles in the ring system.

\item
From separate orbit fits to the ring edge measurements obtained from square-well model fits to individual ring profiles, we determined the width-radius relations for nearly all of the detected modes, with positive width-radius slopes for ILR modes (including the $m=1$ elliptical orbits) and negative slopes for most of the detected OLR modes, as predicted by standard self-gravity models. Ring mass and surface density estimates based on standard self-gravity and collisional self-gravity models differ by orders magnitude and predict radically different radial optical depth profiles. These differences are unresolved at present, and warrant additional investigation. We found no convincing evidence for librations of any of the rings.

\item
The Uranus pole direction at epoch TDB 1986 Jan 19 12:00 is $\alpha_P=77.311327\pm 0.000141^\circ$ and $\delta_P=15.172795\pm0.000618^\circ$, where the error bars take into account possible systematic errors in the \Voyager\ ephemeris. The slight pole precession predicted by \cite{Jacobson2023} was not detectable in our orbit fits, and the absolute radius scale is not strongly correlated with the pole direction.

\item
From Monte Carlo fits to the measured apsidal precession and nodal regression rates of the eccentric and inclined rings, we determined the zonal gravitational coefficients 
$J_2=(3509.291\pm0.412)\times 10^{-6}$ and $J_4=(-35.522\pm0.466)\times 10^{-6}$ with a correlation coefficient $\rho(J_2,J_4)=0.9861$ and $J_6$ held fixed at= $0.5\times 10^{-6}$, for a reference radius $R=$25559 km. This result is significantly displaced from previous results \citep{Jacobson2014}, owing to the inclusion of previously neglected systematic effects. The quoted errors account for systematic differences (and their uncertainties) between the fitted semimajor axes of the ring centerlines (COR) and their estimated centers of opacity (COO), a surrogate for the center of mass (COM) semimajor axes that should be used when computing the radial gradient in the apse and node rates. The minor satellites Cordelia and Ophelia contribute significantly to the precession rate of the $\epsilon$ ring, affecting the final solution for $J_2$ and $J_4$. Although we cannot set useful independent limits on $J_6$, we obtain strong joint constraints on combinations of $J_2,J_4,$ and $J_6$ that are consistent with our measurements. These can be used to limit the range of realistic models of the planet's internal density distribution and wind profile with depth.

\item
The measured anomalous apsidal and nodal precession rates of the $\alpha$ and $\beta$ rings are consistent with the presence of unseen moonlets with masses and orbital radii predicted by \cite{Chancia2016}.

\item
From the amplitudes and resonance radii of normal modes forced by moonlets, we determined the masses of Cressida, Cordelia, and Ophelia. Their estimated densities vary systematically with orbital radius and generally follow the radial trend of the Roche critical density for a shape parameter $\gamma=1.6$. This may provide an important clue to the formation process and possible orbital migration of these small moons.

\end{itemize}
%
\section{Acknowledgements}
We are grateful to two anonymous reviewers for their helpful suggestions. We also thank Robert Jacobson and Ryan Park of JPL for incorporating the Uranus ring occultation data into the updated planet, satellite, and Voyager ephemerides used for our orbit fits. Robert French generously shared unpublished orbital elements for Cressida and Ophelia from \hst\ satellite astrometry. Naor Movshovitz and Benno Neuenschwander provided helpful assessments of the implications for Uranus internal models of the new constraints on $J_2, J_4,$ and $J_6$. Matthew Tiscareno provided helpful insight into the interpretation of the trend of small satellite densities with orbital radius. This work was supported by NASA PDART grant NNX15AJ60G: ``Restoration and submission of Uranus ring occultation observations to the Planetary Data System'' and NASA Solar System Workings grant NNX15AH45G: ``Uranian ring dynamics and constraints on Uranus’ internal structure from occultation data."
This work has made use of data from the European Space Agency (ESA) mission
{\it Gaia} (\url{https://www.cosmos.esa.int/gaia}), processed by the {\it Gaia}
Data Processing and Analysis Consortium (DPAC,
\url{https://www.cosmos.esa.int/web/gaia/dpac/consortium}). Funding for the DPAC
has been provided by national institutions, in particular the institutions
participating in the {\it Gaia} Multilateral Agreement.

%

%

\restartappendixnumbering
\appendix

 \section{{\it Voyager\ 2} occultations and square-well model evaluation}
 \label{appendix:sqwell}
The {\it Voyager\ 2} radio and stellar occultations provide our current best view of the fine-scale radial internal structure of the narrow Uranian rings. (For a detailed description of the \Voyager\ observations and of the shapes and observed structure of each ring, see \cite{French1991}.) 
The high-SNR $\sigma$ Sgr occultation chord intersected only the $\delta$, $\lambda$, and $\epsilon$ rings, while the lower-SNR $\beta$ Per event spanned the entire ring system on ingress and egress. The viewing geometry of the two stellar occultations was quite different from all other Uranus ring occultations, providing additional constraints on the Uranus pole direction. 

The \Voyager\ RSS data present an excellent opportunity to assess the accuracy of the square-well model approximation to ring structure because the raw observations are affected by diffraction with a Fresnel scale $F\sim 1.6-2.3$ km, comparable to typical values for Earth-based stellar occultations, while the diffraction-reconstructed data reveal the true ring structure at 50~m resolution \citep{Gresh1989}. {\bf Figure~\ref{fig:RSSgesqw}} shows the diffraction pattern visible in the uncorrected RSS observations of the egress $\gamma$ ring profile, along with the best-fitting square-well model to the measured intensity. It closely resembles the diffraction pattern in Fig.~\ref{fig:U36AATgisqw}. In this instance, the best fitting square-well model is 1.37 km wide and nearly opaque, with a corresponding diffraction pattern computed for the event geometry that matches the observations for about seven fringes on either side.

\begin{figure}
\centerline{\resizebox{4in}{!}{\includegraphics[angle=90]{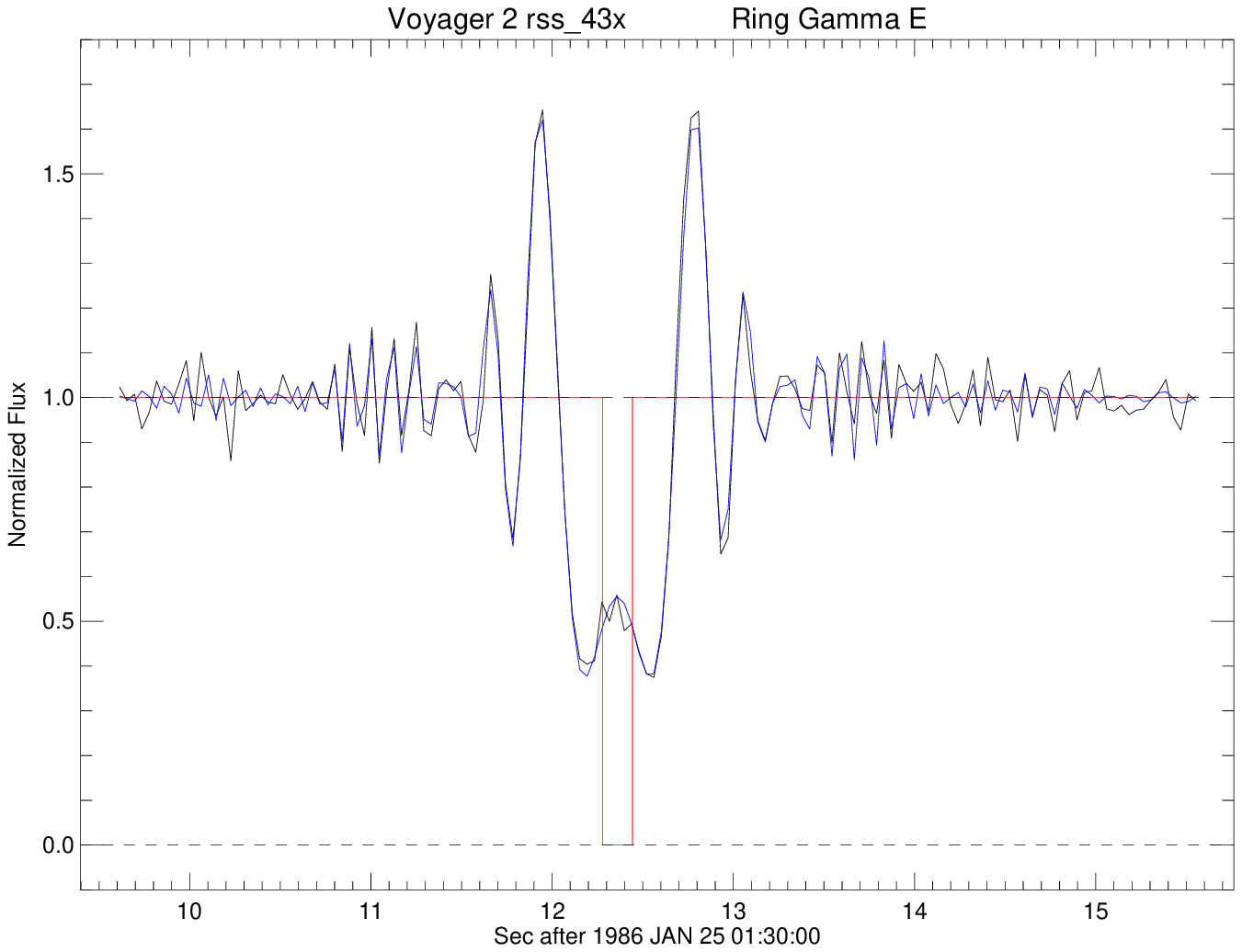}}}
\caption{Raw intensity profile of \Voyager\ 2 $\gamma$ ring egress profile, uncorrected for diffraction. The best-fitting square-well model is overplotted, and the box function square-well model is shown in red, with a width of 1.37 km.}
\label{fig:RSSgesqw}
\end{figure}

Unlike stellar occultation observations, however, the availability of phase information in the recorded RSS signal enables the diffraction effects to be greatly reduced using Fresnel inversion \citep{Marouf1986}. The diffraction-corrected profile for this profile is shown in {\bf Fig.~\ref{fig:VgrGErecon}}, at a processing resolution of 50~m. The upper panel shows the optical depth profile, with sharp edges and perhaps some evidence of noise-limited internal structure. The lower panel shows the reconstructed intensity profile, which is a more appropriate representation for comparison with a square-well model. In this instance, the ring does indeed resemble a square-well with sharp edges and a measured width of 1.43 km, determined from Eq.~(\ref{eq:widthcomparison}) below (shown dashed), very close to the 1.37 km width of the square-well fit to the diffraction-limited data. Note that the radius scale is from the data in the PDS archive file, and not from the current orbit determination.\footnote{\tt VG\_2803/U\_RINGS/EASYDATA/KM00\_025/RU1P2XGE.TAB}

\begin{figure}
\centerline{\resizebox{4in}{!}{\includegraphics[angle=0]{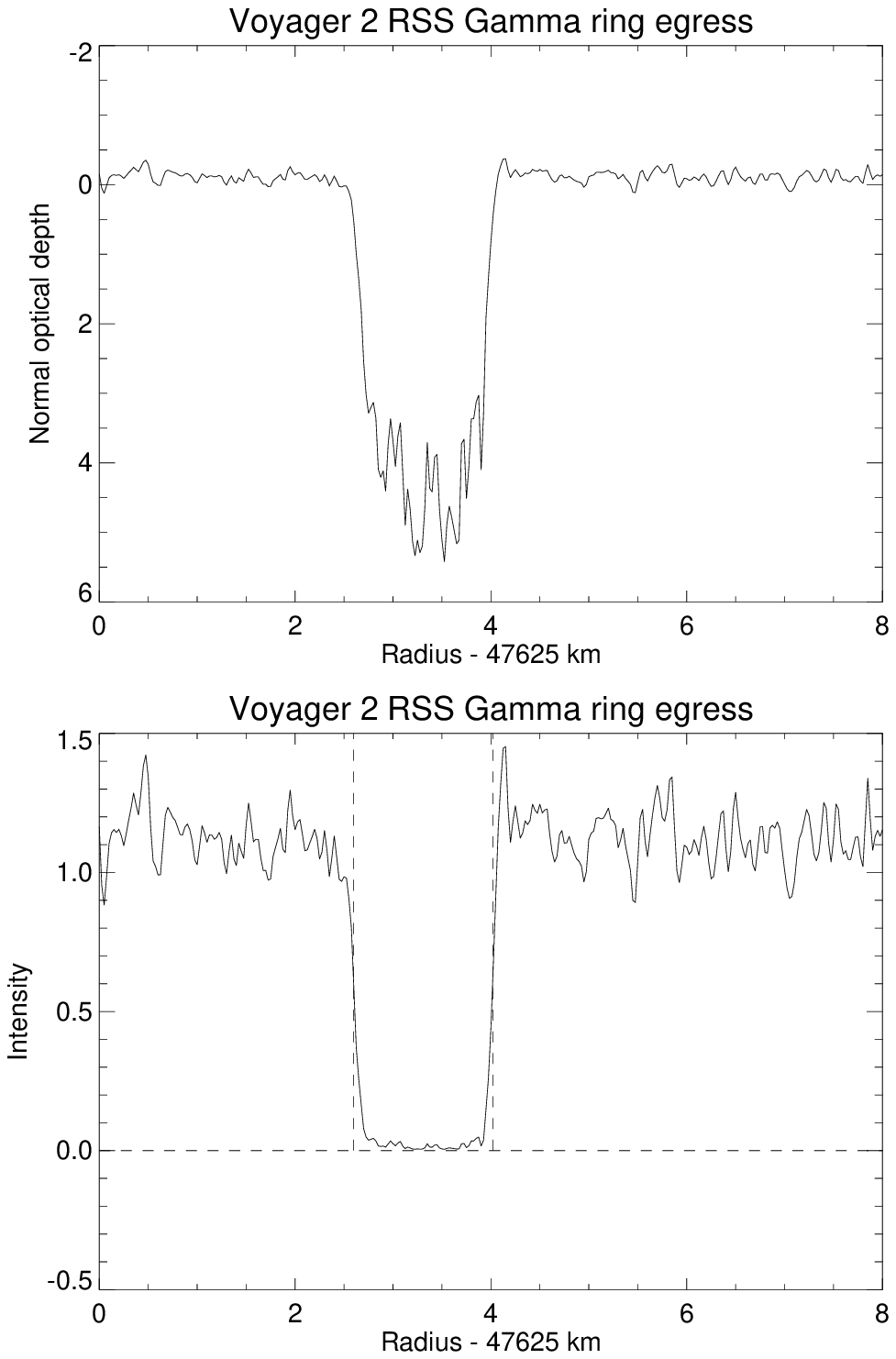}}}
\caption{Diffraction-reconstructed RSS egress $\gamma$ ring profile at an effective resolution of 50~m, plotted as a function of normal optical depth (upper panel) and normalized intensity (lower panel). The width of the reconstructed profile determined from Eq.~(\ref{eq:widthcomparison}) is 1.427 km, shown by the vertical dashed lines, very similar to the 1.370 km width obtained from the square-well model fit to the diffraction-limited profile shown in Fig.~\ref{fig:RSSgesqw}. }
\label{fig:VgrGErecon}
\end{figure}

The close agreement between the measured width and the width obtained from the square-well model demonstrates its accuracy and applicability for actual observations of an intrinsically sharp-edged ring with an intensity profile that resembles a box function. As shown by \cite{Gresh1989}, however, several of the Uranian rings have radially variable internal structure, and not all of them have sharp edges. We repeated the exercise described above and compared the ring widths determined from the square-well fits of the diffraction-limited RSS observations to the observed ring widths of the diffraction-corrected profiles. To provide an objective measure of the ring width, we define the edges as the locations where

\beq
I = 1-f(1-\min I),
\label{eq:widthcomparison}
\eeq
\noindent where $I$ is the normalized diffraction-corrected signal intensity and $f=0.364$, a value chosen to minimize the mean offset between the square-well model widths obtained from fits to the complete set of diffraction-limited RSS ring profiles and the actual ring widths estimated from the diffraction-corrected profiles as given by Eq.~(\ref{eq:widthcomparison}). Here, $f$ corresponds to the chosen fractional drop in the signal (from unit intensity to min $I$, the minimum observed intensity of a ring profile) that formally defines the edge of the ring.
The results of these comparisons are given in {\bf Table~\ref{table:sqwvgr}}, which lists the ingress and egress width estimates and their differences for each ring. The mean difference in width $dW=0.002$ km between the square-well fits and the measured widths is very small, and the standard deviation $\sigma(dW)=0.27$ km.

\begin{table*} [ht] 
\scriptsize 
\begin{center} 
\caption{Comparison of square-well and {\it Voyager} ring widths} 
\label{table:sqwvgr} 
\begin{threeparttable}
\centering 
	\begin{tabular}{l l r r r }\hline 
	Ring & Dir\tnote{a} &  $W$  (km)\tnote{b}& $W$ (km)\tnote{c}& $dW$ (km)\\
		 &   &  (square-well)  &  {\it Voyager} & \\
\hline 
              6 & I & $ 0.899$ & $ 1.141$ & $-0.242$ \\
              6 & E & $ 1.408$ & $ 1.337$ & $ 0.071$ \\
              5 & I & $ 2.615$ & $ 2.376$ & $ 0.239$ \\
              5 & E & $ 1.169$ & $ 1.279$ & $-0.111$ \\
              4 & I & $ 1.405$ & $ 1.295$ & $ 0.110$ \\
              4 & E & $ 1.880$ & $ 1.663$ & $ 0.217$ \\
       $\alpha$ & I & $ 9.423$ & $ 8.676$ & $ 0.747$ \\
       $\alpha$ & E & $ 3.866$ & $ 4.120$ & $-0.254$ \\
        $\beta$ & I & $ 5.069$ & $ 5.081$ & $-0.012$ \\
        $\beta$ & E & $ 8.334$ & $ 8.099$ & $ 0.235$ \\
         $\eta$ & I & $ 1.371$ & $ 1.479$ & $-0.108$ \\
         $\eta$ & E & $ 1.105$ & $ 1.220$ & $-0.115$ \\
       $\gamma$ & I & $ 3.648$ & $ 3.745$ & $-0.097$ \\
       $\gamma$ & E & $ 1.371$ & $ 1.427$ & $-0.056$ \\
       $\delta$ & I & $ 5.182$ & $ 5.777$ & $-0.595$ \\
       $\delta$ & E & $ 2.229$ & $ 2.272$ & $-0.044$ \\
     $\epsilon$ & I & $22.247$ & $22.315$ & $-0.068$ \\
     $\epsilon$ & E & $74.855$ & $74.776$ & $ 0.079$ \\
 	 \hline
	\end{tabular}
	\begin{tablenotes}
		\item[a] I: ingress E:egress
		\item[b] Square-well fits provide the COR midtimes of {\it Voyager} RSS profiles used in ring orbit fits.
		\item[c] From Eq.~(\ref{eq:widthcomparison}) with $f=0.364$.
	\end{tablenotes}
  \end{threeparttable}
\end{center} 
\end{table*}

Shown graphically in 
{\bf Fig.~\ref{fig:sqwvgr}}, it is evident from these results that, except for two outliers (the ingress $\alpha$ and $\delta$ ring profiles), all of the measured width differences $dW$ are within 1-$\sigma$ of zero. 
Of particular relevance for the measured normal mode distortions of selected ring edges in Section \ref{sec:normalmodes}, we find that the RSS $\eta$ ring profiles have measured width differences $|dW|\le0.115$~km, the $\gamma$ ring profiles have $|dW|\le0.097$~km, and $\epsilon$ ring profiles have $|dW|\le0.079$~km, suggesting that the square-well fitted widths to these intrinsically sharp-edged rings are reliable, at least for ring profiles that are not significantly smoothed by the finite angular diameters of the occultation stars or by instrumental time constants.

\begin{figure}
\centerline{\resizebox{5in}{!}{\includegraphics[angle=90]{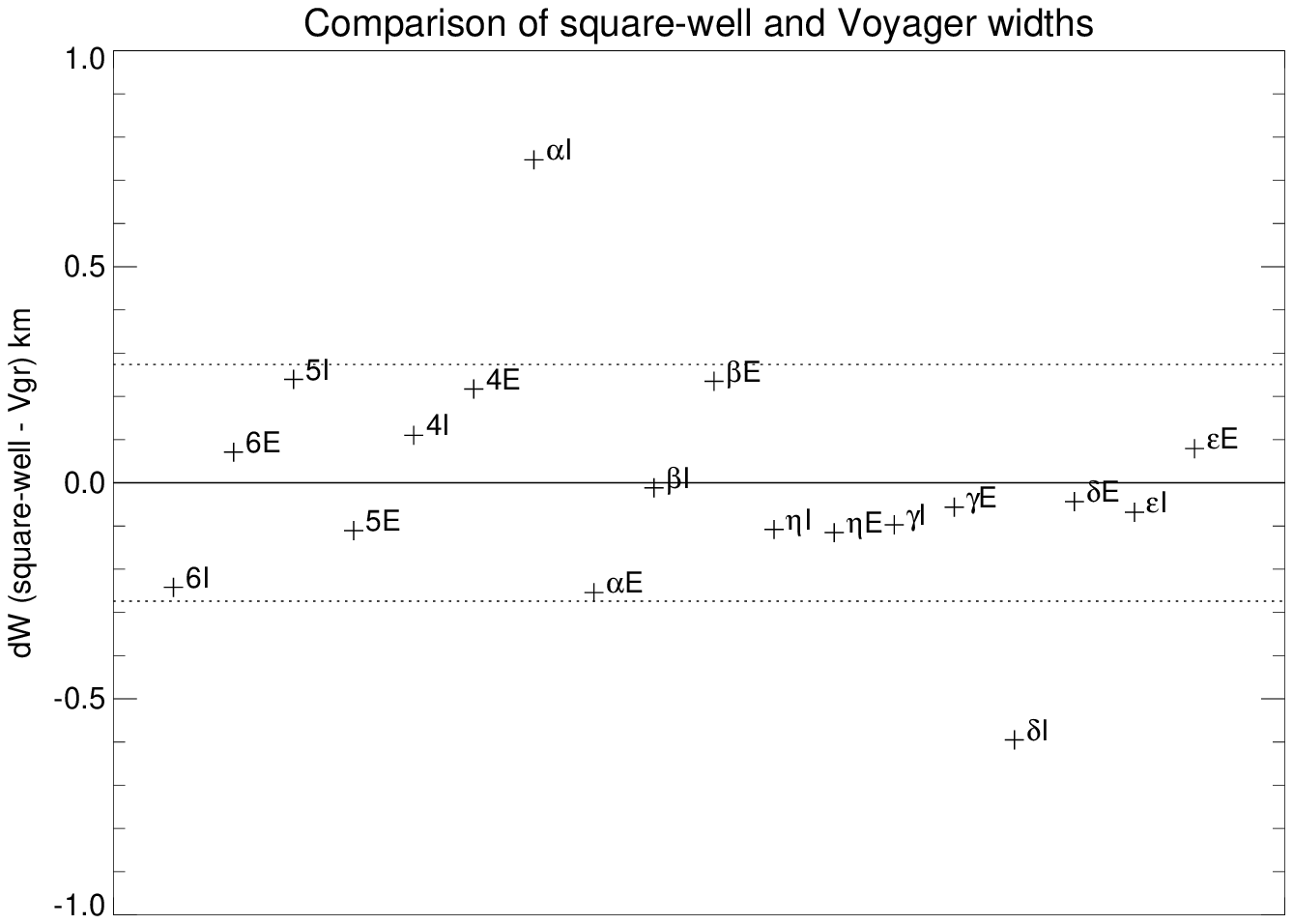}}}
\caption{Differences $dW$ between the widths of \Voyager\ RSS ring profiles obtained from square-well model fits to the uncorrected diffraction patterns in the raw data and the measured widths of the diffraction-corrected profiles using Eq.~(\ref{eq:widthcomparison}) with $f=0.364$. The dotted lines mark the $\pm 1$-$\sigma$ values computed from the full set of residuals, including the two outlier points ($\alpha$I and $\delta$I). The barely visible dashed line shows the mean offset in $dW$ of the measurements.}
\label{fig:sqwvgr}
\end{figure}

The two exceptions noted above for the $\alpha$ and $\delta$ rings illustrate some of the challenges of defining a metric for ring widths that reflects their actual internal structure. 
{\bf Figure~\ref{fig:vgrAlphaI}} shows the RSS diffraction-reconstructed $\alpha$ ingress ring profile, with the vertical dashed lines marking the ring edges as defined by Eq.~(\ref{eq:widthcomparison}). The marked outer edge reasonably matches the moderately sharp outer edge of the ring, but the marked inner edge is clearly a compromise between the abrupt drop in intensity near 44734 km and the gradual variation in ring opacity between 44729--44734 km. (Note that the radius scale is from the data in the PDS archive file, and not from the current orbit determination.\footnote{\tt VG\_2803/U\_RINGS/EASYDATA/KM00\_025/RU1P2XAI.TAB}) The best-fitting square-well model fit is shown in {\bf Fig.~\ref{fig:sqwAlphaI}}, where the underlying structure of the $\alpha$ ring's inner edge (seen at right in this time-ordered ingress profile) is resolved at the level of a few km and resembles the observed structure near the inner edge of the \Voyager\ RSS diffraction-corrected profile. The square-well model accurately matches the outer edge but not the inner edge, resulting in likely systematic errors in the fitted midtimes and widths for at least some Earth-based $\alpha$ ring profiles. 

\begin{figure} \centerline{\resizebox{4in}{!}{\includegraphics[angle=90]{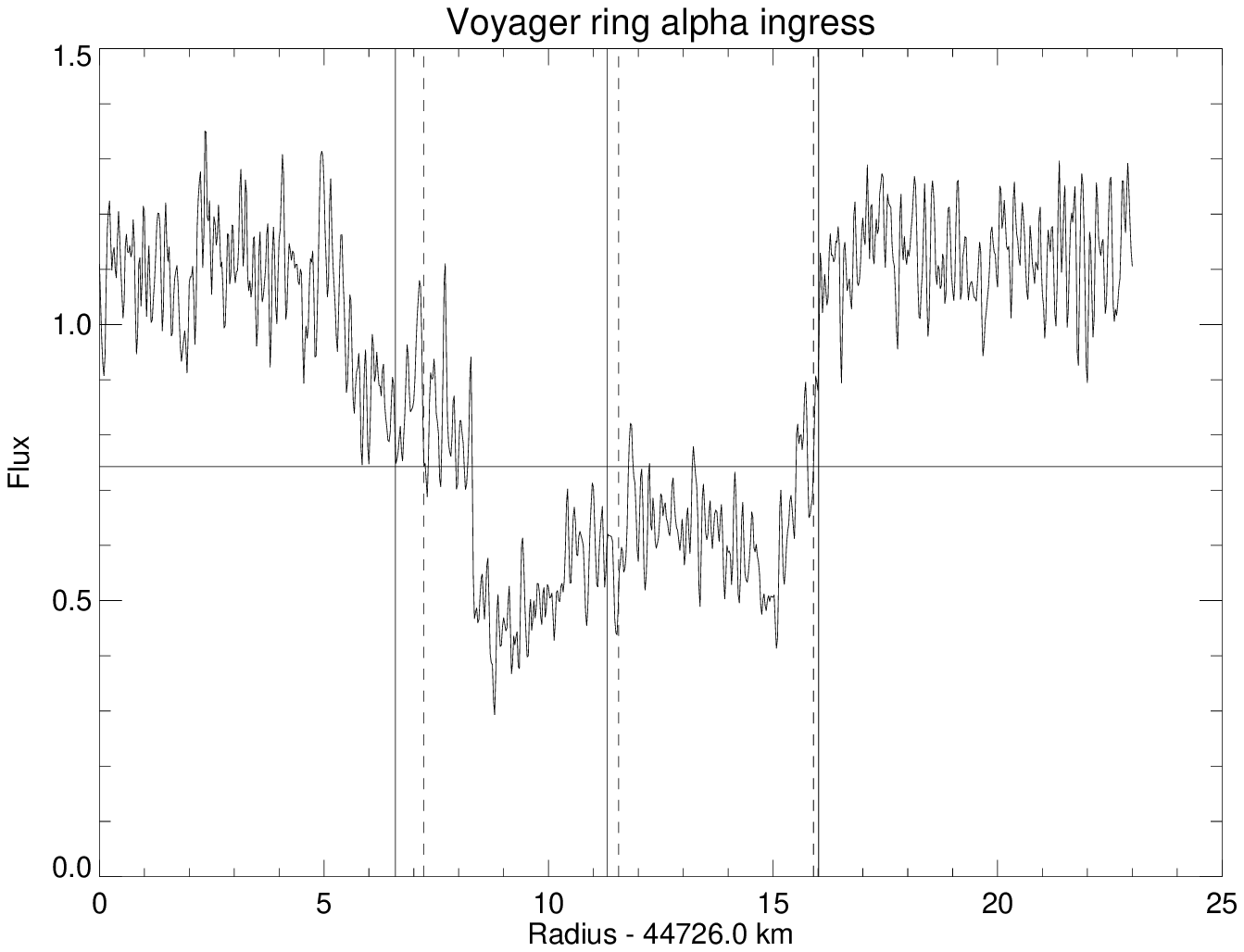}}}
\caption{Radial profile of the diffraction-corrected \Voyager\ RSS observations of the $\alpha$ ring during ingress. The dashed vertical lines mark the midline and the inner and outer edges of the ring as estimated using Eq.~(\ref{eq:widthcomparison}) with $f=0.364$, corresponding to the flux level marked by the solid horizontal line. The solid vertical lines marks the corresponding radii as determined from the square-well model fit to the diffraction-limited RSS profile, based on the radius scale of the archived profiles on the PDS, which differs slightly from the radius scale of our adopted orbit solution.}
\label{fig:vgrAlphaI}
\end{figure}

\begin{figure} 
\centerline{\resizebox{4in}{!}{\includegraphics[angle=90]{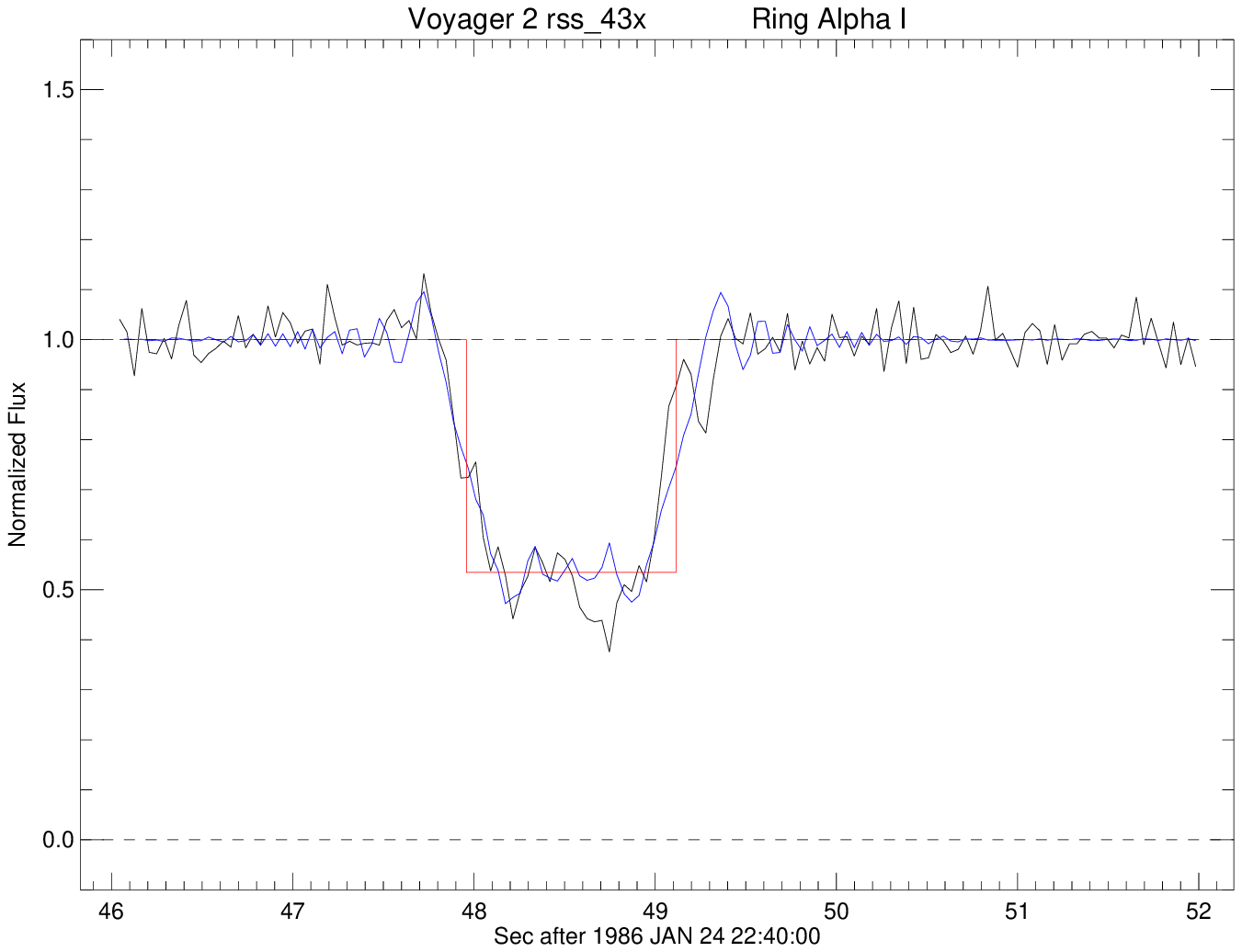}}}
\caption{Square-well model fit to the uncorrected diffraction-limited \Voyager\ RSS $\alpha$ ring ingress profile. Note the good fit of the square-well model to the sharp outer edge of the ring (at left) and the poor fit to the more gradual inner edge at right.}
\label{fig:sqwAlphaI}
\end{figure}

A similar situation prevails for the $\delta$ ingress event, whose RSS diffraction-reconstructed profile is shown in 
{\bf Fig.~\ref{fig:vgrDeltaI}}. Again, the outer edge of the ring is seen to be intrinsically quite sharp, but the inner edge is indistinct, with the ring's opacity increasing gradually over a radial range of $\sim3$ km before reaching a maximum near mid-ring. The vertical dashed lines mark the ring edges as defined by Eq.~(\ref{eq:widthcomparison}). (Note that the radius scale is from the data in the PDS archive file, and not from the current orbit determination.\footnote{\tt VG\_2803/U\_RINGS/EASYDATA/KM00\_025/RU1P2XDI.TAB})
In this instance, the square-well model fit, shown in {\bf Fig.~\ref{fig:sqwDeltaI}}, underestimates the measured with of the high-resolution profile. Notice that the diffraction fringes at the left (earlier in time, corresponding to the outer edge of this ingress profile) match the data for several fringes, whereas the diffraction-limited RSS observations of the inner edge show much more disorder, resulting from the lack of a sharp inner ring edge.

\begin{figure} 
\centerline{\resizebox{4in}{!}{\includegraphics[angle=90]{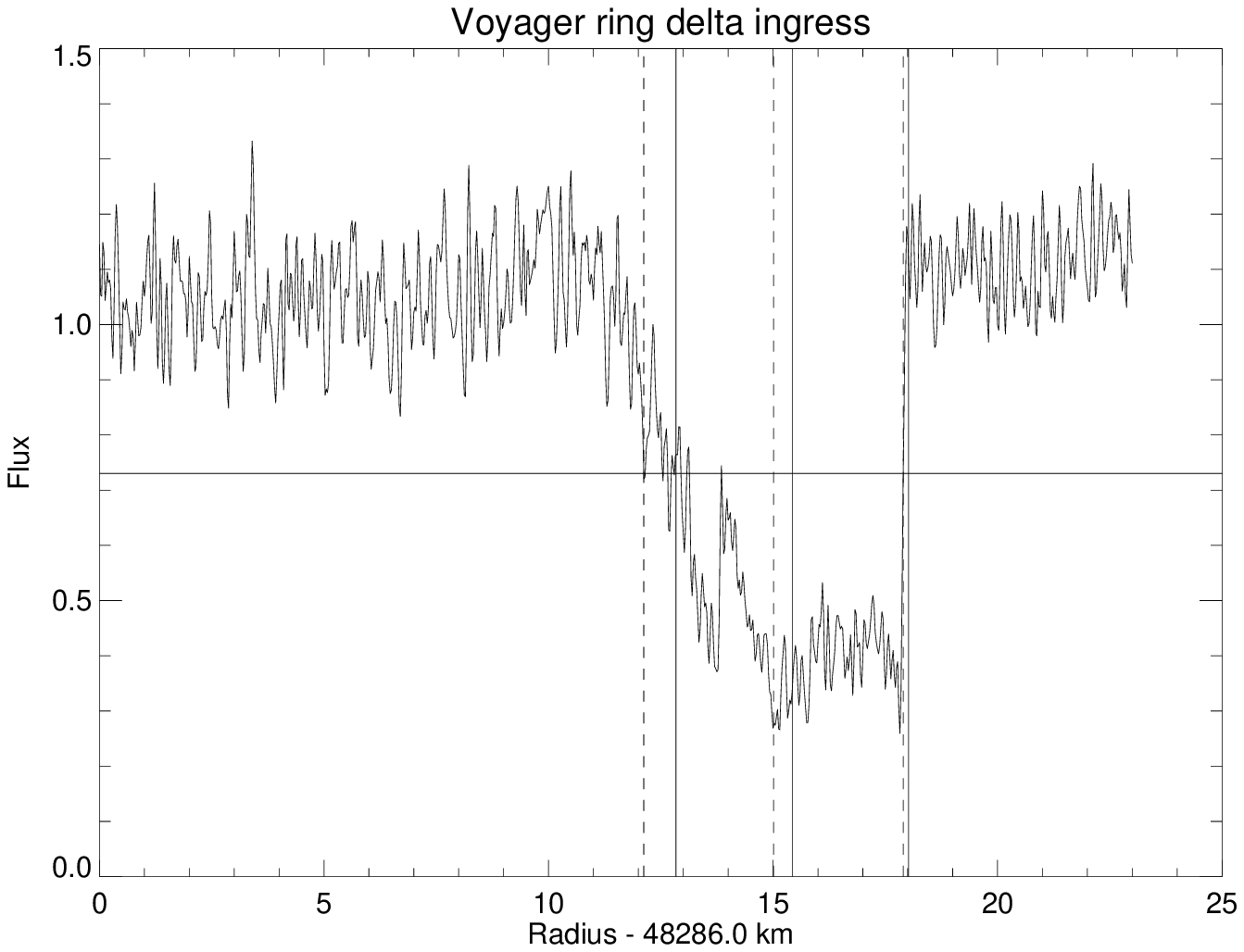}}}
\caption{Radial profile of the diffraction-corrected \Voyager\ RSS observations of the $\delta$ ring during ingress. The dashed vertical lines mark the midline and the inner and outer edges of the ring as estimated using Eq.~(\ref{eq:widthcomparison}) with $f=0.364$, corresponding to the flux level marked by the solid horizontal line. The solid vertical lines marks the corresponding radii as determined from the square-well model fit to the diffraction-limited RSS profile, based on the radius scale of the archived profiles on the PDS, which differs slightly from the radius scale of our adopted orbit solution.}
\label{fig:vgrDeltaI}
\end{figure}

\begin{figure}  
\centerline{\resizebox{4in}{!}{\includegraphics[angle=90]{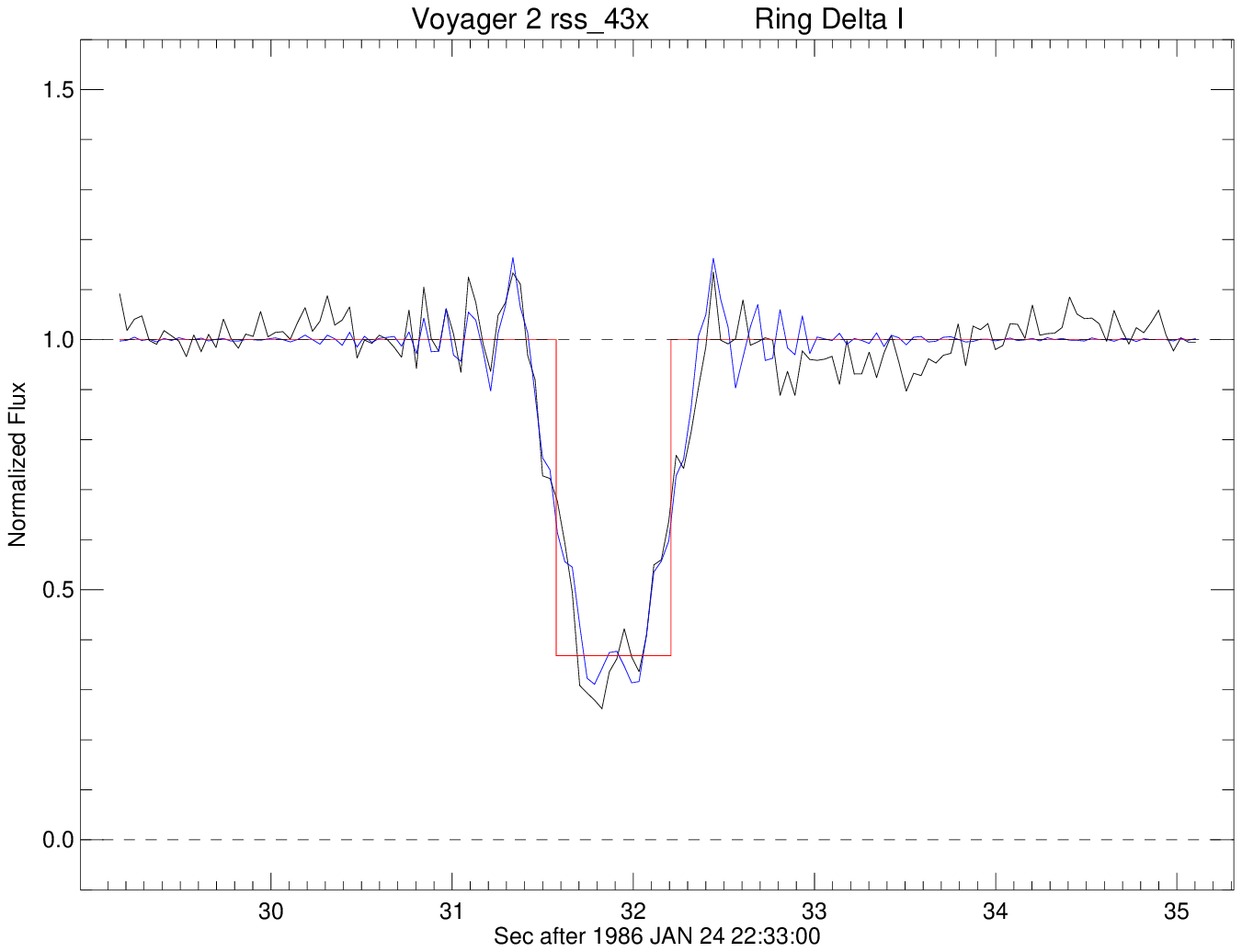}}}
\caption{Square-well model fit to the uncorrected diffraction-limited \Voyager\ RSS $\delta$ ring ingress profile. The sharp outer edge of the ring at left is well-matched by the square-well model, but the modeled diffraction fringes for the inner edge at right are a poor match to the observed diffraction pattern of the gradual inner edge.}
\label{fig:sqwDeltaI}
\end{figure}

Given this level of disagreement between the square-well results and the actual ring observations for rings with non-sharp edges, one could envision a generalization of the simple square-well picture to a multi-well model that fitted the data to a more realistic radial opacity profile. Unfortunately, the radial optical depth profiles of the narrow Uranian rings at different ring longitudes do not follow a simple, discernible pattern, and thus it has not proven possible to construct an $N$-streamline model for any of the rings that accurately matches the observed structure at all longitudes.
%
%

\section{Center of Opacity (COO) from Voyager and Earth-based ring occultation profiles}
\label{appendix:COO}
In Section \ref{sec:COO}, we distinguished between the observed radius of a ring's centerline (COR) and its radius derived from its radially-integrated optical depth (COO), using the \Voyager\ ring 6 occultation profiles as an example in Fig.~\ref{fig:coo6}. We apply this same method to \Voyager\ ring occultation profiles for rings 5, 4, $\alpha, \beta,\gamma,$ and $\epsilon$ in {\bf Figs.~\ref{fig:coo5}--\ref{fig:cooE}}. For the $\epsilon$ ring, we have made additional use of the \Voyager\ $\sigma$ Sgr optical depth profiles, in part because the RSS ingress profile sampled the ring near periapse ($f=30.0^\circ$) with noise-affected high optical depths. (Note that RSS optical depths are twice as large as comparable  spacecraft or Earth-based stellar occultation profiles, due to diffraction scattering signal loss.) The four \Voyager\ profiles yield values of $\Delta a_{\rm COR}$ ranging from 0.46 to 1.53 km.
The ring is also radially resolved in several of the Earth-based stellar occultations. We have selected thirteen of the highest SNR profiles from this set. To minimize the effects of uncertain normalization that afflict estimates of high optical depth when the observed normalized stellar flux is near zero, we limited our selection to profiles more than $90^\circ$ from periapse, shown in {\bf Fig.~\ref{fig:cooEearth}}. They yielded values of $\Delta a_{\rm COR}$ ranging from 0.54 to 1.95 km, comparable in average and range to the \Voyager\ results. 
The mean $\epsilon$ ring value for $\Delta a_{\rm COR}=1.134\pm0.488$ km. 

The derived centers of opacity for all the eccentric ringlets are given in {\bf Table \ref{tbl:rCOO}}, which lists
 $a_{\rm COR}$, the unweighted average values for 
$\Delta a_{\rm COO}$ and their RMS variation, the statistical significance $\Delta a_{\rm COO}/\sigma(\Delta a_{\rm COO})$, and finally of $a_{\rm COO}$, with an uncertainty given by the sum in quadrature of $\sigma(a_{\rm COR})$ and $\sigma(\Delta a_{\rm COO})$.

\begin{figure}
\centerline{\resizebox{5in}{!}{\includegraphics[angle=90]{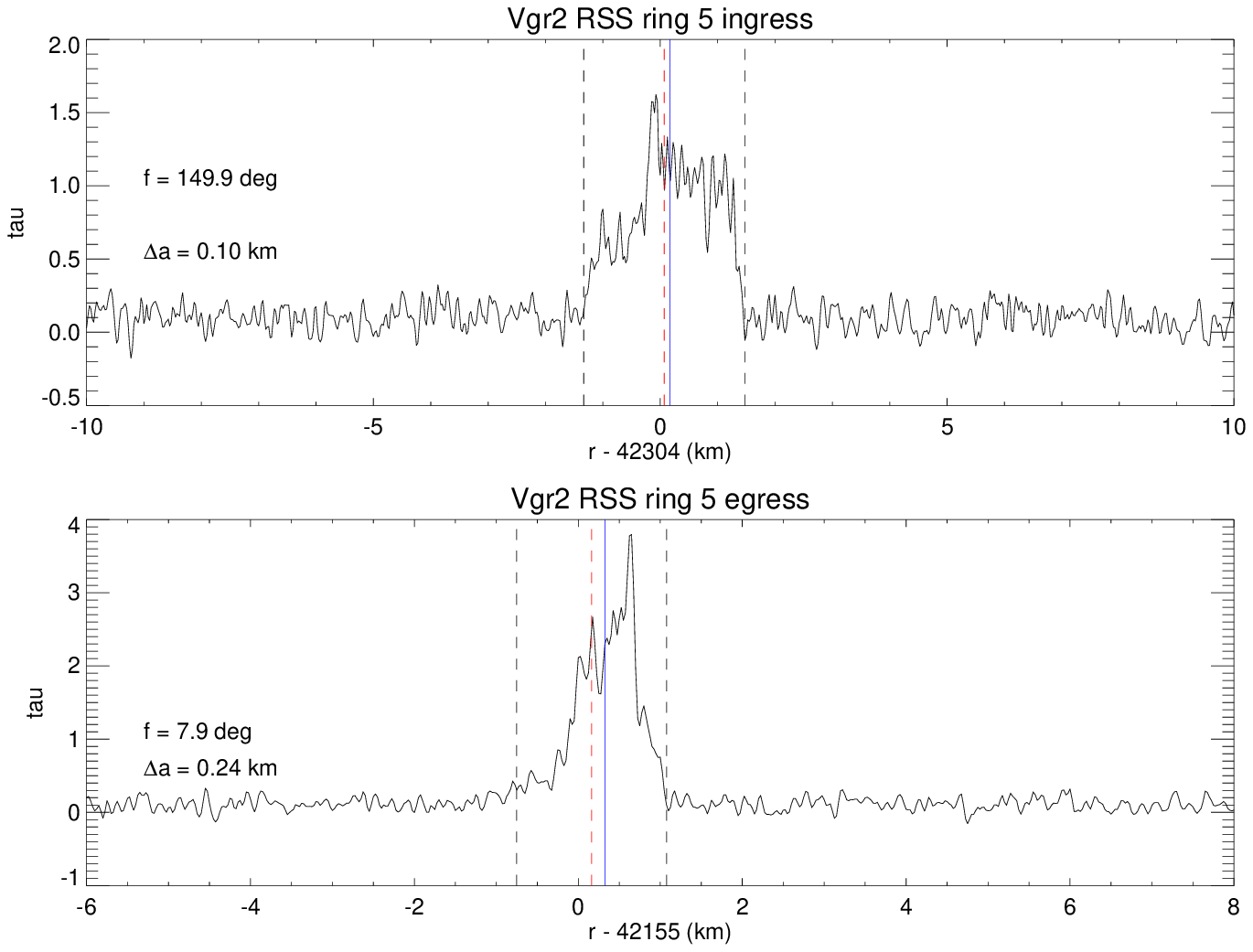}}}
\caption{Ring 5 center of opacity from \Voyager\ RSS optical depth profiles.}
\label{fig:coo5}
\end{figure}

\begin{figure}
\centerline{\resizebox{5in}{!}{\includegraphics[angle=90]{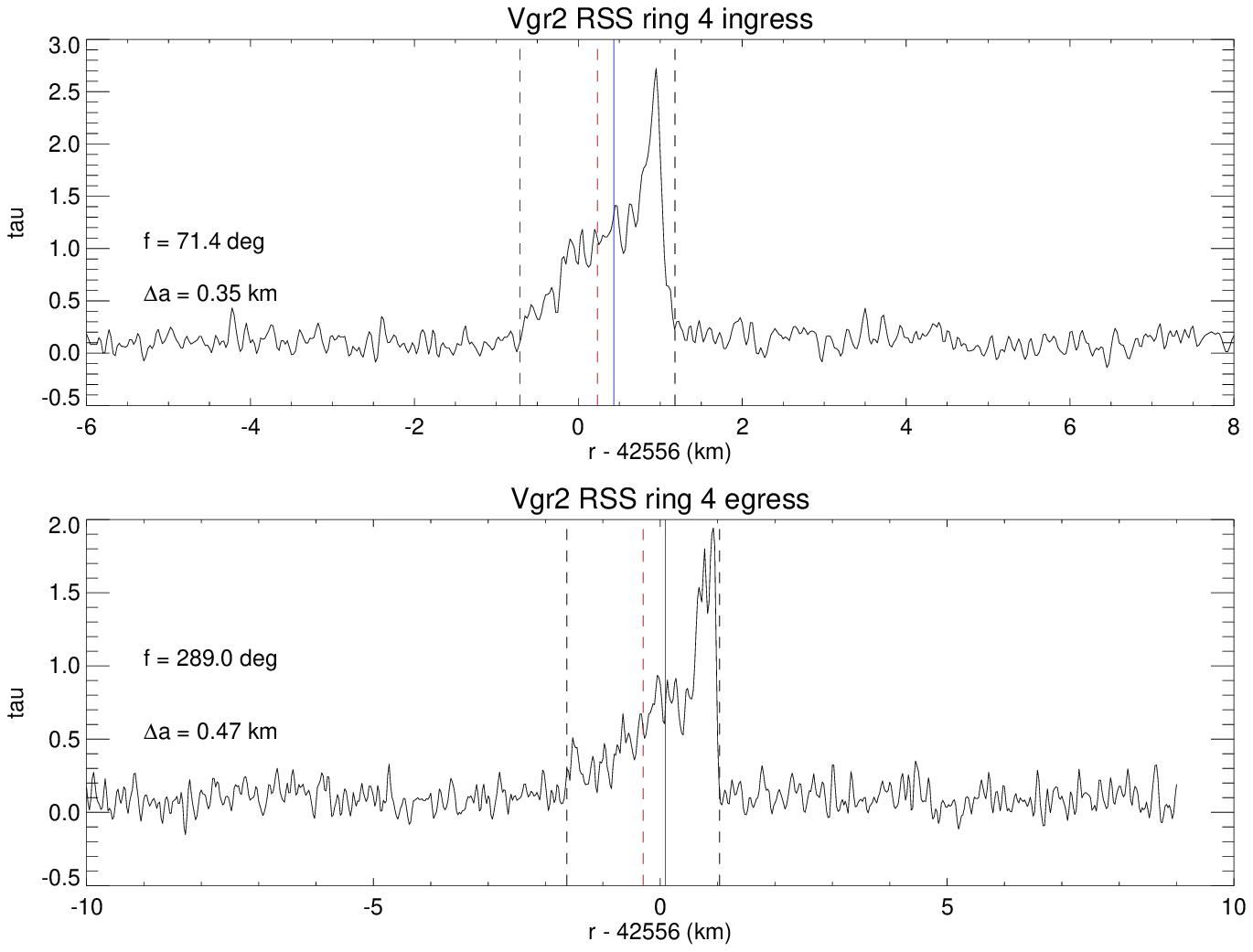}}}
\caption{Ring 4 center of opacity from \Voyager\ RSS optical depth profiles.}
\label{fig:coo4}
\end{figure}

\begin{figure}
\centerline{\resizebox{5in}{!}{\includegraphics[angle=90]{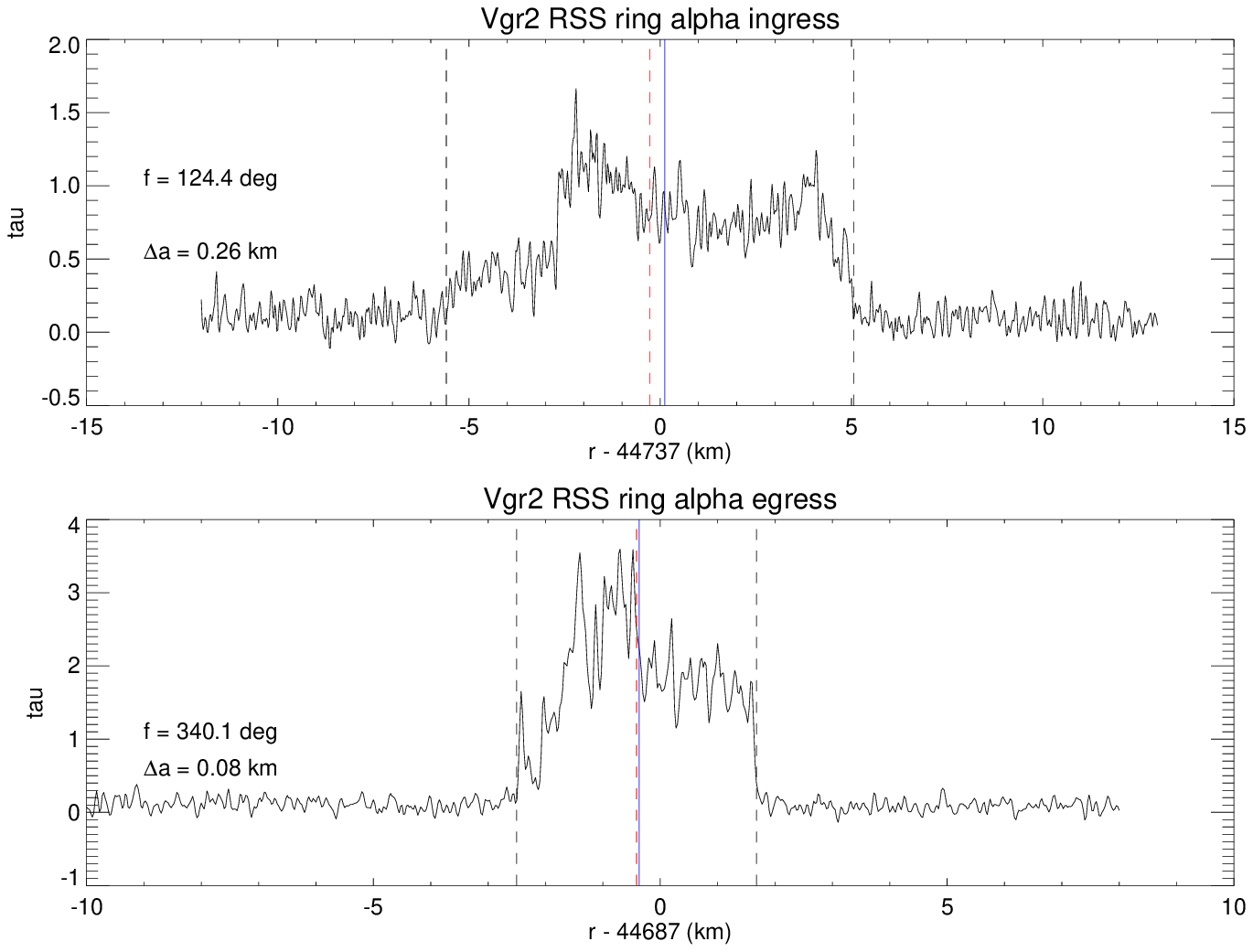}}}
\caption{Ring $\alpha$ center of opacity from \Voyager\ RSS optical depth profiles.}
\label{fig:cooA}
\end{figure}

\begin{figure}
\centerline{\resizebox{5in}{!}{\includegraphics[angle=90]{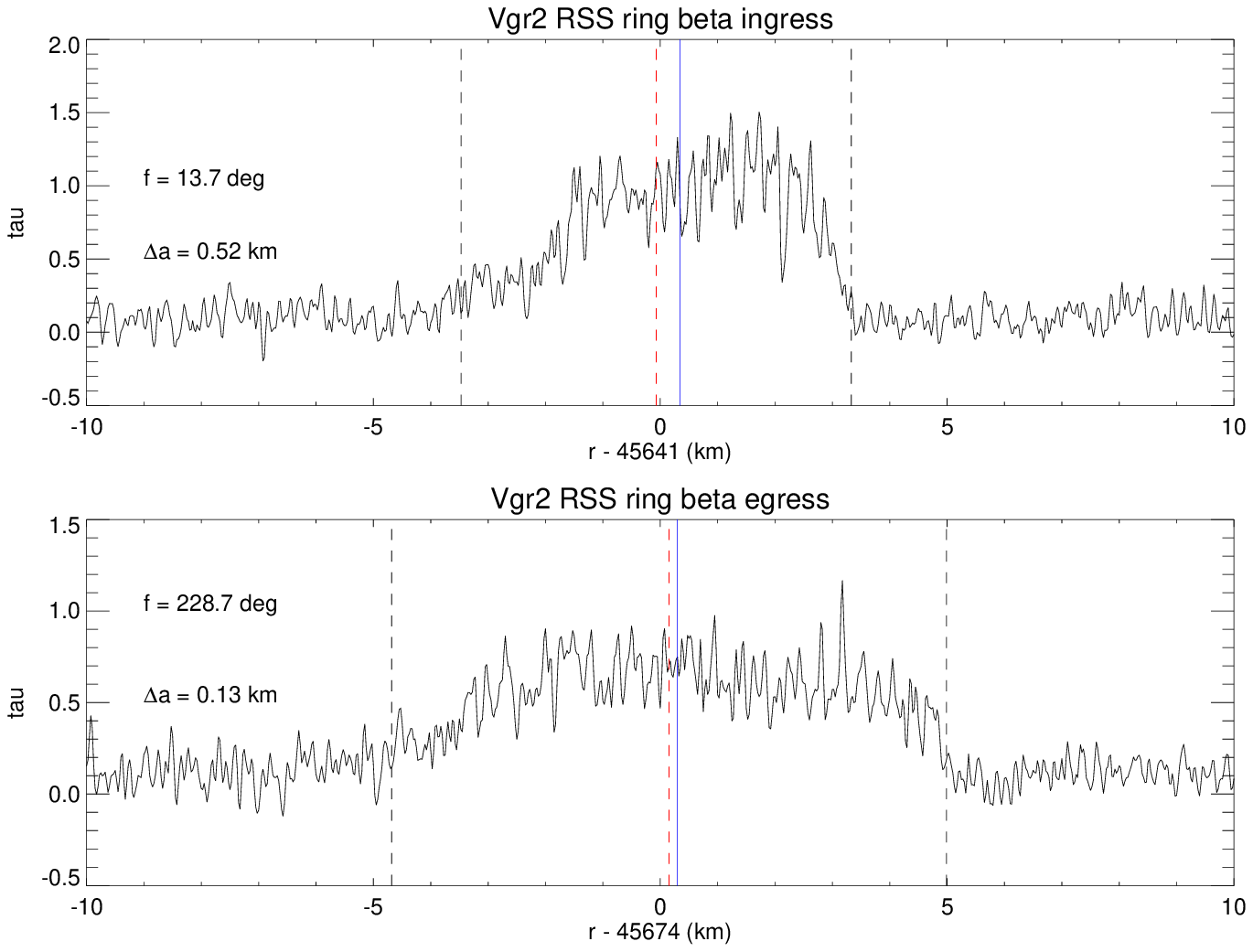}}}
\caption{Ring $\beta$ center of opacity from \Voyager\ R/coS optical depth profiles.}
\label{fig:cooB}
\end{figure}

\begin{figure}
\centerline{\resizebox{5in}{!}{\includegraphics[angle=90]{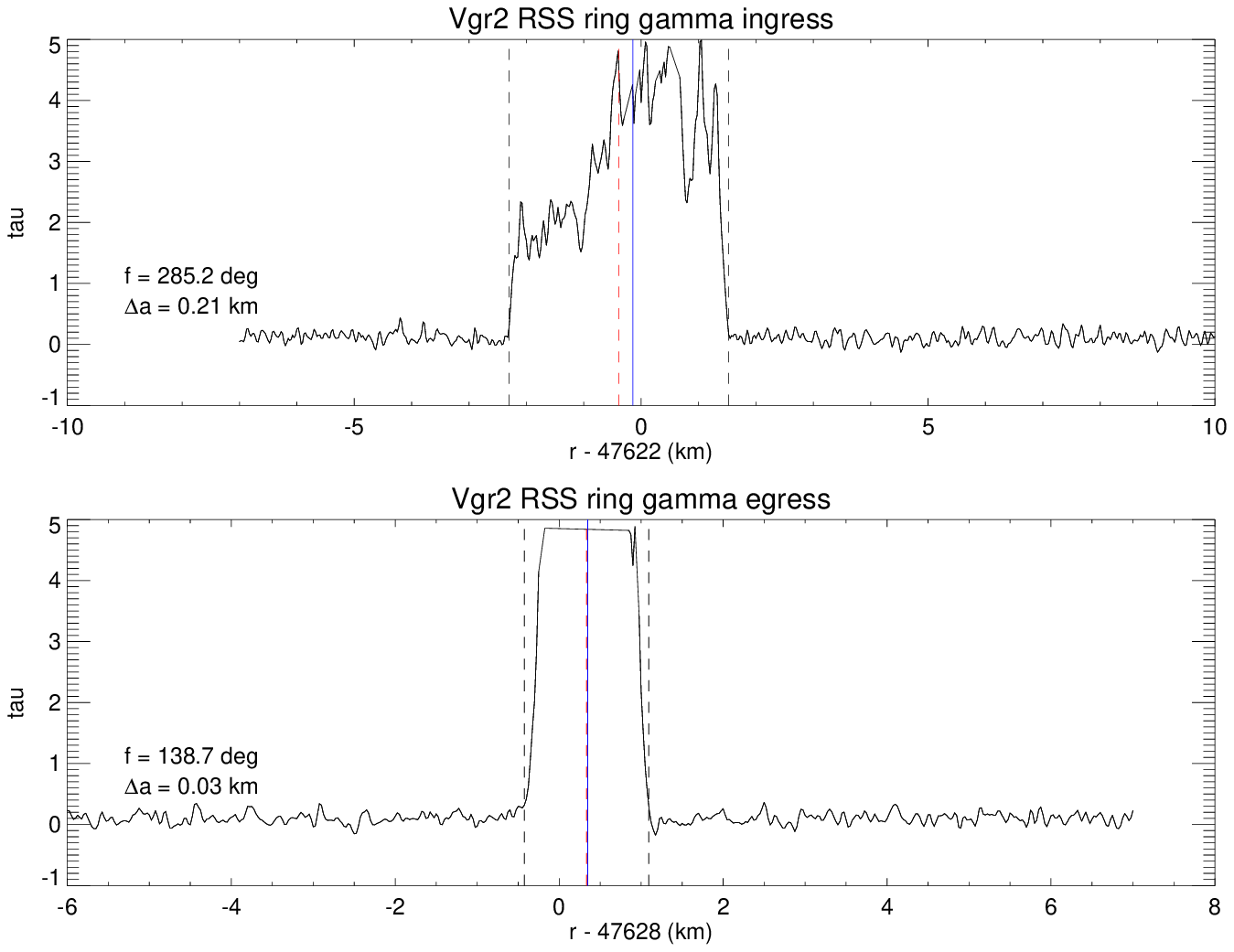}}}
\caption{Ring $\gamma$ center of opacity from \Voyager\ RSS optical depth profiles.}
\label{fig:cooG}
\end{figure}

\begin{figure}
\centerline{\resizebox{5in}{!}{\includegraphics[angle=90]{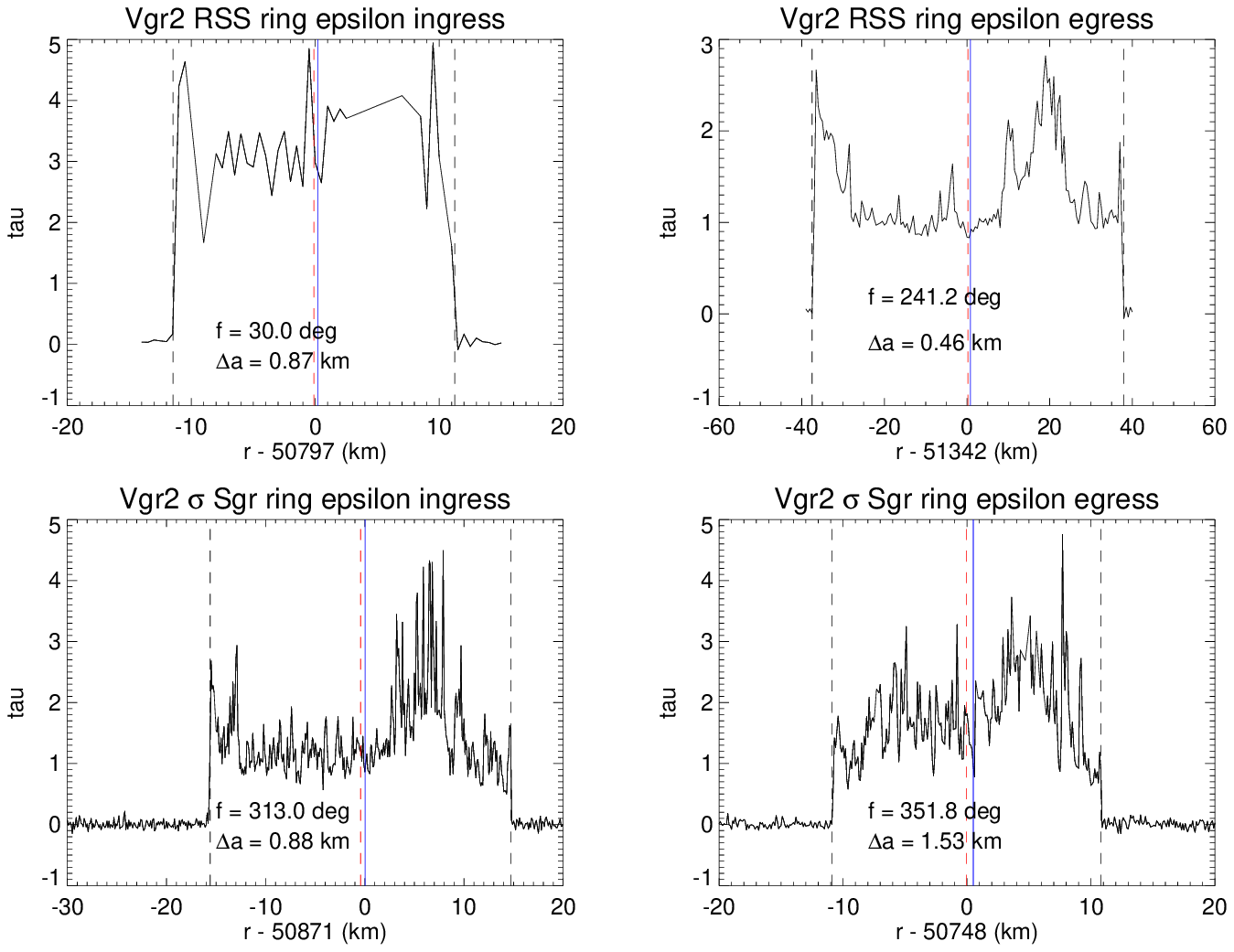}}}
\caption{Ring $\epsilon$ center of opacity from \Voyager\ RSS and $\sigma$ Sgr optical depth profiles. Note that stellar occultation geometric optical depths should be multiplied by a factor of two before comparison with RSS normal optical depths.}
\label{fig:cooE}
\end{figure}

\begin{figure}
\centerline{\resizebox{7in}{!}{\includegraphics[angle=90]{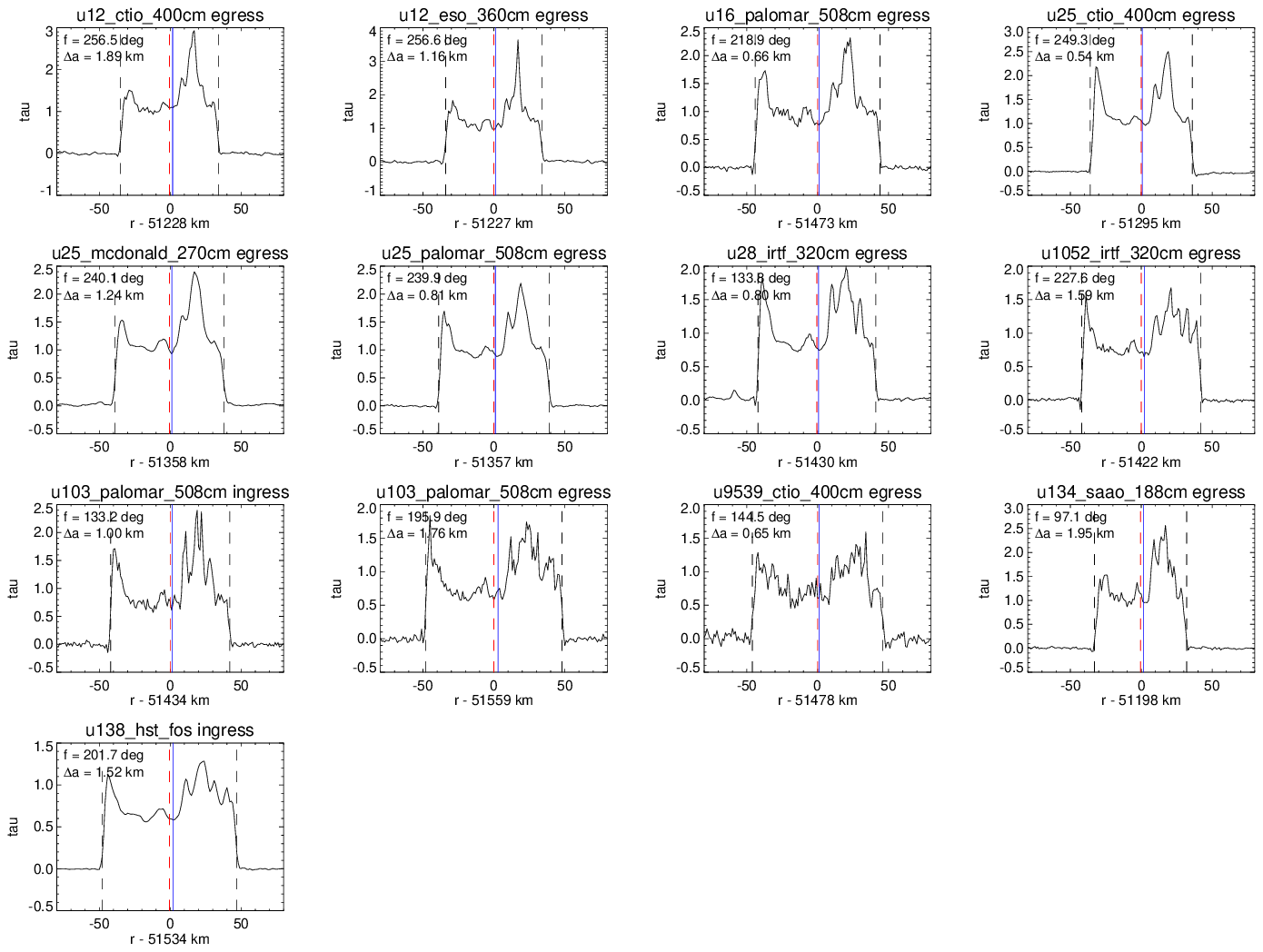}}}
\caption{Estimates of the $\epsilon$ ring center of opacity (COO) from Earth-based occultation profiles. Note that stellar occultation geometric optical depths should be multiplied by a factor of two before comparison with RSS normal optical depths. See text for details.}
\label{fig:cooEearth}
\end{figure}


\section{The compatibility of a model gravity field $\{J'_2,J'_4,J'_6\}$ with the adopted solution from ring occultations}
\label{appendix:grav}
In Section \ref{sec:gravity}, we noted the strong correlations between $J_2, J_4,$ and $J_6$ in the gravity field determination from the Uranus ring occultation data. Here, we use the observed correlations to provide a prescription for determining the compatibility of model values $\{J'_2,J'_4,J'_6\}$ with the adopted gravity field.

First, we quantify the approximate relationship between $J_2$ and $J_4$. Given the strong correlation between $J_2$ and $J_4$ (reflected in the narrowness of the error ellipse for Fit 15 shown in Fig.~\ref{fig:J2J4}), it is possible to construct an approximate relationship between the two leading terms in the gravitational field. Under the restricted circumstances that $J_6=0$, the local value of $dJ_4/dJ_2$ can be estimated by setting the differential of the leading terms in $\dot\varpi$ to zero:
\beq
d \dot\varpi = \biggl(\frac{GM_{\rm Ur}}{a^3}\biggr)^{1/2}\biggl[\frac{3}{2}d J_2\biggl(\frac{R}{a}\biggr)^2 - \frac{15}{4}d J_4\biggl(\frac{R}{a}\biggr)^2\biggr] =0,
\eeq 
from which we find
\beq
dJ_4/dJ_2 = \frac{2}{5}\biggl(\frac{a}{R}\biggr)^2 = 1.24 \biggl(\frac{a}{45000\ \rm{km}}\biggr)^2,
\label{eq:dj4dj2}
\eeq
Setting $a$ in Eq.~(\ref{eq:dj4dj2}) to 44695 km, the mean of the semimajor axes of rings 6, 5, 4, $\alpha$, $\beta$, and $\epsilon$ (those with accurate apse rates), and letting $R=25559$ km, we obtain
$ dJ_4/dJ_2\simeq1.223$. This is close to the slope of the long axis of the error ellipse for our adopted fit in Fig.~\ref{fig:J2J4}, given by $ dJ_4/dJ_2=\tan \theta=1.134$, where $\theta$ is defined in Eq.~(\ref{eq:theta}) below. Under a linear approximation, a model value $J'_4$ that is consistent with the observed values of $J_2$ and $J_4$ and an assumed model value $J'_2$ is given by
\beq
J'_4(J'_2) \simeq J_4 + \frac{dJ_4}{dJ_2} (J'_2 - J_2).
\eeq
Evaluated numerically:
\beq
J'_4(J'_2) \simeq [-35.533 + 1.134\times (J'_2 - 3509.281)]\times10^{-6}.
\eeq

Next, we assume that the partial derivatives in Table \ref{tbl:Jn} accurately represent the sensitivity of the fitted values for $J_2(J_6,\Delta a)$ and $J_4(J_6,\Delta a)$ in our adopted solution (Fit 15) to small differences from our assumed values of $J_6=0.5\times10^{-6}$ and $\Delta a=0$. Then the following relations specify the values of $J_2$ and $J_4$ predicted to result from a fit in which $J_6$ and the systematic radius scale offset have the fixed {\it a priori} values
 $J'_6$ and $\Delta a$:

\beq
J_2(J'_6,\Delta a) = J_2(0.5\times10^{-6},0) + \frac{dJ_2}{dJ_6} (J'_6 -0.5\times 10^{-6}) + \frac{dJ_2}{d\Delta a}\Delta a
\label{eq:Cj2func}
\eeq
\beq
J_4(J'_6,\Delta a) = J_4(0.5\times10^{-6},0) + \frac{dJ_4}{dJ_6} (J'_6 -0.5\times 10^{-6}) + \frac{dJ_4}{d\Delta a}\Delta a.
\label{eq:Cj4func}
\eeq

Numerically, our final values for $J_2$ and $J_4$ map onto the corresponding values for non-zero $J'_6$ and $\Delta a$ according to:
\beq
J_2(J'_6,\Delta a) = 3509.281\times 10^{-6}  + 0.39909 \times (J'_6-0.5\times 10^{-6}) +0.130 \times 10^{-6} \Delta a({\rm km})
\label{eq:Cj2funcnum}
\eeq
\beq
J_4(J'_6,\Delta a) = -35.533\times 10^{-6}  + 1.07570 \times (J'_6-0.5\times 10^{-6}) -0.180 \times 10^{-6} \Delta a( {\rm km}).
\label{eq:Cj4funcnum}
\eeq
To determine whether any planetary gravity model 
$\{J'_{2}, J'_{4}, J'_{6},\Delta a\}$ falls within the 1-$\sigma$ error ellipse of our adopted solution, the Mahalanobis distance $r$ must be $\le 1$, where 

\beq 
r= \sqrt{\frac{x'^2}{\lambda_1} + \frac{y'^2}{\lambda_2}},
\label{eq:Cr}
\eeq
 and $\lambda_1$ and $\lambda_2$ are the eigenvalues of the ($J_2,J_4$) covariance matrix: 
\beq
{\rm covar}(J_2,J_4) = 
\begin{bmatrix}
a & b\\
b & c
\end{bmatrix},
\eeq
with
\beq
a = \sigma^2(J_2),
\eeq
\beq
b = \rho(J_2,J_4)\sigma(J_2)\sigma(J_4),
\eeq
and
\beq
c= \sigma^2(J_4), 
\eeq
and $x'$ and $y'$ are defined below.\footnote{\url{https://cookierobotics.com/007/}}
The halfwidths of the two axes of the error ellipse are given by $\sqrt{\lambda_1}$ and $\sqrt{\lambda_2}$, where
\beq 
\lambda_1= \frac{a+c}{2} + \sqrt{\biggl(\frac{a-c}{2}\biggr)^2 + b^2},
\eeq
and
\beq 
\lambda_2= \frac{a+c}{2} - \sqrt{\biggl(\frac{a-c}{2}\biggr)^2 + b^2}.
\eeq
The following transformation rotates the error ellipse by $\theta$: 
\beq
x' =x \cos \theta + y \sin \theta
\label{eq:Cxp}
\eeq
\beq
y' = -x \sin \theta + y \cos \theta,
\label{eq:Cyp}
\eeq
where
\beq
x = J'_{2} -J_2(J'_6,\Delta a),
\label{eq:Cx}
\eeq
\beq
y = J'_{4} -J_4(J'_6,\Delta a),
\label{eq:Cy}
\eeq
and the rotation angle $\theta$ is specified by 
\beq
\theta = \tan^{-1}(\lambda_1,b),
\label{eq:theta}
\eeq
where the arguments of the arctangent are the column entries of the covariance matrix eigenvectors, simplified by multiplying by their common divisor.

From our adopted gravity parameters (Fit 15 in Table \ref{tbl:Jn}), $\sigma(J_2) = 0.412\times10^{-6}, \sigma(J_4)=0.466\times10^{-6},$ and $ \rho(J_2,J_4)=0.9861$, yielding the numerical values
$\lambda_1=3.84252\times 10^{-13},\lambda_2=2.64830\times 10^{-15}, a=1.697440\times 10^{-13}, b= 1.893233\times 10^{-13}, c=2.17156\times 10^{-13},$ and $\theta=48.5686^\circ$.

The complementary cumulative probability CCP$(r)$ that a given sample has a standard deviation $\ge r$ by chance for a given $(J'_2,J'_4,J'_6,\Delta a)$ is given by
\beq
{\rm CCP}(r) = e^{-r^2(J'_2,J'_4,J'_6,\Delta a) /2},
\label{eq:CCP}
\eeq
 appropriate for a two-dimensional gaussian distribution, where $r$ is given by Eq.~(\ref{eq:Cr}). Note that CCP$(1) = 0.6065$, CCP$(2)=0.1353$, CCP$(3)=0.0111$, and CCP$(4)=0.0003$. In comparison, the complementary cumulative probability for a one-dimensional normal distribution is given by $\rm{erfc} (r/\sqrt{2})$, yielding the more familiar corresponding values $1\sigma=0.3173, 2\sigma=0.0455, 3\sigma=0.0027,$ and $4\sigma=0.0001$.
 
 The inverse relationship
 \beq
 r=\sqrt{-2\ln {\rm CCP}(r)}
 \label{eq:rinverse}
 \eeq
 provides the factor $r$ for a given confidence level CCP$(r)$. For example, the values of $r$ corresponding to CCP=30\%, 10\% and 1\% cutoff values for acceptable interior models that are consistent with our gravity measurements are 1.55, 2.15 and 3.03, respectively.

The prescription for computing the difference in standard deviations $r(J'_2,J'_4,J'_6,\Delta a)$ between a model with assumed values $\{J'_2,J'_4,J'_6,\Delta a\}$ and our adopted solution is:
 \begin{itemize}
 \item Specify values for $J'_2,J'_4,J'_6,$ and $\Delta a$.
  \item Compute $J_2(J'_6,\Delta a) $ and $J_4(J'_6,\Delta a)$ from Eqs.~(\ref{eq:Cj2funcnum}) and (\ref{eq:Cj4funcnum}).
 \item Compute $x$ and $y$ from Eqs.~(\ref{eq:Cx}) and (\ref{eq:Cy}).
  \item Compute $x'$ and $y'$ from Eqs.~(\ref{eq:Cxp}) and (\ref{eq:Cyp}), using the given numerical value for $\theta$.
  \item Compute $r$ from Eq.~(\ref{eq:Cr}) and the given numerical values for $\lambda_1$ and $\lambda_2$.
\end{itemize}


\section{Alignment of ILR and OLR Ring Edge Mode Apses Relative to Forcing Satellite} 

In the main body of the paper, we have assumed the following properties of ring edge modes force by a satellite:
\begin{itemize}
\item In the absence of dissipation, the forced mode (periapse/apoapse) is aligned with the satellite mean longitude for an (ILR/OLR), respectively.
\item In the presence of dissipation, the relevant apse is (lagging/leading) the satellite mean longitude in angle for an (ILR/OLR), respectively.
\end{itemize}

Here, we justify these assumptions on dynamical grounds by drawing on well-known results for ILR edge modes and making use of an exact formal symmetry between inner and outer edge modes to determine the corresponding OLR edge mode properties. 

ILR edge modes are to date much more extensively studied than OLR ones; in particular, detailed numerical solutions are available only for ILR edge modes. It turns out that a few properties, including simple geometric features, can only be obtained through detailed solutions of the dynamical equations. OLR edge mode properties must therefore also be obtained from such detailed solutions.

In the large $|m|$ limit, the dynamical equations display an exact formal symmetry between inner and outer edge modes. This symmetry allows us to deduce a number of properties of inner edge modes from outer edge ones.\footnote{The large $m$ limit applies quantitatively as soon as $|m|$ is larger than a few; the approximation is also semi-quantitatively relevant even for $|m|=2$.}

Let us show this explicitly. In the streamline formalism, the mode structure is controlled by two dynamical equations, one for the mode eccentricity profile and one for its apsidal shift profile. In a discrete, $N$-streamlines formulation, these equations read (see \citealt{Longaretti2018} and \citealt{L23} for details):
\begin{align}
	& \sum_{j\neq i} \frac{\bar{n}}{\pi}\frac{m_j}{M_p} a^2
	H(q_{ij}^2)\frac{\epsilon_i-\epsilon_j}{(\Delta a_{ij})^2} 
	- [\kappa_i - m(\Omega_i-\Omega_P)]\epsilon_i
	= - \bar{n}\frac{a\Psi_{m,k}}{2 G M_p},\label{sat}\\
	& \sum_{j\neq i} \frac{\bar{n}}{\pi}\frac{m_j}{M_p} a^2
	H(q_{ij}^2)\frac{\epsilon_i\sin m\Delta_i-\epsilon_j\sin m\Delta_j}{(\Delta a_{ij})^2} 
	- [\kappa_i - m(\Omega_i-\Omega_P)] \epsilon_i\sin m\Delta_i\nonumber\\
	&\hspace{7.5truecm} = \frac{2\pi}{\overline{n} m_i} \Delta^{\pm}\left( \frac{a}{q}\frac{d\epsilon}{da}t_1 \right). \label{dissip}
\end{align}
In these equations, $m$ is the azimuthal wavenumber (number of lobes of the edge mode), $a_i$, $\epsilon_i$ and $\Delta_i$ are the semi-major axis, eccentricity and apsidal shift of streamline $i$ ($\Delta_i = -\delta_i$ in the notation of Eq. 8 of the present paper), $m_j$ the mass of streamline $j$, $\Delta a_{ij} = a_j - a_i$, $\bar{n}$ is the mean motion at the mode resonance and $a$ the resonance location, $q_{ij}= a(\epsilon_j - \epsilon_i)/\Delta a_{ij}$, $\kappa_i$ and $\Omega_i$ the epicyclic radial frequency and angular rotation velocity of streamline $i$, $M_p$ the planet mass (here Uranus), $\Omega_P$ the mode pattern speed, $\Psi_{m,k}$ the satellite forcing term, $H(q_{ij})$ a dimensionless function of order unity which arises from the nonlinearity of the ring self-gravity, and $t_1$ is an effective stress tensor quantity, characterizing the ring viscous dissipation. The viscous coefficient $t_1$ is a function of the nonlinearity parameter $q = |ad\epsilon/da|$; $t_1 < 0$ in the absence of viscous overstability, and $|t_1|\rightarrow 0$ when $q\rightarrow 0$. Finally, $\Delta^\pm X$ is the variation of $X$ across the current streamline (outer boundary minus inner boundary value of $X$). The azimuthal wavenumber $m$ is positive at an inner Lindblad resonance and negative at an outer Lindblad resonance.

The first term on the left-hand side of both equations represents the effect of the ring self-gravity, the second term the frequency drift with respect to the pattern speed, as by definition $\kappa - m(\Omega-\Omega_P)=0$ at the resonance. On the right-hand side, one finds the satellite forcing term in the first equation and the ring internal (collisional) dissipation in the second one. The two equations are very similar; in particular the left-hand sides are formally identical under the substitution $\epsilon \leftrightarrow \epsilon\sin m\Delta$. 

Furthermore, one can write
\begin{align}
	\kappa_i - m(\Omega_i-\Omega_P) = \frac{3\Omega(m-1)}{2a}(a_i-a) \simeq \frac{3\Omega m}{2a}(a_i-a),\label{freq}
\end{align}
for $|m| > 1$. The mode exists mostly outside the resonance at an ILR and inside the resonance at an OLR. The second equality applies in the large $|m|$ limit.

\begin{figure}[ht]
	\centering
		\includegraphics[width=0.5\columnwidth]{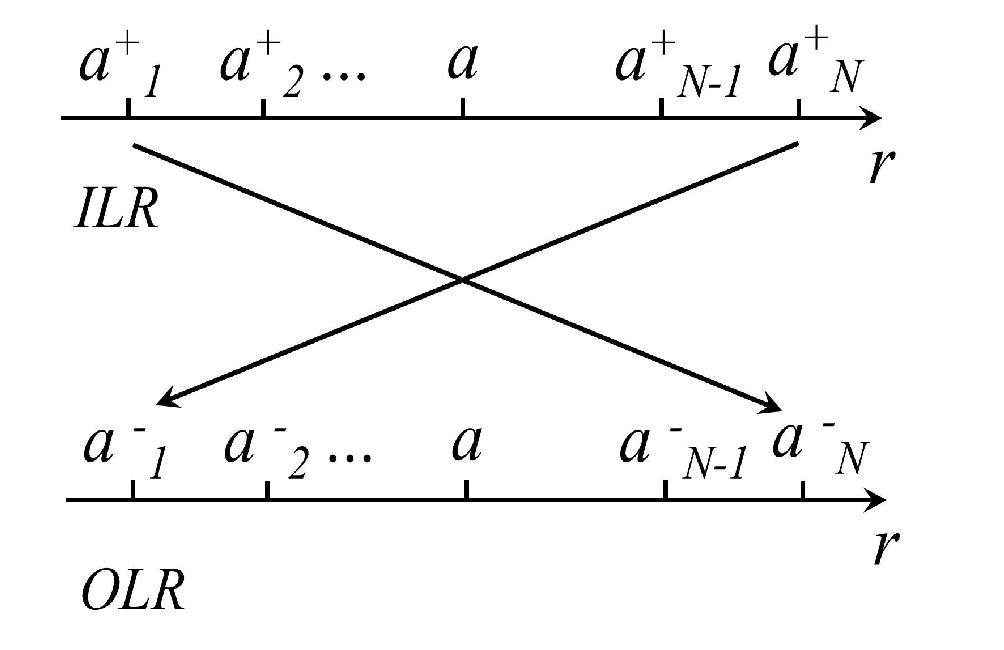}
	\caption{\small{Sketch of the ILR/OLR streamline symmetry: streamline semimajor axes $a_i$ are symmetrized with respect to the resonance location $a$, while keeping their index order. As as consequence, they are ordered in increasing semimajor axis order at ILR but decreasing semimajor axis at OLR.}}
	\label{fig:sym}
\end{figure}

This large $|m|$ limit reveals an interesting symmetry property between an ILR and an OLR equations with the same but opposite $m$ wavenumbers (see Fig.~\ref{fig:sym}). To see this, let us make a formal correspondence between an ILR and an OLR. We assume for now that the two resonance radii are equal.\footnote{This is hypothetical, and only for the sake of displaying the OLR/ILR edge mode symmetry. This does not imply that such a configuration of degenerate ILR and OLR resonances actually exists in rings.} We specify the streamlines semi-major axes of an ILR edge mode by a `$+$' superscript as $a_i^+$. We also define a mirror set of streamlines with respect to the resonance location, with semi-major axes $a_i^- = a - (a_i^+ - a)$; by construction, $a_i^-$ is the symmetric location of $a^+_{N+1-i},$ with respect to the resonance location $a$. Calling $m^+ >0$ the ILR azimuthal wavenumber and $m^-=-m^+ < 0$ the OLR one, in the large $|m|$ limit, 
\begin{align}
	\kappa^+_i - m^+(\Omega^+_i-\Omega_P) = \kappa^-_i - m^-(\Omega^-_i-\Omega_P),\label{freqsym}
\end{align}
with the same meaning for the superscripts $+$ and $-$ on the frequencies, e.g. $\kappa^+_i =\kappa(a^+_i)$. Defining similarly $\epsilon^\pm_i$ and $\Delta^\pm_i$, this symmetry implies that the left-hand sides of Eqs.~\eqref{sat} and \eqref{dissip} are unchanged under the substitution $m^+ \leftrightarrow m^-$, $a_i^+  \leftrightarrow a_i^-$ if one assumes $\epsilon_i^- = \epsilon_i^+$ and $\Delta_i^- = \Delta_i^+$. 

This would also provide the correct OLR mode solution from the ILR one if the right-hand sides were unchanged. However, this is not the case, due to changes in the right-hand sides under the ILR/OLR symmetry.

Let us start with Eq.~\eqref{sat}. In the large $m$ limit, and for leading order inner Lindblad resonances ($k=0$, the only ones of interest here):
\begin{align}
\Psi_{m,0}(a) & \simeq -\frac{G M_s}{a_s}  \frac{2m}{\pi}\Big[2K_0(2/3) + K_1(2/3)\Big] \nonumber\\
& \simeq - \frac{G M_s}{a_s} \frac{5.04m}{\pi} \simeq - \frac{G M_s}{a_s}\frac{5m}{3},\label{psi+}	
\end{align}
where $M_s$ is the satellite mass, $a_s$ its semimajor axis, and $K_0$ and $K_1$ are modified Bessel functions (see \citealt{GT80} for this approximation). The same expression holds for a leading order OLR of same $|m|$. Note that $\Psi_{m,k} < 0$ at an ILR and $>0$ at an OLR due to the change of sign of $m$.

These remarks have the following implication: \textit{under the substitution of an ILR by an OLR as described above and assuming for simplicity that $\Psi_{m^-,k} = - \Psi_{m^+,k}$ --- which is nearly true in the $|m| \gg 1$ limit --- the ILR solution of Eq.~\eqref{sat} is also a solution of the OLR form of the equation under the substitution $\epsilon^- = -\epsilon^+$}. In other words, the eccentricity profile of an ILR forced mode is also the eccentricity profile of the associated OLR (with the meaning specified above for this association), once this profile is antisymmetrized with respect to the resonance location.

This sign change has a notable consequence. From known numerical solutions of the forced mode equations at an ILR \citep{L23}, dissipationless ILR modes have their periapse aligned with the origin of phases of the forcing potential (i.e., the mean longitude of the satellite for leading order Lindblad resonances). The change of sign implies that, at an OLR, the apoapse is aligned with the satellite mean longitude instead, as $a(1-\epsilon)$ (periapse) becomes $a(1+\epsilon)$ (apoapse).

Let us now consider Eq.~\eqref{dissip}. The symmetry just described leaves the nonlinearity parameter profile $q=|ad\epsilon/da|$ unchanged but symmetrized with respect to the resonance location. Relatedly $ad\epsilon/da$ is also symmetric with respect to the resonance location (due to the change of sign of the eccentricity, the derivative remains positive). This implies that $(a/q)(d\epsilon/da)t_1$ is symmetric as well, so that, for any given streamline, the right-hand side of Eq.~\eqref{dissip} changes sign under the symmetry discussed here, but is otherwise left unchanged. Therefore, $\Delta^-_i = -\Delta^+_i$ is the correct solution at an OLR under the same symmetry (as $\epsilon$ has changed sign according to the preceding discussion of the eccentricity profile and $m$ changes sign as well).
 
To sum up: under the ILR/OLR symmetry described above, the ILR solution for a forced edge mode is also a solution at an OLR under antisymmetrization of the eccentricity $\epsilon$ and apsidal shift $\Delta$ profiles with respect to the resonance location. As a consequence, the mode apoapse is aligned with the satellite mean longitude in the absence of dissipation. When dissipation is added, the apoapse is expected to be leading (in angle) with respect to the satellite mean longitude because the mode global phase $m\phi + m\Delta$ vanishes for some positive $\phi$ ($\phi$ is the azimuthal angle of the mode in the rotating frame). At an ILR, the mode periapse is nearly aligned with and lagging with respect to the satellite instead. 

The conclusions just stated on the nature of the apse aligned with the satellite for an ILR/OLR forced mode are robust, but the sign of the lag applies for the simplest physical context only, and is therefore subject to various qualifications. These are discussed in the body of the text. 

Note finally that the sign of the lag is consistent with the requirement that angular momentum exchanges with the satellite flow from the inside to the outside, in order for the satellite to confine a ring edge. This conclusion follows very simply from the following generic torque expression
\begin{align}\label{torquenum}
	T_s = \int da\ \pi ma\sigma_0\Psi_{m,k}\epsilon \sin m\Delta,
\end{align} 
which is valid for inner and outer resonances \citep{L92,Longaretti2018}. For an ILR edge mode, $m > 0$, $\Psi_{m,k} < 0$ and $\epsilon \sin m\Delta > 0$ so that $T_s < 0$, as expected (see \citealt{L23} for actual eccentricity and apsidal shift profiles to substantiate this statement). For an OLR edge mode, and because of the ILR/OLR symmetry shown in this Appendix, $m < 0$, $\Psi_{m,k} > 0$ and $\epsilon \sin m\Delta < 0$ so that $T_s > 0$, again in line with expectations.

The arguments presented here are exact for identical ILR/OLR resonance locations and identical but opposite satellite forcing terms for both an ILR and an OLR. These requirements are not satisfied in reality, but the difference is only quantitative, not qualitative. In particular, for the large $|m|$ resonances of interest in the Uranian rings context, the relative quantitative difference introduced with respect to actual resonance locations and satellite forcing magnitudes is a small fraction of unity. Such a small difference will produce barely noticeable deviations in the OLR eccentricity and apsidal shift profiles from their symmetric ILR counterpart, for any given satellite. Furthermore, the qualitative issues examined here (which apse is nearly aligned with the satellite, and sign of their small lags) are independent of the quantitative details, such as the satellite mass, the azimuthal wavenumber, and the resonance location.



\end{document}